\documentclass[aps,prd,preprint,12pt,superscriptaddress,nofootinbib,a4paper,preprintnumbers
]{revtex4-1}

\usepackage[utf8]{inputenc}
\usepackage{amsmath,amssymb}
\usepackage[separate-uncertainty=true]{siunitx}
\usepackage{graphicx}
\usepackage[usenames,dvipsnames]{xcolor}
\usepackage[caption=false]{subfig}

\usepackage[colorlinks=true
,urlcolor=blue
,anchorcolor=blue
,citecolor=blue
,filecolor=blue
,linkcolor=blue
,menucolor=blue
,linktocpage=true
,pdfproducer=medialab
,pdfa=true
]{hyperref}
\usepackage[capitalize]{cleveref}
\usepackage{booktabs}
\usepackage{tabularx}
\usepackage{xspace}
\usepackage[normalem]{ulem}
\usepackage{slashed}
\usepackage{bbold}
\usepackage{wasysym}
\usepackage{mathrsfs} 
\usepackage{multirow}
\usepackage[compat=1.0.0]{tikz-feynman}
\usepackage{placeins}
\usepackage{tablefootnote}
\usepackage{threeparttable}

\usepackage{appendix}

\allowdisplaybreaks

\graphicspath{{graphics/}}

\newcommand{\eqn}{equation}
\newcommand{\lb}{\left(}
\newcommand{\rb}{\right)}
\newcommand{\be}{\beta}
\newcommand{\al}{\alpha}
\newcommand{\tanb}{\ensuremath{\tan\beta}\xspace}
\newcommand{\sina}{\ensuremath{\sin\alpha}\xspace}
\newcommand{\cosa}{\ensuremath{\cos\alpha}\xspace}
\newcommand{\mH}{\ensuremath{m_{\PS}}\xspace}
\newcommand{\mh}{\ensuremath{m_{\Ph}}\xspace}
\newcommand{\wH}{\ensuremath{\Gamma_{\PS}}\xspace}
\newcommand{\relwH}{\ensuremath{\frac{\Gamma_{\PS}}{m_{\PS}}}\xspace}
\newcommand{\wh}{\ensuremath{\Gamma_{\Ph}}\xspace}
\newcommand{\wSM}{\ensuremath{\Gamma_{\mathrm{SM}}}\xspace}
\newcommand{\pT}{\ensuremath{p_{\mathrm{T}}}\xspace}
\newcommand{\mhh}{\ensuremath{m_{\Ph\Ph}}\xspace}
\newcommand{\HT}{\ensuremath{\mathrm{H}_{\mathrm{T}}}\xspace}

\newcommand{\ie}{i.e.~}
\newcommand{\eg}{e.g.~}

\newcommand{\HiggsBounds}{\texttt{HiggsBounds}\xspace}
\newcommand{\HiggsSignals}{\texttt{HiggsSignals}\xspace}
\newcommand{\HiggsTools}{\texttt{HiggsTools}\xspace}

\newcommand{\GeV}{{\ensuremath\,\rm GeV}\xspace}

\newcommand{\lam}{\lambda}
\newcommand{\MeV}{{\ensuremath\rm MeV}\xspace}
\newcommand{\pb}{{\ensuremath\rm pb}\xspace}
\newcommand{\fb}{{\ensuremath\rm fb}\xspace}
\newcommand{\invfb}{\ensuremath{\mathrm{fb}^{-1}}\xspace}

\newcommand{\Sh}{\ensuremath{\mathrm{S_{h}}}\xspace}
\newcommand{\SH}{\ensuremath{\mathrm{S_{H}}}\xspace}
\newcommand{\ShBox}{\ensuremath{\mathrm{S_{h}}{\text -}\Box}\xspace}
\newcommand{\SHBox}{\ensuremath{\mathrm{S_{H}}{\text -}\Box}\xspace}
\newcommand{\SHSh}{\ensuremath{\mathrm{S_{H}}{\text -}\mathrm{S_{h}}}\xspace}
\newcommand{\POWHEG} {{\textsc{POWHEG}}\xspace}
\newcommand{\PYTHIA} {{\textsc{PYTHIA}}\xspace}
\newcommand{\MADGRAPH} {{\textsc{MadGraph}{}5\_a\textsc{mc@nlo}}\xspace}
\newcommand{\PS}{\ensuremath{H}\xspace}
\newcommand{\Ph}{\ensuremath{h}\xspace}

\newcommand{\kappalam}{\ensuremath{\kappa_{\lambda_{\Ph\Ph\Ph}}}\xspace}
\newcommand{\kappat}{\ensuremath{\kappa_{q}^{\Ph}}\xspace}
\newcommand{\yt}{\ensuremath{y_{t}^{\Ph}}\xspace}
\newcommand{\ytSM}{\ensuremath{y_{t}^{\mathrm{SM}}}\xspace}
\newcommand{\ytH}{\ensuremath{y_{t}^{\PS}}\xspace}
\newcommand{\yb}{\ensuremath{y_{b}^{\Ph}}\xspace}
\newcommand{\ybSM}{\ensuremath{y_{b}^{\mathrm{SM}}}\xspace}
\newcommand{\ybH}{\ensuremath{y_{b}^{\PS}}\xspace}
\newcommand{\lamSM}{\ensuremath{\lambda_{\mathrm{SM}}}\xspace}
\newcommand{\lamhhh}{\ensuremath{\lambda_{\Ph\Ph\Ph}}\xspace}
\newcommand{\lamHhh}{\ensuremath{\lambda_{\PS\Ph\Ph}}\xspace}
\newcommand{\kappalamH}{\ensuremath{\kappa_{\lambda_{\PS\Ph\Ph}}}\xspace}
\newcommand{\kappatH}{\ensuremath{\kappa_{q}^{\PS}}\xspace}
\newcommand{\sigSH}{\ensuremath{\sigma_{\SH}}\xspace}

\newcommand{\intabs}{\ensuremath{\lb\Delta \sigma\rb_\text{abs}}\xspace}
\newcommand{\intabslow}{\ensuremath{\lb\Delta \sigma\rb_\text{abs}^{<}}\xspace}
\newcommand{\intabshig}{\ensuremath{\lb\Delta \sigma\rb_\text{abs}^{>}}\xspace}
\newcommand{\intabssum}{\ensuremath{\lb\Delta \sigma\rb_\text{abs}^{\sum}}\xspace}
\newcommand{\intrel}{\ensuremath{\lb\Delta \sigma\rb_\text{rel}}\xspace}
\newcommand{\intrellow}{\ensuremath{\lb\Delta \sigma\rb_\text{rel}^{<}}\xspace}
\newcommand{\intrelhig}{\ensuremath{\lb\Delta \sigma\rb_\text{rel}^{>}}\xspace}
\newcommand{\intrelsum}{\ensuremath{\lb\Delta \sigma\rb_\text{rel}^{\sum}}\xspace}

\newcommand{\BMa}{BM1\xspace}
\newcommand{\BMb}{BM2\xspace}
\newcommand{\BMc}{BM3\xspace}
\newcommand{\BMd}{BM4\xspace}
\newcommand{\BMe}{BM5\xspace}
\newcommand{\BMf}{BM6\xspace}
\newcommand{\BMg}{BM7\xspace}
\newcommand{\BMh}{BM8\xspace}
\newcommand{\BMi}{BM9\xspace}

\DeclareSIUnit{\pb}{pb}
\DeclareSIUnit{\fb}{fb}
\AtBeginDocument{
\heavyrulewidth=.08em
\lightrulewidth=.05em
\cmidrulewidth=.03em
\belowrulesep=.65ex
\belowbottomsep=0pt
\aboverulesep=.4ex
\abovetopsep=0pt
\cmidrulesep=\doublerulesep
\cmidrulekern=.5em
\defaultaddspace=.5em

\newcolumntype{C}{>{\centering\arraybackslash}X}
\newcolumntype{b}{C}
\newcolumntype{s}{>{\hsize=.6\hsize}C}
\newcolumntype{R}{>{\raggedleft\arraybackslash}X}
}

\begin{document}
\date{\today}
\rightline{RBI-ThPhys-2024-15}
\title{\vspace{6mm}{\Large Interference effects in resonant di-Higgs production at\\ the LHC in the Higgs singlet extension}\vspace{3mm}}

\author{Finn Feuerstake}
\email{finn.feuerstake@stud.uni-hannover.de}
\affiliation{Institut für Theoretische Physik, Leibniz Universität Hannover, Appelstraße 2, 30167 Hannover, Germany}

\author{Elina Fuchs}
\email{elina.fuchs@itp.uni-hannover.de}
\affiliation{Institut für Theoretische Physik, Leibniz Universität Hannover, Appelstraße 2, 30167 Hannover, Germany}
\affiliation{Physikalisch-Technische Bundesanstalt (PTB), Bundesallee 100, 38116 Braunschweig, Germany}

\author{Tania Robens}
\email{trobens@irb.hr}
\affiliation{Ruder Boskovic Institute, Bijenicka cesta 54, 10000 Zagreb, Croatia}

\author{Daniel Winterbottom}
\email{d.winterbottom15@imperial.ac.uk}
\affiliation{Imperial College London,
Department of Physics, Blackett Laboratory,
Prince Consort Rd, London, SW7 2BW, United Kingdom}

\renewcommand{\abstractname}{\texorpdfstring{\vspace{0.5cm}}{} Abstract}

%%%%%%%%%%%%%%%%%%%%%%%%%%%%%%%
{%\setstretch{1.0}
\begin{abstract}
    \vspace{0.5cm}
Interference effects are well founded from the quantum mechanical viewpoint and in principle cannot be ignored in realistic studies of New Physics scenarios.
In this work, we investigate the size of interference effects between resonant and non-resonant contributions to di-Higgs production
in the singlet extension of the Standard Model,  
where the additional heavy scalar provides a resonant channel.
We find these interference contributions to have a non-negligible effect on the cross-sections and differential distributions. In order to allow for a computationally efficient treatment of these effects via reweighting, we introduce a new tool utilising a matrix-element reweighting method: \texttt{HHReweighter}.
In addition to the broadly used di-Higgs invariant mass $m_{hh}$, we analyse the sensitivity to the interference terms for other kinematic variables, such as the Higgs boson transverse momentum, and find that these also can be sensitive to interference effects.
Furthermore, we provide updates on  the latest experimental and theoretical limits on the parameter space of the real singlet extension of the Standard Model Higgs sector.

\end{abstract}

%%%%%%%%%%%%%%%%%%%%%%%%%%%%%%%

\maketitle

\tableofcontents
}
\newpage
\section{Motivation}
\label{sec:intro}

Higgs pair production is a process of high interest at the LHC because of its sensitivity to the triple-Higgs coupling and therefore to the underlying mechanism of electroweak symmetry breaking and possible modifications of the scalar potential with respect to its prediction in the Standard Model (SM). Thus, a meaningful interpretation of the measured total and differential cross sections of this process requires an accurate theoretical description and simulation.
We focus on di-Higgs production stemming from an additional resonance, as \eg present in models with singlet extensions (see \eg Refs.~\cite{Robens:2015gla,Robens:2016xkb,Robens:2022cun} and references therein). Di-Higgs production has been discussed in great detail in Ref.~\cite{DiMicco:2019ngk} (see also the short summary in Ref.~\cite{Brigljevic:2024vuv}), where in particular finite widths and interference effects from Refs. \cite{Dawson:2015haa,Dawson:2016ugw,Lewis:2017dme,Carena:2018vpt} are presented. More recent work on interference effects in di-Higgs production in extended scalar sectors can be found \eg in Refs.~\cite{Basler:2019nas,Heinemeyer:2024hxa}.

The assumption of using factorised cross sections has its own merits. One example is the fact that often it is much easier to calculate higher-order corrections to a single $s$-channel production and decay, without taking the full corrections into account. In some cases, these can be $\mathcal{O}\lb 100\% \rb$, see \eg Refs.~\cite{Ellis:1996mzs,Campbell:2017hsr},
and might easily be more important than a leading order (LO) calculation where the interference effects are correctly described. However, it is not clear a priori that this is the case in all scenarios and realisations of physical models. In particular, for some channels strong relations exist between couplings of new physics particles, \eg via unitarity relations~\cite{Gunion:1990kf} or other theory constraints. In such cases, it is imperative to include the full matrix elements.

Resonant di-scalar searches are currently investigated by both ATLAS and CMS \cite{ATLAS:2024ish,CMS:2024phk}, and correspond to one of the prime channels for testing the scalar sector realised in nature via triple scalar couplings. Therefore, it is of high importance that 
the correct theoretical description of such processes is consistently applied and up-to-date tools are used in the corresponding simulations.
For example, the correct interpretation of results relies on the correct understanding of the simulated process, that might include kinematic effects in differential distributions resulting from irreducible interference terms. 
Similarly, as the finite width of unstable particles is a model prediction, it cannot be set to a value chosen independently of the involved masses and couplings. Note that modifications to the di-Higgs kinematic distributions can also appear in effective field theory interpretations, see \eg Refs.~\cite{Carvalho:2015ttv,Buchalla:2018yce,Capozi:2019xsi,Alasfar:2023xpc}.
First studies showing the possible importance of interference effects in di-Higgs production were presented by the CMS Collaboration in Ref.~\cite{CMS:2024phk}.

In particular, there can be non-negligible interference effects between resonant and non-resonant diagrams that affect kinematic distributions as already discussed in Refs.~\cite{Dawson:2015haa,Carena:2018vpt,DiMicco:2019ngk}. These effects have been neglected by all experimental LHC searches for resonant di-Higgs production to date. In fact, most searches disregard the non-resonant contribution to the di-Higgs spectrum entirely \ie it is neither taken into account as part of the signal or the background. Due to the complicated nature of these analyses, it is then not always possible to reinterpret these experimental results to set limits on realistic physics models. A careful examination of the interference effects is thus essential for exploring realistic new physics models in future LHC runs. 
Moreover, incorporating the non-resonant process and interference effects can even enhance the sensitivity of future searches as it allows for the exploration of deviations in kinematic distributions away from the resonance peak. This provides an additional handle for differentiating between the SM and new physics scenarios.

In this work, we therefore concentrate on the most simple model, adding an additional CP even neutral scalar to the SM, which has already been vastly explored by the experimental collaborations. We show that even for such a simple scenario, a description of the signal as a resonant peak in the invariant mass spectrum does not suffice to correctly describe the behaviour of the experimental distributions in all possible scenarios and that interference effects can also affect other distributions. We see this as a first example that interference effects need to be taken into account even for simple models, and we propose a reweighting method to incorporate such effects.

This paper is organised as follows. 
In Section~\ref{sec:singlet_extension}, we introduce the real singlet model used for our studies, and briefly discuss general features of width and interference effects in Section~\ref{sec:WidthInterference}. In Section~\ref{sec:simulation} and Section~\ref{sec:parscan} we describe the tools used to simulate di-Higgs production and present our method for determining regions of the parameter space with sizeable interference effects. We introduce our tool used to model interference effects using matrix-element  reweighting  in Section~\ref{sec:method}. This is followed by an investigation of interference effects for several benchmark points with distinct features in Section~\ref{sec:benchmarks}. We summarise our investigations in Section~\ref{sec:summary}.

\section{Real singlet extension of the SM}
\label{sec:singlet_extension}

\subsection{Summary of the model}
\label{sec:model}
In this work, we consider the simplest extension of the Standard Model of particle physics that can provide resonance-enhanced di-Higgs production at colliders. We choose a model that extends the SM electroweak sector by an additional real scalar field that transforms as a singlet under the SM gauge group. This model has been widely discussed in the literature, see \eg Refs.~\cite{Schabinger:2005ei, OConnell:2006rsp,Patt:2006fw,Barger:2007im,Bowen:2007ia,Robens:2015gla, Robens:2016xkb, Huang:2016cjm, Dawson:2017vgm, Ilnicka:2018def, Heinemann:2019trx, Fuchs:2020cmm, Robens:2022cun} for details.

The most general renormalizable Lagrangian of the SM extended by one real singlet scalar $S$ is given by
\begin{equation}\label{lag:s}
\mathscr{L}_s = \left( D^{\mu} \Phi \right) ^{\dagger} D_{\mu} \Phi + 
\partial^{\mu} S \partial_{\mu} S - V(\Phi,S ) \, ,
\end{equation}
with the scalar potential
\begin{eqnarray}\label{potential}\nonumber
V(\Phi,S ) &=& -m^2 \Phi^{\dagger} \Phi - \mu ^2  S ^2 +
\left(
\begin{array}{cc}
\Phi^{\dagger} \Phi &  S ^2
\end{array}
\right)
\left(
\begin{array}{cc}
\lambda_1 & \frac{\lambda_3}{2} \\
\frac{\lambda_3}{2} & \lambda _2 \\
\end{array}
\right)
\left(
\begin{array}{c}
\Phi^{\dagger} \Phi \\  S^2 \\
\end{array}
\right) \\
\nonumber \\ 
&=& -m^2 \Phi^{\dagger} \Phi -\mu ^2 S ^2 + \lambda_1
(\Phi^{\dagger} \Phi)^2 + \lambda_2  S^4 + \lambda_3 \Phi^{\dagger}
\Phi S ^2,
\end{eqnarray}
where we already imposed a $\mathbb{Z}_2$ symmetry that is spontaneously broken by the vacuum expectation value (VEV) of the singlet field $S$\footnote{In fact, for models which do not introduce new coloured states or more than one additional scalar boson, the general phenomenology and interference effects are independent of this choice and only depend on the coupling rescaling and the total width for a chosen parameter point. For the real singlet extension, applying symmetries only impacts the available model parameter space, mainly through theoretical constraints.}.

After minimisation, the model contains in total 5 free parameters in the electroweak sector. A physical choice is~\cite{Robens:2015gla,Robens:2016xkb}
\begin{equation}
    m_h,\,m_H,\,v,\,v_s,\,\sin\alpha,
\end{equation}
where the masses of the two physical states are ordered as $m_h\,\leq\,m_H$, and where $v$ and $v_s$ denote the VEVs of the doublet $\Phi$ and singlet $S$, respectively. The mixing angle $\sin\alpha$ corresponds to the angle that rotates the gauge into the mass eigenstates, 
\begin{align}
    \begin{pmatrix}
        h\\H 
    \end{pmatrix}
 =  \begin{pmatrix}
        \cos\alpha & -\sin\alpha\\
        \sin\alpha & \cos\alpha
    \end{pmatrix}
    \begin{pmatrix}
    \tilde h\\ h'
    \end{pmatrix}\,,
\end{align}
where 
$\Phi = \left(0, (\tilde h +v)\right)^T/\sqrt{2}$ and 
$S=\left(h'+v_s\right)/\sqrt{2}$.
Hence, $\sin\al\,=\,0$ denotes complete decoupling of the second scalar resonance $S$ from the SM. 

In this work, we are interested in interference effects in pair production of the discovered SM-like Higgs boson, $h_{125}$, at a mass of $m_{125}=125\,\GeV$. We therefore identify the lighter state $h$ as $h_{125}$.
Similarly, the doublet VEV is fixed to be $v=246\,\GeV$ from electroweak precision measurements. Hence, in total three free parameters remain:
\begin{equation}
    m_H,\,\sin\alpha,\,\tan\beta\,\equiv\,\frac{v}{v_s},
\end{equation}
which we choose as the independent input parameters in this work.

The trilinear Higgs couplings $\lambda_{ijk}$ (of mass dimension one) play a crucial role in the di-Higgs production processes.
They are defined as $V\,\supset\,\sum_{ijk}\,\lam_{ijk}\,h_i\,h_j\,h_k$,  with $\left\{i,j,k\right\}\,\in\,\{1,2\}$ and $h_1\,\equiv\,h,\,h_2\,\equiv\,H$, 
where \lamhhh is expressed in relation to the SM-value as $\kappalam\,\equiv\,\frac{\lamhhh}{\lamSM}$, with $\lamSM\,=\,\frac{m_{125}^2}{2\,v}$ being the triple scalar coupling in the SM-like case. 

The $hhh$ and $hhH$ couplings depend on the model parameters in the following way: 
\begin{eqnarray}
\label{eqn:couplings}
\lamhhh &=& \lamSM \left(\cos^3{\alpha}-\tanb\sin^3{\alpha} \right),\nonumber\\
\lamHhh &=& \lamSM\frac{2\mh^2+m_{\PS}^2}{\mh^2}\,\frac{\sin\lb 2 \alpha\rb}{2}\left(\cos{\alpha} + \tanb\sin{\alpha}\right).
\end{eqnarray}

The total widths of the two scalars are given by
\begin{eqnarray}
\label{eqn:Hwidth}
    \wh &=& \left(1-\sin^2\alpha\right)\wSM,\nonumber\\
    \wH &=& \sin^2\alpha\,\wSM(\mH) + \frac{\lamHhh^2 \sqrt{1-4\mh^2/\mH^2}}{8\pi\mH},
\end{eqnarray}
where $\wSM$ is the width computed for a SM-like Higgs boson of a given mass. 
For small values of $|\sina|$, these formulas can be approximated using a Taylor expansion around $\sina=0$:
\begin{eqnarray}
\label{eqn:couplings_approx}
  \lamhhh &\approx& \lamSM\left(1-\frac{3}{2}\sin^2\alpha\right), \nonumber\\
  \lamHhh &\approx& \lamSM \frac{2\mh^2+m_{\PS}^2}{\mh^2}\left(\sina +\tanb\,\sin^2\alpha\right)\,,
\end{eqnarray}
such that we obtain 
\begin{eqnarray}
\wH &\approx& \,\sin^2\alpha\,\wSM(\mH) + \lamSM^2 \left(\sin^2\alpha+2\tanb\,\sin^3\alpha\right)\left(\frac{2\mh^2+\mH^2}{\mh^2}\right)^2\frac{ \sqrt{1-4\mh^2/\mH^2}}{8\pi\mH}.\nonumber\\
&&
\end{eqnarray}

\subsection{Constraints on the model}

The model discussed here is subject to a large number of theoretical and experimental constraints that have been discussed in detail in the references given above \cite{Pruna:2013bma,Lopez-Val:2014jva,Robens:2015gla,Robens:2016xkb,Ilnicka:2018def}. Here, we focus on the mass region of $m_H\,\geq\,2\,m_h = 250\,\GeV$; in this regime, the most important updates stem from direct searches for di-boson states in the full Run-2 analyses as well as the higher-order contributions to the $W$-boson mass. The latter have first been calculated in \cite{Lopez-Val:2014jva}, with a recent update in \cite{Papaefstathiou:2022oyi}.
We only require agreement with all constraints on the electroweak scale, which means that the bounds do not include constraints using renormalization group equation running of the couplings. 

\begin{figure*}[hbt!]
  \includegraphics[width=0.48\textwidth]{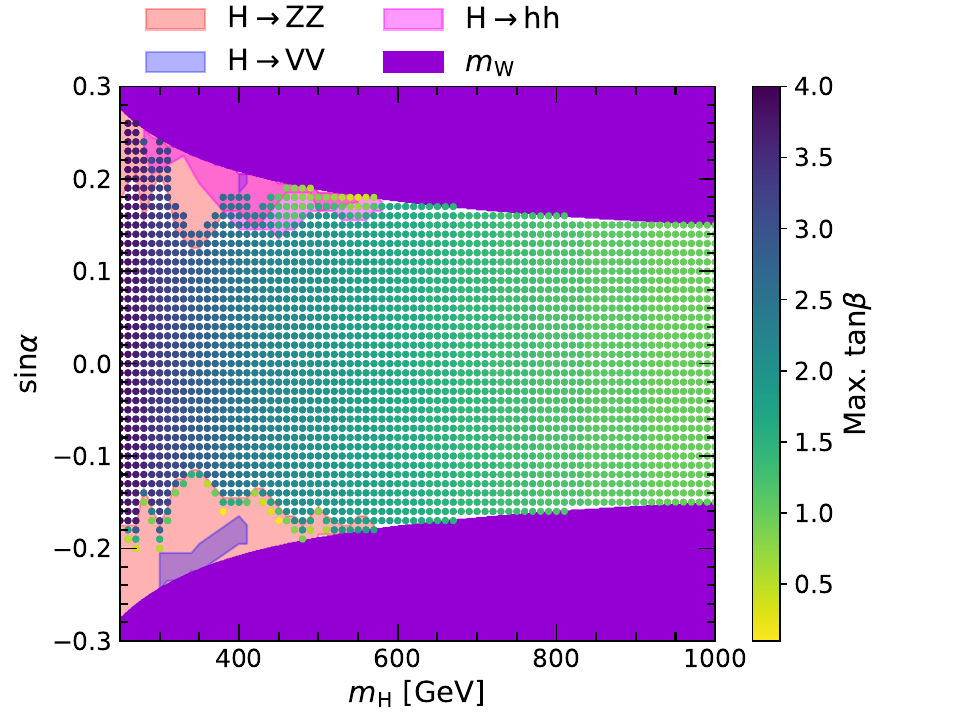}
  \includegraphics[width=0.48\textwidth]{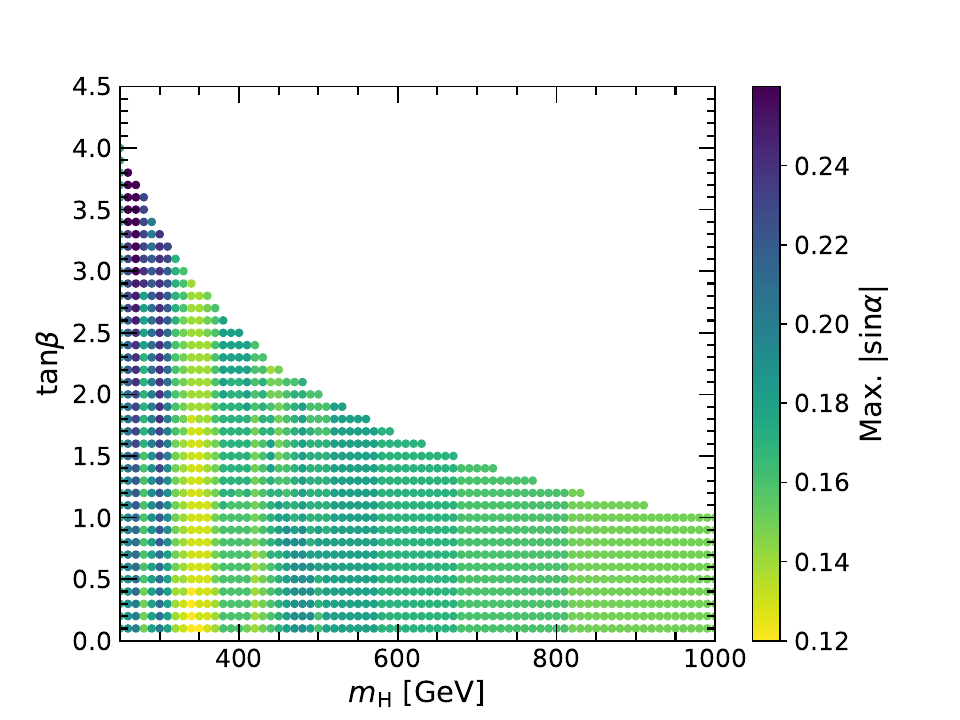}
\caption{
Allowed parameter space in the real singlet extension in the $\lb m_H,\,\sin\alpha\rb$ (left) and $\lb m_H,\,\tan\beta\rb$ (right) planes. The circular dots indicate allowed regions of the parameter space. Maximal values for the third free parameter are shown in colour coding. In addition, on the left side we also display exclusion bounds from the $W$-mass measurement in solid purple. Additionally, regions where LHC resonance searches exclude parts of the parameter space are indicated with colour-shaded areas. We note that these shaded areas indicate that the searches are sensitive to certain portions of the \tanb range, but do not imply a full exclusion of the entire \tanb range. See text for details.
}
\label{fig:allowed_points}
\end{figure*}

We include the most up-to-date  experimental constraints. 
We do this by making use of the publicly available tool \HiggsTools \cite{Bahl:2022igd}, that builds on \HiggsBounds \cite{Bechtle:2008jh,Bechtle:2011sb,Bechtle:2013wla,Bechtle:2020pkv} and \HiggsSignals \cite{Bechtle:2013xfa,Bechtle:2014ewa,Bechtle:2020uwn}. The allowed parameter space in the $\lb m_H,\,\sin\alpha\rb$ and $\lb m_H,\,\tan\beta\rb$ planes is shown in Figure~\ref{fig:allowed_points}.  Direct search results stem from $H\rightarrow\,Z\,Z$ \cite{CMS:2018amk,ATLAS:2020tlo}, $H\,\rightarrow\,V\,V$ \cite{ATLAS:2018sbw,ATLAS:2020fry}, and $H\,\rightarrow\,h\,h$ \cite{CMS:2021yci,ATLAS:2021ifb,ATLAS:2022xzm}.

The parameter scan used here follows the prescription in Refs.~\cite{Pruna:2013bma,Robens:2015gla,Robens:2016xkb,Ilnicka:2018def}.
We first applied all possible theoretical and experimental constraints apart from direct search limits and signal strength measurements. This is due to the fact that most of these can be implemented directly in the scan in a straightforward way. As is standard, we include the following theoretical constraints:
\begin{itemize}
\item{}boundedness from below of the potential,
\item{}limits from perturbative unitarity (PU),
\item{}limits from vacuum stability,
\item{}perturbativity of the couplings.
\end{itemize}

The above bounds impose limits on the potential couplings in the gauge-eigenbasis. In the scan we apply, we require all of these to be fulfilled, \ie any parameter point that does not pass these constraints is no longer processed further. The above bounds have been discussed in great detail in Ref.~\cite{Pruna:2013bma} and have been further evolved on in Ref.~\cite{Robens:2015gla}, which we refer the reader to for more details. In general, one can \eg say that PU puts an upper limit on $\tan\be$ that is inversely proportional to the heavy scalar mass for the mass range considered here, while vacuum stability and perturbativity of the couplings lead to constraints in the $\lb \sin\al, \tan\be\rb$ plane.

The scan then also tests agreement with direct searches and signal strength measurements using the standalone \HiggsBounds and \HiggsSignals version with results up to 2022. The allowed points are then checked against the most up to date constraints using the most recent version of \HiggsTools\footnote{
Exclusion criteria within \HiggsTools that are used here follow the prescription in Ref.~\cite{Robens:2023pax}, \ie we use the observed and not the expected sensitivity.
}. The $W$-boson mass was calculated according to Refs.~\cite{Lopez-Val:2014jva,Papaefstathiou:2022oyi}.

We show the most constraining channel, using the current \HiggsTools version, from this last step in Figure~\ref{fig:allowed_points}. This corresponds to an update of the results presented in previous work, as \eg \cite{Robens:2015gla,Robens:2016xkb,Ilnicka:2018def,Robens:2022cun}.

Concerning the cross section values used for the determination of the rates, \HiggsTools applies a factorised approach, where the production cross sections are obtained by rescaling the respective values given in the Yellow Report 4 of the Higgs working group \cite{LHCHiggsCrossSectionWorkingGroup:2016ypw}, rescaled by $\sin^2\,\alpha$ and $\cos^2\alpha$ for $H$ and $h$, respectively. 
Note that care must be taken to distinguish the validity regimes of the respective calculations:
for $m_H\,>\,m_\text{top}$ the next-to-next-to-LO (NNLO) with next-to-next-to-leading logarithmic resummation value~\cite{Harlander:2009mq,Pak:2009dg,Harlander:2009bw,Harlander:2009my} must be taken, as it contains full top quark mass dependence up to next-to-LO order (NLO), while the next-to-NNLO calculation~\cite{Anastasiou:2014lda,Anastasiou:2014vaa,Anastasiou:2015vya} relies on the effective theory and is therefore viable for $m_H\,\leq\,m_\text{top}$.

\section{Width and interference effects in a nutshell}
\label{sec:WidthInterference}

\subsection{Narrow width approximation}
\label{sec:NWA}

In general, from a quantum field theoretical viewpoint, only stable particles can be used as the incoming and outgoing states of the S-matrix (see \eg Ref.~\cite{Itzykson:1980rh}), implying that all unstable particles can be treated only as intermediate states (see also the discussion in \eg Ref.~\cite{Denner:2019vbn}). In practise, however, it can be advantageous to make use of the narrow width approximation (NWA) \cite{pilkuhn1967interactions,Dicus:1984fu} to factorise a more complicated process into the on-shell production of an intermediate particle and its subsequent decay. This way, one can conveniently include higher-order corrections to the production and decay processes.
We briefly review the assumptions underlying this approximation.

Let us consider a process $a\,b\,\rightarrow\,S\,\rightarrow\,c\,d$, where we now assume $S$ can be produced and decay resonantly and $a,\,b,\,c,$ and $d$ are stable initial and final state particles, respectively. In the case the $s$-channel is the only contributing diagram, the matrix element can be written 
via the resummation of the infinite series of self-energy insertions $\Sigma(p^2)$ between the tree-level propagators $D(p^2)= i/(p^2 - m_S^2)$ (see Ref.~\cite{Denner:2019vbn})
as
\begin{\eqn}\label{eq:mabcd}
  \mathcal{M}_{a\,b\,\rightarrow\,c\,d}\,\,=\, \frac{1}{p^2 - m_{S,0}^2 + \hat{\Sigma}(p^2)}\,\mathcal{F}
  \sim\,\frac{1}{p^2-m_S^2+i\,\Gamma_S\,m_S}\,\mathcal{F},
  \end{\eqn}
where $p$ denotes the four-momentum in the $s$-channel, $\mathcal{F}$ depends on the four-momenta of the external particles as well as other model parameters, and $m_{S,0}, m_S$ denote the tree-level mass and physical mass of the scalar, respectively. 

In Eq.~\eqref{eq:mabcd}, we used the optical theorem to relate the quantum field theoretical renormalized self-energy, $\hat{\Sigma}$, to the physical total width of $S$, $\Gamma_S$, at the pole $p^2=m_S^2$:

\begin{\eqn}
  \text{Im} \hat{\Sigma}\,\lb p^2\,=\,m_S^2 \rb \,=\,\Gamma_S\,m_S,
  \end{\eqn}
 giving rise to the well-known Breit-Wigner propagator in the final step of Eq.~\eqref{eq:mabcd}. For a detailed discussion see \eg Ref.~\cite{Fuchs:2016swt}. The total width $\Gamma$ corresponds to the sum over all possible partial decay widths\footnote{We here for simplicity assume that $m_S$ is real. To ensure gauge invariance, it is helpful to work in the complex mass scheme, see \eg \cite{Denner:1999gp,Denner:2005fg,Denner:2006ic,Denner:2014zga,Denner:2019vbn} for details. 
}.

In the limit of $\Gamma\,\rightarrow\,0$, one obtains
\begin{\eqn}\label{eq:NWA_delta}
  \frac{1}{|p^2-m^2+i\,m\,\Gamma|^2}\,\rightarrow\,\frac{\pi}{m\,\Gamma}\,\delta\lb p^2-m^2 \rb.
  \end{\eqn}
One can also use the Breit-Wigner propagator on the left-hand side of Eq.~\eqref{eq:NWA_delta} for finite widths. 
Further assumptions in the NWA are that the function $\mathcal{F}$ only obtains major contributions around $p^2\,\sim\,m^2$ 
and only varies mildly in the region $p^2\,\in\,\left[ \lb m-\Gamma\rb^2,\lb m+\Gamma \rb^2 \right]$\footnote{This assumption, and the validity of the NWA, can be violated \eg near kinematic thresholds.}. In this case, and making use of phase space factorisation, one obtains the $s$-channel mediated cross section
\begin{align}
\sigma_{a\,b\,\rightarrow\,c\,d}^s\,\simeq\,\sigma_{a\,b\,\rightarrow\,S}\,\times\,\int^{p^2_\text{max}}_{p^2_\text{min}}\,\frac{d\,p^2}{2\,\pi}\,\frac{2\,m}{|p^2-m^2+i\,m\,\Gamma|^2}\,\times\,\Gamma_{S\,\rightarrow\,c\,d}.
\end{align}
Making use of Eq.~\eqref{eq:NWA_delta}, the limit $\Gamma\,\rightarrow\,0$ leads to the well-known factorisation into the on-shell production cross section times the branching ratio, 
\begin{\eqn}\label{eq:fac}
\sigma_{a\,b\,\rightarrow\,c\,d}^s\,\simeq\,\sigma_{a\,b\,\rightarrow\,S}\,\times\,\underbrace{\frac{\Gamma_{S\,\rightarrow\,c\,d}}{\Gamma}}_{\text{BR}\lb S\,\rightarrow\,c\,d \rb}.
\end{\eqn}
For the case that $\Gamma\,\neq\,0$, one can prove that the formal error of the above approximation is $\mathcal{O}\lb \frac{\Gamma}{m} \rb$. For an introduction as well as discussions of the limitations of the above approximation such as off-shell and threshold effects, the impact of nearby resonances and non-factorisable contributions, see \eg Refs.~\cite{Berdine:2007uv,uhlemann_dipl,Uhlemann:2008pm,Cacciapaglia:2009ic, Fuchs:2015jwa,Fuchs:2017wkq,Bagnaschi:2018ofa,Hoang:2024oeq}. See also Ref.~\cite{Denner:2019vbn} for a more general discussion regarding the correct treatment of  unstable particles. A generalisation of the NWA to include interference effects in an on-shell approximation was developed in Ref.~\cite{Fuchs:2014ola}.

The result in Eq.~(\ref{eq:fac}) highlights the importance of using the physical value of the total width in order to correctly describe the underlying physics. 
In particular, using the total width as an input parameter $\Gamma_\text{in}$, \eg determined by the detector resolution, in turn leads to arbitrary branching ratios that can easily exceed 1 if $\Gamma\,\equiv\,\Gamma_\text{in}\,\leq\,\Gamma_{S\,\rightarrow\,c\,d}$, and therefore unphysical rates.
Instead, using a properly calculated total width as a prediction of the model parameters enables a correct physical description.   

\subsection{Interference effects}
Here, we briefly discuss interference effects between resonant and non-resonant contributions to the matrix element of a considered process. In the case of additional contributions to the process $a\,b\,\rightarrow\,c\,d$, namely, diagrams that do not go via the $s$-channel resonance $S$, the total matrix element is given by
\begin{\eqn}
\mathcal{M}_\text{tot}\,=\,\mathcal{M}_S\,+\,\mathcal{M}_\text{rest},
\end{\eqn}
where $\mathcal{M}_\text{rest}$ contains all additional diagrams. For simplicity, let us assume none of these contains an additional resonant channel, \ie here we do not consider interference between two resonances, but between a resonance and the continuum.
The total squared matrix element is thus given by
\begin{\eqn}
|\mathcal{M}_\text{tot}|^2\,=\,|\mathcal{M}_S|^2\,+\,|\mathcal{M}_\text{rest}|^2\,+\,\underbrace{2\,\text{Re}\left[ \mathcal{M}_S\,\mathcal{M}^*_\text{rest} \right]}_\text{Interference}.
\end{\eqn}

In case the process is dominated by an on-shell resonance $S$, very often the process is well-described by the first contribution in the above equation. However, it is not a priori clear that this holds for all kinematic variables in the process, including differential distributions or specific regions of phase space. In such cases, all above contributions need to be taken into account. In particular, for processes with electroweak vector bosons in the initial and final state that include scalar $s$-channel contributions, unitarity links these \cite{Gunion:1990kf} and requires that the full squared matrix element is included for a correct description.

In the following, we will investigate the effect of the interference term in di-Higgs production mediated via an $s$-channel resonance term.
\section{Simulation of di-Higgs events}
\label{sec:simulation}
\begin{figure*}[htbp]
  \raisebox{0.27cm}{\includegraphics{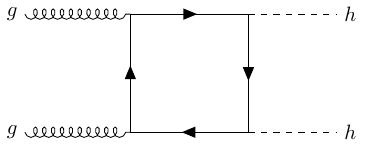}}
  \hspace{1cm}
  \includegraphics{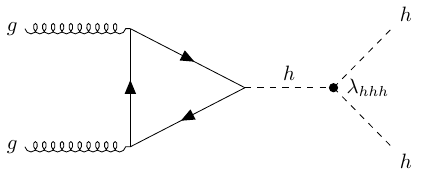} \\
  \centering\includegraphics{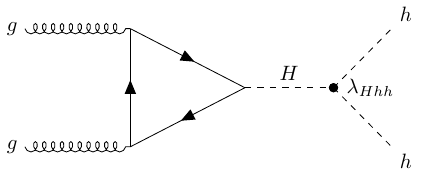}
  
    \caption{Feynman diagrams of the dominant contributions to di-Higgs production in the SM+singlet model at $pp$ colliders. The upper-left plot displays the so-called box diagram that does not depend Higgs trilinear couplings ($\Box$). The upper-right diagram shows the $s$-channel diagram that depends on \lamhhh (\Sh). The resonant contribution to the di-Higgs production due to an additional scalar $H$ (\SH) is displayed in the lower diagram.}
 \label{fig:feynman}

\end{figure*}

To investigate interference effects between resonant and non-resonant contributions to pair production of SM-like Higgs bosons, we simulate 
events of gluon fusion di-Higgs production ($gg\to hh$) for 13 TeV proton-proton collisions both at LO and NLO in perturbative quantum chromodynamics (QCD).
We perform the LO simulations with the \MADGRAPH~\cite{Alwall:2014hca} event generator, using an implementation of the Two Real Singlet Extension of the SM~\cite{Papaefstathiou:2020lyp} with the model file from \cite{gitlabandreas}. 
While this model contains three neutral Higgs bosons, we decouple one of these in our implementation.
We denote the two non-decoupled scalar fields as \Ph and \PS, where \Ph is identified as the observed Higgs boson with a mass of $\mh=125\,\GeV$, and \PS denotes the additional heavier Higgs boson with a variable mass \mH.
The model leads to contributions to the di-Higgs amplitude from both non-resonant and resonant Feynman diagrams, including interference terms. Among the non-resonant diagrams, the box diagram, $\Box$, (Figure~\ref{fig:feynman}, upper-left), is independent of the Higgs trilinear couplings, $\lambda_{i}$, whereas the $s$-channel diagram, \Sh, depends on the \lamhhh coupling (Figure~\ref{fig:feynman}, upper-right).
The resonant $s$-channel contribution, \SH, depends on the \lamHhh trilinear coupling (Figure~\ref{fig:feynman}, bottom).

The NLO simulations are performed using the \POWHEG 2.0~\cite{Nason:2004rx,Frixione:2007vw,Alioli:2010xd} event generator. 
Currently there is no \POWHEG model that simulates di-Higgs events including both non-resonant and resonant diagrams. Instead we simulate these contributions separately neglecting the interference between resonant and non-resonant diagrams, and in Section~\ref{sec:method} we will employ a reweighting method to account for the missing interference terms. The non-resonant simulations follow the implementation described in Refs.~\cite{Heinrich:2017kxx,Heinrich:2019bkc,Heinrich:2020ckp}. The resonant simulation is performed by generating single \PS events using the 
implementation presented in \POWHEG~\cite{Bagnaschi:2011tu}, followed by the $\PS\rightarrow\Ph\Ph$ decay implemented in \PYTHIA 8.310~\cite{Bierlich:2022pfr}.

All samples are generated with the NNPDF3.1~\cite{Ball_2017} NNLO parton distribution functions. Parton showering and hadronization are modelled using \PYTHIA with the {CP5} tune~\cite{Sirunyan_2020_CP5}. 

We also apply a smearing to the outgoing
\Ph bosons to study the impact of detector effects on the final distributions. 
This smearing aims to describe the typical resolution of the four $b$-jet final state. The smearing is performed as follows:
\begin{enumerate}
    \item The \Ph are first decayed into $b$-quark pairs using \PYTHIA.
    \item The $\phi$ and $\eta$ direction of the $b$-quarks are smeared independently by shifting the truth-level $\phi$ and $\eta$ on an case-by-case basis by an amount determined from randomly sampling a Gaussian function with a mean of 0 and an uncertainty of $\sigma_\phi=\sigma_\eta =0.05$, which is the typical angular resolution for $b$-jets with $\pT\approx 100\,\GeV$ as estimated using simulations of the CMS detector. 
    \item Finally, the energy/momentum of each $b$-quark is smeared by scaling its 4-vector by $y$, where $y$ is determined by randomly sampling a Gaussian function with a mean of 1 and a width of $\sigma=0.15$. The 15\% value is chosen to ensure that the overall resolution on the reconstructed \Ph mass matches the resolution quoted by the CMS Collaboration in Ref.~\cite{CMS:2019uxx}.
\end{enumerate}

We will refer to the smeared $b$-quarks as ``$b$-jets'' in the following. 
The \Ph candidates are reconstructed by pairing the $b$-jets. 
There are three possible combination of pairs, and we choose the combination that minimises 
$(m_{\Ph_{1}}-125\,\GeV)^2+(m_{\Ph_{2}}-125\,\GeV)^2$, 
where $m_{\Ph_{1}}$ and $m_{\Ph_2}$ 
are the reconstructed masses of the two \Ph candidates. 
We estimate the di-Higgs invariant mass, \mhh, following the method in Ref.~\cite{ATLAS:2022hwc}. This method improves the \mhh resolution by scaling
the four-momenta of each of the reconstructed \Ph such that $\mh=125\,\GeV$.  
Following this method, for signal events with $\mH=600\,\GeV$, the \mhh resolution is about 6\%.

When we study the reconstruction-level distributions in subsequent sections we also apply a loose set of kinematic selections. 
These selections are chosen to mimic the CMS trigger requirements deployed in Run-3 of the LHC, as described in Ref.~\cite{CMS-DP-2023-050}. The $b$-jets are required to have $\pT>30\,\GeV$ and $|\eta|<2.5$. The scalar sum of \pT of the four $b$-jets, \HT, is also required to be larger than $280\,\GeV$.
\section{Numerical investigation of interference effects}
\label{sec:parscan}

In this section, we analyse the parameter dependence of the LO QCD simulations performed with \MADGRAPH. In the first part, we discuss in detail different observables and their interference behaviour at two example parameter points. 
The second part contains a parameter scan, where we are mostly interested in the effect of the interference on the cross-section.
 
To evaluate the impact of the interference term, we split the diagrams describing the total process $P_{\text{tot}}=pp\rightarrow \Ph\Ph$ into two subprocesses, defined as
\begin{itemize}
    \item[1.] $P_{\SH}=pp\rightarrow \PS\rightarrow \Ph\Ph$,
    \item[2.] $P_{\text{woH}}=pp\rightarrow \Ph\Ph$ without $\PS$.
\end{itemize} 
To generate the $P_{\text{woH}}$ subprocesses, which include the $\Box$ and \Sh diagrams as shown in Figure~\ref{fig:feynman}, we set the coupling constant $\lamHhh$ to zero in the simulation. For the widths of the Higgs bosons \PS and \Ph, we use an estimated tree-level total width, consisting of the partial widths for $\Ph\rightarrow VV^{\lb*\rb}$, $\Ph\rightarrow ff$, and $\PS\rightarrow \Ph\Ph$ and an effective $\Ph\rightarrow\gamma \gamma$ partial width\footnote{The total width as the sum of these partial widths is calculated using the ``AutoWidth'' feature of \MADGRAPH.}.
As the current  model file does not include the gluonic decay mode automatically, we also sum this decay width by hand, where we use the full top quark mass dependence and perform the calculation at LO.
All partial widths were calculated with \MADGRAPH. Some example values are shown in Table~\ref{hgg_contributions_table}.

\begin{table}[]
    \begin{center}   
   \resizebox{\textwidth}{!}{\begin{tabular}{c|cccccccccc}
    \hline
    Higgs mass $m_H$\,[GeV] & 260 &320 & 380 & 440 & 500 & 560 & 620 & 680 & 740 & 800 \\ \hline
    $\Gamma_{\PS\rightarrow gg}^{\text{SM}}$ [MeV] & 2.111 & 4.895 & 13.55 & 21.64 & 27.64 & 32.24 & 35.87 & 38.79 & 41.17 & 43.14 \\
    \wH [GeV] & 0.043 & 0.111 & 0.214 & 0.370 & 0.570 & 0.817 & 1.115 & 1.472 & 1.891 & 2.380 \\   \hline
    \end{tabular}}
    \end{center}
    \caption{For a Higgs mass as given in the first row, the second row shows the \PS decay widths into $gg$ as in the SM ($\Gamma_{\PS\rightarrow gg}^{\text{SM}}$) for that given mass.  
     The third row shows the total decay width of \PS, calculated as the sum of all partial widths (all decays apart from $H\to gg$ are automatically taken into account by \MADGRAPH and $\Gamma_{\PS\rightarrow gg}= \sin^2\alpha\,\Gamma_{\PS\rightarrow gg}^{\text{SM}}$ is added), evaluated at the example parameter point of $\sina=0.08$ and $\tanb=1$. The maximal branching ratio into the digluon final state for the points in the table is about 0.048\,$\%$.
    }
    \label{hgg_contributions_table}
    \end{table}

\subsection{Illustration of interference Effects}\label{subsec:parscan_interference}

For illustrative purposes, we show the results for two parameter points, namely the points $\mH=300\,\GeV$, $\sina=0.17$, and $\tanb=3.3$; and $\mH=600\,\GeV$, $\sina=0.17$, and $\tanb=1.6$. For these points, we simulated $10^6$ events for each subprocess, $P_{\text{tot}}$, $P_{\SH}$ and $P_{\text{woH}}$. We investigate the interference effects for different observables directly by evaluating the difference between the coherent sum $P_{\text{tot}}$ of all contributions to $gg\to hh$ and the incoherent sum $P_{\SH} + P_{\text{woH}}$. In particular, we analyse the \mhh, \pT, and $\Delta\phi_{\Ph\Ph}$ distributions, where $\Delta\phi_{\Ph\Ph}$ is the angle between the final state Higgs bosons. Here, \pT is shown for both final state Higgs bosons. 
While the behaviour of the di-Higgs invariant mass is well known, and the corresponding interference effects have been discussed in detail in the literature (see \eg Ref.~\cite{DiMicco:2019ngk} and references therein), other variables have not been investigated yet. However, all of these can in principle serve as input for both cut-based as well as machine learning based analyses. Therefore, it is imminent to equally be aware of interference effects in as many as possible differential distributions.

\begin{figure}[htbp]
    \centering
    \subfloat[]{\includegraphics[width=0.46\textwidth]{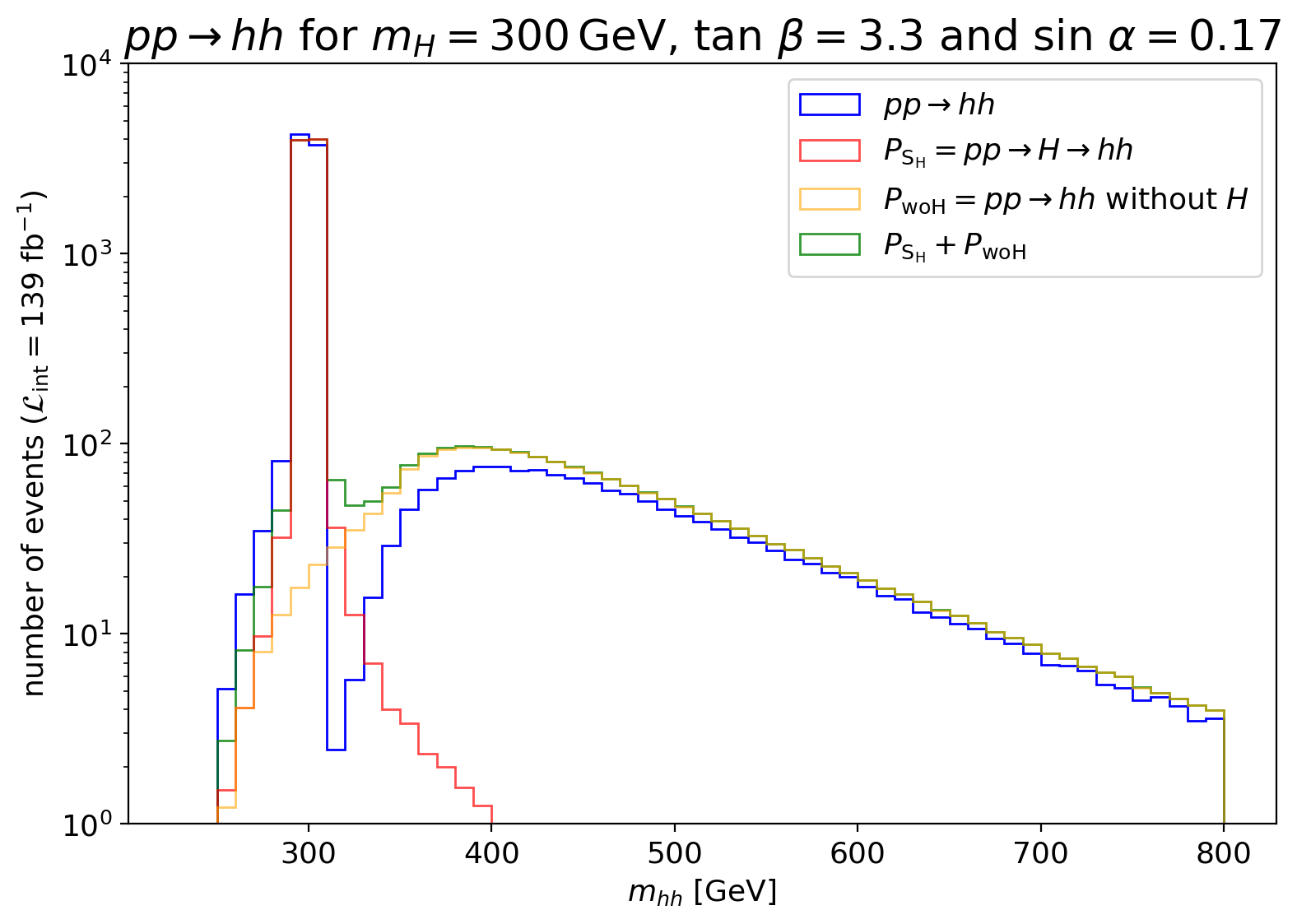}\label{figure_300_benchmark_mhh}}\quad
    \subfloat[]{\includegraphics[width=0.46\textwidth]{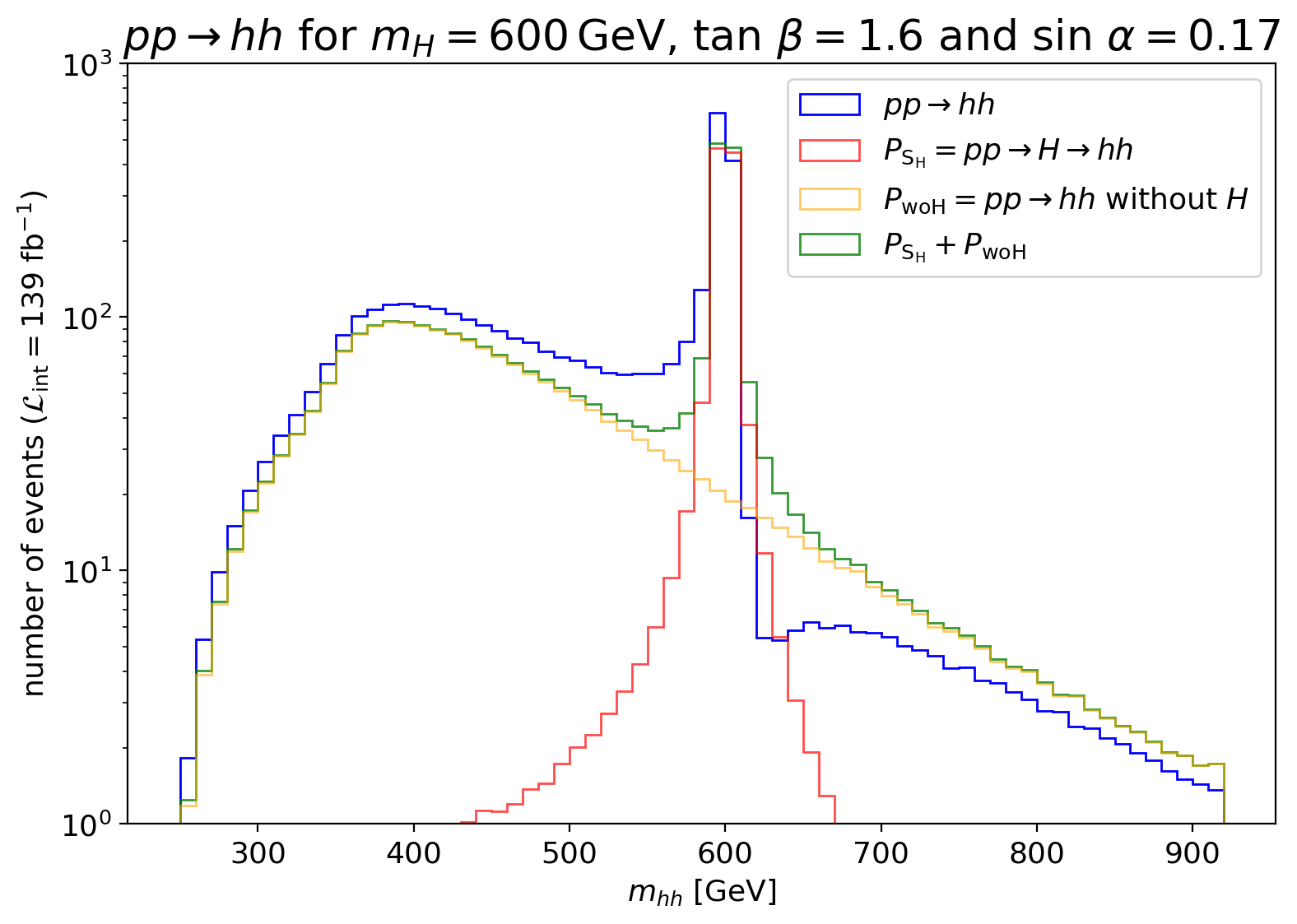}\label{figure_600_benchmark_mhh}}\quad
    \subfloat[]{\includegraphics[width=0.46\textwidth]{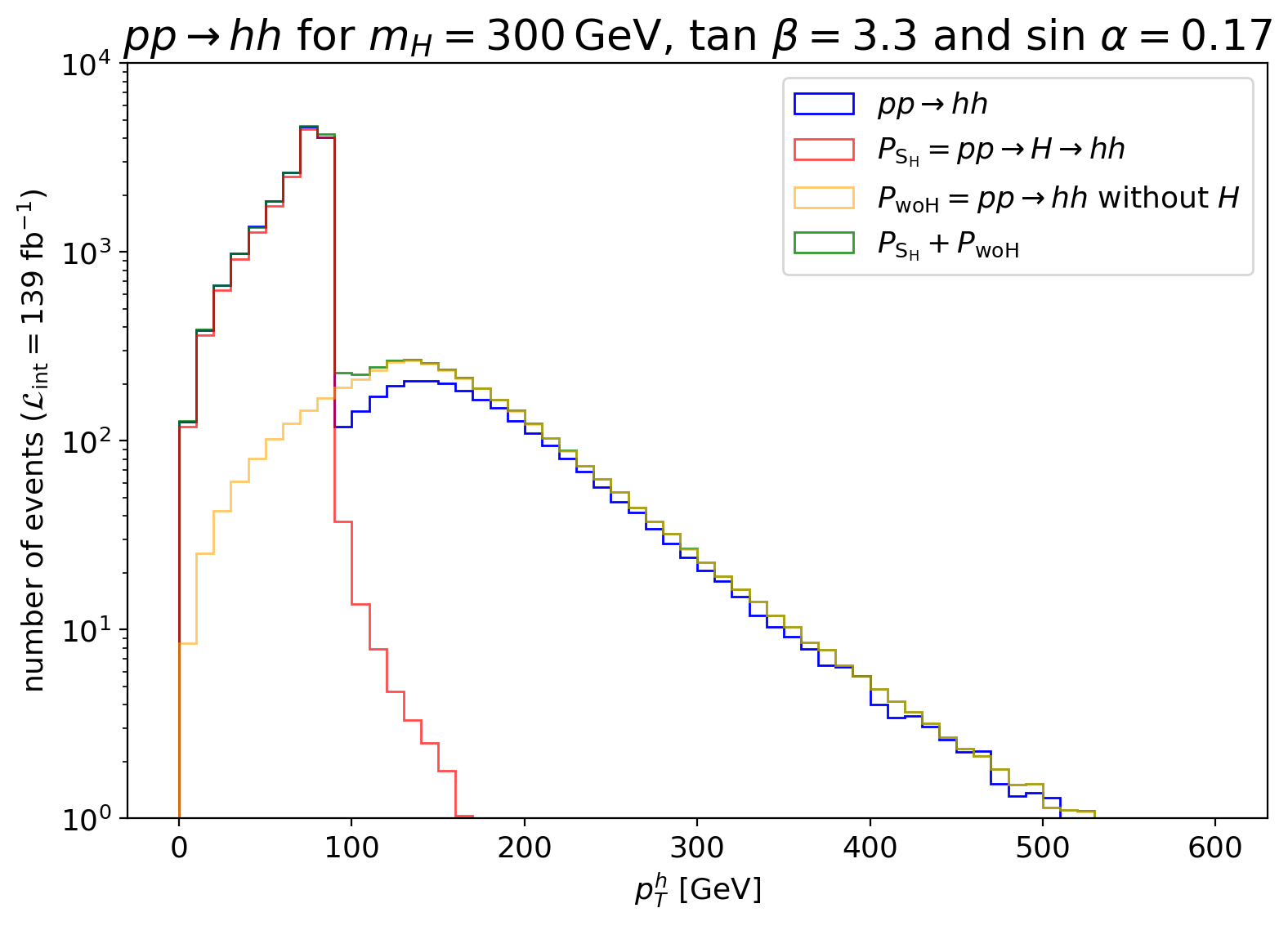}\label{figure_300_benchmark_pt}}\quad
    \subfloat[]{\includegraphics[width=0.46\textwidth]{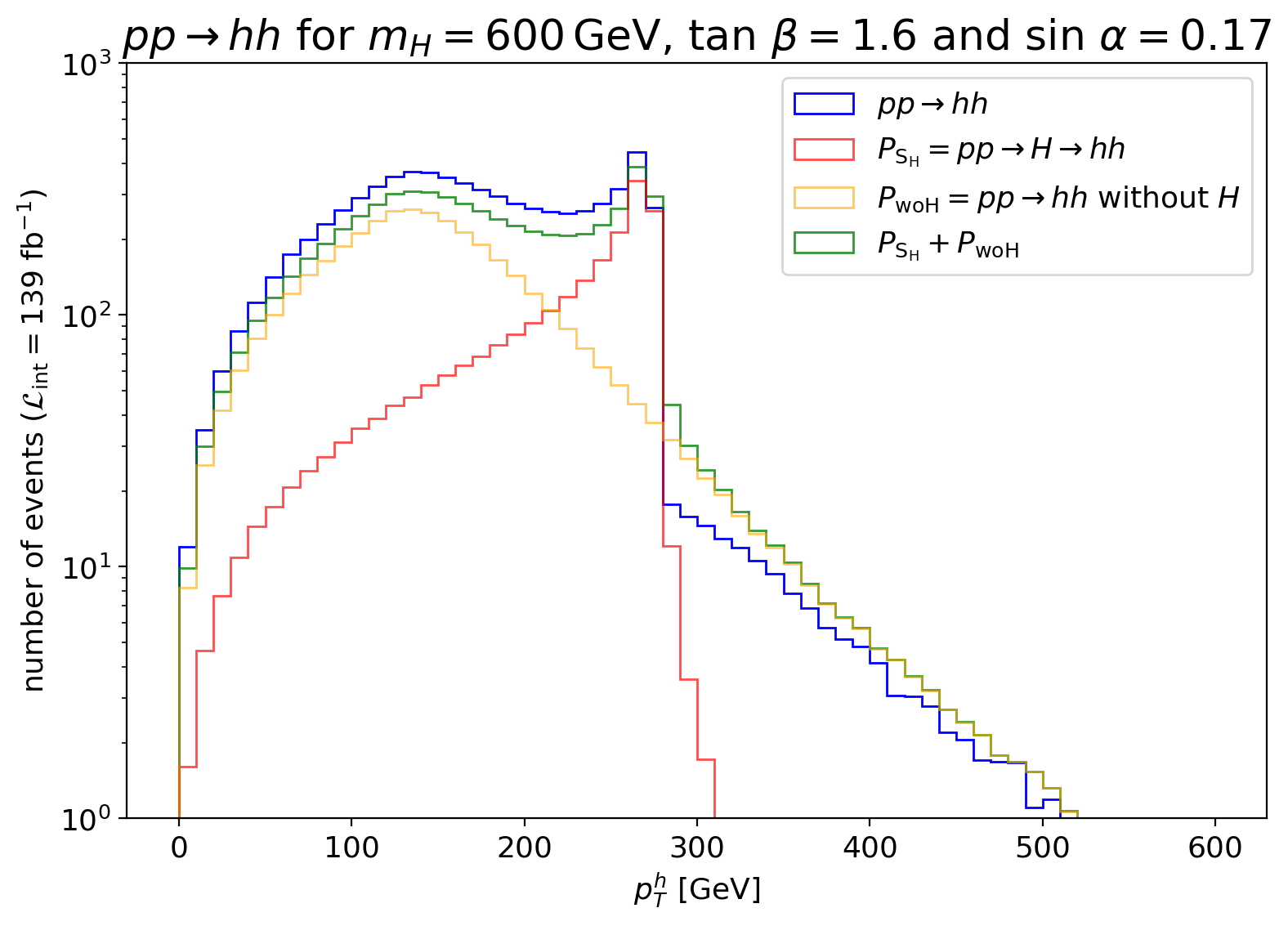}\label{figure_600_benchmark_pt}}
    \caption{The upper plots show the invariant mass \mhh and the lower plots show the transverse momenta \pT of both final state Higgs bosons for the two parameter points. The blue line corresponds to $N_{\text{tot}}$, the red line corresponds to $N_{\SH}$, the orange line corresponds to $N_{\text{woH}}$ and the green line is the sum $N_{\SH}+N_{\text{woH}}$ neglecting interference effects. The event numbers $N_i$ corresponding to process $P_i$ were simulated assuming full Run-2 luminosity. The bin size is $10\,\GeV$.}
    \label{figure_Observable_Plots_two_benchmarks}
\end{figure}

As we will discuss in the following, \mhh and \pT show interesting interference effects, as can be seen in Figure~\ref{figure_Observable_Plots_two_benchmarks}, which shows the number of total events expected using full Run-2 luminosity $\int\mathcal{L}\,=\,139\,\fb^{-1}$. In contrast, $\Delta\phi_{\Ph\Ph}$ is not significantly affected for these parameters points, and we chose to defer the corresponding discussion to  Appendix~\ref{sec:AppendixA}.
We observe the well-known peak-dip structure around $\mhh\,\sim\,m_H$, that has already been discussed in the literature. For \mhh slightly larger than the mass of the heavy scalar, there is a negative interference that leads to a smaller number of events than expected from incoherently adding the different distributions. 
A similar effect appears for \mhh slightly below \mH, although the interference effect has a different sign. These are therefore regions in the \mhh distributions where interference effects can become important.
The interference behaves similarly for the \pT distributions, with positive interference for \pT lower than the peak and negative interference for \pT higher than the peak. 

\begin{figure}[htbp]
    \centering
    \subfloat[]{\includegraphics[width=0.46\textwidth]{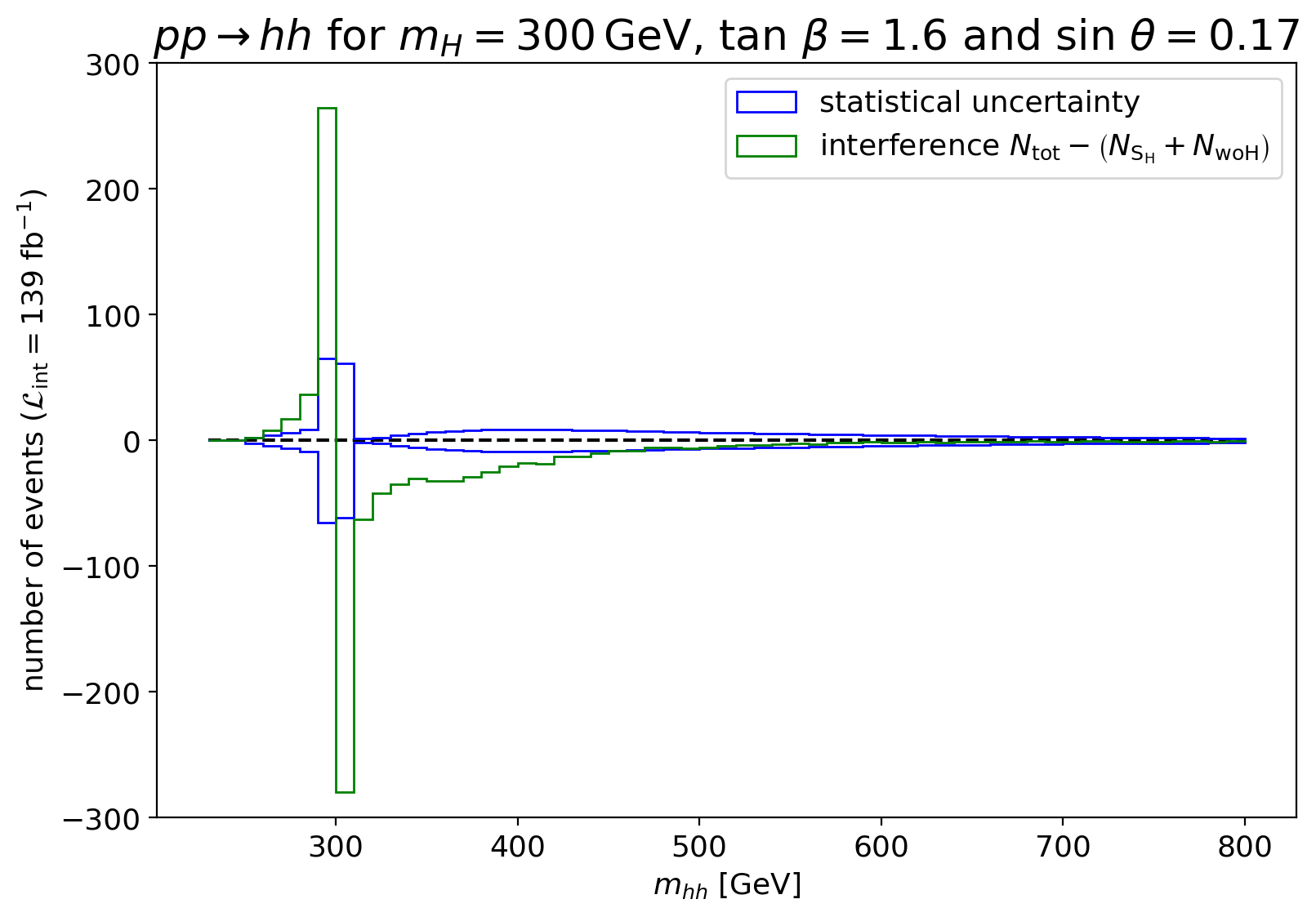}\label{figure_300_benchmark_mhh_diff}}\quad
    \subfloat[]{\includegraphics[width=0.46\textwidth]{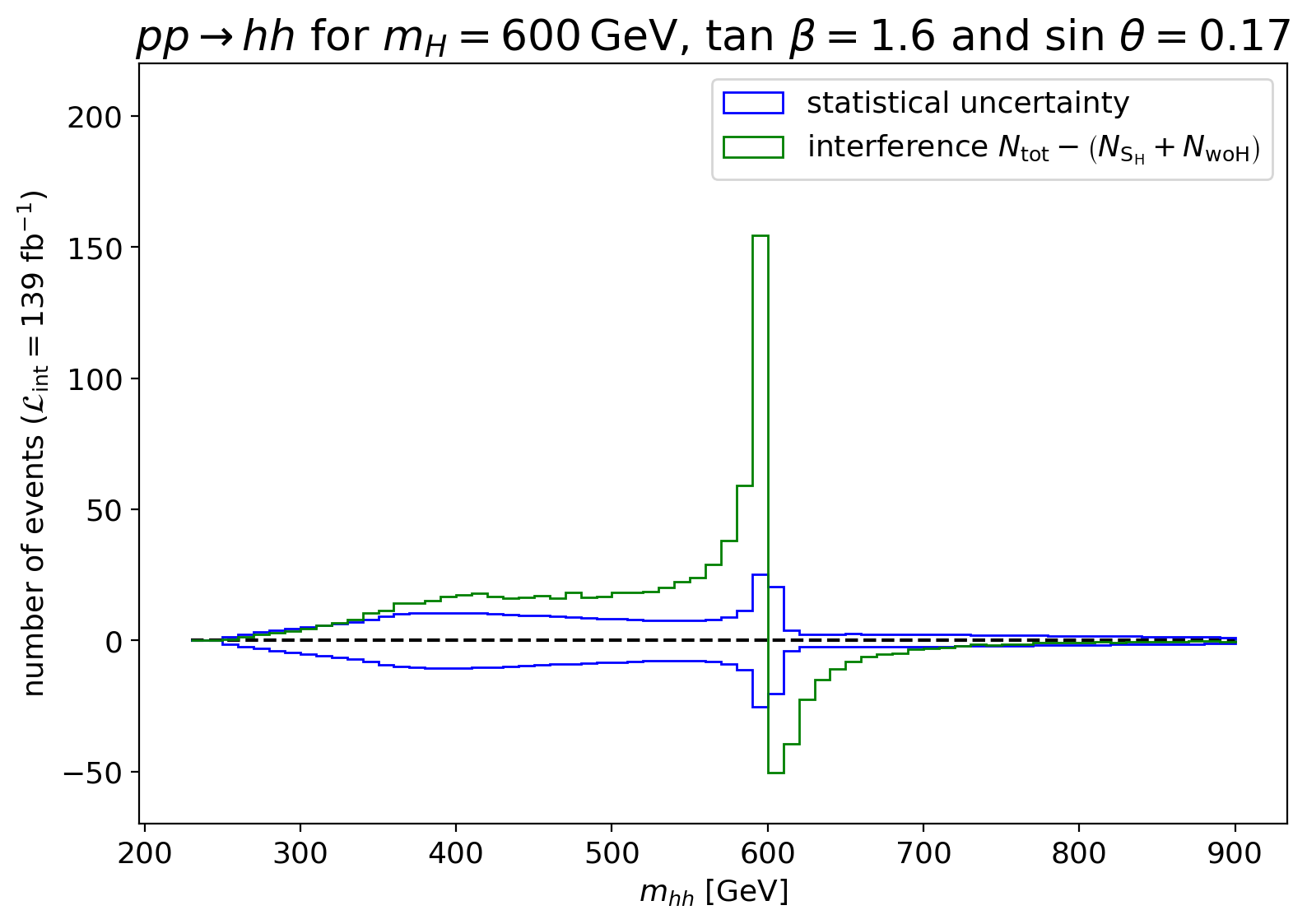}\label{figure_600_benchmark_mhh_diff}}\quad
    \subfloat[]{\includegraphics[width=0.46\textwidth]{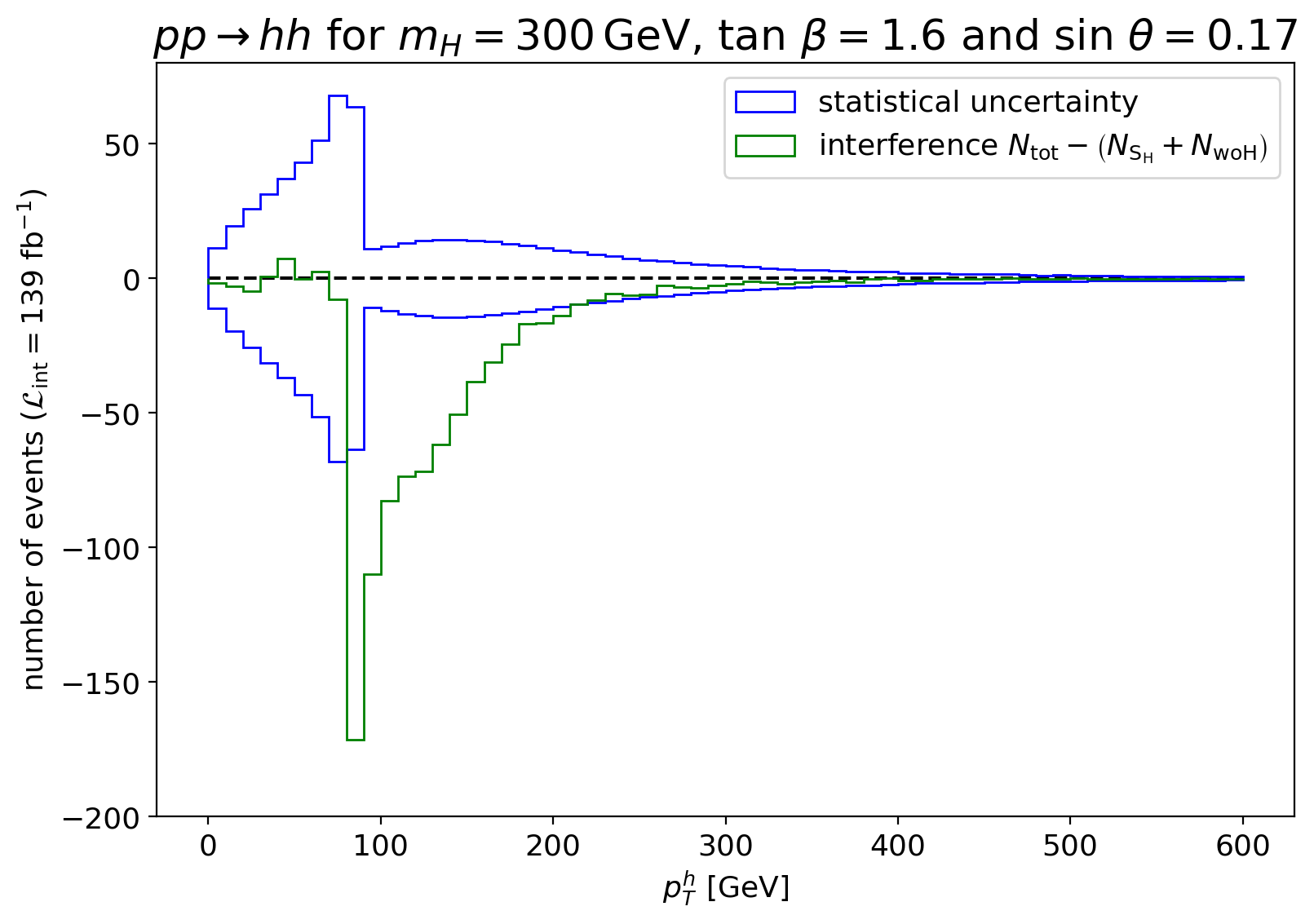}\label{figure_300_benchmark_pt_diff}}\quad
    \subfloat[]{\includegraphics[width=0.46\textwidth]{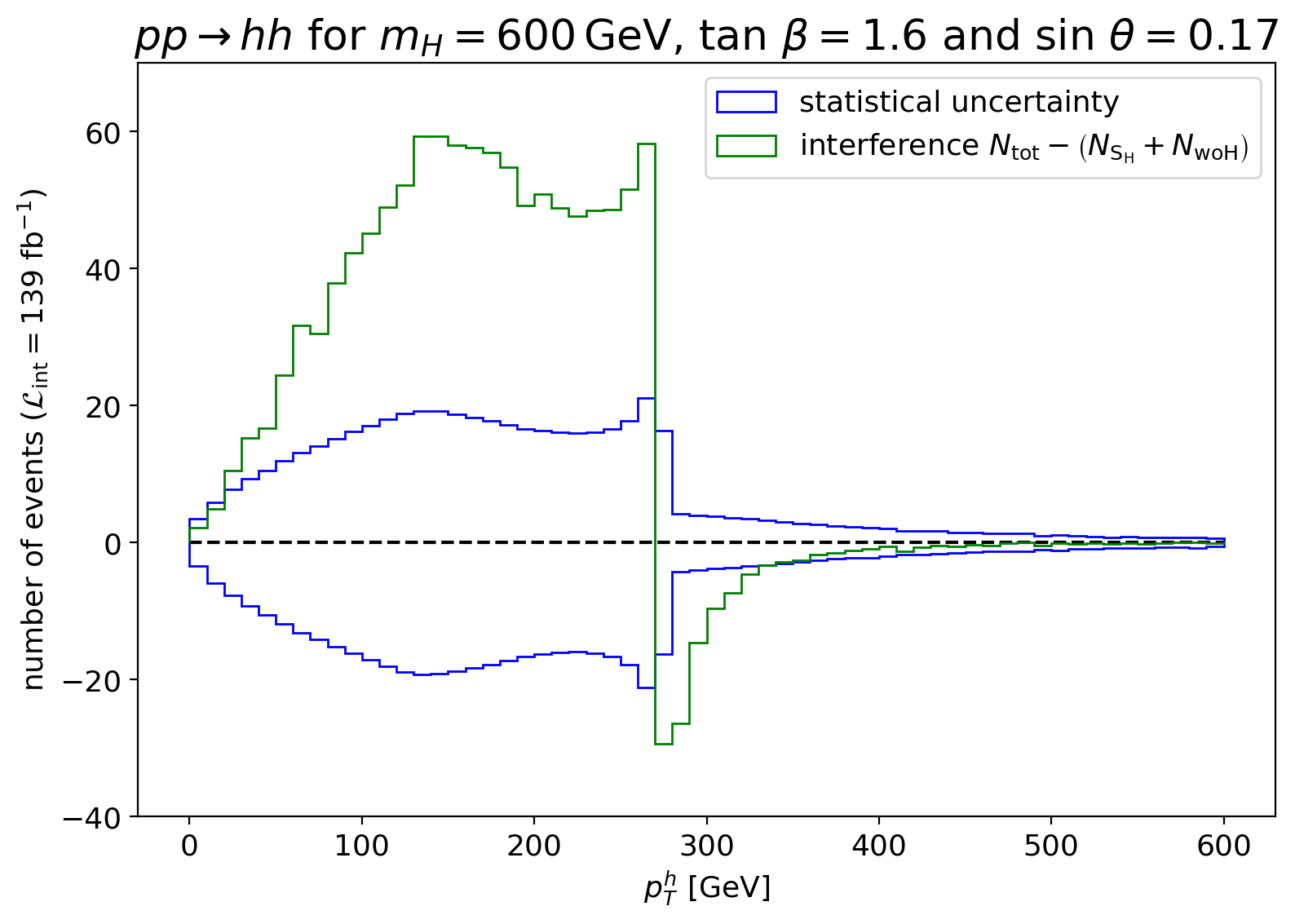}\label{figure_600_benchmark_pt_diff}}
    \caption{These plots show the interference calculated from the difference between the blue line and the green line seen in Figure~\ref{figure_Observable_Plots_two_benchmarks} as the green line and a statistical uncertainty, for which we used the $\sqrt{N_\text{tot}}$ of the $P_{\text{tot}}$ process assuming full Run-2 luminosity. The upper plots show this for the invariant mass \mhh and the lower plots show this for the transverse momentum \pT for the two parameter points.
    }
    \label{figure_Observable_Plots_two_benchmarks_difference}
\end{figure}

Another important question is whether these effects are statistically significant. In Figure~\ref{figure_Observable_Plots_two_benchmarks_difference}, we therefore show the relative difference for the event numbers for the full Run-2 luminosity. As a simple estimate, we only take statistical errors into account. We see that the interference effects are non-negligible over large regions of parameter space, and exceed the statistical uncertainty estimate we applied.

\subsection{Parameter scan}

To investigate the effects of interference between the resonant and non-resonant diagrams, we examine the cross-section corresponding to the interference term for a broad range of parameters. 
For this purpose, we perform a parameter scan for a range of the relevant parameters in the parameter volume of $[260\,\GeV,800\,\GeV]\times[-0.24,0.24]\times[0.5,4.0]$ of the 
$\lbrace{\mH, \sina, \tanb \rbrace}$ 
space 
where we scan $\mH$ in steps of $60$\,GeV, $\sina$ is steps of $0.08$ and $\tanb$ in steps of $0.5$. 
The resulting parameter points are displayed in Table~\ref{parameter_values_table}. We omitted the points with $\sina=0$ as they represent the SM case where the second Higgs decouples.
    \begin{table}[tbh!]
    \begin{center}
    \begin{tabular}{c|cccccccccc}
    \hline 
    parameter &  & & & & values & & & & & \\ \hline
    \mH [GeV] & 260 & 320 & 380 & 440 & 500 & 560 & 620 & 680 & 740 & 800 \\
    \sina & & & -0.24 & -0.16 & -0.08 & 0.08 & 0.16 & 0.24 & & \\
    \tanb & & 0.5 & 1.0 & 1.5 & 2.0 & 2.5 & 3.0 & 3.5 & 4.0 & \\   
    \noalign{\smallskip}\hline
    \end{tabular}
    \end{center}
    \caption{Parameters used in the interference scan.}
     \label{parameter_values_table}
\end{table}

\begin{figure}[htbp]
    \centering
    \subfloat[\intabs]{\includegraphics[width=0.46\textwidth]{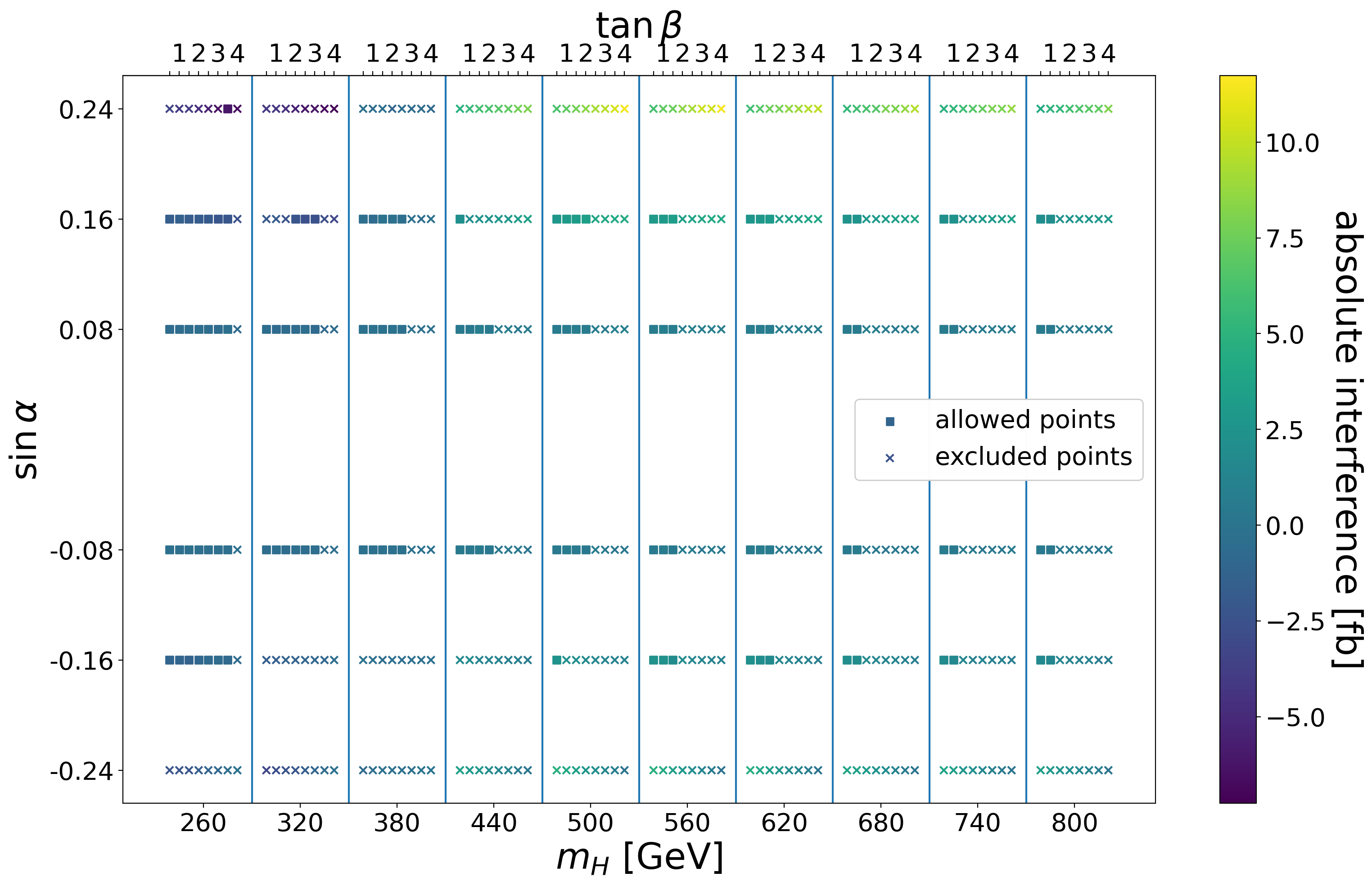}\label{figure_abs_interference_Plot}}\quad
    \subfloat[\intrel]{\includegraphics[width=0.46\textwidth]{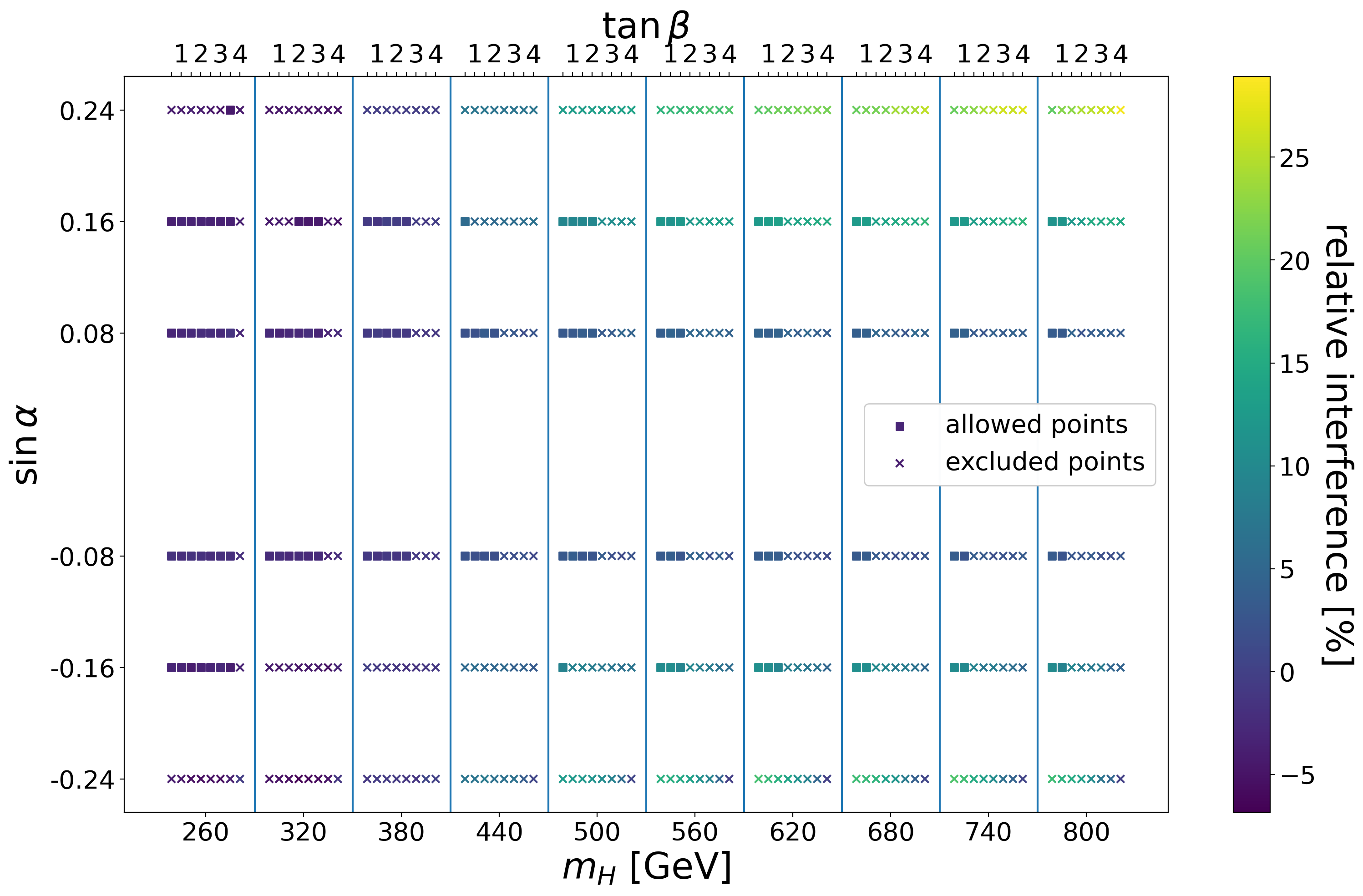}\label{figure_rel_interference_Plot}}
    \caption{The different bins on the x-axis show the investigated values of \mH. The points inside the bins are ordered by the corresponding \tanb values from left to right. The y-axis shows \sina, where the SM case of $\sina=0$ is omitted. Squares represent allowed points and crosses represent the points that do not pass the scan constraints described in Section~\ref{sec:singlet_extension}. The maximal relative interference for allowed points is around $13\%$. The colours of the points indicate the size of \intabs and \intrel for the left and right plots, respectively.}
    \label{figure_undivided_interference_Plots}
\end{figure}

We then investigated the absolute and the relative interference, defined as
\begin{align}
    \intabs &=\sigma_{P_{\text{tot}}}-\sigma_{P_{\SH}}-\sigma_{P_{\text{woH}}}~\text{and}\\
   \intrel &=\frac{\intabs}{\sigma_{P_{\text{tot}}}}\label{eq:reldef},
\end{align}
respectively. In  Figures~\ref{figure_abs_interference_Plot} and ~\ref{figure_rel_interference_Plot}, we display the absolute and relative interference, where we mark all points in the volume which are excluded by the constraints in Section~\ref{sec:model} 
as crosses, while allowed points are represented by squares. 
We observe that the interference is largest for high values of \sina, while the size of the interference decreases for  small values of \mH and/or \sina. Furthermore, the absolute values of \intrel are larger for positive than for negative interference,  reaching 12.6 $\%$ and -4.6 $\%$ for the allowed points, respectively. We checked that we generated sufficient events such that the interference terms are more significant than the statistical uncertainty at the respective parameter point\footnote{For some parameter points with relative interference effects $\leq\,1\,\%$, the integration errors can be larger, effectively leading to effects compatible to 0 within integration errors. As such points are not of particular interest for our study, we chose not to rerun such scenarios with larger statistics.}.

\begin{figure}[htbp]
    \centering
    \subfloat[The total interference \intabslow for $\mhh<\mH$.]{\includegraphics[width=0.46\textwidth]{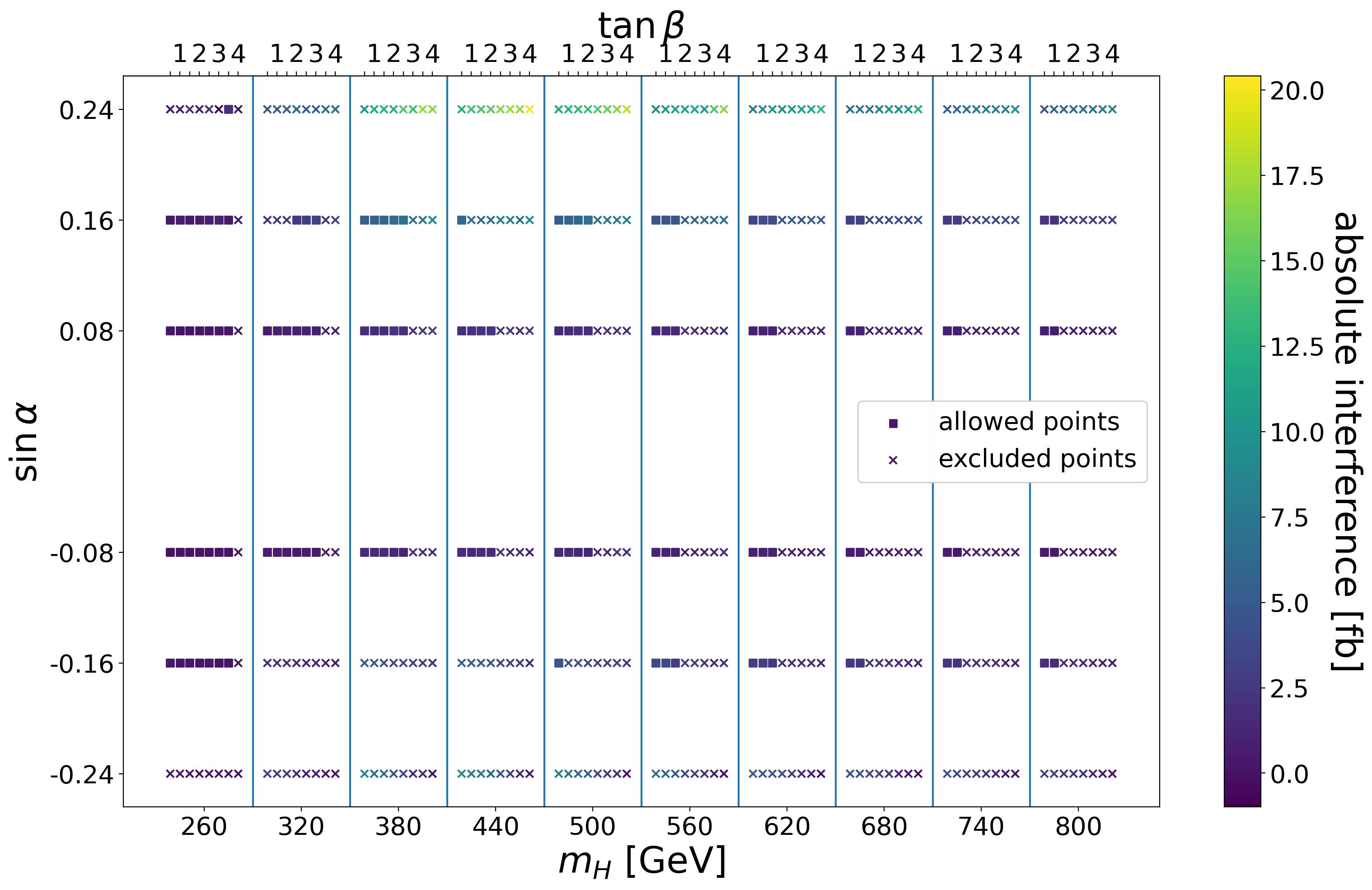}\label{figure_abs_low_interference_Plot}}\quad
    \subfloat[The relative interference \intrellow for $\mhh<\mH$.]{\includegraphics[width=0.46\textwidth]{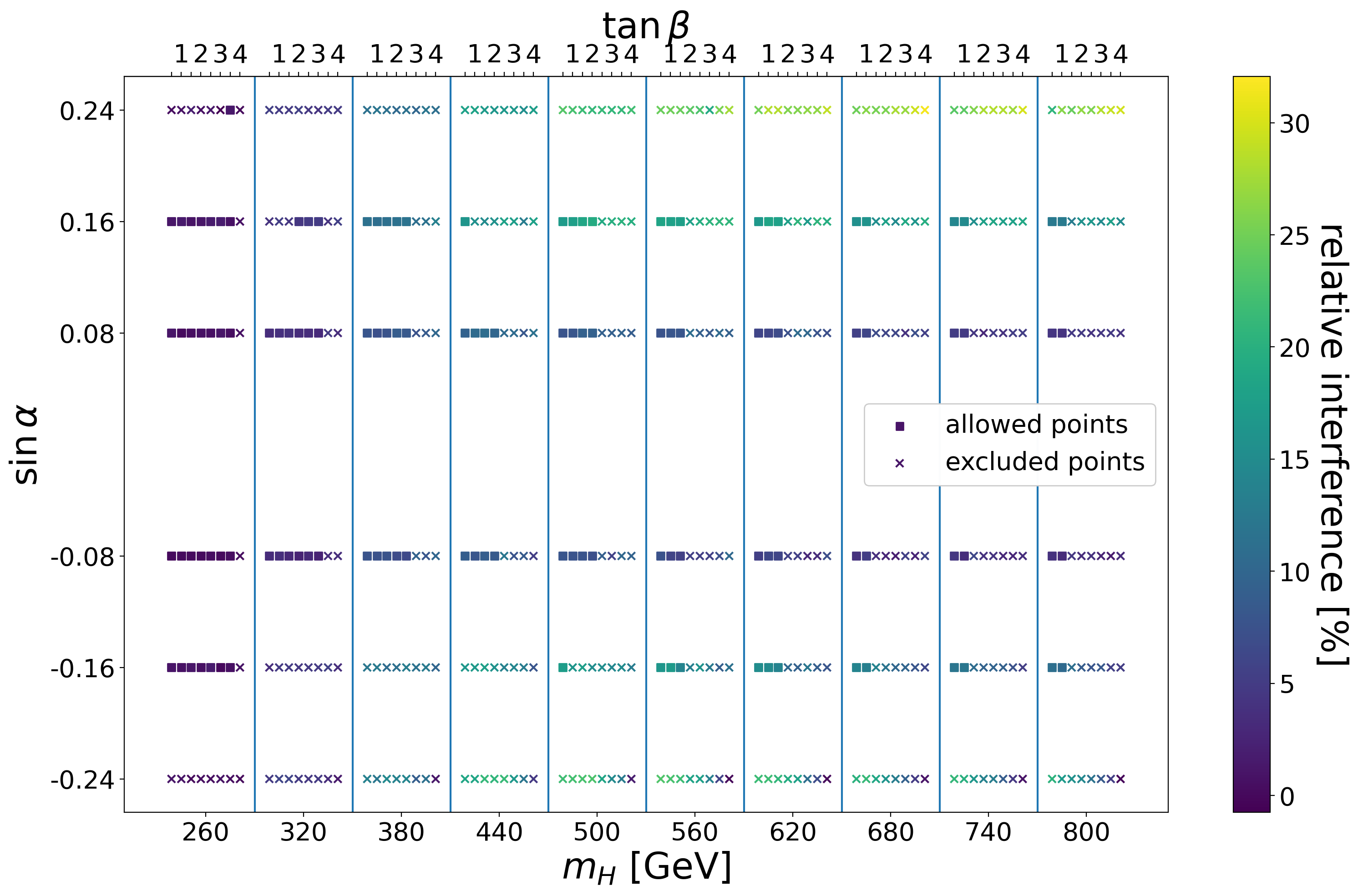}\label{figure_rel_low_interference_Plot}}\quad
    \subfloat[The total interference \intabshig for $\mhh>\mH$.]{\includegraphics[width=0.46\textwidth]{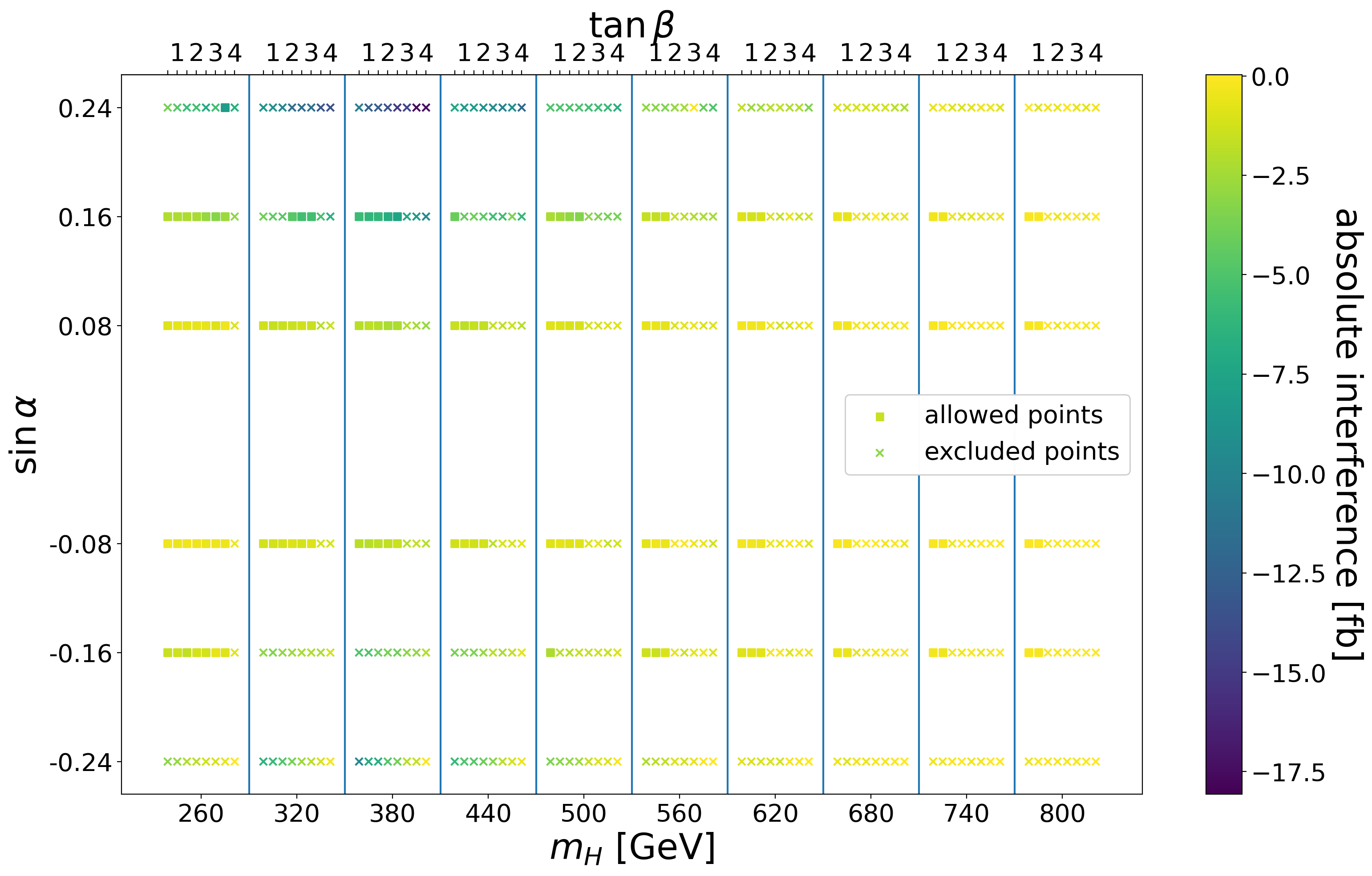}\label{figure_abs_hig_interference_Plot}}\quad
    \subfloat[The relative interference \intrelhig for $\mhh>\mH$.]{\includegraphics[width=0.46\textwidth]{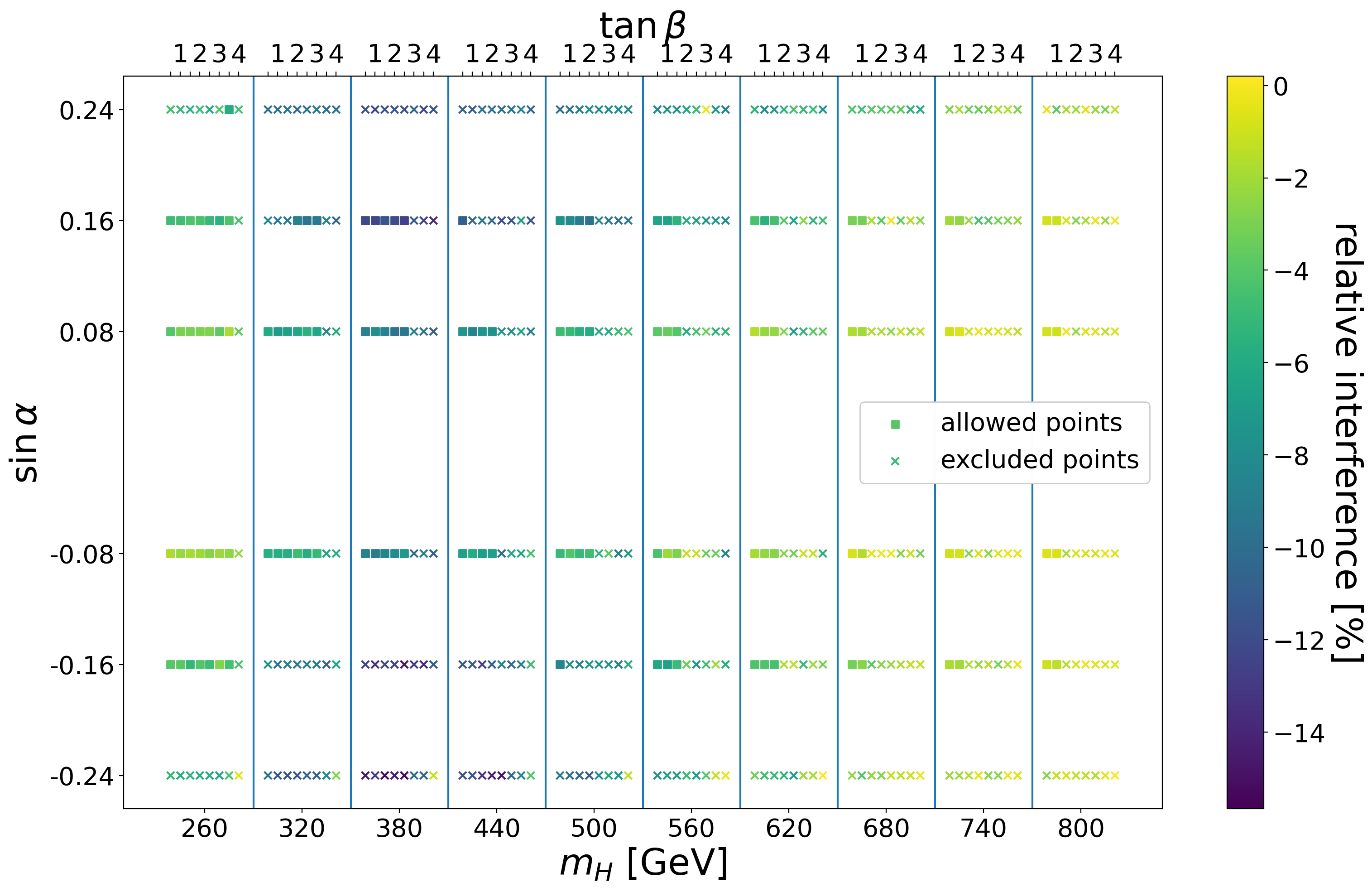}\label{figure_rel_hig_interference_Plot}}\quad
    \subfloat[The total interference sum $\intabssum=\left|\intabslow\right|+\left|\intabshig\right|$.]{\includegraphics[width=0.46\textwidth]{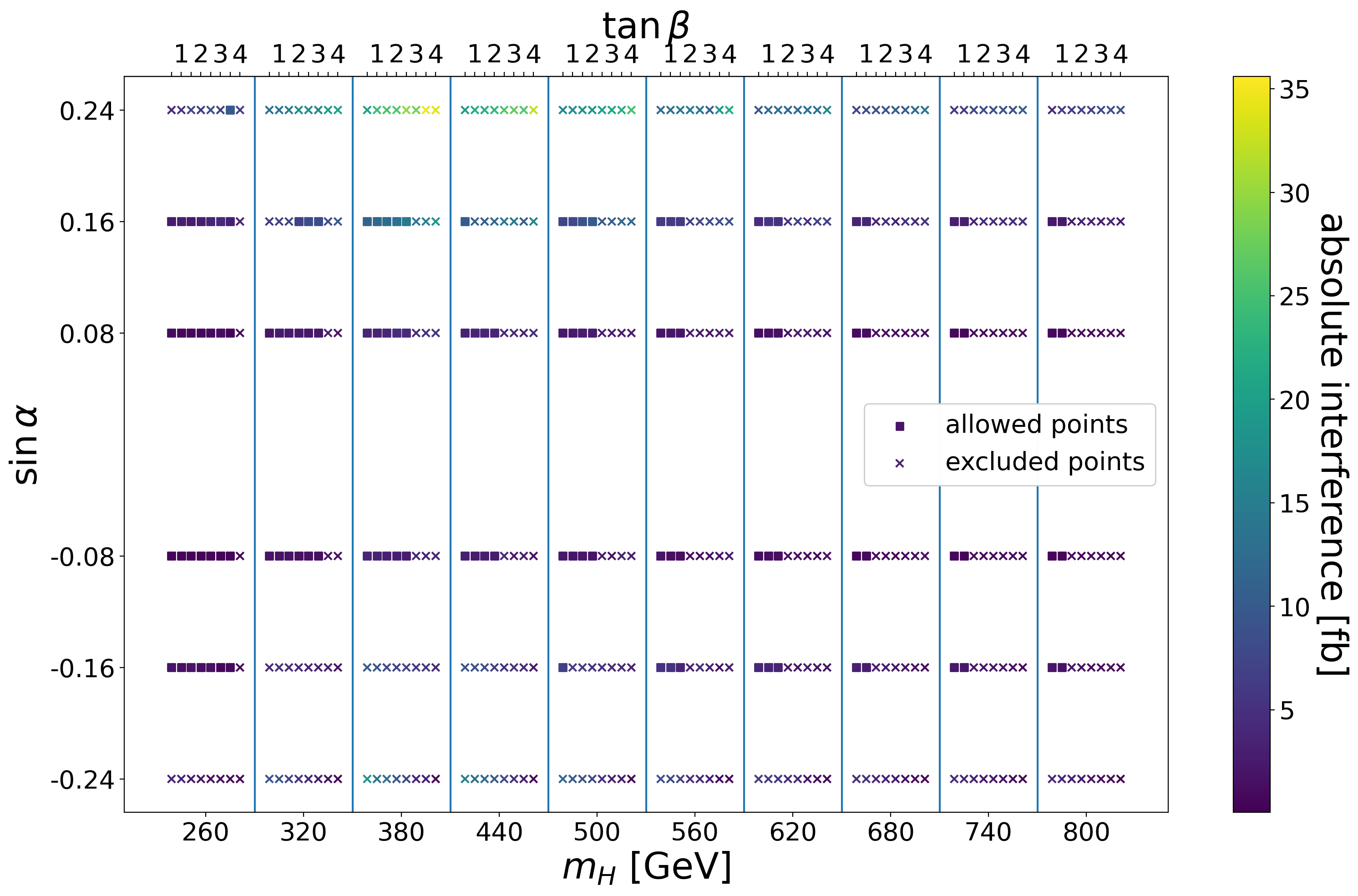}\label{figure_abs_sum_interference_Plot}}\quad
    \subfloat[The relative interference sum $\intrelsum=\left|\intrellow\right|+\left|\intrelhig\right|$.]{\includegraphics[width=0.46\textwidth]{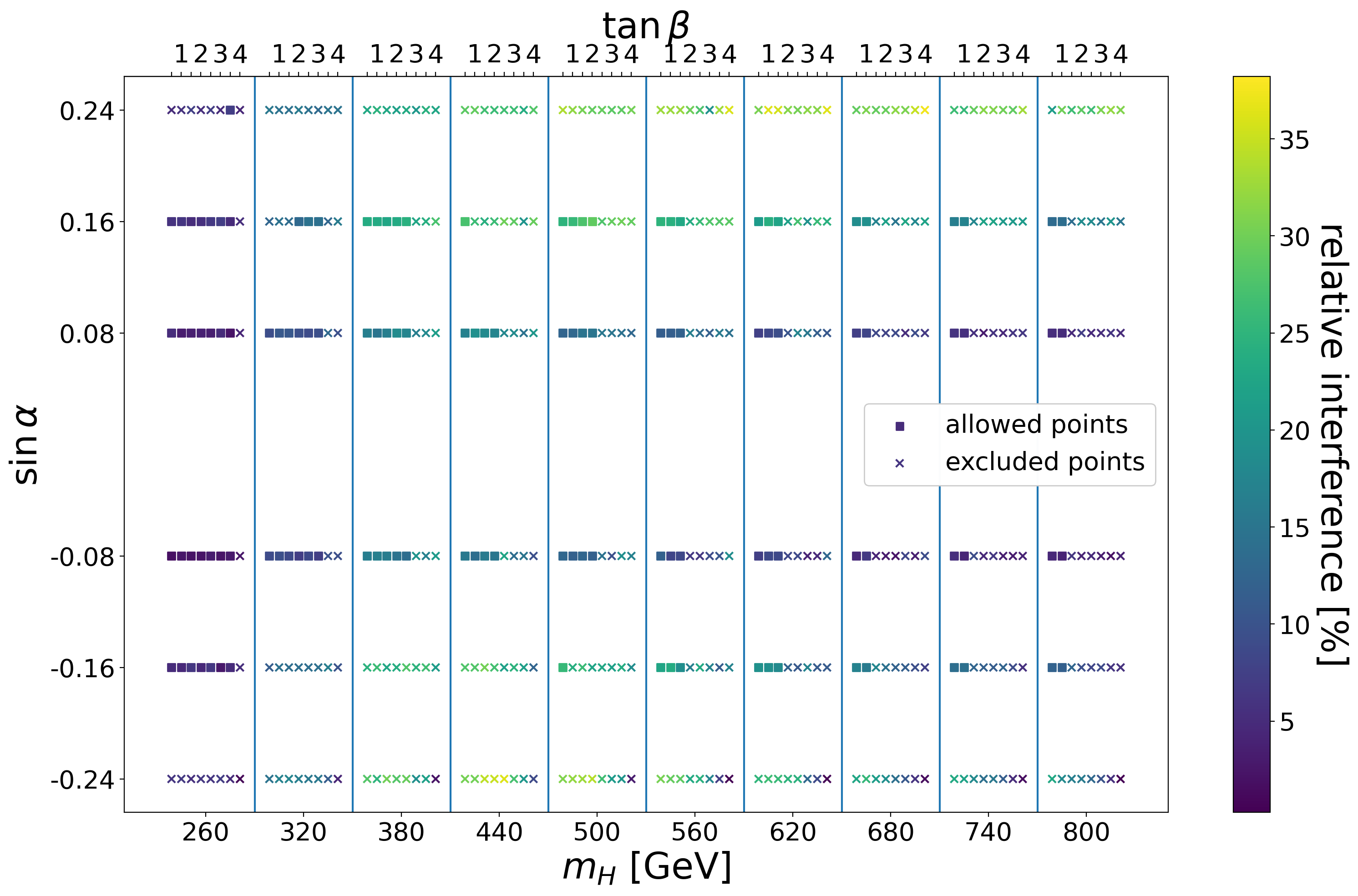}\label{figure_rel_sum_interference_Plot}}
    \caption{Interferences for invariant di-Higgs 
    masses $\mhh$ above and below the heavy scalar mass \mH.
    Bins on the x-axis represent different \mH values, ordered by increasing \tanb within each bin. The y-axis shows \sina, excluding $\sina=0$ (SM case). Squares represent allowed points and crosses represent 
    the points that do not pass the scan constraints described in Section~\ref{sec:singlet_extension}. 
    The colours of the points indicate the size of \intabslow, \intrellow, \intabshig, \intrelhig, \intabssum, and \intrelsum for the upper-left, upper-right, middle-left, middle-right, lower-left, and lower-right, plots, respectively.}
    \label{divided_interference_Plots}
\end{figure}

We investigated the interference effects for events with invariant mass \mhh above or below \mH. For this we used the \texttt{pylhe}~\cite{pylhe} python package to divide our generated events into those with higher and those with lower invariant mass than the respective \mH. 
Accordingly, we split the cross-sections into
the lower invariant mass range, \intabslow, and for the higher invariant mass range, \intabshig, respectively, and define their absolute sum as
\begin{equation}
    \intabssum\,\equiv\,\left|\intabslow\right|+\left|\intabshig\right|
\end{equation}
and the corresponding relative interference \intrellow is defined by normalization to the total cross-section over the whole mass range,
\begin{equation}
    \intrelsum\,\equiv\,\frac{\intabssum}{\sigma_{P_{\text{tot}}}}.\label{eq:relintsumdef}
\end{equation}
The relative interferences \intrellow, \intrelhig, are also defined equivalent to Eq.~\eqref{eq:relintsumdef}. The results are displayed in Figure~\ref{divided_interference_Plots}. As shown in  Figures~\ref{figure_300_benchmark_mhh} and especially \ref{figure_600_benchmark_mhh}, the interference seems to have a positive value for $\mhh < \mH$ and a negative value for $\mhh > \mH$\footnote{In other beyond the SM extensions, one can also find scenarios where this behaviour is reversed, see \eg discussion in Ref.~\cite{Arco:2022lai}  in the context of a two-Higgs-doublet model.}. Therefore, although the total interference integrated over the whole mass range might be small in some scenarios, effects in differential distributions could still be large. We therefore also provide the corresponding information on the different mass regions discussed above as well as the absolute value of deviations between naive summing of the different contributions and the total process including interference. 
Figure~\ref{figure_abs_low_interference_Plot} shows that a higher positive mixing angle enhances both \intabslow and \intrellow, see Figure~\ref{figure_rel_low_interference_Plot}.
Furthermore, Figure~\ref{figure_abs_hig_interference_Plot} shows that \intabshig has the largest negative 
values for large mixing angles in the \mH region around $380\,\GeV$, see also Figure~\ref{figure_rel_hig_interference_Plot}. Figure~\ref{figure_abs_sum_interference_Plot} and especially Figure~\ref{figure_rel_sum_interference_Plot} show that some points, which appeared to have negligible interference based on the total cross-sections (\intabs and \intrel) displayed in Figures~\ref{figure_abs_interference_Plot} and~\ref{figure_rel_interference_Plot}, actually experience two interference effects of opposite sign, which partially cancel when the total cross-section is computed. In Section~\ref{sec:benchmarks_bandc} we will introduce benchmark (BM) points that exhibit this feature (benchmarks \BMb and \BMc).

\begin{figure}[htbp]
    \centering
    \subfloat[\kappalam]{\includegraphics[width=0.46\textwidth]{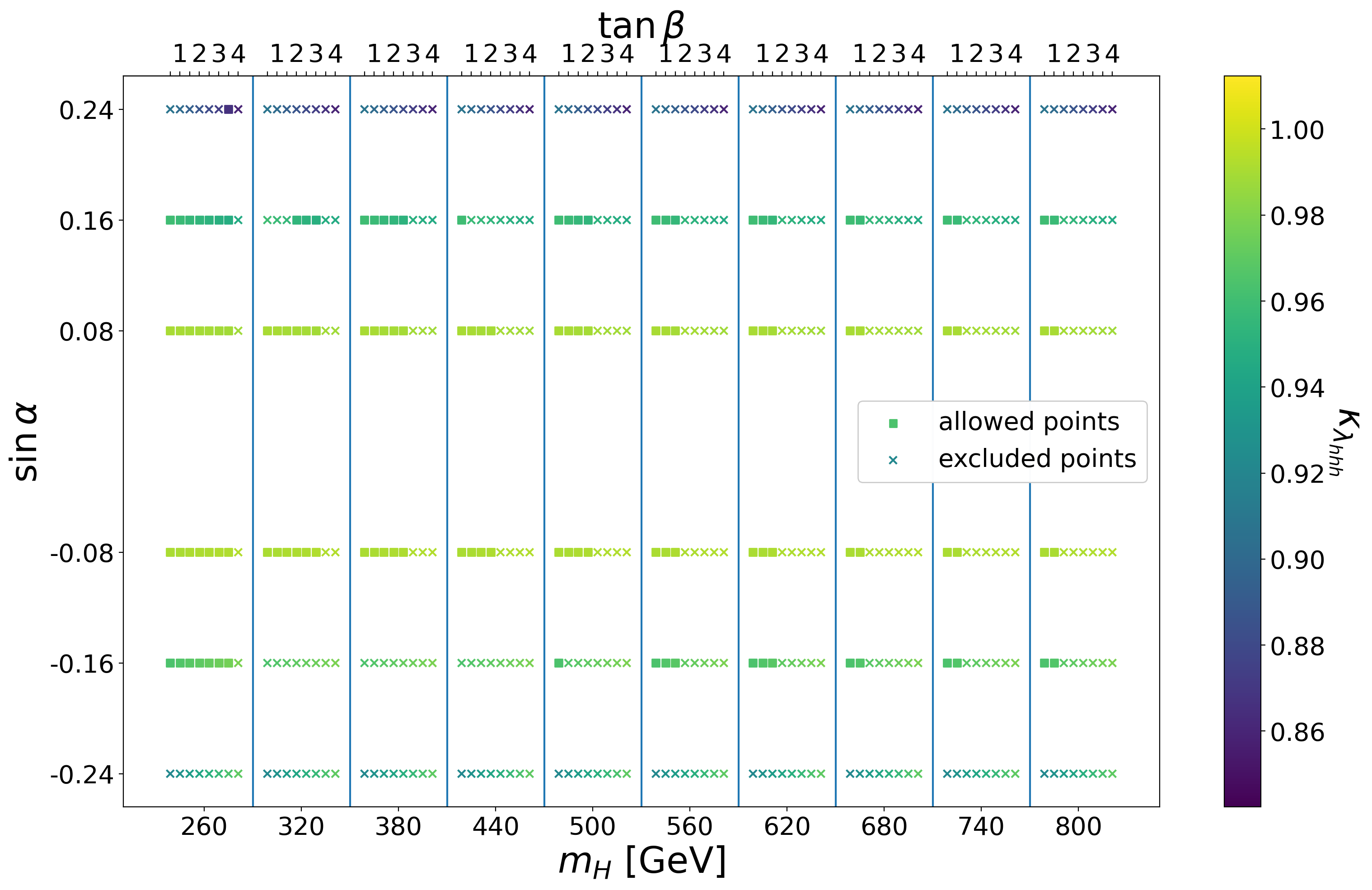}\label{figure_kappa_lambda_Plot}}\quad
    \subfloat[\lamHhh]{\includegraphics[width=0.46\textwidth]{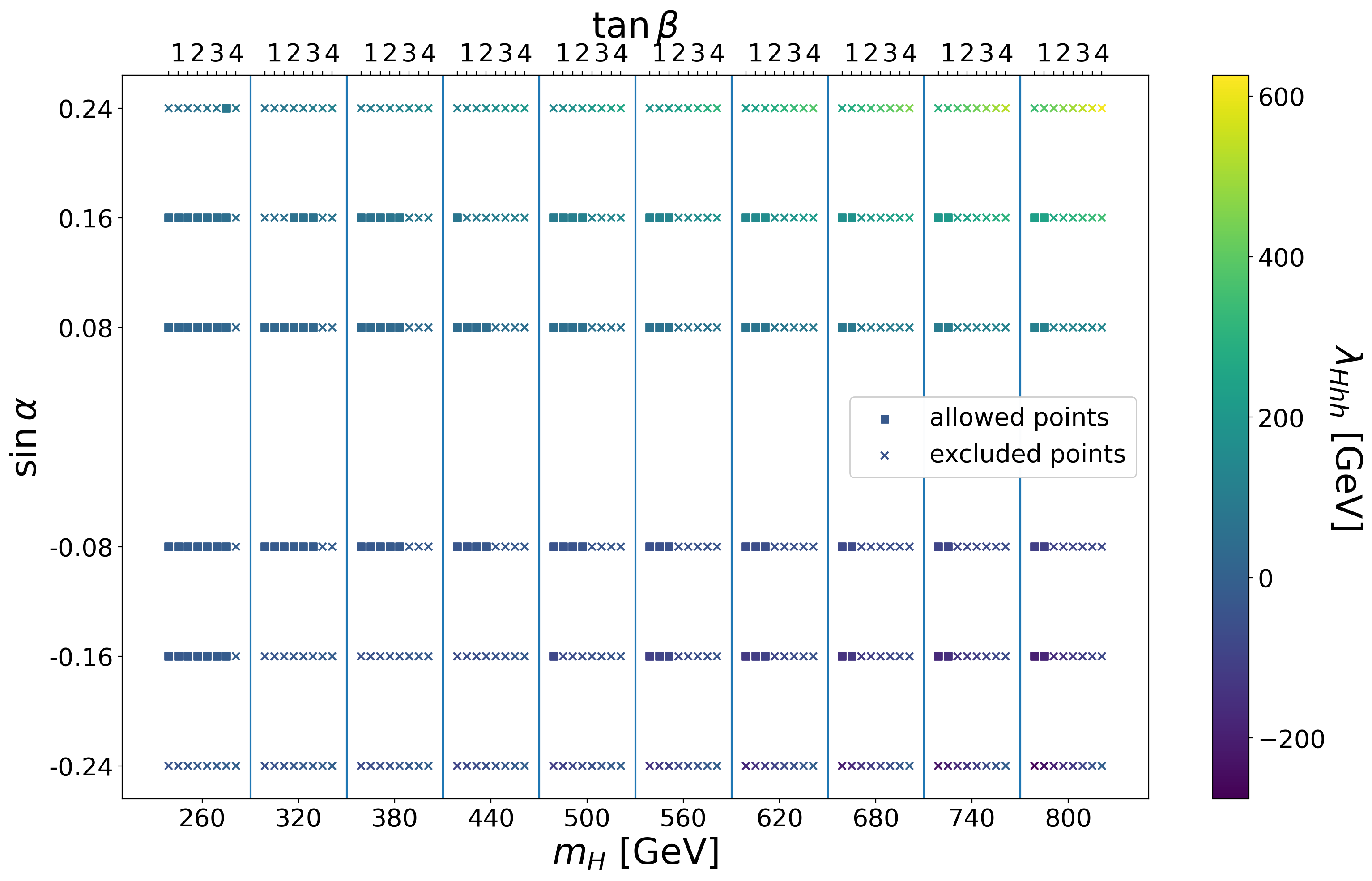}\label{figure_lambda_Hhh_Plot}}\quad
    \subfloat[\wh]{\includegraphics[width=0.46\textwidth]{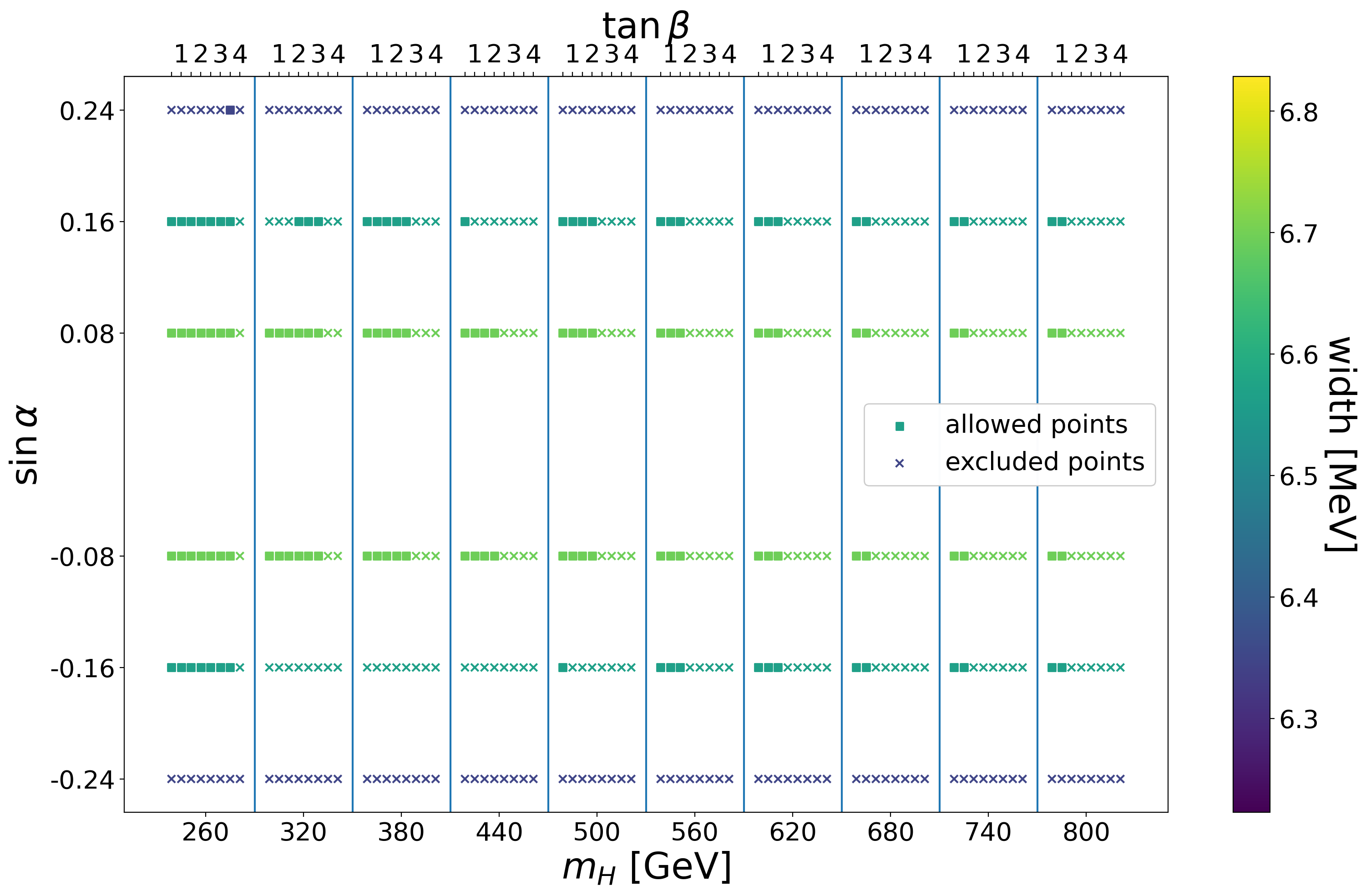}\label{figure_width_h_Plot}}\quad
    \subfloat[\wH]{\includegraphics[width=0.46\textwidth]{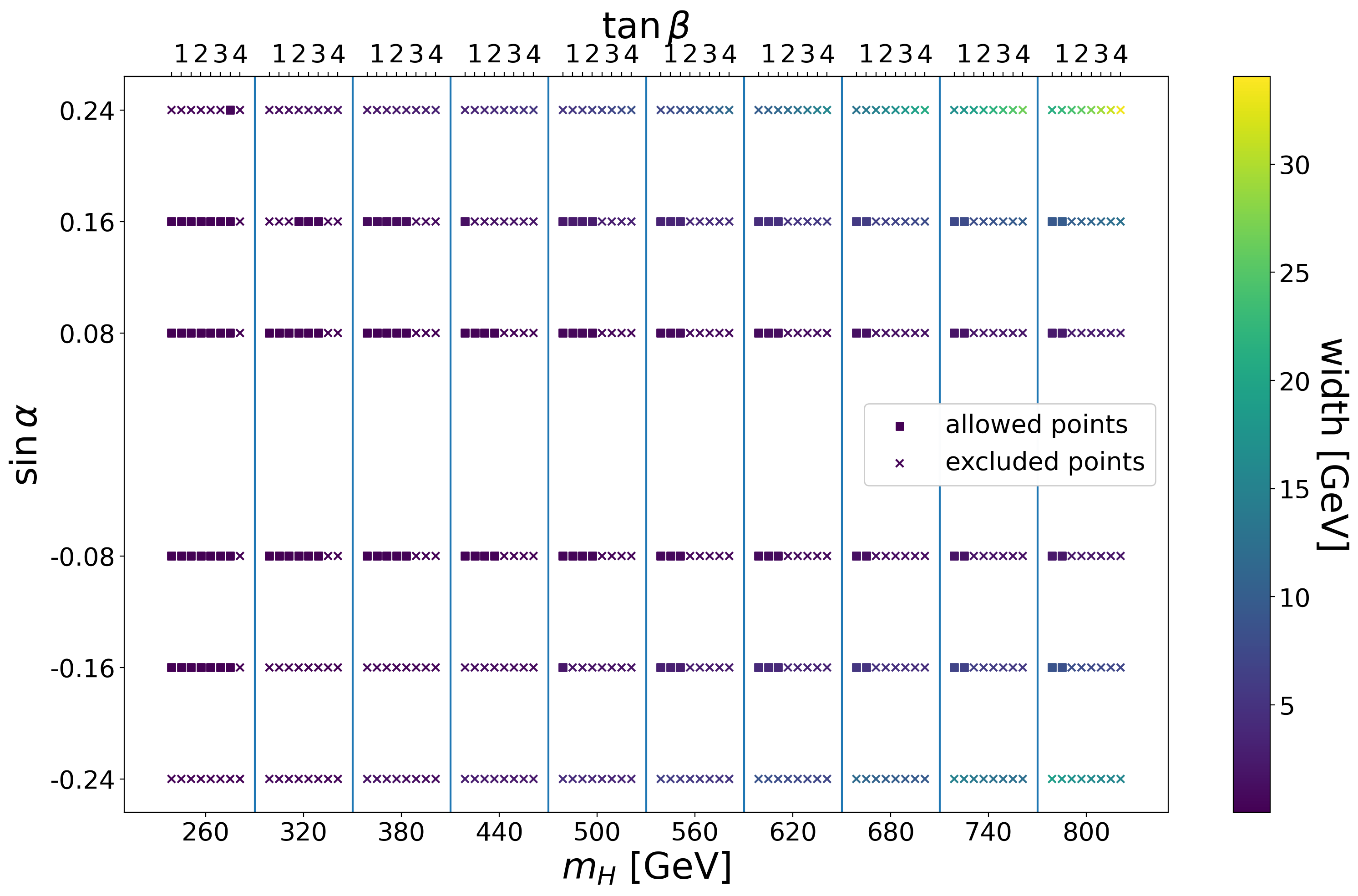}\label{figure_width_H_Plot}}
    \caption{The different bins on the x-axis show the different values of \mH investigated and inside the bins the points are ordered by the corresponding \tanb values from left to right. The \sina value is shown on the y-axis, where $\sina=0$ is left out, as it corresponds to the SM-case. Squares represent allowed points and crosses represent the points that do not pass the scan constraints described in Section~\ref{sec:singlet_extension}. The colours of the points indicate the size of \kappalam, \lamHhh, \wh, and \wH for the upper-left, upper-right, lower-left, and lower-right plots, respectively.}\label{lambda_&_width_Plots}
\end{figure}

In order to relate the dependence of the cross-sections to the dependence of the involved couplings, we also display the triple Higgs couplings \lamhhh and \lamHhh defined in Eq.~\eqref{eqn:couplings}, as well as the total widths of the two Higgs bosons, \wh and \wH 
in Figure~\ref{lambda_&_width_Plots}. From Figure~\ref{figure_kappa_lambda_Plot}, we observe that \kappalam has a larger deviation from the SM for larger $|\alpha|$ values, but never exceeds 1\footnote{Several works have investigated one loop electroweak effects on the triple scalar coupling from new physics scenarios, see \eg \cite{Kanemura:2002vm,Aiko:2023xui,Bahl:2023eau,Heinemeyer:2024hxa}.
}.
Figure~\ref{figure_lambda_Hhh_Plot} shows that \lamHhh has the sign of the mixing angle for the points considered here.  
These patterns reflect the analytic behaviour of the couplings for small values of $\sin\alpha$, see Eq.~\eqref{eqn:couplings_approx}.
Figure~\ref{figure_width_h_Plot} demonstrates that \wh only depends the mixing angle and Figure~\ref{figure_width_H_Plot} shows that \wH increases with both the absolute value of the mixing angle and \tanb, and reaches maximal ratios of $\Gamma_H/m_H\,\sim\,0.012$ for the points still allowed by current constraints. 

A small comment is in order: the reader may notice that the width of the SM-like scalar is around $\sim\,6\,\MeV$, which is in principle compatible with the current experimental values of $3.2^{+2.4}_{-1.7}\,\MeV$~\cite{CMS:2022ley} and $4.5^{+3.3}
_{-2.5}\,\MeV$~\cite{ATLAS:2023dnm} from off-shell $H\rightarrow ZZ$ measurements by CMS and ATLAS, respectively,  but is not in agreement with the most up to date theoretical prediction of $\Gamma_{125}\,\sim\,4\,\MeV$ as \eg documented in Ref.~\cite{LHCHiggsCrossSectionWorkingGroup:2016ypw}. 
This is mainly due to the fact that in naive \textsc{MadGraph} calculations, the bottom mass in the $h\,\rightarrow\,b\,\bar{b}$ decay width is defined as the on-shell and not the $\overline{\text{MS}}$ mass, which takes a much smaller value at the scale $\mu\,\equiv\,125\,\GeV$\footnote{We thank M. Spira for useful discussions on this point.}. This in principle also manifests itself in discrepancies of the branching ratios of the $125\,\GeV$ scalar. However, as in our scenario the $125\,\GeV$ particle only appears in highly off-shell scenarios, we did not modify the total width but keep the LO calculated value from \textsc{MadGraph} in our simulations.
\section{Modelling interference effects using reweighting}
\label{sec:method}

There are numerous examples of extended Higgs sector models that include an additional scalar that can impact the di-Higgs cross-section and differential distributions, see \eg Ref.~\cite{wg3ext} for an overview of models currently investigated by the LHC experiments. The final di-Higgs spectrum depends on the parameters of the model. For example, in our investigated singlet model there are three free parameters that affect the relative contributions of the diagrams shown in Figure~\ref{fig:feynman} to the total cross-section, and there are countless other models that may contain even more free parameters. While in principle it is possible to generate Monte Carlo (MC) events for different combinations of parameters, and simulate the detector responses, the computational resources required would be immense. We have therefore developed a tool, \texttt{HHReweighter}, utilising a matrix-element (ME) reweighting method to reduce the total CPU time required to model the full parameter space.
The MEs are computed using \MADGRAPH (version 2.6.7). Our tool is publicly available in GitLab~\cite{reweight_repo}.

\subsection{Reweighting LO samples}

LO-accurate ME reweighting methods work by reweighting events of a reference sample by a weight, $w$, given by 
\begin{equation}\label{eq:reweight}
w = \frac{|\mathcal{M}_{\mathrm{target}}|^2}{|\mathcal{M}_{\mathrm{ref}}|^2},    
\end{equation} 
where $\mathcal{M}_{\mathrm{ref}}$ is the ME used in the generation of the reference sample and $\mathcal{M}_{\mathrm{target}}$ is the ME for the target model/model-parameters. 
While it is possible to compute $|\mathcal{M}_{\mathrm{target}}|^2$ for all interesting models/parameters, and this is certainly less CPU intensive than performing full event simulations, it would still require substantial computing resources to obtain all the final distributions in fits to data.

To mitigate this limitation, our strategy is to first decompose the squared ME describing di-Higgs production into several terms corresponding to the three diagrams shown in Figure~\ref{fig:feynman}, as well as the corresponding interference terms. This includes three squared terms corresponding to the $\Box$, \Sh, and \SH diagrams, plus three  cross-terms describing the interferences between these contributions: \ShBox, \SHBox, and \SHSh. The MEs depend on the  Yukawa couplings of the \Ph and \PS to the top quark, denoted \yt and \ytH, respectively, the Yukawa couplings to the bottom quark, denoted \yb and \ybH, respectively, and the \lamhhh and \lamHhh trilinear couplings. For convenience, we compute each of the MEs setting $\yt=\ytH=\ytSM$, $\yb=\ybH=\ybSM$, and  $\lamhhh=\lamHhh=\lamSM$, where \ytSM (\ybSM) and \lamSM are the top (bottom) Yukawa and trilinear couplings of the SM Higgs sector. We denote the squared MEs $\mathcal{M}_{i}^2$ (where $i=\Box,\Sh,\SH,\ShBox,\SHBox,\SHSh$) in the following. 
For the interference terms, we use the shorthand notation
\begin{equation}
    \widetilde{\mathcal{M}}^2_{a\text{-}b} \equiv 2 \text{Re}\left[ \mathcal{M}_a\,\mathcal{M}^*_b \right].
\end{equation}

The total squared ME for any model can then be constructed by a linear combination of these terms according to 
\begin{equation}
\begin{split}
|\mathcal{M}_{\mathrm{total}}|^2 =& 
\mathcal{M}_{\Box}^2\cdot (\kappat{})^4 
+\mathcal{M}_{\Sh}^2\cdot (\kappat{})^2\kappalam^2 
+ \widetilde{\mathcal{M}}_{\ShBox}^2\cdot (\kappat{})^3\kappalam
+\mathcal{M}_{\SH}^2(\mH,\wH)\cdot (\kappatH{})^2\kappalamH^2 \\
& +\widetilde{\mathcal{M}}_{\SHBox}^2(\mH,\wH)\cdot (\kappat{})^2\kappatH{}\kappalamH
 +\widetilde{\mathcal{M}}_{\SHSh}^2(\mH,\wH)\cdot \kappat{}\kappatH{}\kappalam\kappalamH,
\end{split}
\label{eqn:decomposition}
\end{equation}
where $\kappatH\equiv\frac{\ytH}{\ytSM}=\frac{\ybH}{\ybSM}$, $\kappat\equiv\frac{\yt}{\ytSM}=\frac{\yb}{\ybSM}$, and $\kappalamH\equiv\frac{\lamHhh}{\lamSM}$\footnote{In our current implementation, all quark Yukawa couplings are modified by universal $\kappa_q^h$ and $\kappa_q^H$. A generalisation to individual Yukawa coupling modifiers will be addressed in future work, \eg to accommodate also the coupling structure of the two-Higgs-doublet model.}. We note that \SH, \SHBox, and \SHSh terms also depend on the assumed values of \mH and \wH.

\begin{figure}[htbp]
  \centering\includegraphics[width=0.48\textwidth]{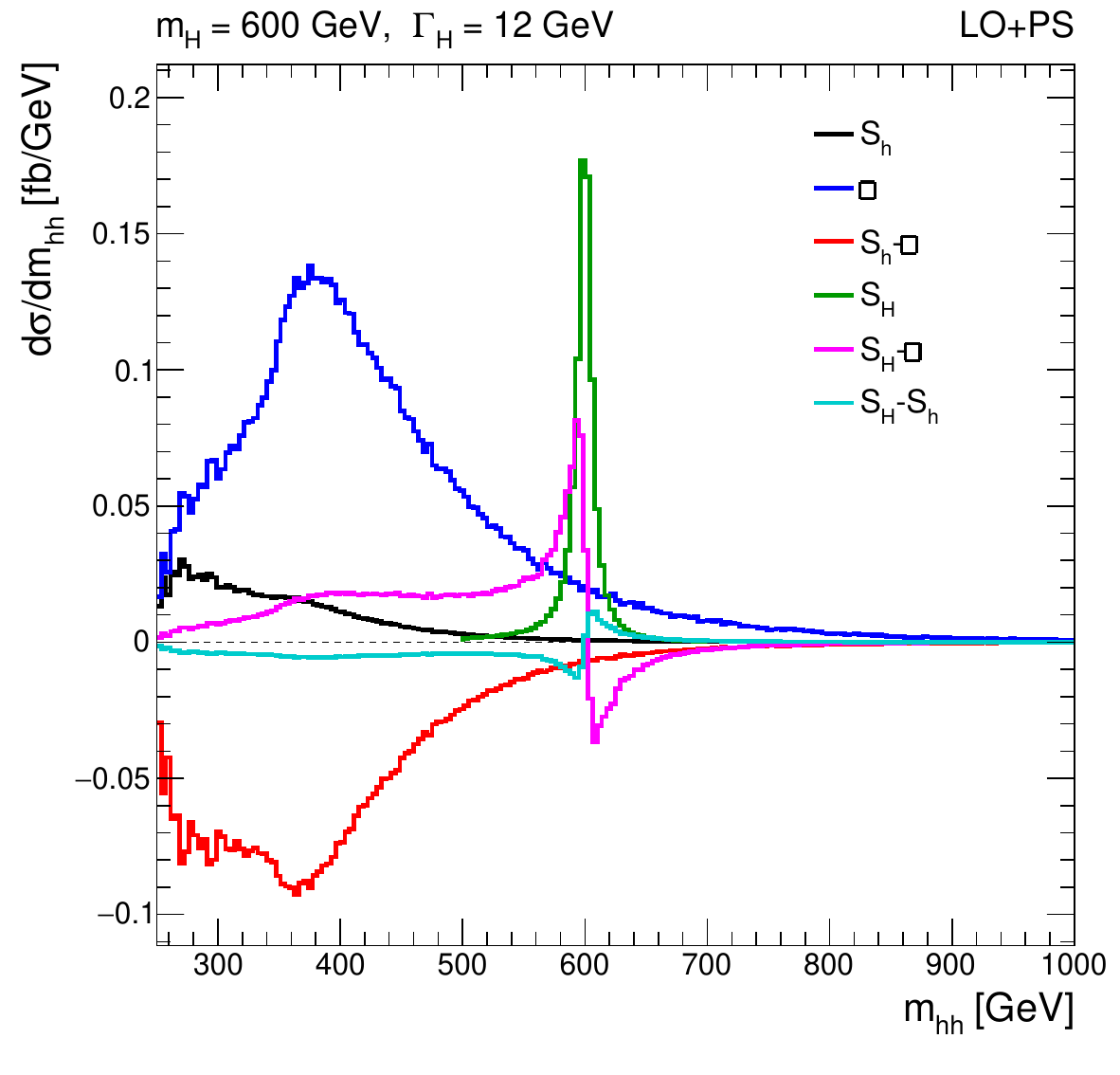}
  \caption{The di-Higgs mass distributions split into the individual contributions corresponding to the three diagrams in Figure~\ref{fig:feynman} and the interference terms. The non-resonant SM-like contributions from the $\Box$,  \Sh, and \ShBox terms are shown in blue, black, and red, respectively. The \SH resonant contributions is displayed in green. The interference between the \SH and the $\Box$ (\Sh) diagrams is shown in magenta (cyan). All distributions are shown for $\kappat=\kappatH=\kappalam=\kappalamH=1$.
}
\label{fig:contributions}
\end{figure}

The $\mathcal{M}_{i}^2$ values can be used to compute weights for the individual components allowing the di-Higgs distributions to be decomposed into the individual terms. An example of this is shown in Figure~\ref{fig:contributions}, for $\mH=600\,\GeV$ and $\wH=12\,\GeV$. All distributions are displayed for $\kappat=\kappatH=\kappalam=\kappalamH=1$.

We briefly summarise the procedure we used to obtain this figure:
\begin{itemize}
\item We generate an event sample $S_\text{SM}$ for the pure SM process $pp\,\rightarrow\,h\,h$, that includes $\Box$, \Sh, and \ShBox contributions. From this sample, all but the $\SH$ distributions are obtained via Eq.~\eqref{eq:reweight}, with $|\mathcal{M}_\text{target}|^2$ given by the respective term of Eq.~\eqref{eqn:decomposition}, and where $|\mathcal{M}_\text{ref}|^2$ corresponds to the squared matrix element for the SM process. 
\item A second sample $S_2$ is generated that corresponds to the resonant $s$-channel contribution. This is used for the distribution denoted by \SH\footnote{In case the sample was generated with a different total width, we also apply a reweighting here, following again the prescription of Eq.~\eqref{eq:reweight}.}.
\end{itemize}
We will comment on the validation of this procedure below. 

\begin{figure}[htbp]
  \centering\includegraphics[width=0.48\textwidth]{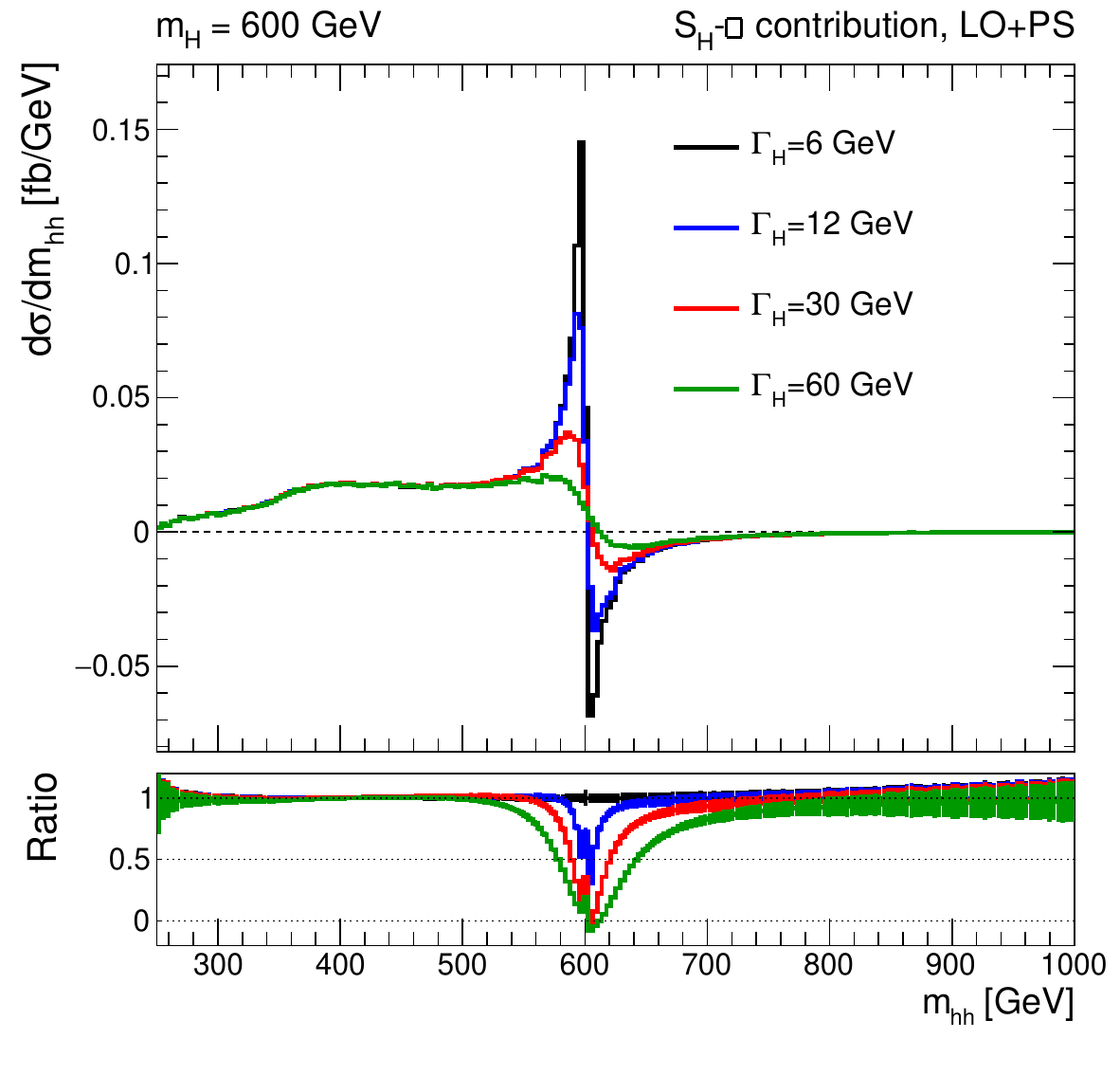} 
\caption{The di-Higgs mass distributions for the \SHBox interference term for different values of \wH. All distributions are shown for $\kappat=\kappatH=\kappalamH=1$. The lower panel shows the ratio of the distributions with respect to $\wH=6\,\GeV$. 
}
\label{fig:width_comp}
\end{figure}

In Figure~\ref{fig:width_comp}, we also illustrate the effect of \wH on one of the interference terms, where again contributions have been obtained as described above. All distributions are shown for $\kappat=\kappatH=\kappalamH=1$. 
\vfill

After the decomposition into the individual terms in Eq.~\eqref{eqn:decomposition}, it is possible for the experimental searches to be conducted in such a way as to allow for reinterpretations in different models, as follows:
\begin{itemize}
    \item Three distributions are produced for the $\Box$, \Sh, and \ShBox contributions, which do not depend on \mH or \wH.
    \item Distributions for the \SH, \SHBox, and \SHSh terms can be produced for several values of \mH and \wH. The number of \wH and \mH values to be considered depends on the experimental resolution. For example, in Section~\ref{sec:simulation} we estimated  the \mhh resolution in the 4$b$ final state to be $\sim6\%$; in this case we consider $\sim 4$ values of \wH for each mass point as sufficient to take into account widths effects for $\frac{\wH}{\mH}<20\%$. The usual interpolation methods can be use to obtain distributions for intermediate points.
    \item The physics model in question predicts the values \kappat, \kappatH, \kappalam, \kappalamH,  and \wH depending on the model parameters. 
    For example, in the singlet model these are computed as a function of the \tanb, \sina, and \mH parameters as
\begin{equation}
\begin{split}
\kappat &= \cosa,  \\
\kappatH &= \sina, \\
\kappalam &= \cos^3{\alpha}-\tanb\sin^3{\alpha}, \\
\kappalamH &= \frac{2\mh^2+m_{\PS}^2}{\mh^2}\,\frac{\sin\lb 2 \alpha\rb}{2}(\cos{\alpha} + \tanb\sin{\alpha}),
\end{split}
\end{equation}

and the width is given in Eq.~\eqref{eqn:Hwidth}.
Using these predictions, the total signal distribution is then constructed by combining the individual distributions according to Eq.~\eqref{eqn:decomposition}.
\end{itemize}  

Our tool gives the user the option to reweight either the SM non-resonant di-Higgs process or a resonant \SH sample. In principle, both options should give the same distributions within statistical uncertainties. However, in practise one needs to be careful to make sure the MC samples being reweighted are sufficiently populated in  phase-space regions where $|\mathcal{M}_{\mathrm{target}}|^2$ is non-vanishing~\cite{Mattelaer:2016gcx}, to ensure statistical uncertainties do not become too large. With this in mind, we take the following approach to obtain distributions for each of the terms in Eq.~\eqref{eqn:decomposition}. We obtain the $\Box$, \Sh, \ShBox, \SHBox, and \SHSh terms by reweighting the SM non-resonant MC samples. To obtain the \SH distribution for a $\mH^{\mathrm{target}}$-$\wH^{\mathrm{target}}$ point, we reweight an \SH MC sample generated with the same \mH, and we ensure that the \wH is not too far from $\wH^{\mathrm{target}}$\footnote{If the widths of the reference and target are too dissimilar then it would result in regions of the parameter space receiving very large weights which would impact the statistical precision of the reweighted sample \eg in the tails of the  distributions.}.

\begin{figure}[htbp]
  \includegraphics[width=0.48\textwidth]{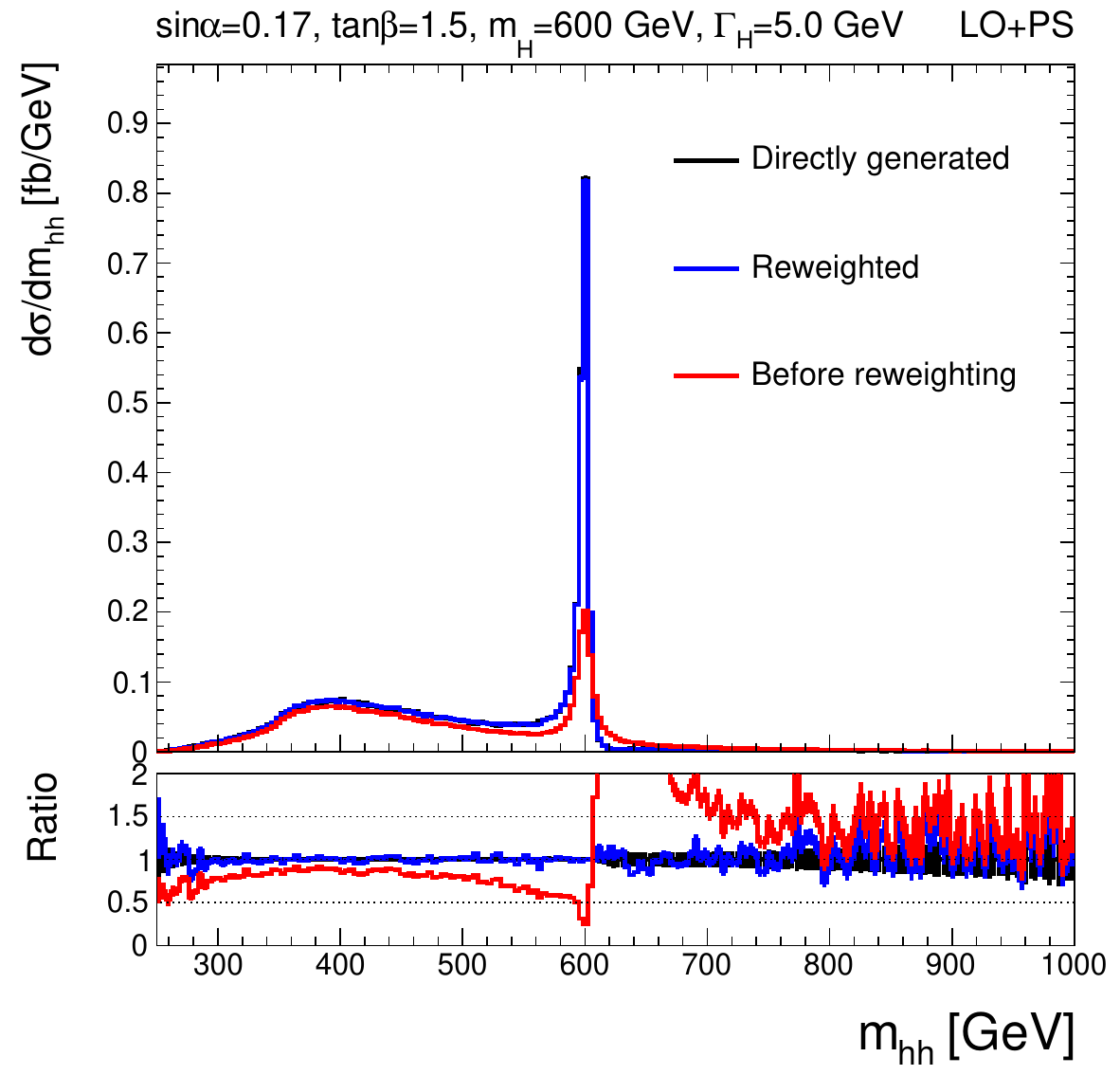}
  \includegraphics[width=0.48\textwidth]{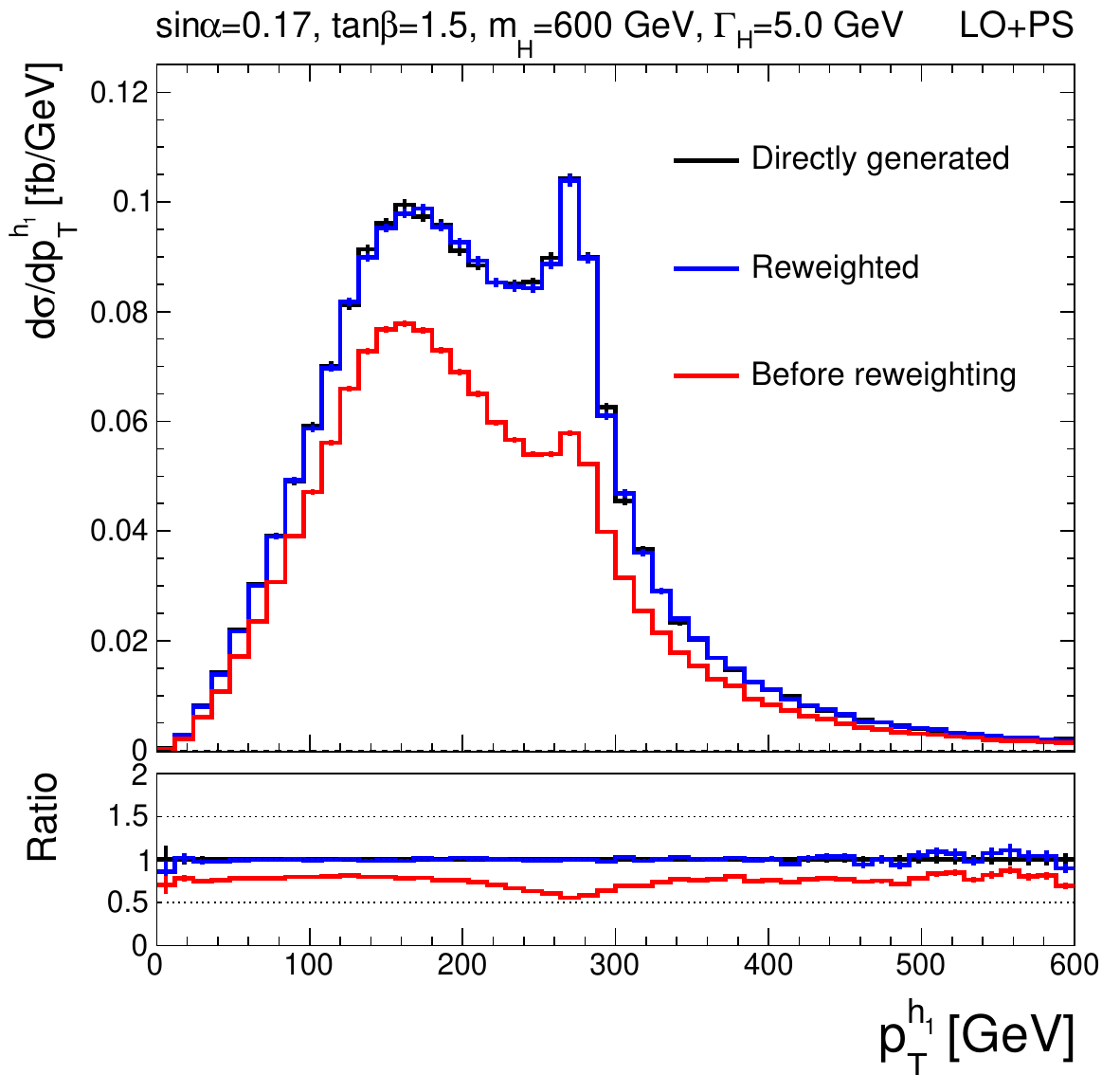}
\caption{The reweighting method is validated by comparing the reweighted di-Higgs mass (left) and leading \Ph \pT (right) distributions (blue) to the distributions obtained by directly simulating the di-Higgs events for a set of parameter points (black). The red distribution corresponds to the MC events used for the reweighting, before the weights have been applied. On the left, the blue line overlays exactly the black points including error bars.}
\label{fig:reweight_validations}
\end{figure}

Figure~\ref{fig:reweight_validations} shows examples of the reweighted  distributions (blue) compared to the distributions obtained from a MC sample generated directly for the target model parameters (black). The distributions are shown for \mhh and \pT of the leading \Ph. However, several other kinematic and angular variables were analysed and found to have the same level of agreement.

In this case the target parameter points are $\sina=0.17$, $\tanb=1.5$, and $\mH=600\,\GeV$, which yields a predicted width of $5\,\GeV$. We obtain the \SH component by reweighting an \SH MC sample with $\wH=12\,\GeV$ and $\kappalamH=\kappatH=1$, while a sample of SM di-Higgs events is used to obtain the other terms, as described above.
The red line in Figure~\ref{fig:reweight_validations} shows the \mhh spectrum for the MC events used in the reweighting procedure before the reweighting is applied for comparison.
The distributions obtained for the reweighted events match the directly generated MC sample within the statistical uncertainties. The method is additionally validated for several other benchmark points as shown in Appendix~\ref{sec:AppendixB}.

\subsection{Including higher-order corrections}

The discussions so far have focused on MC events generated at LO. However, it is well known that the LO cross-section for gluon-fusion is significantly different compared to higher-order predictions (see \eg Ref.~\cite{LHCHiggsCrossSectionWorkingGroup:2016ypw} and references therein). It is therefore necessary to correct the LO cross-section through the application of K-factors to account for this. 
The NNLO cross-sections for the $\Box$, \Sh, and \ShBox components~\cite{deFlorian:2017qfk,Buchalla:2018yce,Grazzini:2018bsd,Amoroso:2020lgh} are used to define K-factors for these terms, which we call $K_{\Box}$, $K_{\Sh}$, and $K_{\ShBox}$, respectively.
For the \SH, \SHBox, and \SHSh components, no dedicated NNLO predictions are provided. However, NNLO predictions are available for the production of a single undecayed SM-like \PS ($gg\rightarrow\PS$)~\cite{LHCHiggsCrossSectionWorkingGroup:2016ypw}, and we use these predictions to define the K-factor for the \SH term:
\begin{equation}
K_{\SH} = \frac{\sigma_{\mathrm{NNLO}}(gg\rightarrow\PS)}{\sigma_{\mathrm{LO}}(gg\rightarrow\PS)},
\end{equation}
where $\sigma_{\mathrm{LO}}$ is also computed for single-Higgs production for consistency using \textsc{MadGraph}{}5. 
To define K-factors for the \SHBox and \SHSh terms we propose the ansatz:
\begin{equation}
    \begin{split}
        K_{\SHBox} &= \sqrt{K_{\SH}K_{\Box}}~\mathrm{and} \\
        K_{\SHSh} &= \sqrt{K_{\SH}K_{\Sh}}.
    \end{split}
    \label{eqn:ansatz}
\end{equation}
Table~\ref{tab:kfactors} lists the LO and NNLO cross-sections and the computed K-factors. All cross-sections are computed for $\kappat=\kappatH=\kappalam=\kappalamH=1$. The cross-sections displayed for the \SH, \SHBox, and \SHSh terms are computed for $\mH=600\,\GeV$ and $\wH=5\,\GeV$. 

\begin{figure*}[htbp]
  \includegraphics[width=0.48\textwidth]{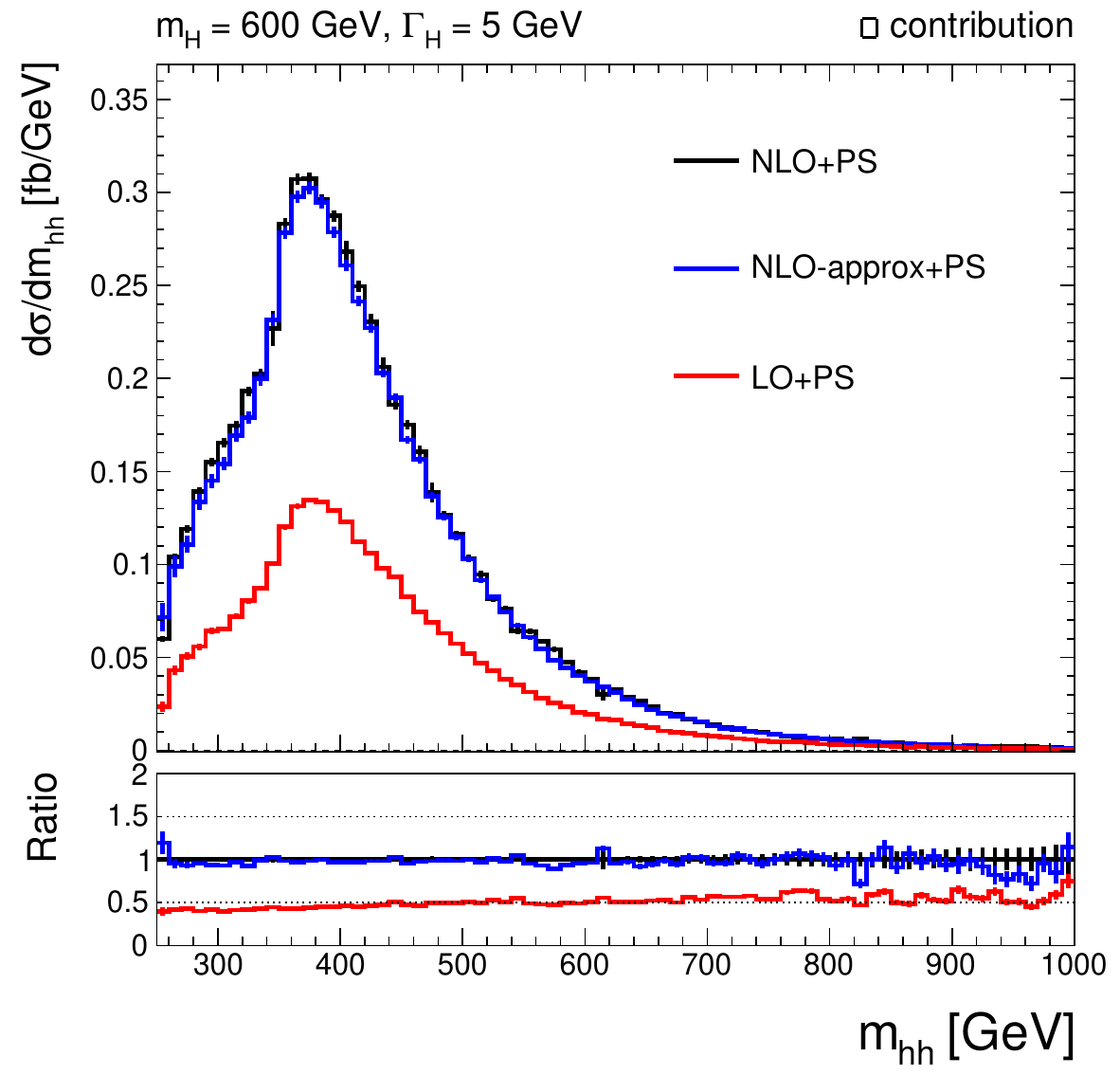}
  \includegraphics[width=0.48\textwidth]{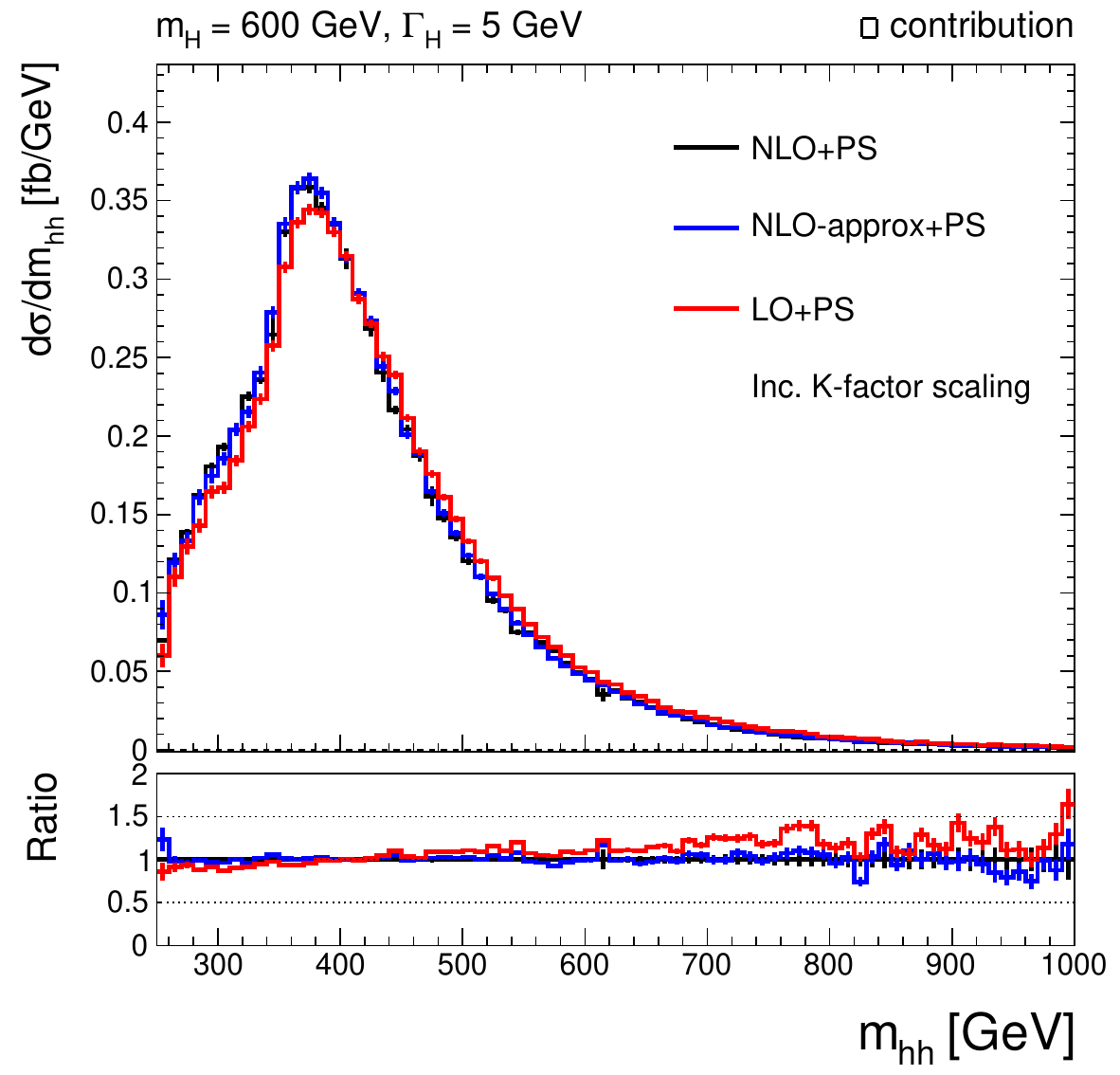}
\caption{The di-Higgs mass distributions obtained for the $\Box$ component. The NLO prediction (black) is compared to the approximate NLO distribution obtained by reweighting the NLO SM di-Higgs MC (blue) and the LO distribution (red). The plot on the left (right) shows the distributions before (after) the NNLO K-factor scaling is applied.   
}
\label{fig:nlo_vs_lo_box}
\end{figure*}

\begin{figure*}[htbp]
  \includegraphics[width=0.48\textwidth]{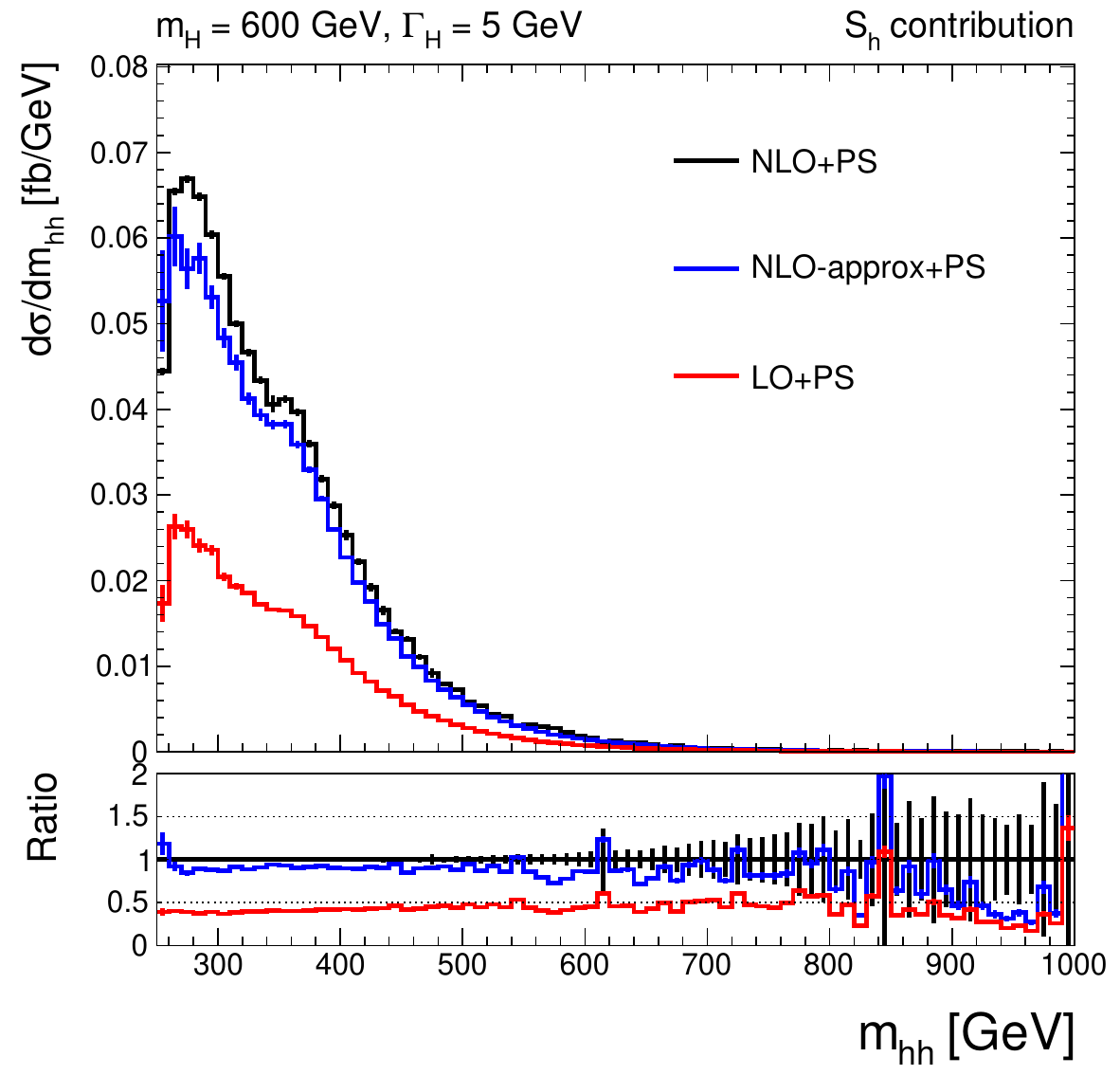}
  \includegraphics[width=0.48\textwidth]{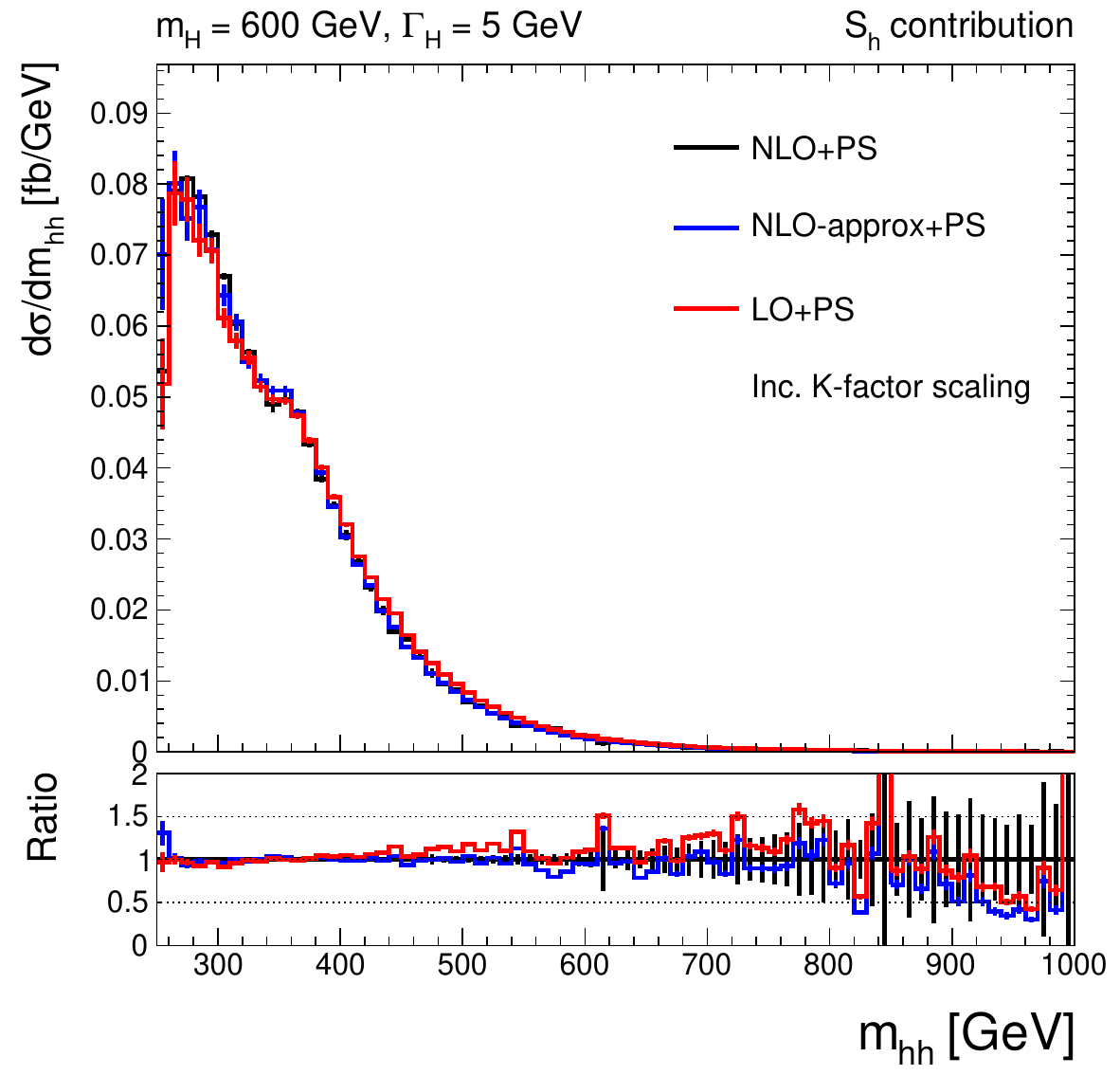}
\caption{The di-Higgs mass distributions obtained for the \Sh component. The NLO prediction (black) is compared to the approximate NLO distribution obtained by reweighting the NLO SM di-Higgs MC (blue) and the LO distribution (red). The plot on the left (right) shows the distributions before (after) the NNLO K-factor scaling is applied.  }
\label{fig:nlo_vs_lo_Sh}
\end{figure*}

\begin{figure*}[htbp]
  \includegraphics[width=0.48\textwidth]{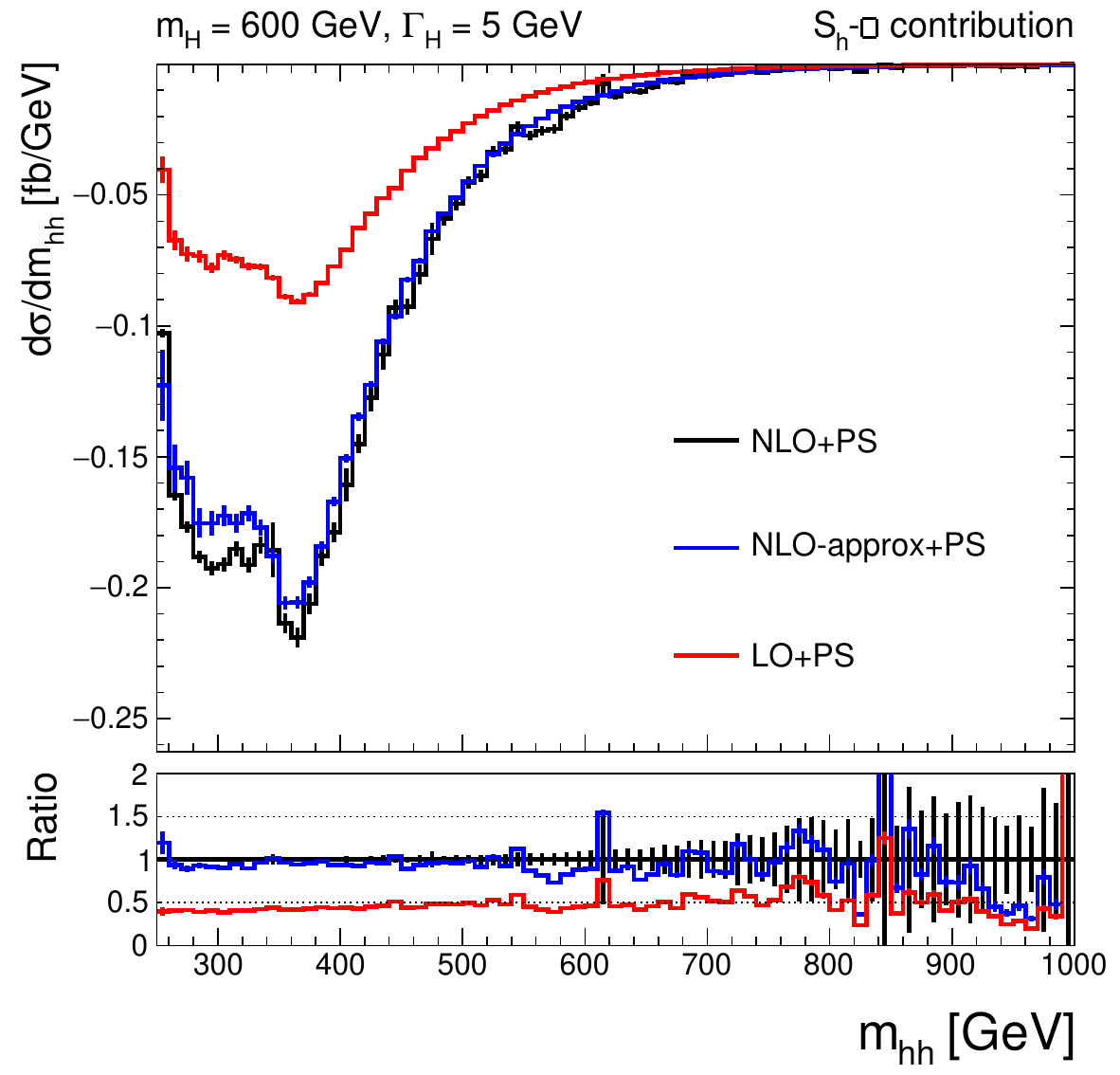}
\includegraphics[width=0.48\textwidth]{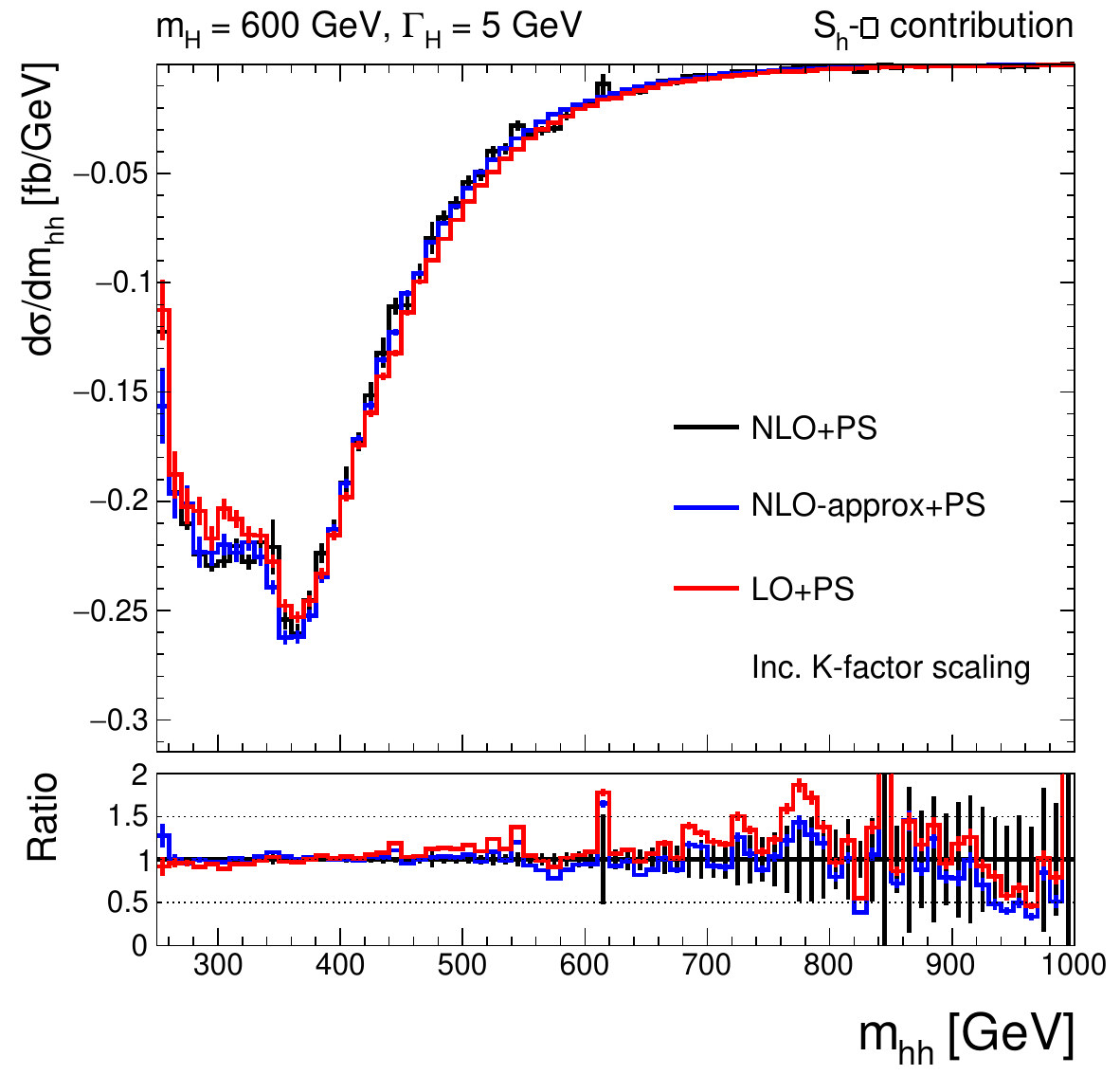}
\caption{The di-Higgs mass distributions obtained for the \ShBox component. The NLO prediction (black) is compared to the approximate NLO distribution obtained by reweighting the NLO SM di-Higgs MC (blue) and the LO distribution (red). The plot on the left (right) shows the distributions before (after) the NNLO K-factor scaling is applied. }
\label{fig:nlo_vs_lo_box_Sh}
\end{figure*}

\begin{figure*}[htbp]
  \includegraphics[width=0.48\textwidth]{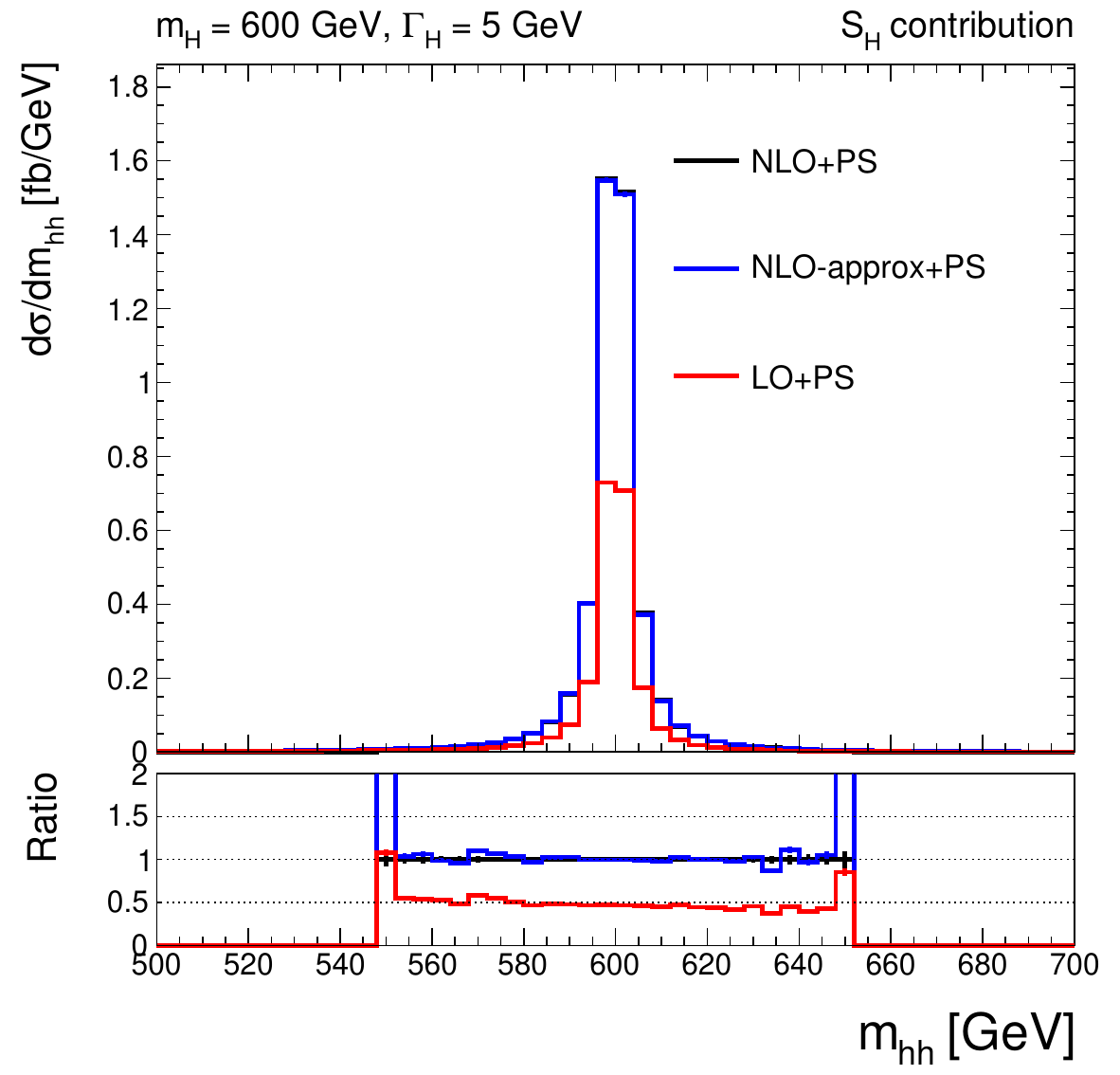}
  \includegraphics[width=0.48\textwidth]{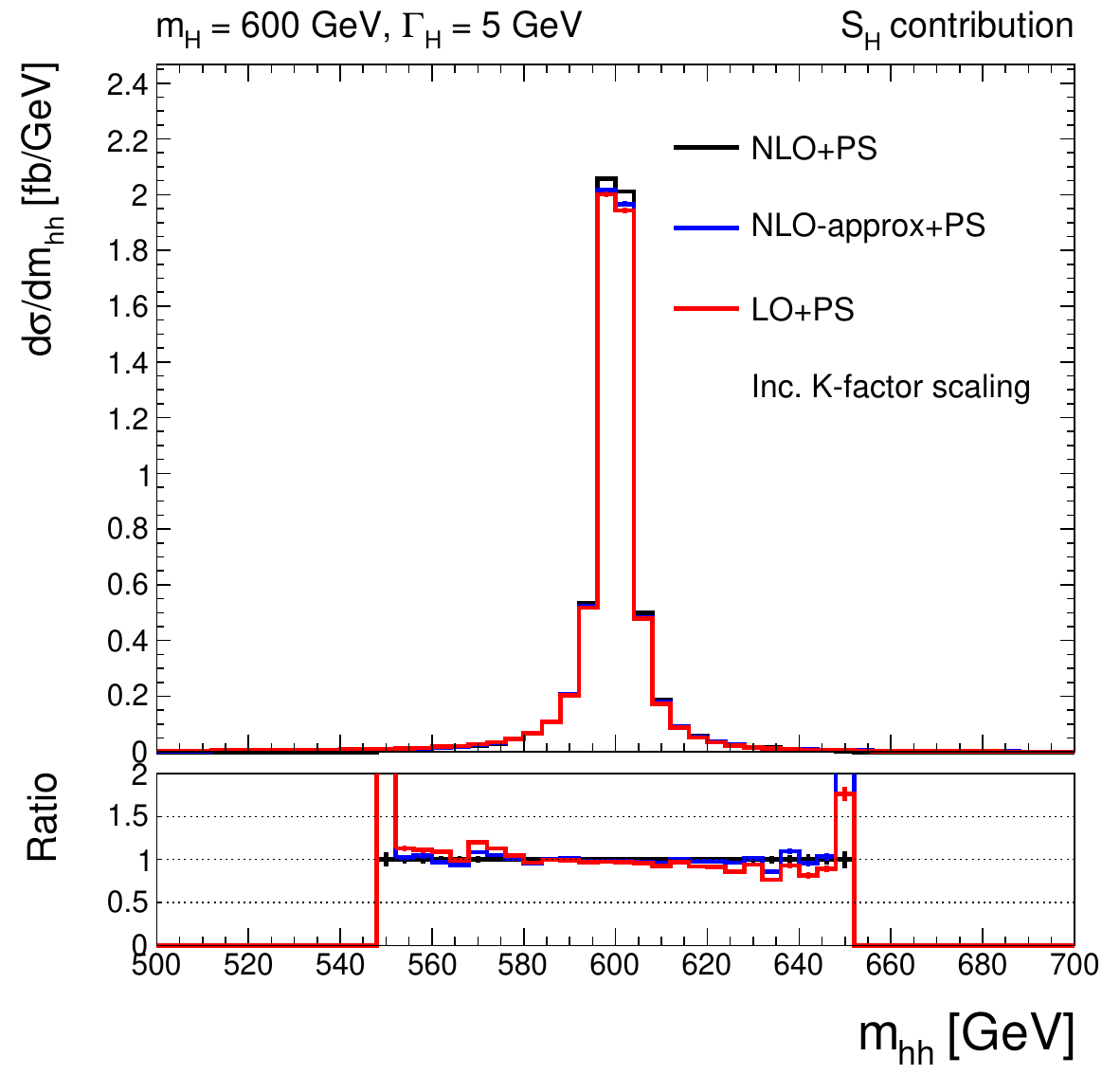}
\caption{The di-Higgs mass distributions obtained for the \SH component. The NLO prediction (black) is compared to the approximate NLO distribution obtained by reweighting the NLO \SH MC sample (blue) and the LO distribution (red). The plot on the left (right) shows the distributions before (after) the NNLO K-factor scaling is applied. }
\label{fig:nlo_vs_lo_SP}
\end{figure*}

\begin{figure*}[htbp]
  \includegraphics[width=0.48\textwidth]{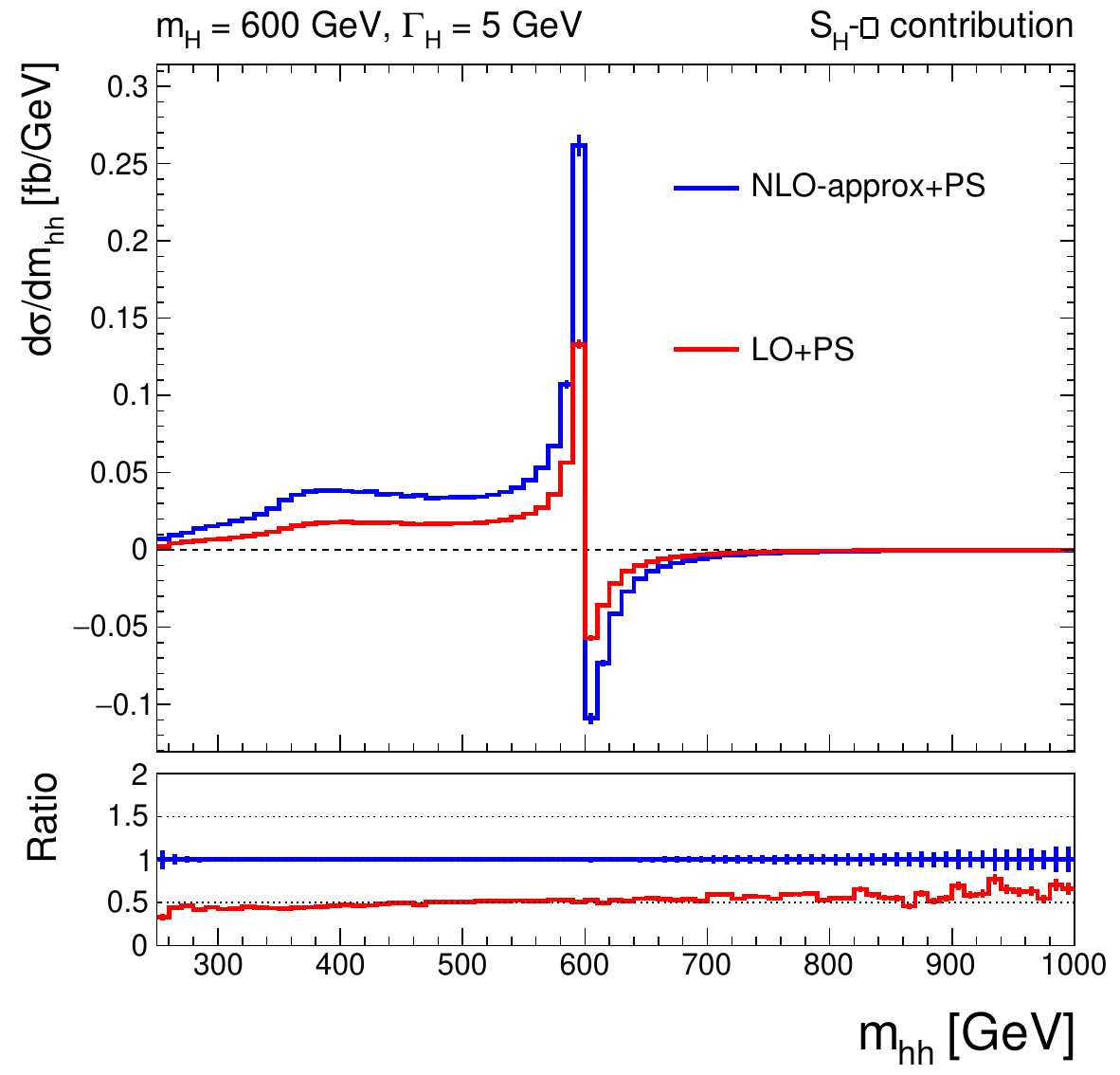}
  \includegraphics[width=0.48\textwidth]{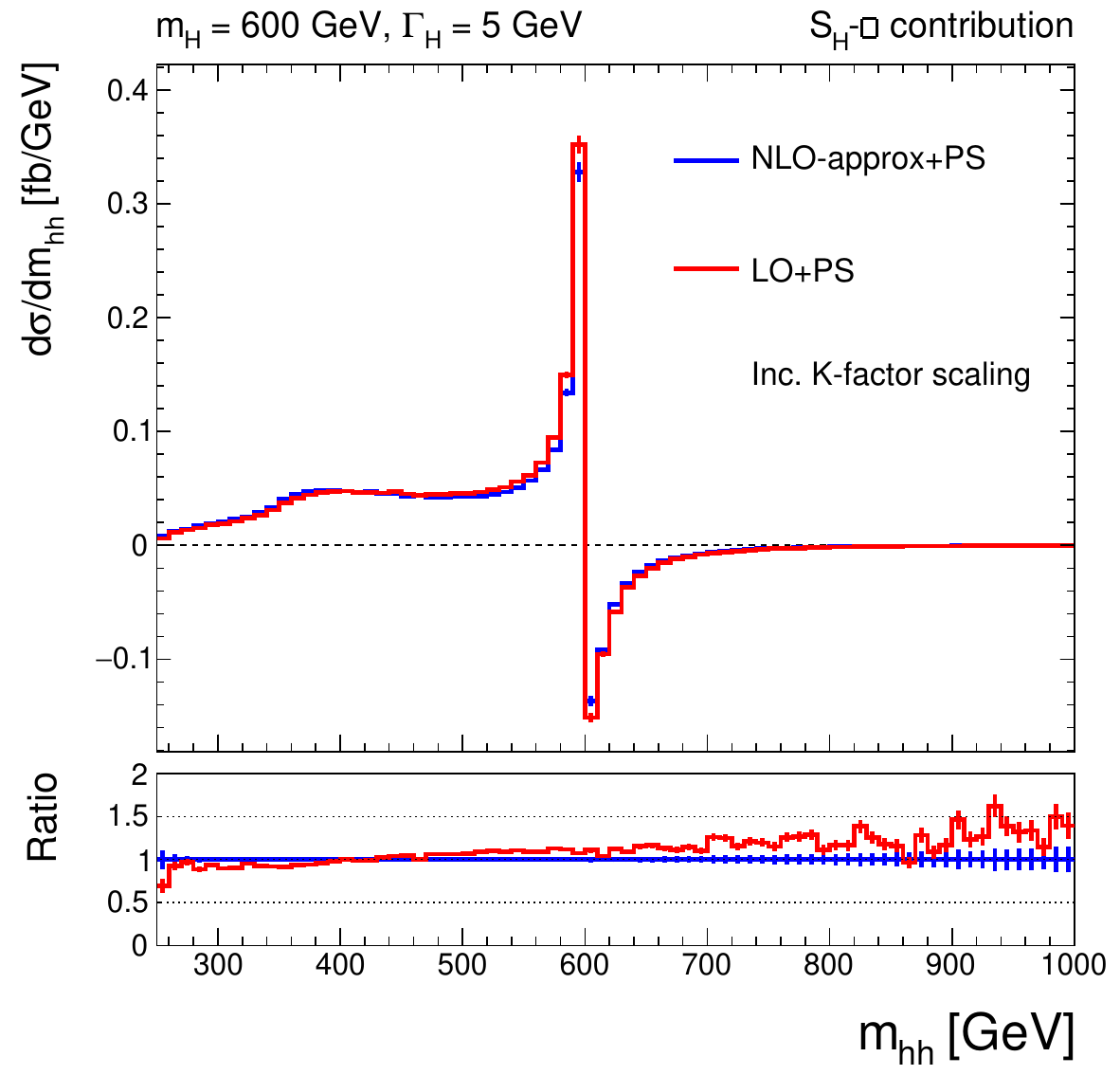}
\caption{The di-Higgs mass distributions obtained for the \SHBox component. The NLO prediction (black) is compared to the approximate NLO distribution obtained by reweighting the NLO SM di-Higgs MC (blue) and the LO distribution (red). The plot on the left (right) shows the distributions before (after) the NNLO K-factor scaling is applied. }
\label{fig:nlo_vs_lo_box_SP}
\end{figure*}

\begin{figure*}[htbp]
  \includegraphics[width=0.48\textwidth]{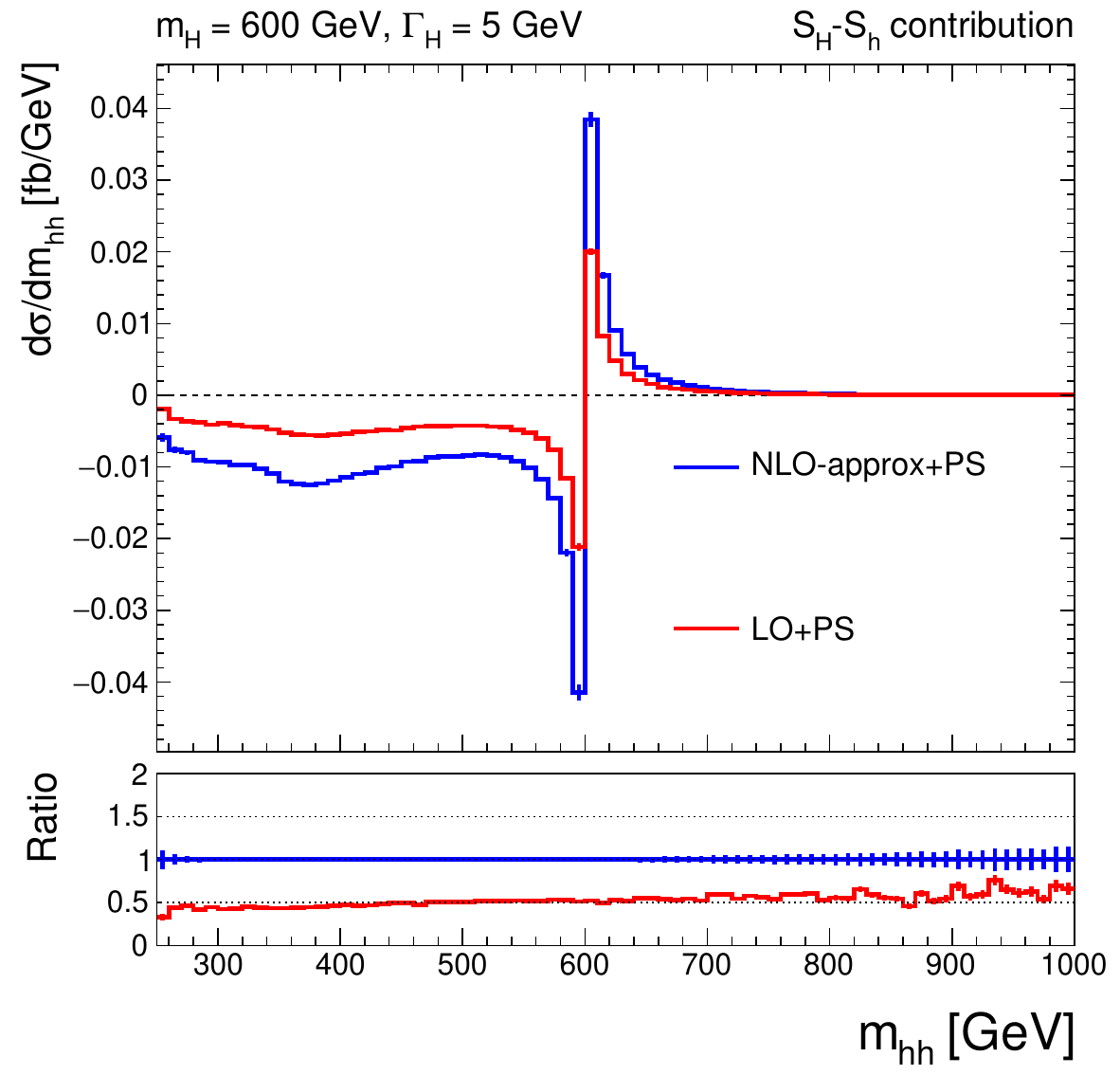}
  \includegraphics[width=0.48\textwidth]{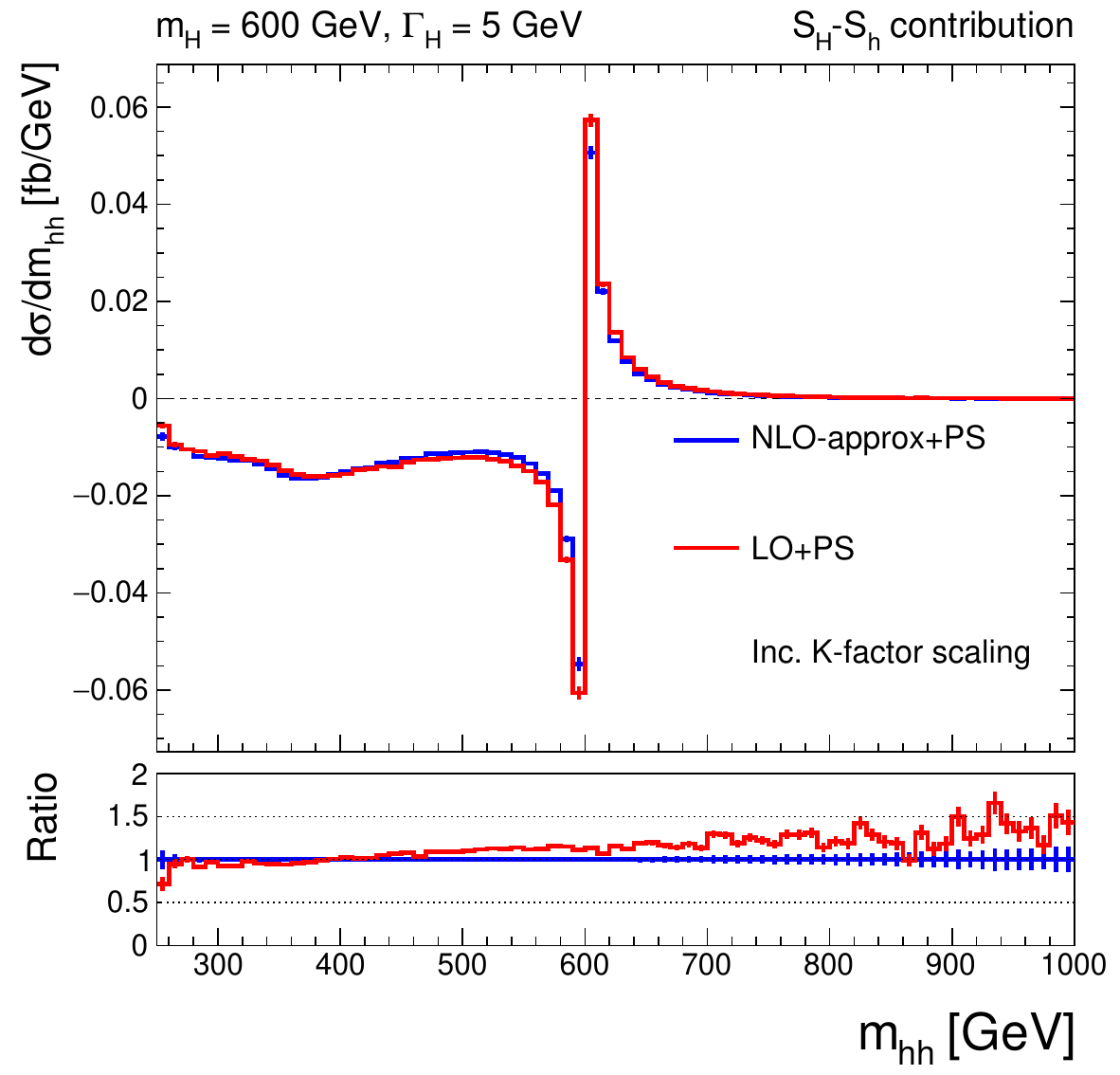}
\caption{The di-Higgs mass distributions obtained for the \SHSh component. The NLO prediction (black) is compared to the approximate NLO distribution obtained by reweighting the NLO SM di-Higgs MC (blue) and the LO distribution (red). The plot on the left (right) shows the distributions before (after) the NNLO K-factor scaling is applied. }
\label{fig:nlo_vs_lo_SP_Sh}
\end{figure*}

We produce di-Higgs mass distributions for each of the terms in Eq.~\eqref{eqn:decomposition} generated at LO before and after the application of the K-factors. 
These distributions are shown by the red lines in Figures~\ref{fig:nlo_vs_lo_box}--\ref{fig:nlo_vs_lo_SP_Sh}. The left and right plots show the distributions before and after the K-factor applications, respectively. We also compare the LO distributions to NLO predictions for cases where MC production is possible in \POWHEG, namely the $\Box$, \Sh, \ShBox, and \SH terms. These predictions are shown by the black lines in the figures.

While the pure resonant contributions as well as all non-resonant contributions are readily available at NLO, there is currently no MC on the market that simulates the full contribution including all interference terms at that order. This motivates us to define an approach that tries to remedy this by mimicking the expected contributions, to be confirmed once the according tools become available in the future. Furthermore, some transverse variables, such as $p_T^{hh}$, typically vary widely between LO and NLO and in principle require the description via NLO kinematics, including real radiation.
We therefore define an approximate scheme for NLO results (labelled ``NLO-approx''), which we obtain by reweighting NLO MC events using LO ME weights. 

In the \POWHEG description \cite{Frixione:2007nu}, the largest part of the events is simulated with NLO kinematics, giving rise to an additional hard real emission parton\footnote{Born type kinematics are Sudhakov suppressed. We thank P. Nason and E. Re for useful comments concerning this topic.}. In order to map this to the LO matrix elements used for reweighting, we apply the following procedure:
We ``ignore'' the additional parton when we perform the reweighting.
Technically, this is obtained in the following way. We take the two outgoing \Ph four-momenta from the ME and boost to the di-Higgs rest frame. We then obtain the four-momenta for the incoming gluons by requiring both gluons to have zero transverse momentum, and equal and opposite longitudinal momentum, also requiring four-momentum conservation between incoming and outgoing particles. When estimating the MEs we also average over all possible spin-states of the incoming gluons.
An NLO SM di-Higgs sample is reweighted to obtain the NLO-approx $\Box$, \Sh, and \ShBox, \SHBox, and \SHSh terms, while an NLO \SH sample with $\wH=12\,\GeV$ is reweighted to obtain the NLO-approx \SH sample for a different width ($\wH=5\,\GeV$ in this case). The NLO-approx results are shown by the blue lines in the figures. The total rate in the NLO-approx method is obtained using the newly determined weights, derived from the LO matrix elements, for each separate contribution and phase space point\footnote{In practise this leads to weighted events. We have ensured that we sampled the corresponding contributions with sufficient statistical accuracy.}.
Furthermore, in order to include corrections to the overall rate to NNLO, we define an equivalent set of K-factors accounting for the scaling of the NLO and NLO-approx cross-sections to the NNLO predictions. The NLO(-approx) cross-sections and K-factors are also displayed in Table~\ref{tab:kfactors}. 

We remind the reader that our method for simulating \SH at NLO generates undecayed $gg\rightarrow\PS$ events using \POWHEG and then performs the $\PS\rightarrow\Ph\Ph$ decays using \PYTHIA (as described in Section~\ref{sec:simulation}). For this reason, the cross-section determined from \POWHEG does not include the $\PS\rightarrow\Ph\Ph$ branching ratio. To obtain the correct \SH cross-section we thus multiply the undecayed $gg\rightarrow\PS$ cross-section by the branching ratio which is determined using the above settings for the couplings, that leads to $\Gamma(H\,\rightarrow\,h\,h)\,=\,0.061\,\GeV$ using \MADGRAPH at LO, and then dividing by the respective width of 5 \GeV, leading to a BR of 0.012\footnote{For the $\sim$NLO \SH contribution, prior to reweighting we first use the branching ratio that would be obtained for a 12 \GeV width. The rate in Table~\ref{tab:kfactors} is then obtained using reweighting as described in the text.}.
\begin{table}[hb!]
\begin{center}
\resizebox{\textwidth}{!}{
\begin{tabular}{p{3cm}|w{c}{1.7cm}w{c}{1.7cm}w{c}{1.7cm}w{c}{1.7cm}w{c}{1.7cm}w{c}{1.7cm}w{c}{1.7cm}}
\hline
Term & $\sigma_{\mathrm{LO}}$ & $\sigma_{\mathrm{NLO}}$ & $\sigma_{\sim\mathrm{NLO}}$ & $\sigma_{\mathrm{NNLO}}$ & $K_{i}^{\mathrm{LO}}$ & $K_{i}^{\mathrm{NLO}}$ & $K_{i}^{\sim\mathrm{NLO}}$ \\

\hline
$\Box$ & 27.50~\fb & 60.37~\fb & 58.45~\fb & 70.39~\fb & 2.56 & 1.17 & 1.20 \\

\Sh & 3.70~\fb & 9.17~\fb & 8.31~\fb & 11.06~\fb & 2.99 & 1.21 & 1.33 \\ 

\ShBox & -18.08~\fb & -42.35~\fb & -39.57~\fb & -50.41~\fb & 2.79 & 1.19 & 1.27 \\

\SH (undecayed) & 0.73~\pb & 1.51~\pb & -- & 2.00~\pb & 2.74 & 1.33 & 1.30 \\
\SH & 8.91~\fb & 18.43~\fb & 18.78~\fb & -- & 2.74 & 1.33 & 1.30 \\

\SHBox & 5.10~\fb & -- & 10.68~\fb & -- & 2.65 & -- & 1.25 \\

\SHSh & -1.40~\fb & -- & -3.02~\fb & -- & 2.87 & -- & 1.32 \\
\hline
\end{tabular}}
\end{center}
\caption{\label{tab:kfactors} The K-factors used to correct the LO, NLO, and NLO-approx samples to the NNLO cross-sections. The cross-sections used in the calculations of the K-factors are also displayed. The NLO-approx values are indicated by the ``$\sim\mathrm{NLO}$'' sub/super-scripts. The K-factors are defined as $K^j \equiv \frac{\sigma_{\mathrm{NNLO}}}{\sigma_j}$, where $j$ denotes the perturbative order. The above table is computed for $m_H\,=\,600\,\GeV$, $\Gamma_H\,=\,5\,\GeV$, and $\kappat=\kappatH=\kappalam=\kappalamH=1$. Please see text for details on the branching ratios used in the decayed \SH cross section calculation.}
\end{table}

As expected, Figures~\ref{fig:nlo_vs_lo_box}--\ref{fig:nlo_vs_lo_box_Sh} show a significant difference between the LO and NLO yields due to the differences in the cross-sections. However, after application of the K-factors the yields agree (by construction) and we observe a good agreement in the shapes of the distributions compared to the NLO, indicating that LO samples are adequate for describing the \mhh spectrum. The shapes of the NLO-approx distributions also agree well with the NLO distributions, and in this case the yields/cross-section before application of the K-factors are also noticeably quite similar to the NLO results. 
We note that the agreement of the NLO-approx results with the NLO are slightly better for the $\Box$ contribution compared to the \Sh contribution. This is because we are reweighting a SM di-Higgs sample where the $\Box$ diagram contributes more strongly to the total ME than the \Sh and hence the weights used to obtain the $\Box$ contribution will tend to be closer to unity and thus less sensitive to the approximations of the method. 

Figures~\ref{fig:nlo_vs_lo_box_SP} and \ref{fig:nlo_vs_lo_SP_Sh} show the LO and NLO-approx distributions for the \SHBox and \SHSh terms, respectively. While no comparisons to the full NLO results are possible in this case, the agreement between the shapes and the yields (after K-factor scaling) of the LO and NLO-approx distributions gives further confidence in the description of the \mhh spectrum by the LO MC samples after reweighting.

\begin{figure*}[htbp]
  \includegraphics[width=0.48\textwidth]{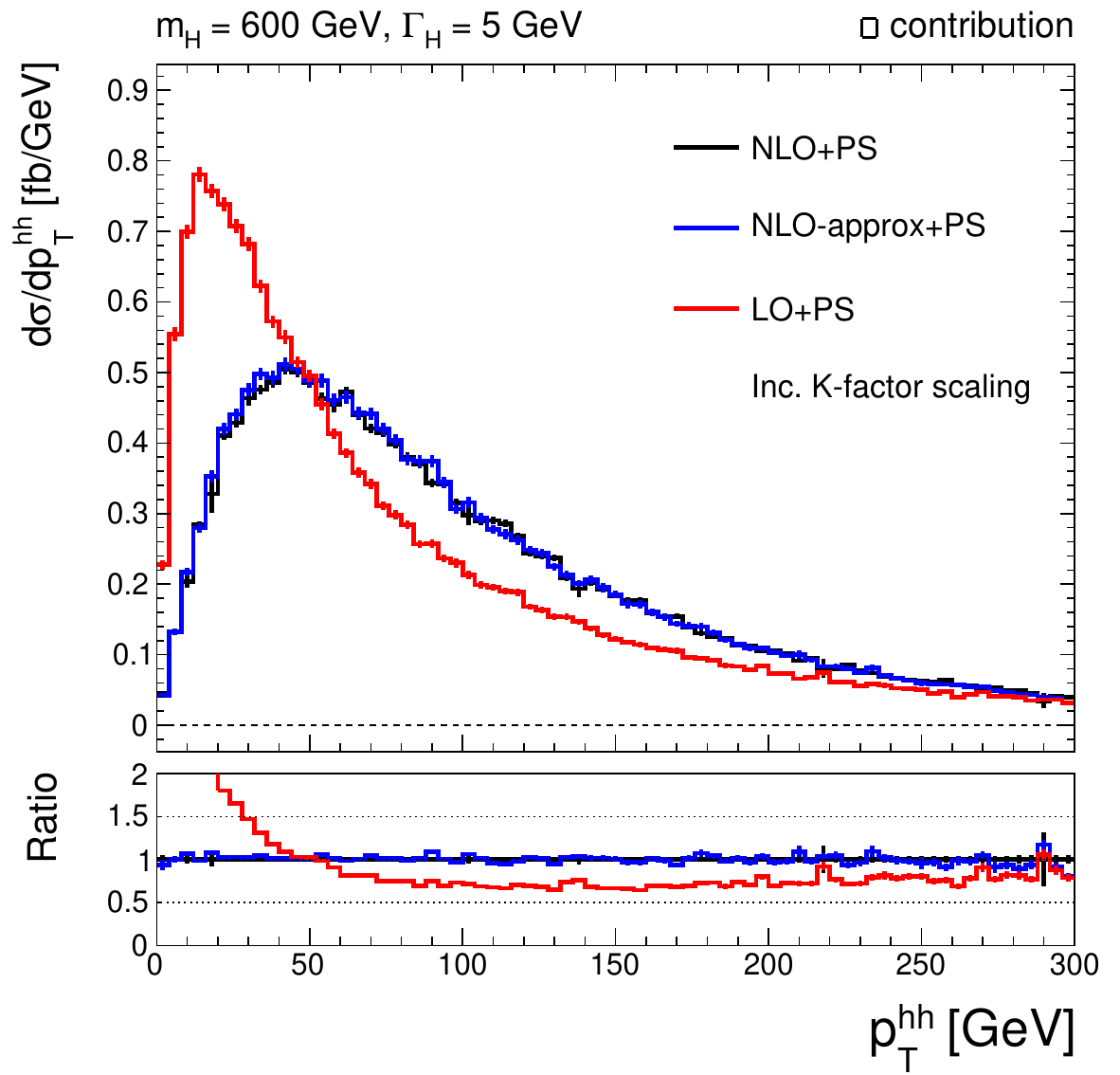}
  \includegraphics[width=0.48\textwidth]{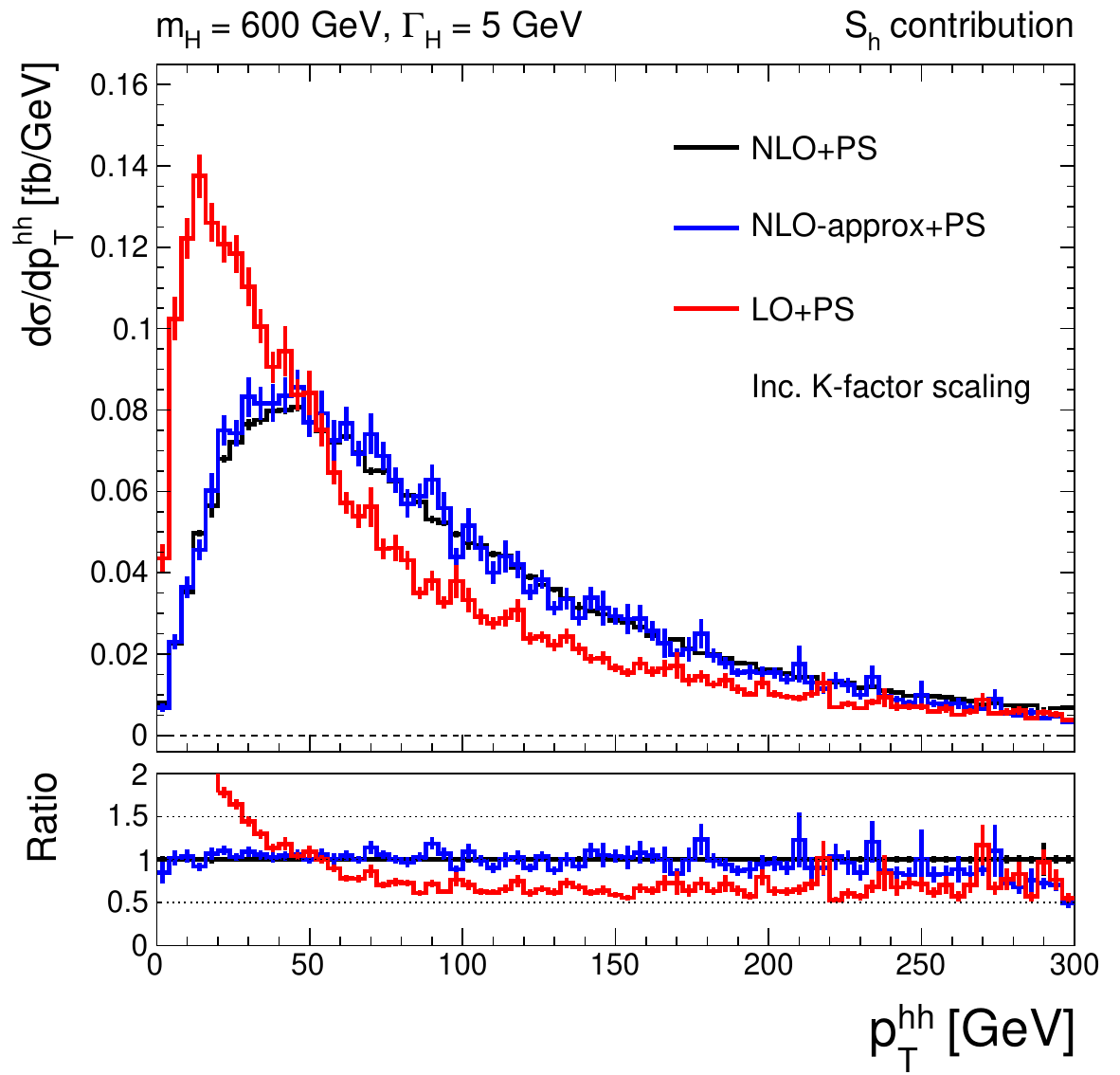} \\
  \includegraphics[width=0.48\textwidth]{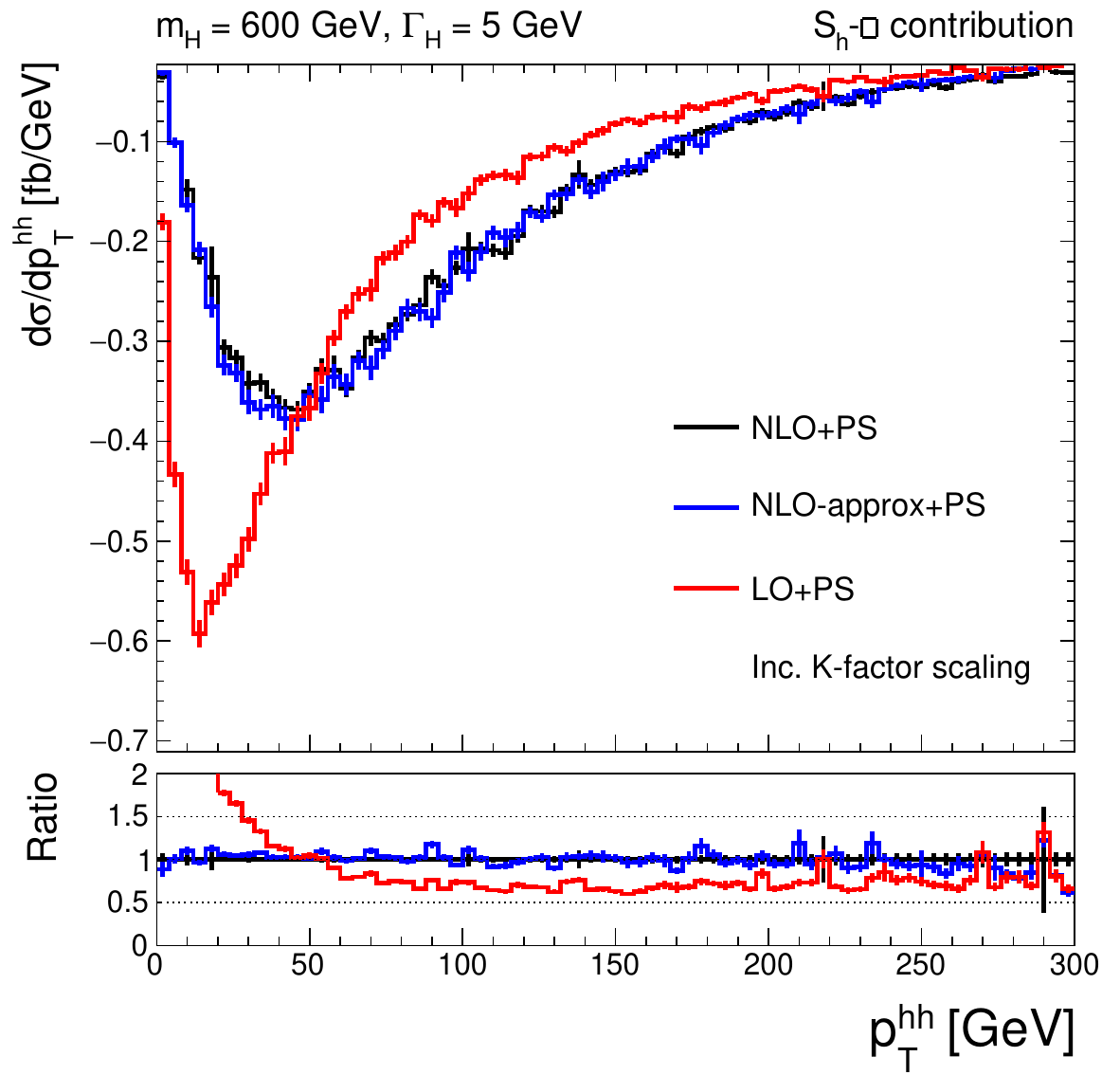}
  \includegraphics[width=0.48\textwidth]{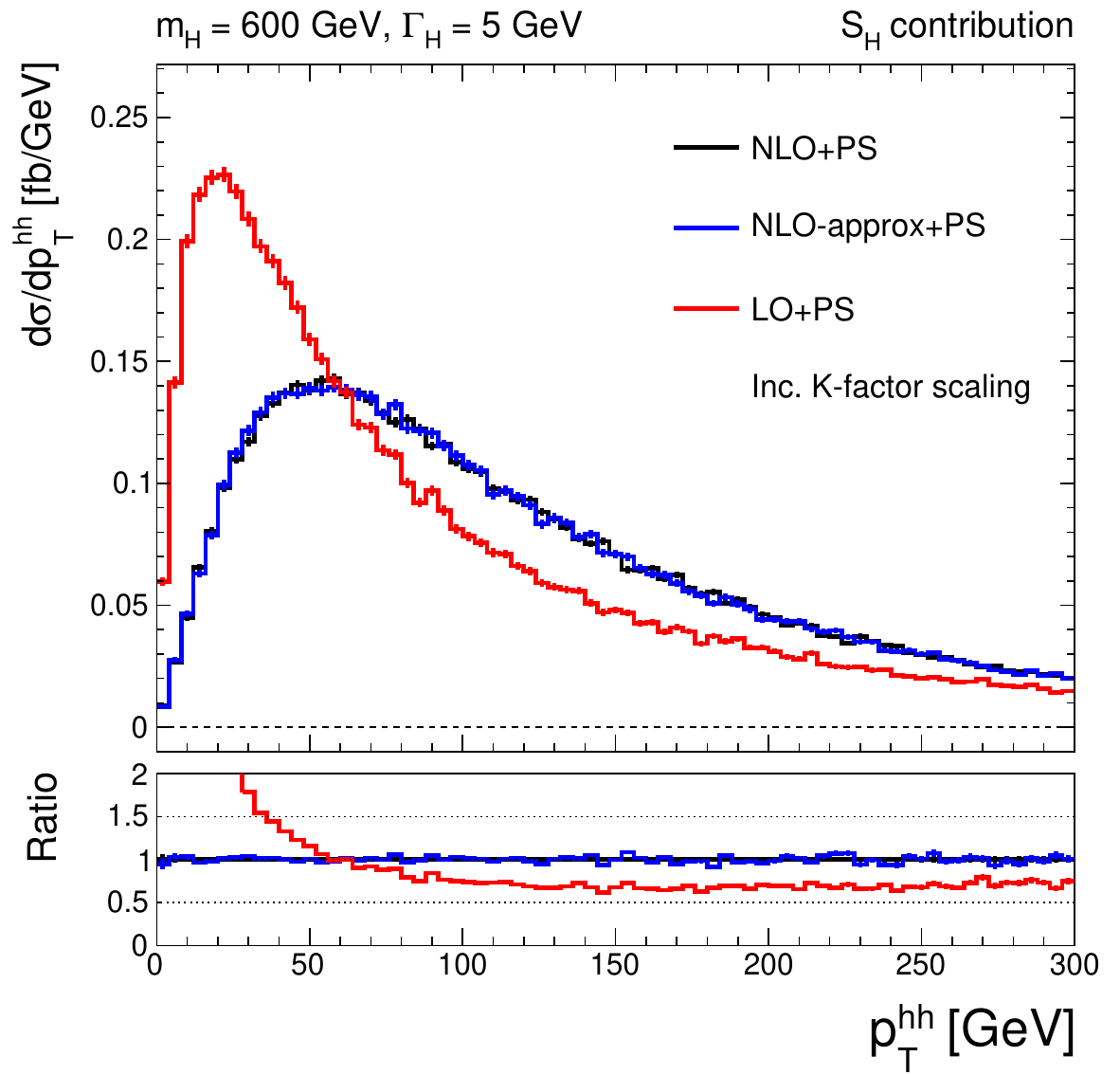}
\caption{The di-Higgs \pT distributions obtained for the $\Box$ component (upper left), \Sh component (upper right), \ShBox component (lower left), and \SH component (lower right). The NLO prediction (black) is compared to the approximate NLO distribution obtained by reweighting the NLO di-Higgs MC (blue) and the LO distribution (red). The NNLO K-factor scaling is applied in all cases. }
\label{fig:nlo_vs_lo_pT}
\end{figure*}

While the LO MC gives a good description of the \mhh spectrum, it is well know that the modelling of variables that are sensitive to the additional radiation are more problematic, as this is approximated by the parton shower. In Figure~\ref{fig:nlo_vs_lo_pT} we show an example of such a variable, the di-Higgs \pT, $\pT^{\Ph\Ph}$. The LO samples clearly tend to predict a softer \pT distribution compared to the NLO. In contrast, the agreement of the NLO-approx method is almost exact.

Finally, in support of the ansatz we made in Eq.~\eqref{eqn:ansatz}, we make the following two observations:
\begin{enumerate}
    \item As shown in Figures~\ref{fig:nlo_vs_lo_box_SP} and ~\ref{fig:nlo_vs_lo_SP_Sh}, the predicted \SHBox and \SHSh yields (and equivalently the cross-sections) for the LO and NLO-approx methods after the K-factor scaling are in good agreement. 
    The cross-sections predicted for the \SHBox term for LO and NLO-approx are 13.52~\fb and 13.35~\fb, respectively. While the cross-sections for the \SHSh term are -4.02~\fb and -4.00~\fb. 
    This demonstrates that the method is, at least, robust regardless of whether it is applied to LO or NLO samples.
    \item We can apply the same ansatz to estimate the K-factors for the \ShBox term and compare to the K-factors computed directly from the NNLO cross-sections. The K-factors predicted via the ansatz are 2.77, 1.19, and 1.26 for the LO, NLO, and NLO-approx methods, respectively, which agree with the values presented in Table~\ref{tab:kfactors} within 1\%.
\end{enumerate}
\section{Interference effects on differential distributions}
\label{sec:benchmarks}

We use the scans shown in Section~\ref{sec:parscan} to define several interesting BM points.
We select the BM points such that they exhibit interesting features that we will describe below. Additionally, we ensure that all points are accessible to experimental searches at the LHC, by requiring that the $gg\rightarrow\PS\rightarrow\Ph\Ph$ (\SH-only) cross-section \sigSH, scaled by the K-factors as described in Section~\ref{sec:method}, is close to the experimental limits presented by the ATLAS Collaboration in Ref.~\cite{ATLAS:2021tyg}.
To achieve this we require $\sigSH>\sigma_{\mathrm{ATLAS}}\times \sqrt{\frac{140}{L[\invfb]}}$, where $\sigma_{\mathrm{ATLAS}}$ is the expected limits from Ref.~\cite{ATLAS:2021tyg}, and $L$ is a target luminosity\footnote{We chose to use the ATLAS limits as no Run-2 CMS combination was published at the time. Since then, the CMS Collaboration has presented a combined result in Ref.~\cite{CMS:2024phk}. We have checked whether taking the best expected limit from either CMS or ATLAS has any influence on the defined BM points, and confirmed that the same BMs are selected in this case.}.
We consider two values of $L$ equal to 400~\invfb and 3000~\invfb, which approximately correspond to the expected luminosity recorded by each of the ATLAS/CMS detectors at the end of Run-3 (assuming Run-2+Run-3 are combined) and at the end of the LHC program, respectively. 

Table~\ref{tab:benchmarks} summarises the BM points, which will be described in more detail in the following sections. All points in the table satisfy the \sigSH requirement for $L=3000~\invfb$, and we additionally indicate in the second to last column if the requirement is also satisfied for $L=400~\invfb$.
In the following sections, we will state the parameters of the singlet model for the chosen benchmark scenarios and show distributions produced from MC events generated at LO in QCD. The K-factor scaling described in Section~\ref{sec:method} is also applied.  
We remind the reader than the shapes of the \mhh distributions generated at LO were shown to agree very well with the NLO distributions, and thus we deem the LO samples to be sufficient for these comparisons. We note that the $\pT^{\Ph\Ph}$ spectrum at LO does not accurately match the NLO predictions, therefore, any analysis relying on these or similar kinematic observables should be generated at NLO, as argued above.

\begin{table*}[htb!]
\begin{center}
\resizebox{\textwidth}{!}{
\begin{tabular}{cccccccccl}
\hline\noalign{\smallskip}
\multirow{2}*{Benchmark} & \multirow{2}*{\sina} & \multirow{2}*{\tanb} & \mH & \wH & \multirow{2}*{\kappalam} & $\sigma$ & \sigSH & Accessible  & \multicolumn{1}{c}{\multirow{2}*{Feature}} \\
 &  &  & [\GeV{}] & [\GeV{}] & & [\fb{}] & [\fb{}] & in Run-3  &  \\
\noalign{\smallskip}\hline\noalign{\smallskip}
\BMa & 0.16 & 1.0 & 620 & 4.6 & 0.96 & 50.5 & 13.5 & \checkmark & Max \intrel \\
\BMb  & 0.16 & 0.5 & 440 & 1.5 & 0.96 & 91.6 & 56.4 & \checkmark & Max \intrelsum  \\
\BMc  & 0.16 & 0.5 & 380 & 0.8 & 0.96 & 119.8 & 90.1 & \checkmark & Max \intrelsum with $\intrel < 1\%$  \\
\BMd  & -0.16 & 0.5 & 560 & 3.0 & 0.96 & 51.4 & 15.5 & \checkmark & Max non-res. within $\mH\pm 10\%$ \\
\BMe  & 0.08 & 0.5 & 500 & 0.6 & 0.99 & 40.6 & 8.1 & & Max non-res. within $\mH\pm 10\%$ \\
\BMf  & 0.16 & 1.0 & 680 & 6.1 & 0.96 & 44.8 & 8.4 & \checkmark & Max \mH \\
\BMg  & 0.15 & 1.1 & 870 & 9.5 & 0.96 & 36.8 & 2.3 & & Max \mH \\
\BMh  & 0.24 & 3.5 & 260 & 0.6 & 0.87 & 374.2 & 357.3 & \checkmark & Max $|\kappalam-1|$ \\
\BMi  & 0.16 & 1.0 & 800 & 9.8 & 0.96 & 38.9 & 3.6 & & Max \relwH \\
\noalign{\smallskip}\hline
\end{tabular}}
\end{center}
\caption{Overview of the investigated benchmark points. The values of the \sina, \tanb, and \mH model parameters are shown. The values of \wH, \kappalam, $\sigma$, and \sigSH, which are fixed by the model parameters, are also displayed. The cross-sections are computed at LO and scaled by NNLO K-factors as described in Section~\ref{sec:method}. The second to last column indicates whether the points are expected to be accessible at the LHC by the end of Run-3. The last column gives a summary of the feature exhibited by the benchmark point.
\label{tab:benchmarks}}

\end{table*}

\subsection{\BMa}

Our first BM point, \BMa, is chosen as one of the points where the influence of the interference on the di-Higgs cross-section is maximal. Accordingly, we take a point with a relatively large value of \intrel (equal to 13\%) from the points that are still allowed according to our scan specifications, which is found for $\mH=620\,\GeV$, $\sina=0.16$, and $\tanb=1.0$.

\begin{figure*}[htbp]
  \includegraphics[width=0.48\textwidth]{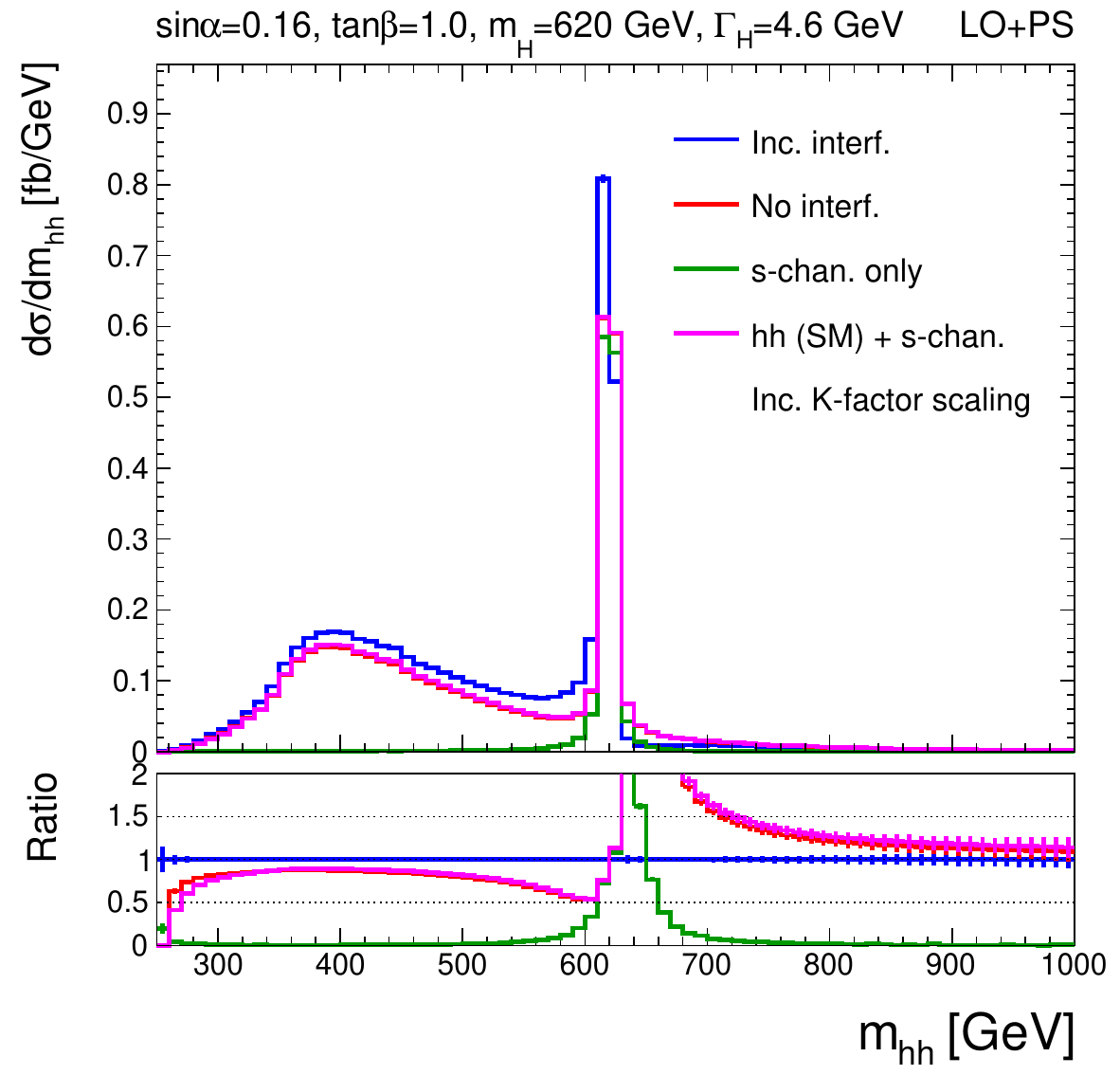}
  \includegraphics[width=0.48\textwidth]{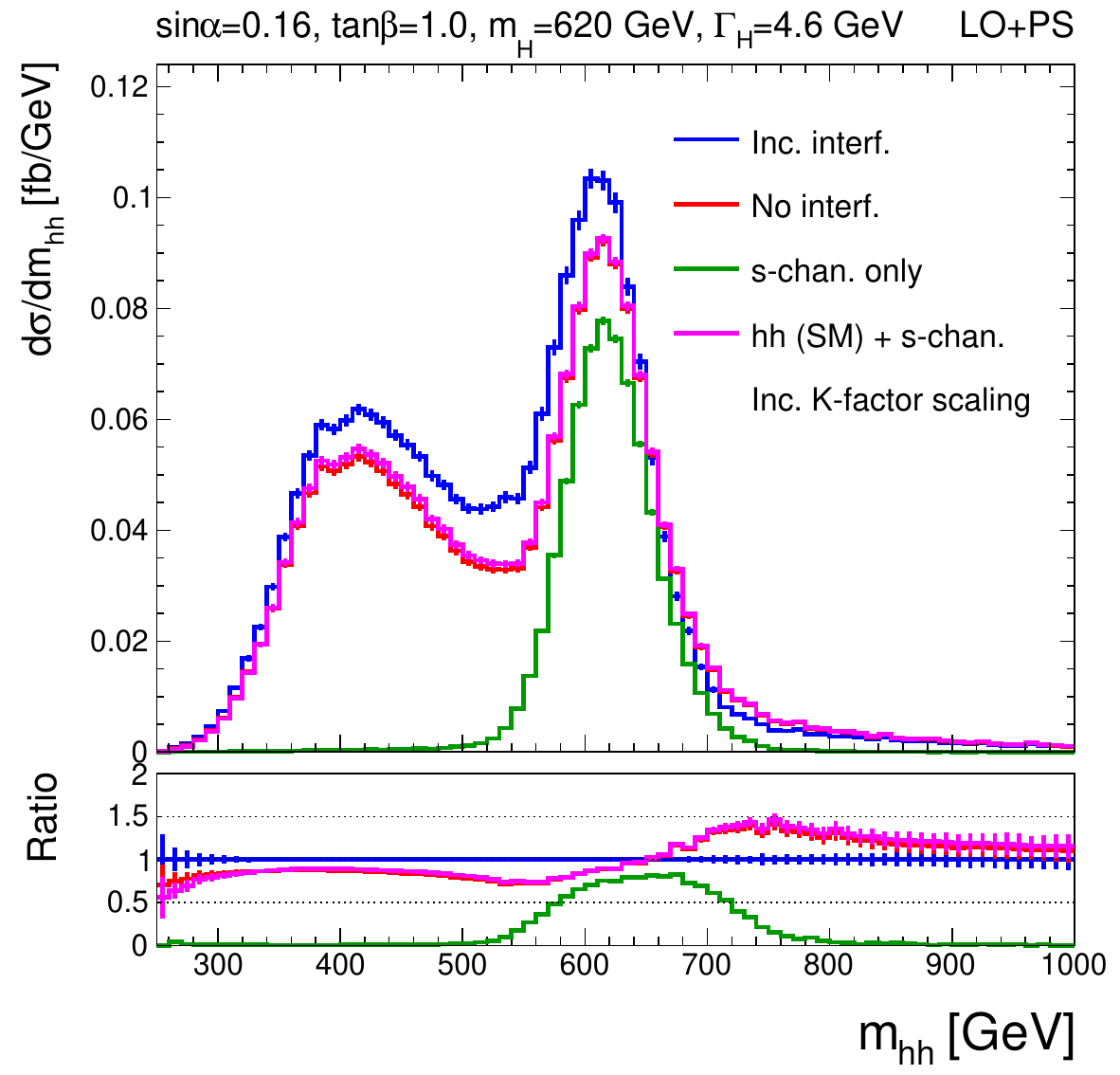}
  
\caption{The di-Higgs mass distributions for \BMa before (left) and after (right) experimental smearing. The blue lines show the full di-Higgs spectrum including all interference terms. This is compared to the spectrum excluding the \SHBox and \SHSh interference terms shown in red, the \PS $s$-channel spectrum (\SH) shown in green, and the incoherent sum of \SH and the SM di-Higgs spectrum in magenta.  }
\label{fig:mass_BMa}
\end{figure*}

The \mhh distributions for this BM with and without detector smearing are displayed in Figure~\ref{fig:mass_BMa}. Smearing in this and all subsequent figures is performed according to the prescription discussed in Section~\ref{sec:simulation}. The figure compares the distributions for four different treatments of the di-Higgs  modelling. The full di-Higgs distribution including both resonant and non-resonant diagrams and interference terms is shown in blue, the distribution neglecting the \SHBox and \SHSh interference terms in shown in red (referred to as ``no-interference'' in the following), the \SH-only distribution is shown in green, and the distribution obtained by fixing the non-resonant di-Higgs spectrum to the SM expectation and summing with the \SH distribution is shown in magenta (referred to as ``SM+\SH''). 

We note that most of the current experimental searches neglect the non-resonant contributions to the spectrum entirely, which corresponds to the green line in Figure~\ref{fig:mass_BMa} and equivalent figures that will follow. However, a small number of searches fix the non-resonant di-Higgs process to the SM expectation and neglect the interference effects, which corresponds to the magenta line in the figures.  

The interference tends to increase (decrease) the broader non-resonant-like continuum above (below) the resonant peak at $\mH\sim 620\,\GeV$. After detector smearing, the overall distribution exhibits an interesting double-peak structure. The interference also causes a shift in the position of the mass peak to slightly lower values, however, after smearing the shift in the peak position is not very noticeable. It is clear from comparing to the \SH-only spectrum that neglecting the non-resonant terms altogether results in a failure to describe the double-peak structure, and it also results in an underestimation of the number of events close to the peak of around 35\%. Including the non-resonant terms while neglecting only the \SHBox and \SHSh interference terms clearly brings an improvement in the overall description of the spectrum. However, it still results in an underestimate of the height of the \SH peak by about 15\%, and an under(over)-estimate of the non-resonant continuum of $\approx$ 15--30\% (15--40\%) below (above) the resonant peak. This clearly motivates the proper consideration of the interference effects when performing resonant searches. We also do not observe a significant difference between no-interference and SM+\SH distributions. This is because only small values of \sina are still allowed by the experimental constraints and this results in \kappalam and \kappat values that are typically close to unity at LO. As will be shown in the subsequent sections, this feature is common to all BM points.

\begin{figure*}[htbp]
  \includegraphics[width=0.48\textwidth]{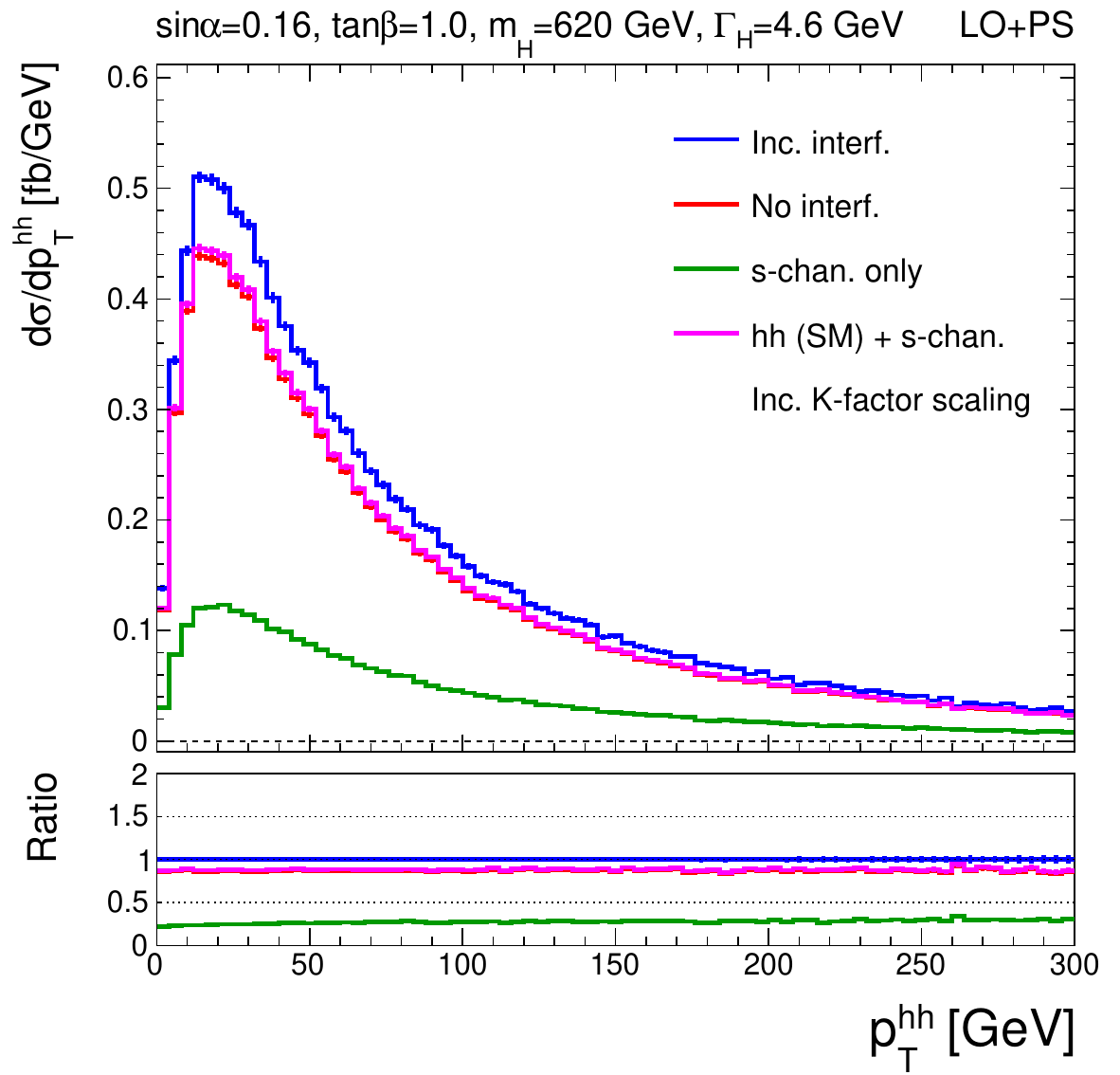}
  \includegraphics[width=0.48\textwidth]{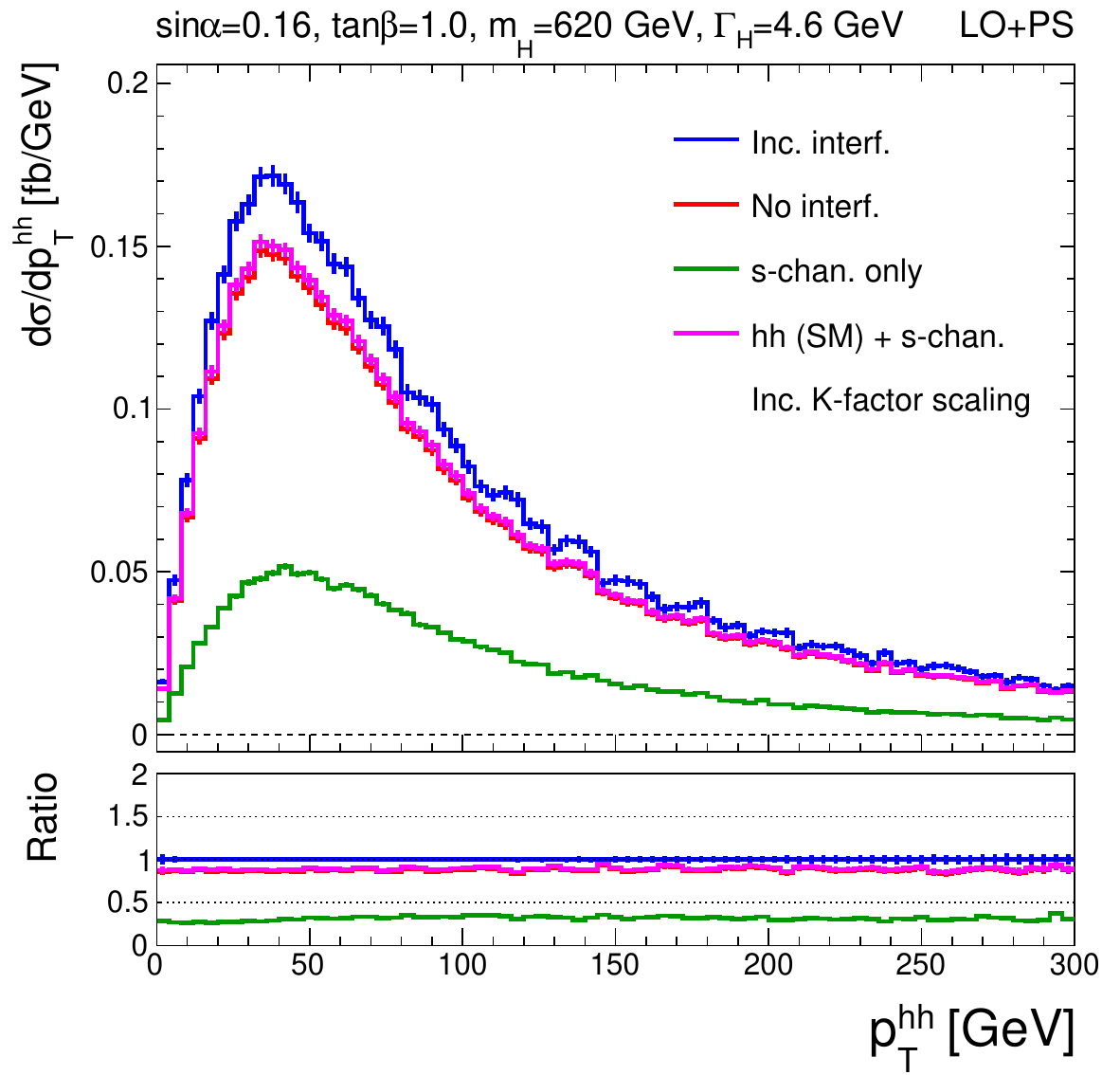}
  
\caption{The di-Higgs \pT distributions for \BMa before (left) and after (right) experimental smearing.
The blue lines show the full di-Higgs spectrum including all interference terms. This is compared to the spectrum excluding the \SHBox and \SHSh interference terms shown in red, the \PS $s$-channel spectrum (\SH) shown in green, and the incoherent sum of \SH and the SM di-Higgs spectrum in magenta.}
\label{fig:hh_pt_BMa}
\end{figure*}

\begin{figure*}[htbp]
  \includegraphics[width=0.48\textwidth]{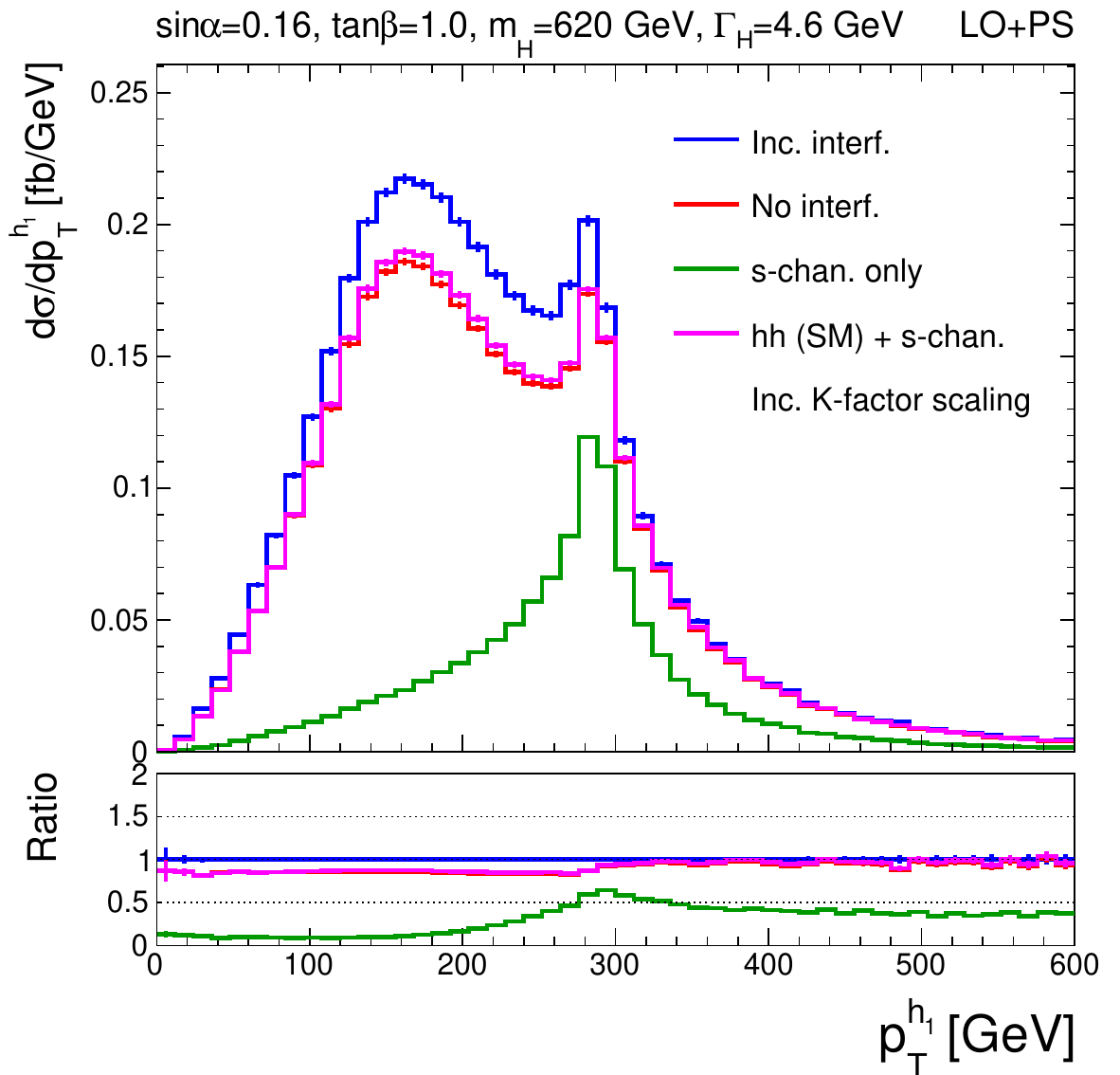}
  \includegraphics[width=0.48\textwidth]{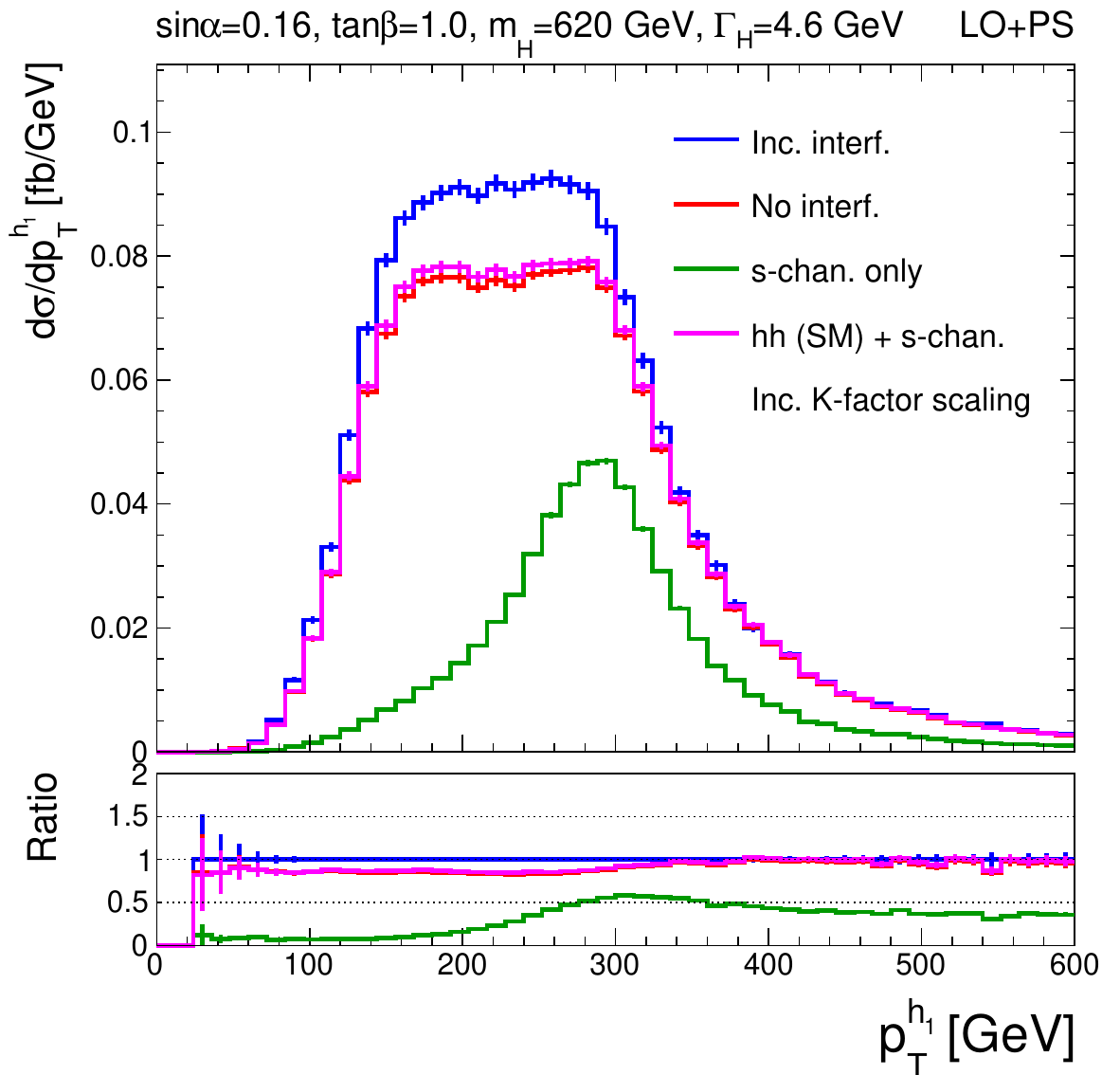}
  
\caption{The leading \Ph \pT for \BMa before (left) and after (right) experimental smearing.
The blue lines show the full di-Higgs spectrum including all interference terms. This is compared to the spectrum excluding the \SHBox and \SHSh interference terms shown in red, the \PS $s$-channel spectrum (\SH) shown in green, and the incoherent sum of \SH and the SM di-Higgs spectrum in magenta.}
\label{fig:h1_pt_BMa}
\end{figure*}

\begin{figure*}[htbp]
  \includegraphics[width=0.48\textwidth]{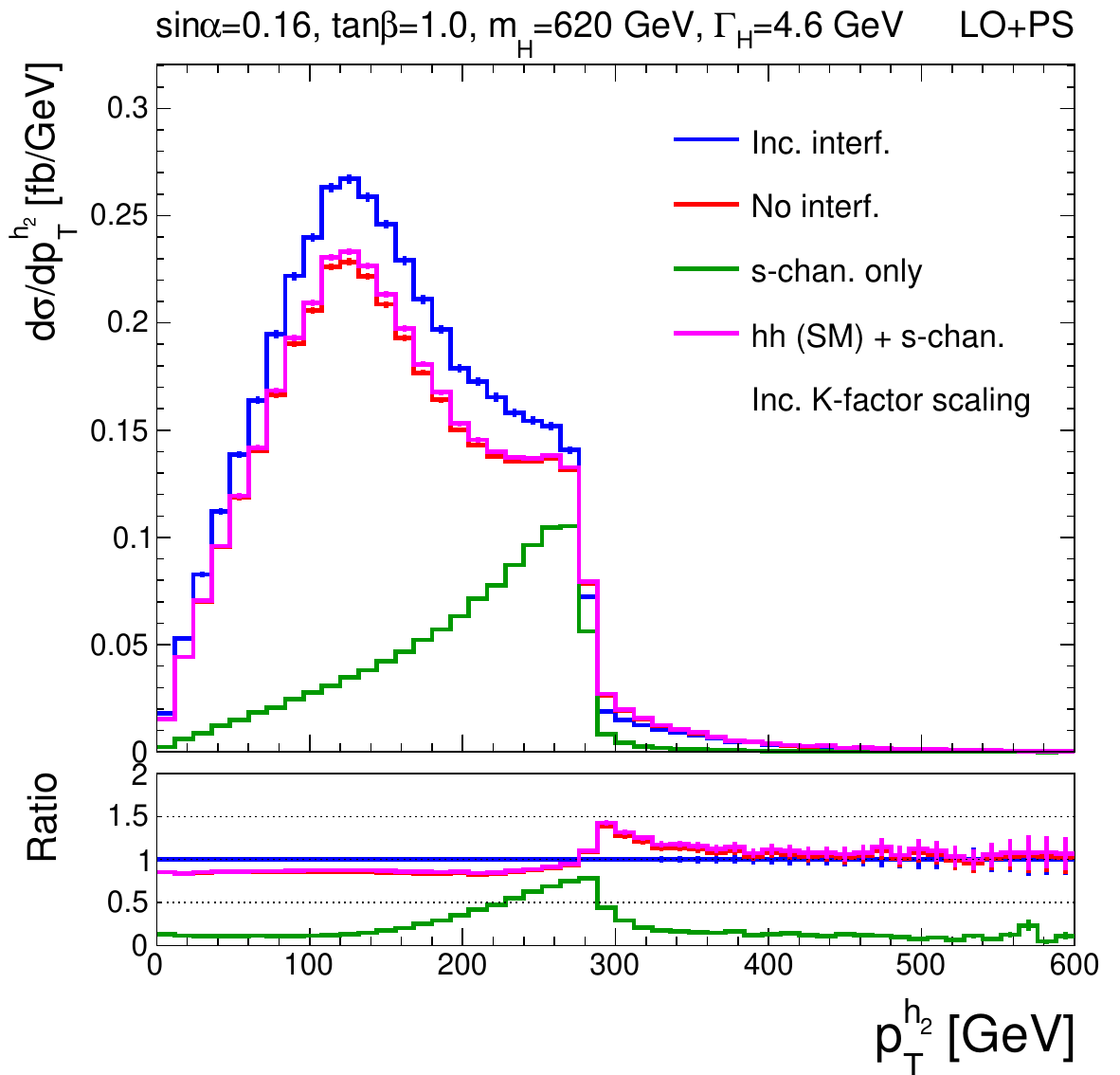}
  \includegraphics[width=0.48\textwidth]{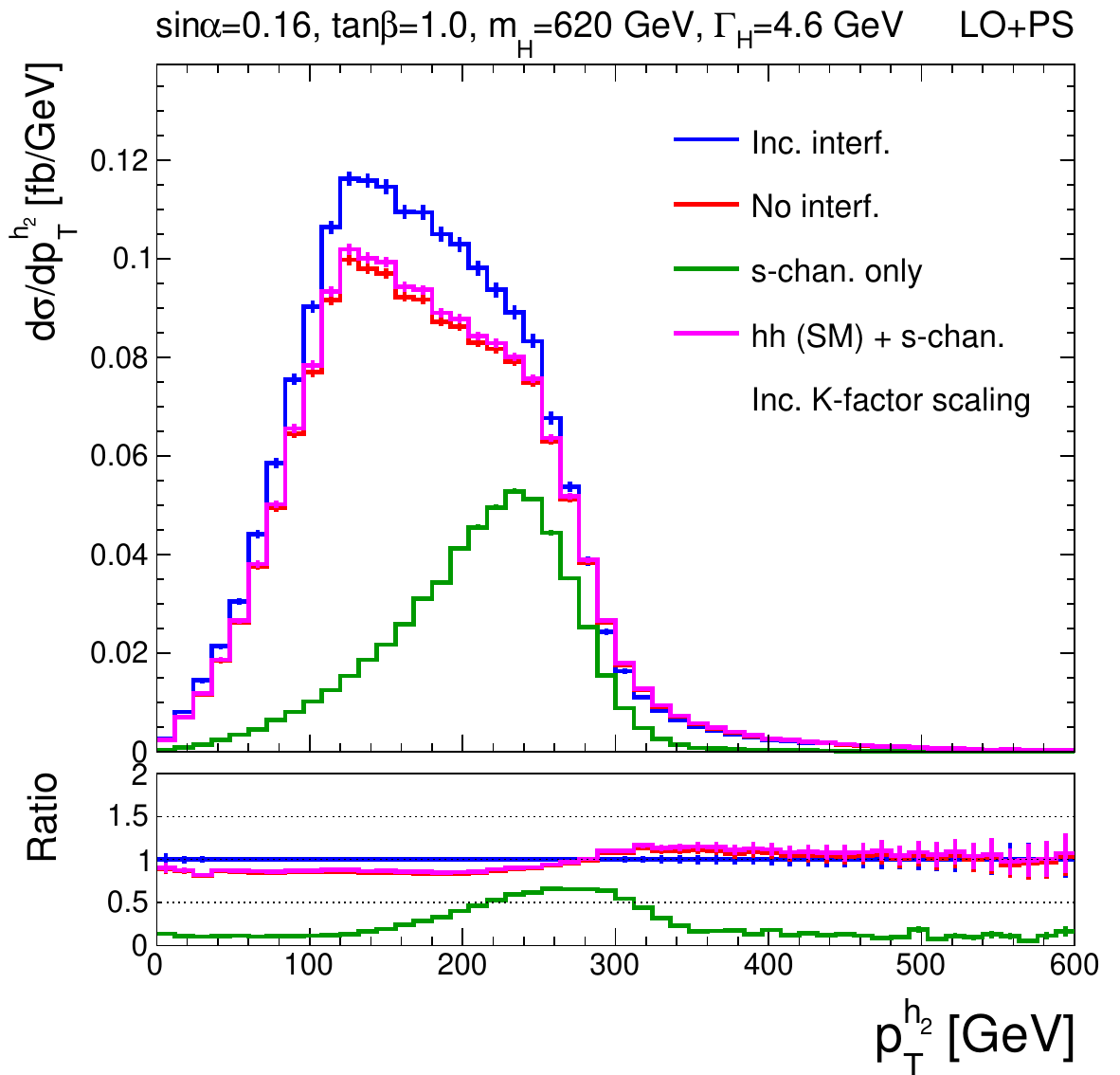}
\caption{The sub-leading \Ph \pT before (left) and after (right) experimental smearing. 
The blue lines show the full di-Higgs spectrum including all interference terms. This is compared to the spectrum excluding the \SHBox and \SHSh interference terms shown in red, the \PS $s$-channel spectrum (\SH) shown in green, and the incoherent sum of \SH and the SM di-Higgs spectrum in magenta.}
\label{fig:h2_pt_BMa}
\end{figure*}

\begin{figure*}[htbp]
  \includegraphics[width=0.48\textwidth]{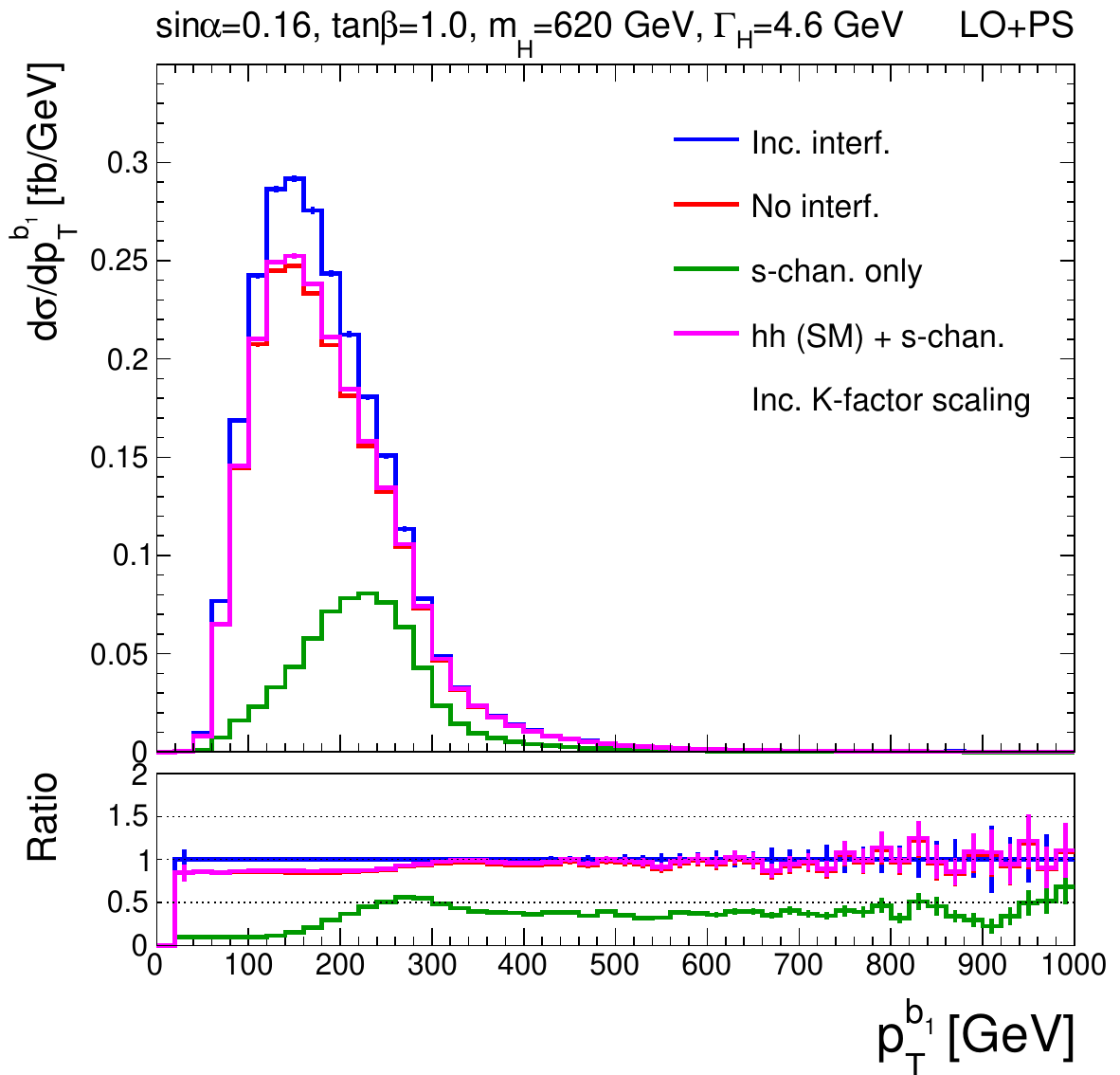}
  \includegraphics[width=0.48\textwidth]{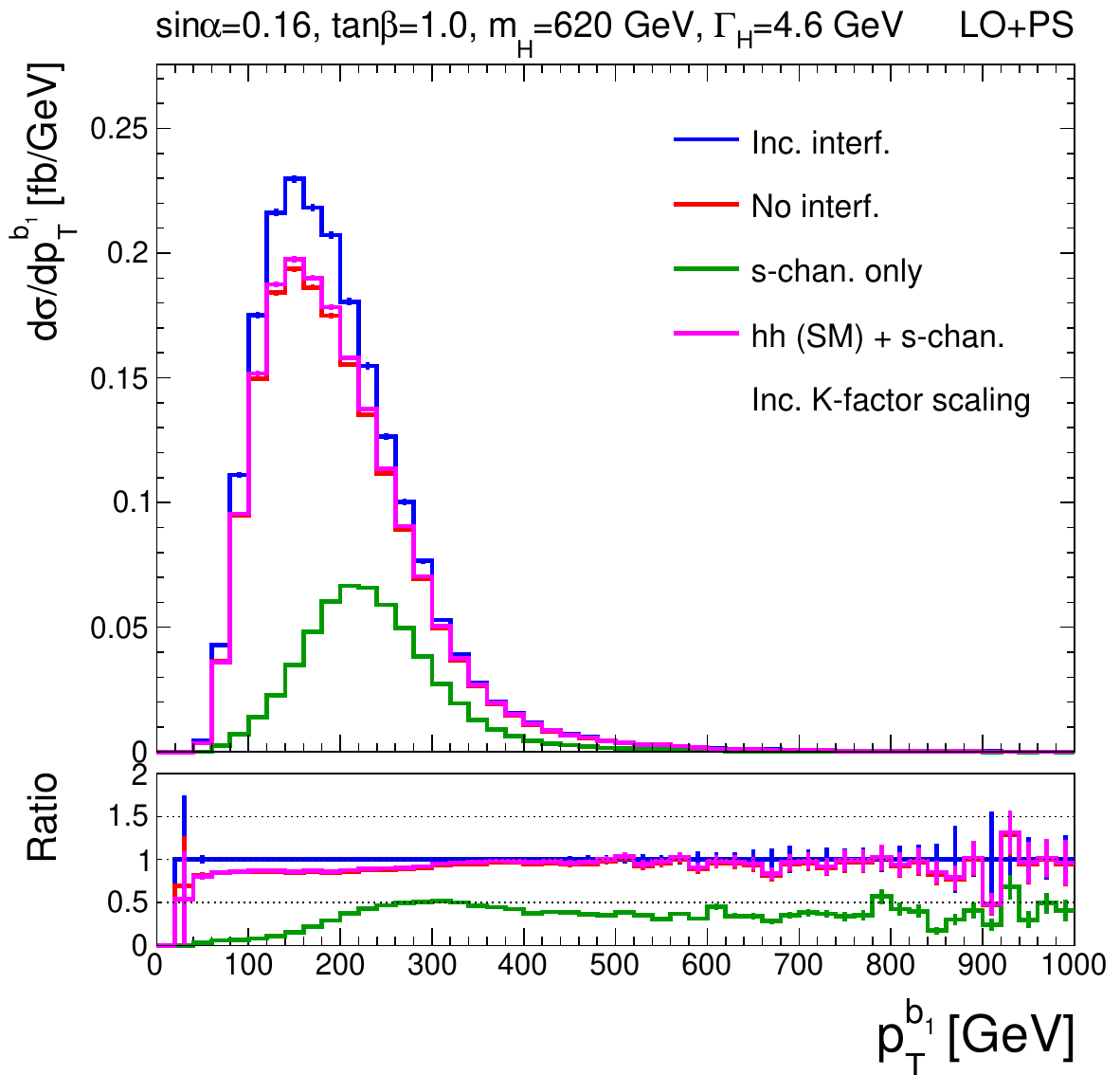}
\caption{The \pT of the leading $b$-jet before (left) and after (right) experimental smearing. 
The blue lines show the full di-Higgs spectrum including all interference terms. This is compared to the spectrum excluding the \SHBox and \SHSh interference terms shown in red, the \PS $s$-channel spectrum (\SH) shown in green, and the incoherent sum of \SH and the SM di-Higgs spectrum in magenta. No \pT or \HT cuts have been applied for the smeared distributions in this case.}
\label{fig:b1_pt_BMa}
\end{figure*}

\begin{figure*}[htbp]
  \includegraphics[width=0.48\textwidth]{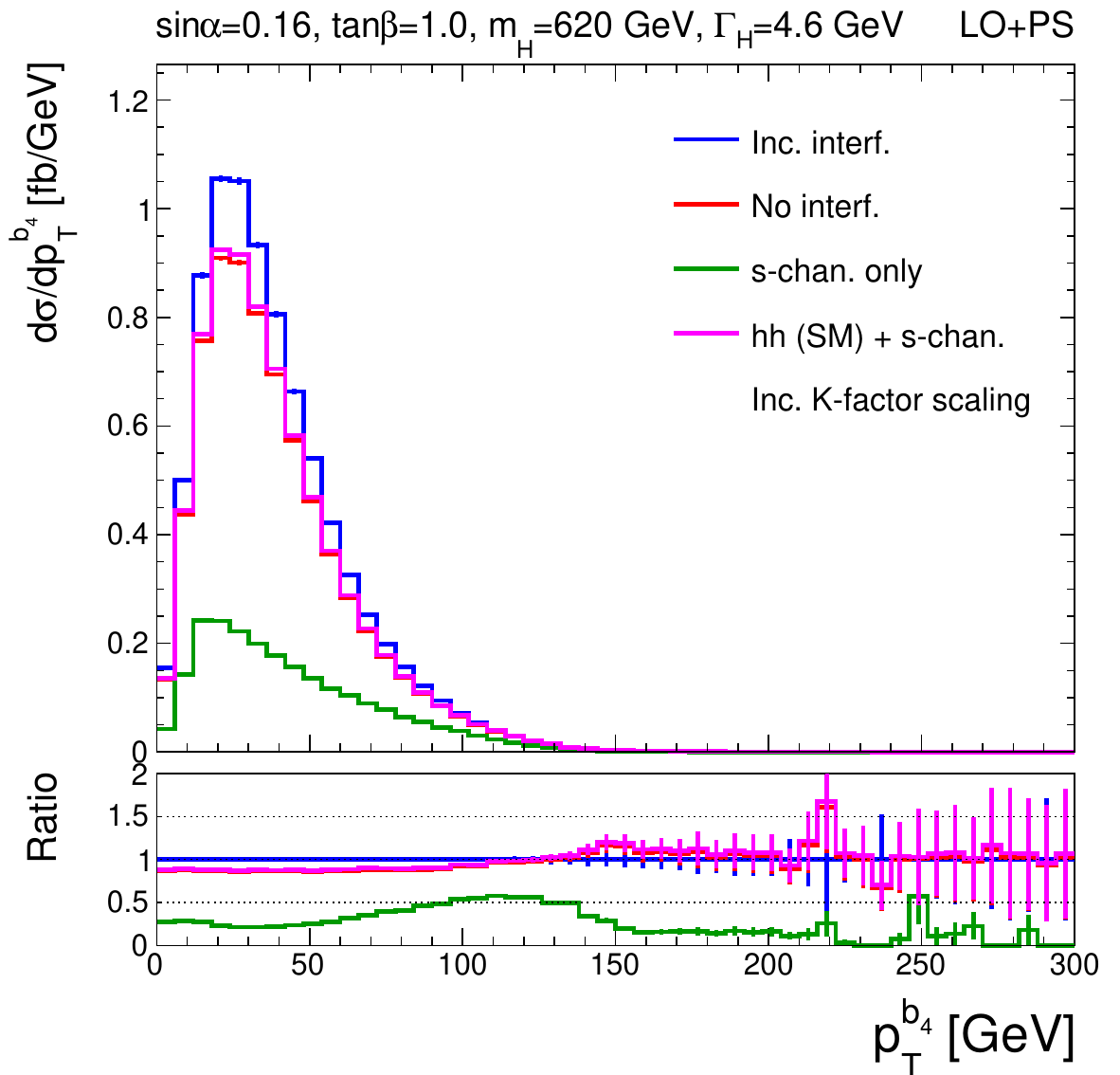}
  \includegraphics[width=0.48\textwidth]{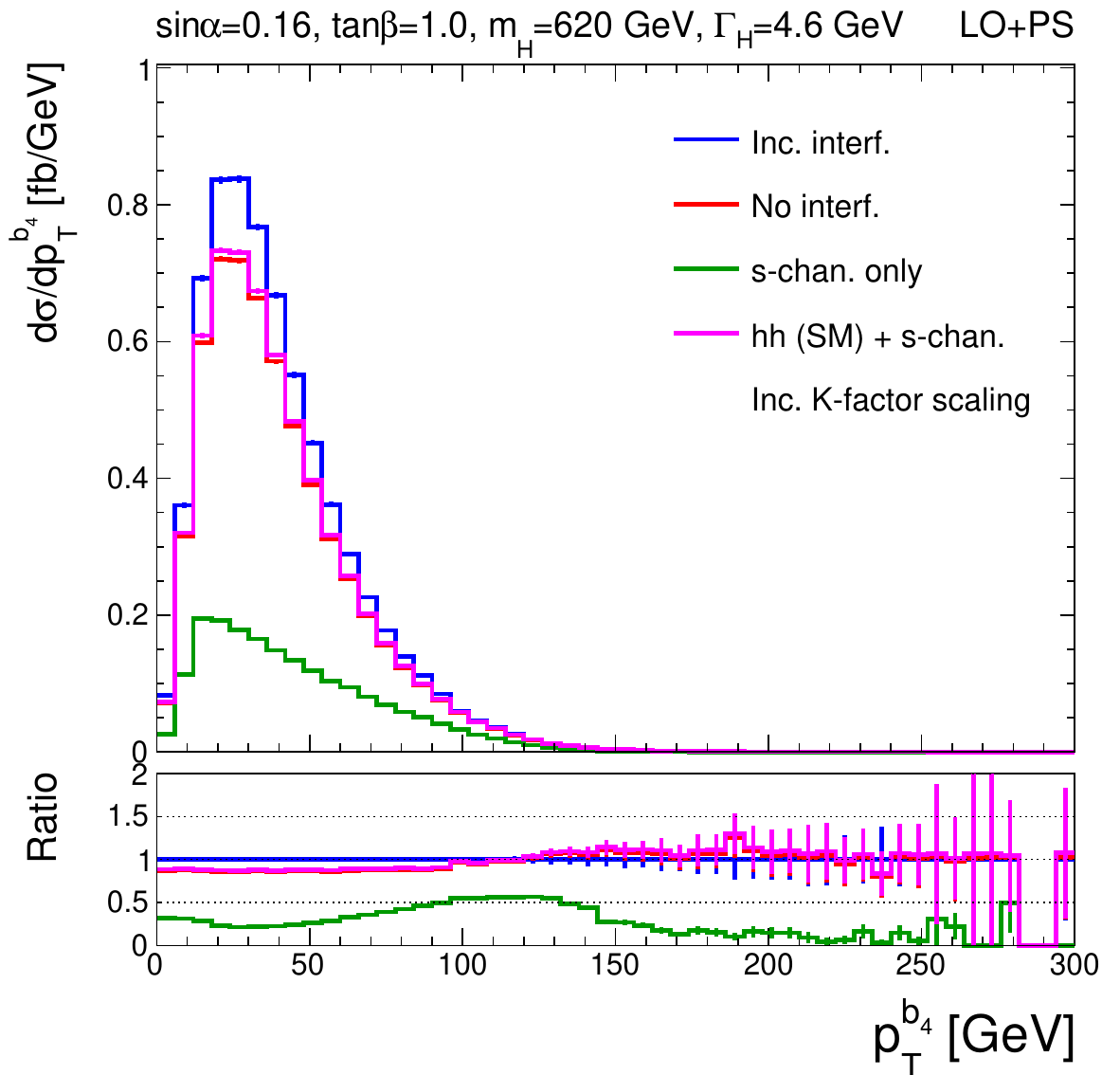}
\caption{The \pT of the softest $b$-jet before (left) and after (right) experimental smearing. 
The blue lines show the full di-Higgs spectrum including all interference terms. This is compared to the spectrum excluding the \SHBox and \SHSh interference terms shown in red, the \PS $s$-channel spectrum (\SH) shown in green, and the incoherent sum of \SH and the SM di-Higgs spectrum in magenta. No \pT or \HT cuts have been applied for the smeared distributions in this case.}
\label{fig:b4_pt_BMa}
\end{figure*}

\begin{figure*}[htbp]
  \includegraphics[width=0.48\textwidth]{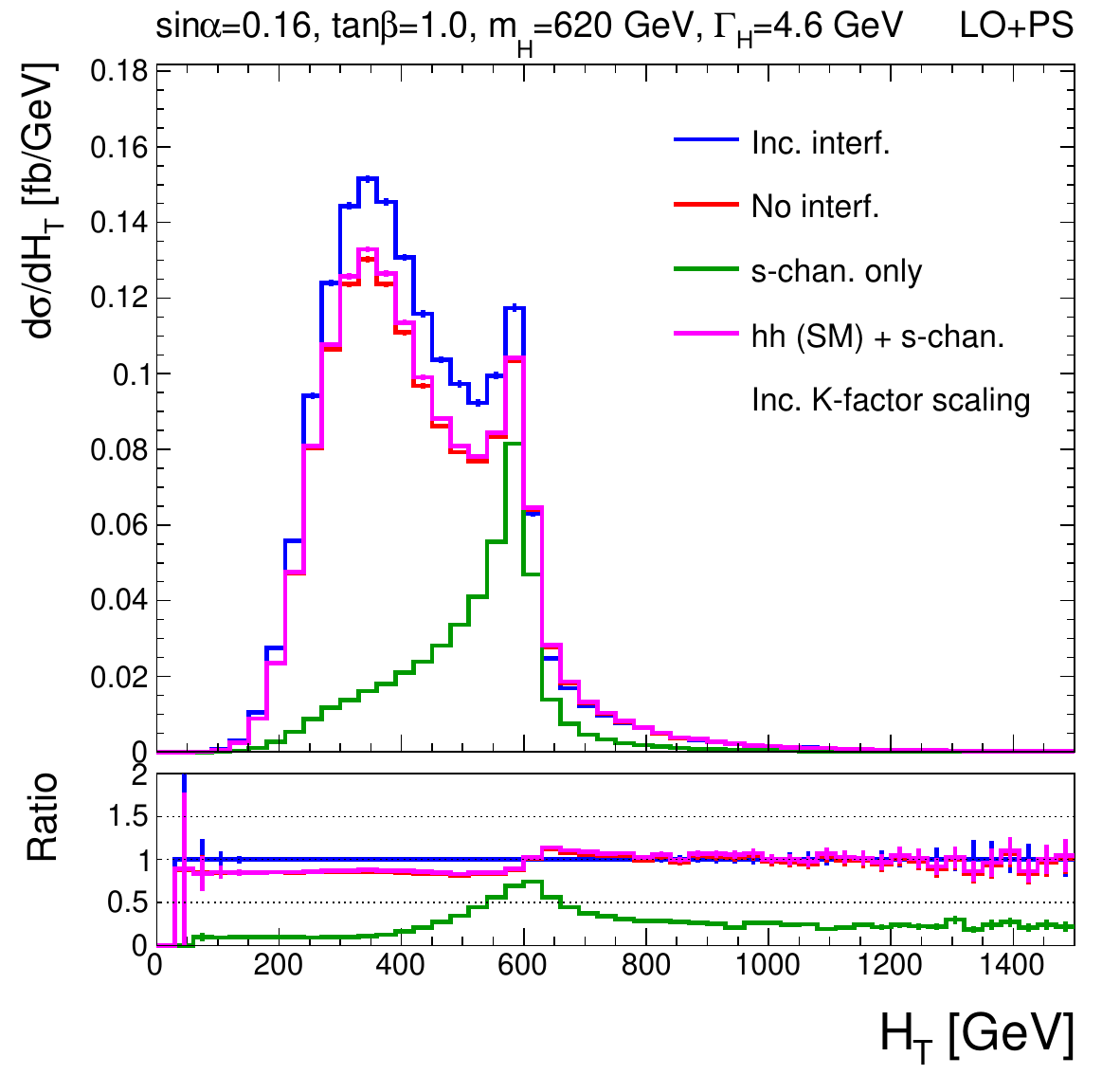}
  \includegraphics[width=0.48\textwidth]{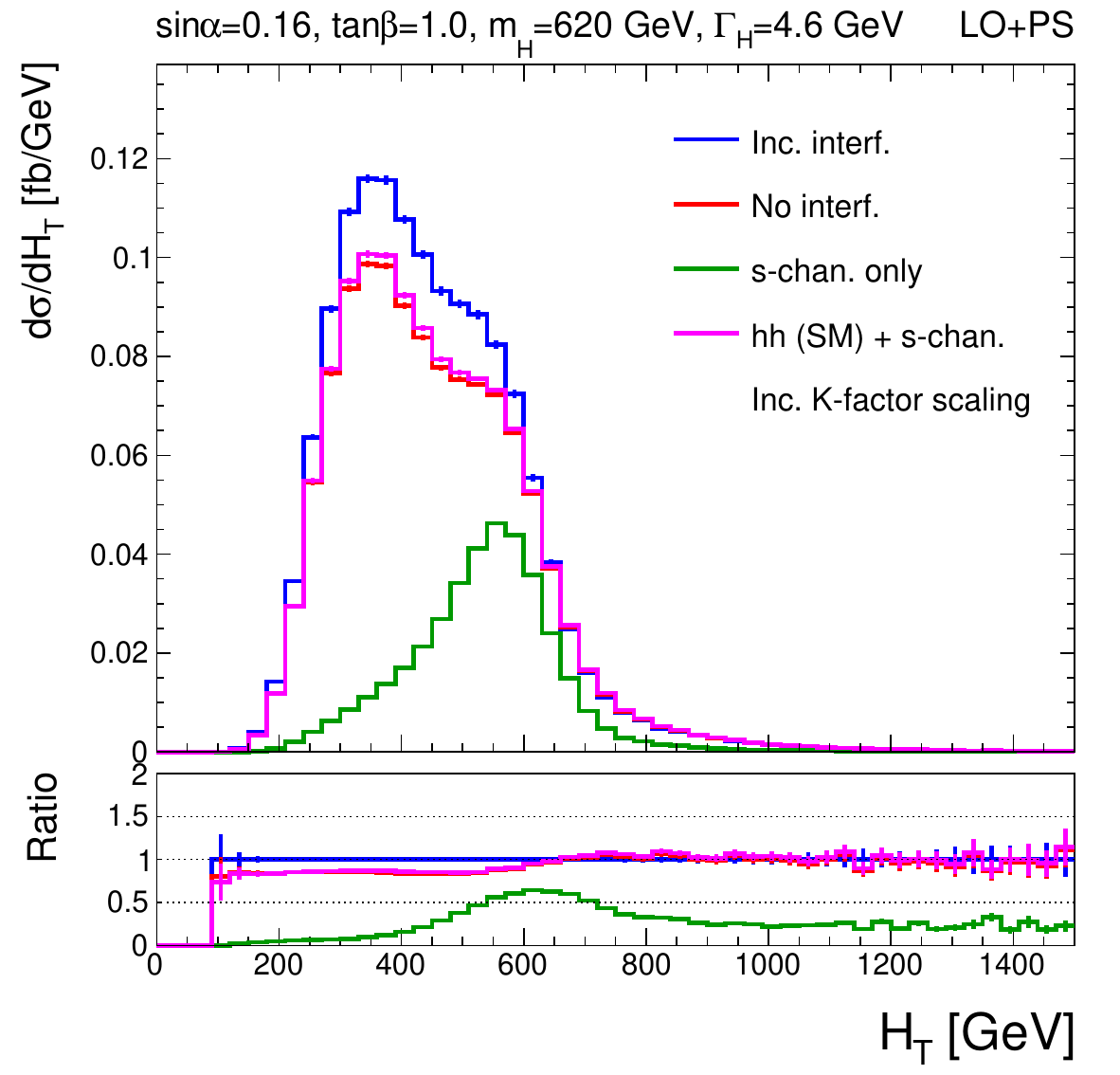}
\caption{The \HT of the $b$-jets before (left) and after (right) experimental smearing. 
The blue lines show the full di-Higgs spectrum including all interference terms. This is compared to the spectrum excluding the \SHBox and \SHSh interference terms shown in red, the \PS $s$-channel spectrum (\SH) shown in green, and the incoherent sum of \SH and the SM di-Higgs spectrum in magenta. No \pT or \HT cuts have been applied for the smeared distributions in this case.}
\label{fig:b1_Ht_BMa}
\end{figure*}

\begin{figure*}[htbp]
  \includegraphics[width=0.48\textwidth]{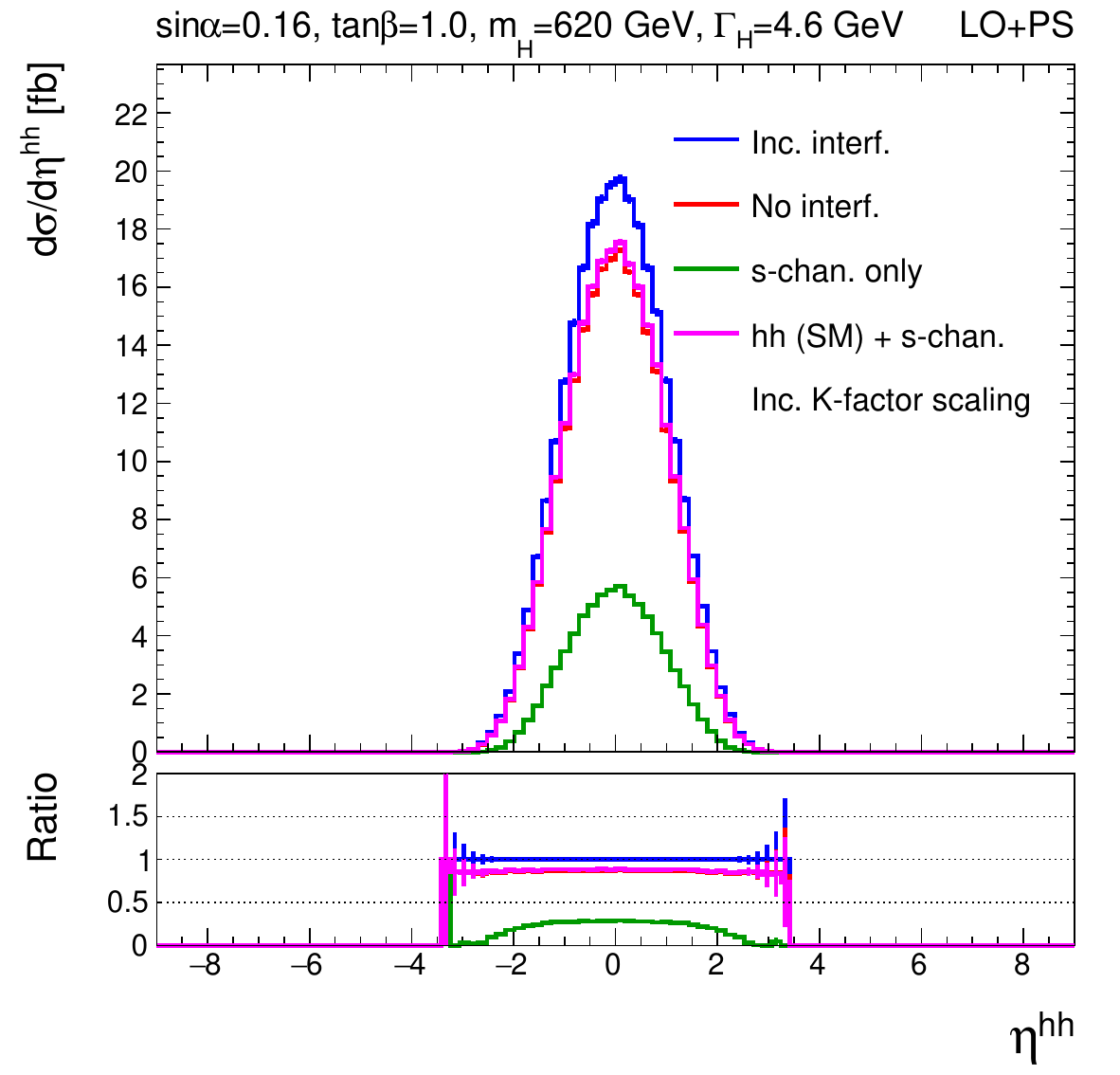}
  \includegraphics[width=0.48\textwidth]{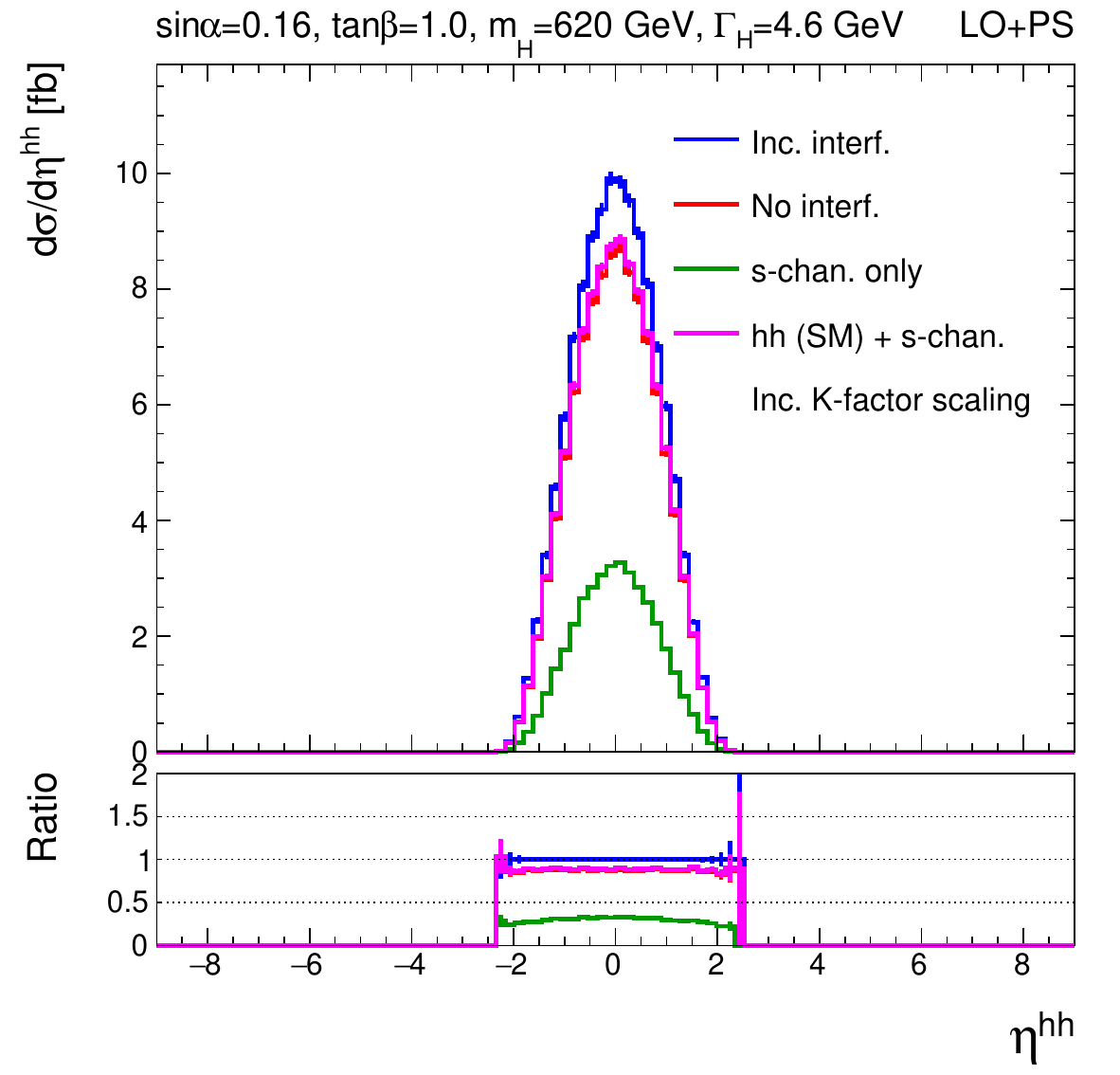}
\caption{The di-Higgs $\eta$ distributions for \BMa before (left) and after (right) experimental smearing.
The blue lines show the full di-Higgs spectrum including all interference terms. This is compared to the spectrum excluding the \SHBox and \SHSh interference terms shown in red, the \PS $s$-channel spectrum (\SH) shown in green, and the incoherent sum of \SH and the SM di-Higgs spectrum in magenta.}
\label{fig:eta_BMa}
\end{figure*}

\begin{figure*}[htbp]
  \includegraphics[width=0.48\textwidth]{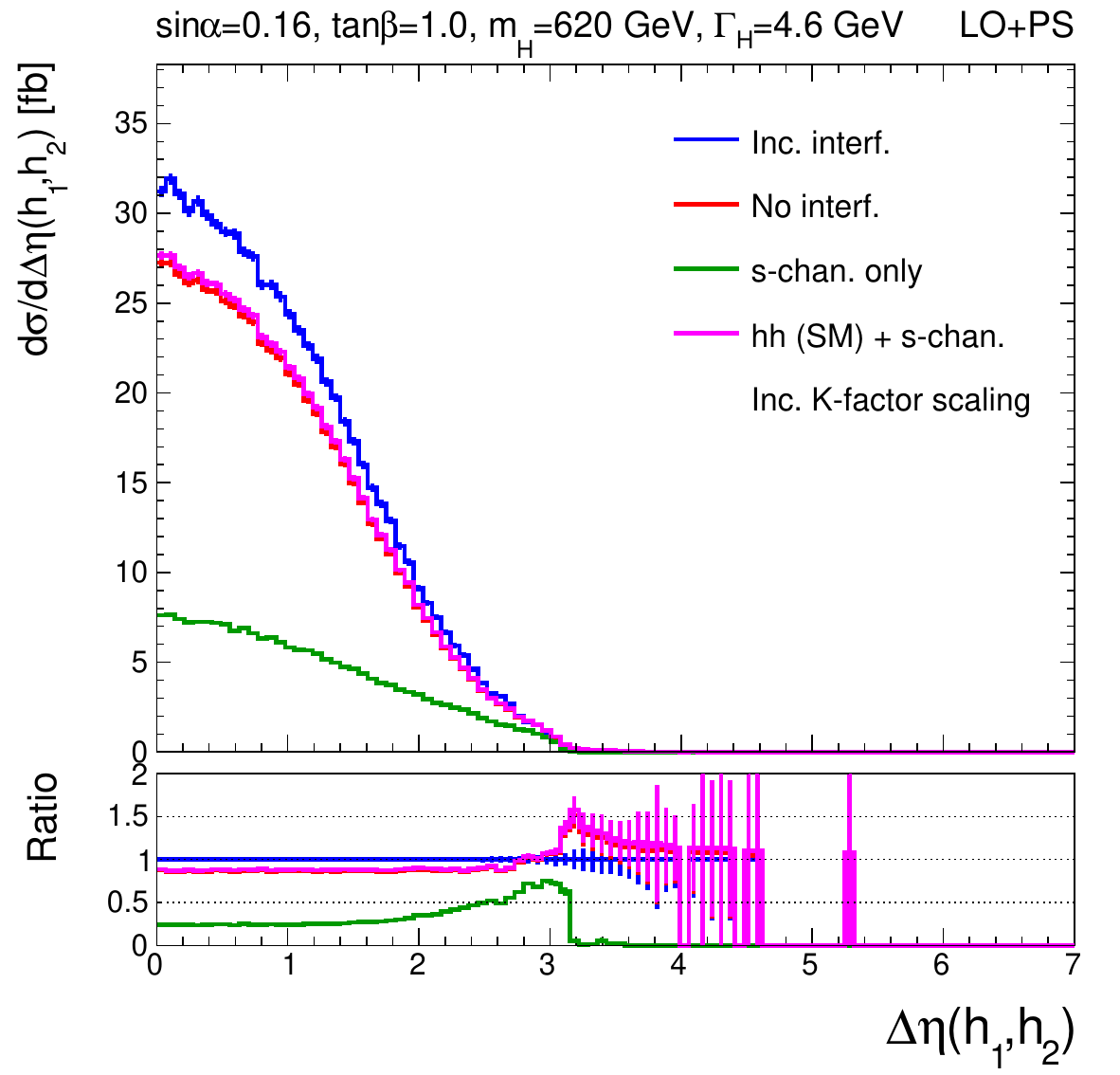}
  \includegraphics[width=0.48\textwidth]{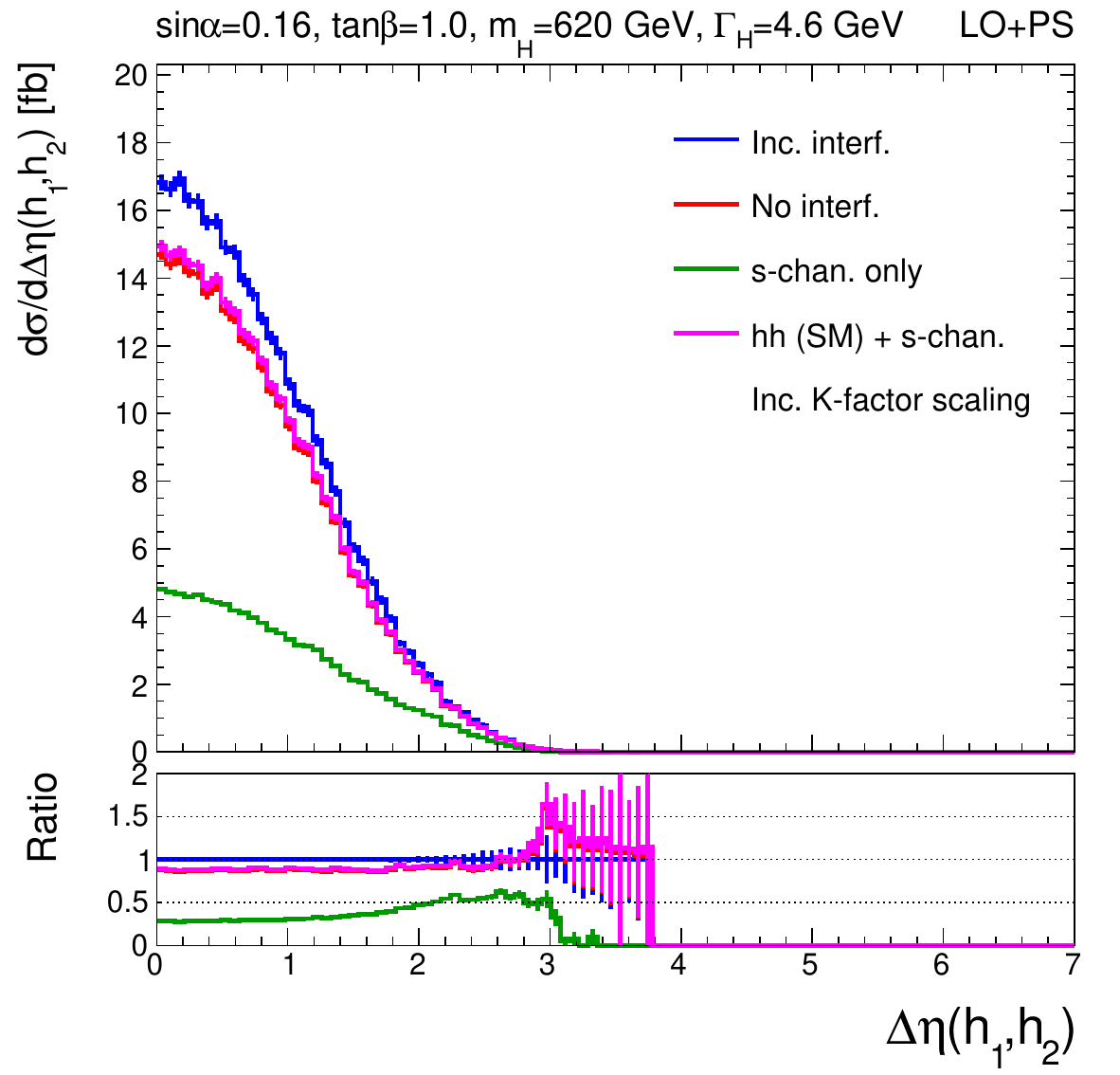}
\caption{The $\Delta\eta$ between the two \Ph for \BMa before (left) and after (right) experimental smearing.
The blue lines show the full di-Higgs spectrum including all interference terms. This is compared to the spectrum excluding the \SHBox and \SHSh interference terms shown in red, the \PS $s$-channel spectrum (\SH) shown in green, and the incoherent sum of \SH and the SM di-Higgs spectrum in magenta.}
\label{fig:deta_BMa}
\end{figure*}

\begin{figure*}[htbp]
  \includegraphics[width=0.48\textwidth]{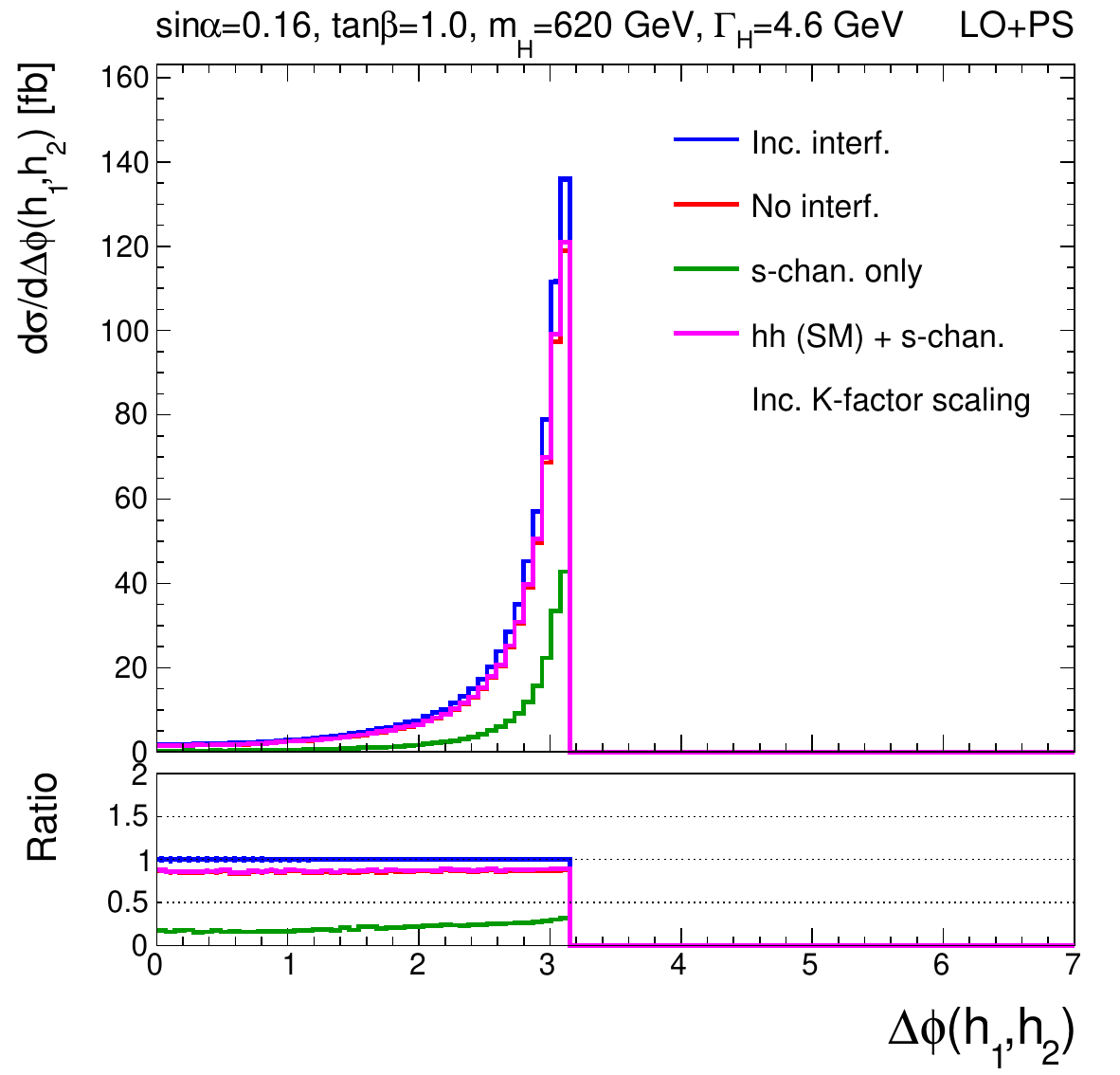}
  \includegraphics[width=0.48\textwidth]{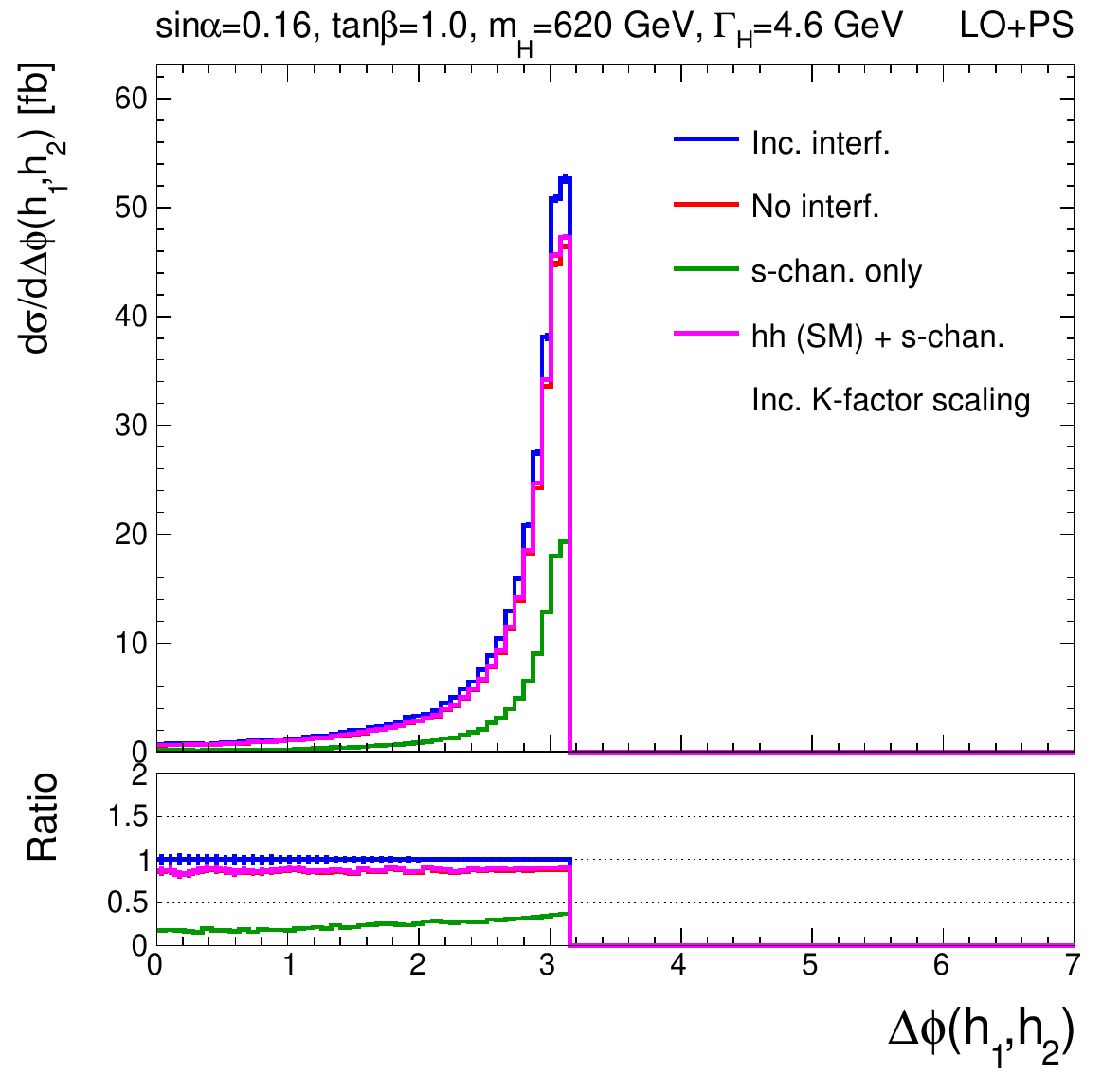}
\caption{The $\Delta\phi$ between the two \Ph before (left) and after (right) experimental smearing.
The blue lines show the full di-Higgs spectrum including all interference terms. This is compared to the spectrum excluding the \SHBox and \SHSh interference terms shown in red, the \PS $s$-channel spectrum (\SH) shown in green, and the incoherent sum of \SH and the SM di-Higgs spectrum in magenta.}
\label{fig:dphi_BMa}
\end{figure*}

\begin{figure*}[htbp]
  \includegraphics[width=0.48\textwidth]{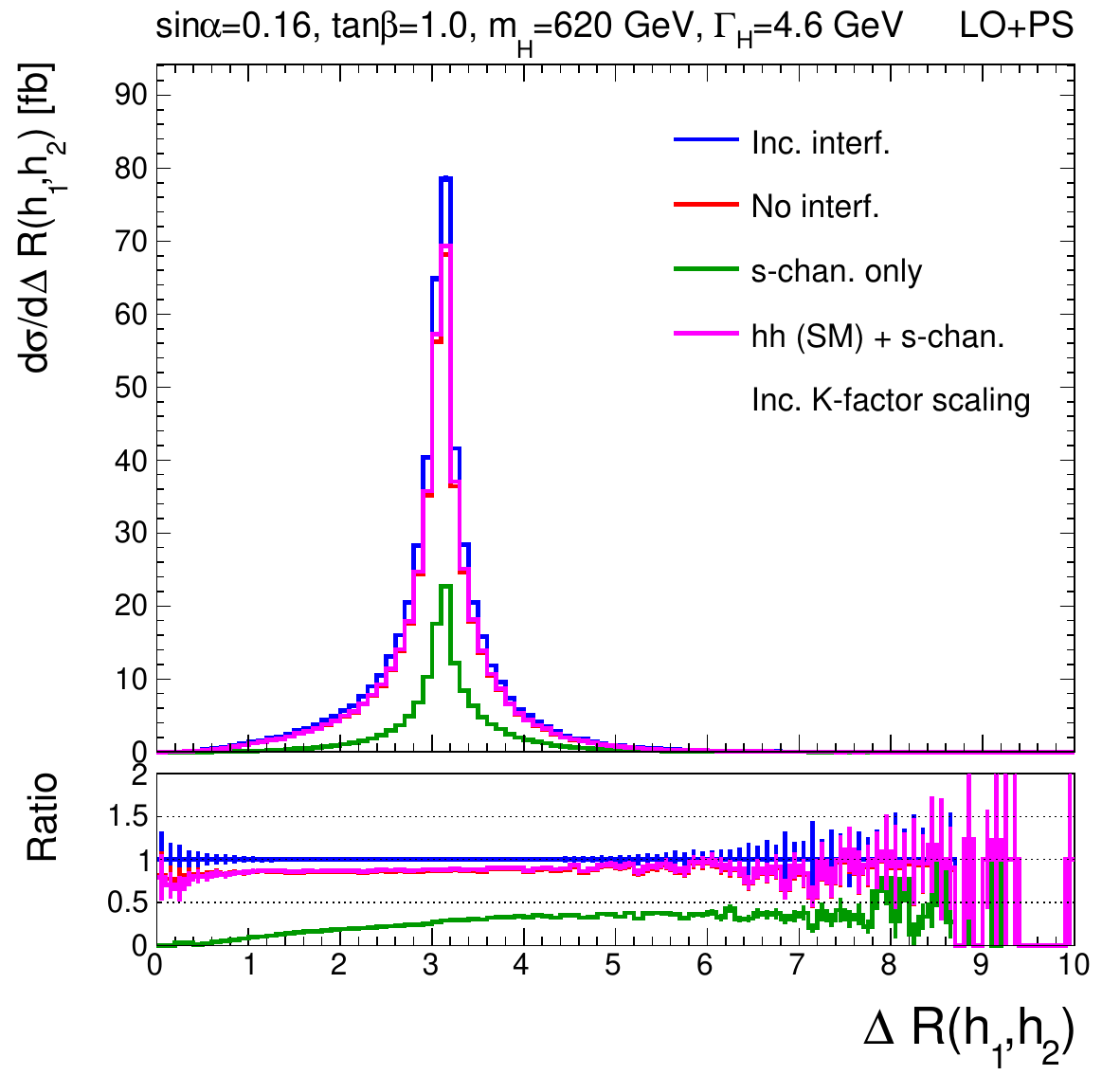}
  \includegraphics[width=0.48\textwidth]{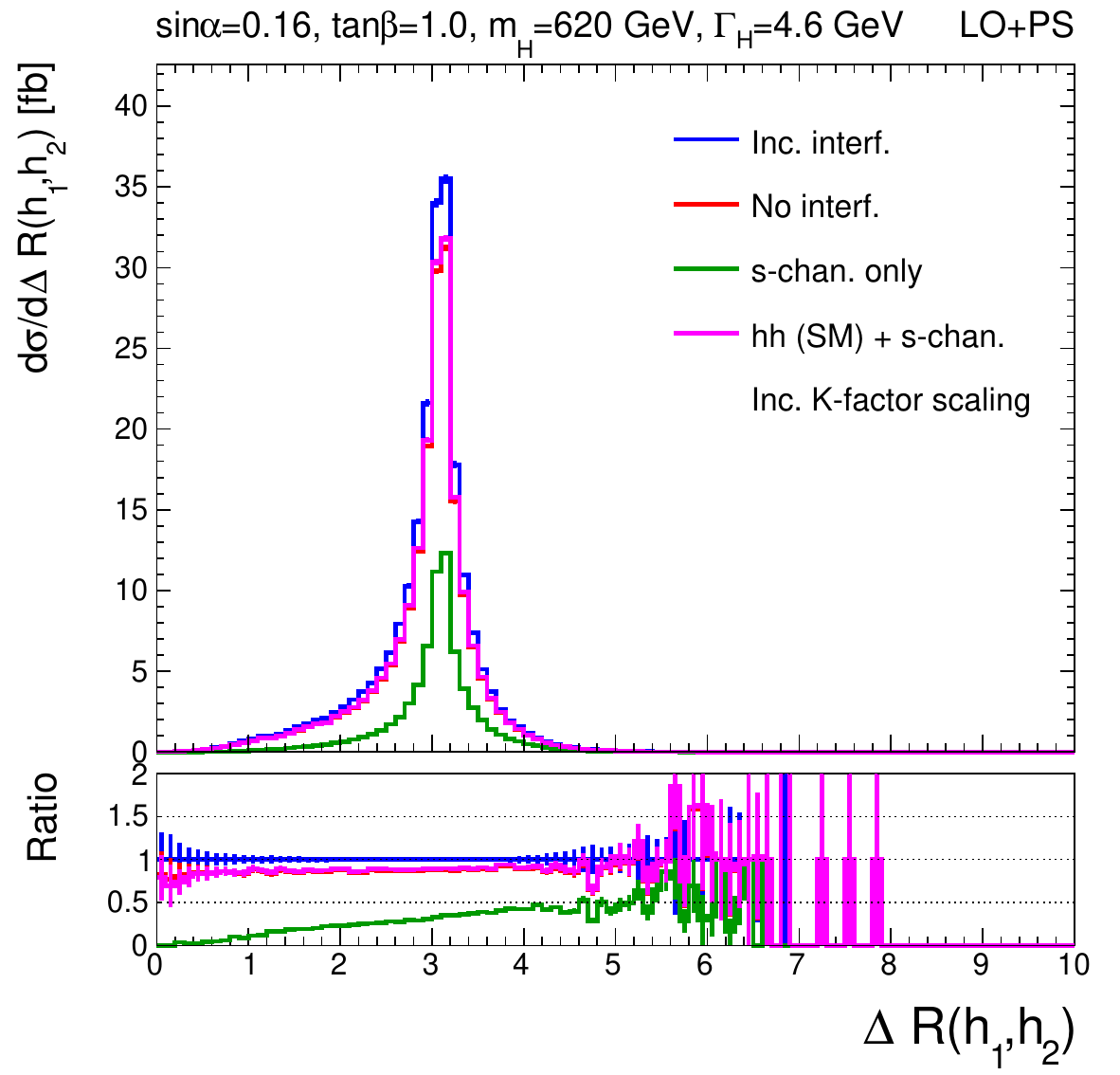}
  
\caption{The $\Delta R$ between the two \Ph before (left) and after (right) experimental smearing.
The blue lines show the full di-Higgs spectrum including all interference terms. This is compared to the spectrum excluding the \SHBox and \SHSh interference terms shown in red, the \PS $s$-channel spectrum (\SH) shown in green, and the incoherent sum of \SH and the SM di-Higgs spectrum in magenta.}
\label{fig:dR_BMa}
\end{figure*}

We also show the distributions for several other variables in Figures~\ref{fig:hh_pt_BMa}-\ref{fig:dR_BMa}. The variables displayed are: $\pT^{\Ph\Ph}$, the \pT of the leading and sub-leading \Ph, $\pT^{\Ph_{1}}$ and $\pT^{\Ph_{2}}$, the \pT of the leading and softest $b$-jets, $\pT^{b_{1}}$ and $\pT^{b_{4}}$, the \HT of the four $b$-jets, the $\eta$ of the di-Higgs system, $\eta^{\Ph\Ph}$, the separation in $\eta$ of the two \Ph, $\Delta\eta(\Ph_{1},\Ph_{2})$, the separation of the $\phi$ between the two \Ph, $\Delta\phi(\Ph_{1},\Ph_{2})$, and the $\Delta R$ (where $\Delta R^2 = \Delta\phi^2+\Delta\eta^2$) between the two \Ph, $\Delta R(\Ph_{1},\Ph_{2})$. In general, the effect of the interference on the shapes of these variables is less significant than was observed for \mhh. The $\pT^{\Ph\Ph}$ and $\eta^{\Ph\Ph}$ distributions in particular do not show any noticeable differences in shapes when neglecting the interference terms. There are more noticeable differences in the other distributions, however, the regions which have largest differences tend to occur in the tails and therefore probably do not have a very large impact on the experimental searches. The \SH-only $\pT^{\Ph_{1}}$, $\pT^{\Ph_{2}}$, $\pT^{b_{1}}$, $\pT^{b_{4}}$, and \HT distributions on the other hand do show very obvious differences compared to the full spectrum which further discourages neglecting the non-resonant contribution when performing experimental searches. 

\subsection{\BMb and \BMc}
\label{sec:benchmarks_bandc}

We define two BM points, \BMb and \BMc, to investigate scenarios where the interference may affect the shapes of the distributions without necessarily changing the total cross-section. \BMb is taken as the point with the largest value of \intrelsum\footnote{We note that a large value of \intrelsum does not necessarily indicate that interference effects are more significant compared to scenarios with larger values of \intrel, such as in the case of \BMa. Rather, \intrelsum serves as an alternative metric for identifying scenarios with non-negligible interference effects to the kinematic distributions which is less sensitive to accidental cancellations that may occur when integrating over \mhh.} (equal to 29.0\%), which is found for $\mH=440\,\GeV$, $\sina=0.16$, and $\tanb=0.5$\footnote{We note that \BMb predicts a \sigSH slightly larger than the observed limits presented by CMS in Ref.~\cite{CMS:2024phk} that are not yet included in \HiggsTools.  
The observed limit is $\sim 52~\fb$ compared to the predicted value of $56~\fb$. However, we still consider this a useful benchmark since it is very close to the experimental limit and CMS neglected the non-resonant di-Higgs background and interference effects when setting their limits, which may have some influence on the exclusions.}.
We define \BMc by additionally requiring $\intrel<1\%$ when determining the maximum \intrelsum value. This allows us to explicitly select a scenario where the effect on the total cross-section is negligible. The values of \intrelsum and \intrel in this scenario are 24.5\% and -0.8\%, respectively. For comparison, the value of \intrel in \BMb is 6.0\%. The values of the model parameters for this scenario are $\mH=380\,\GeV$, $\sina=0.16$, and $\tanb=0.5$.

\begin{figure*}[htbp]
  \includegraphics[width=0.48\textwidth]{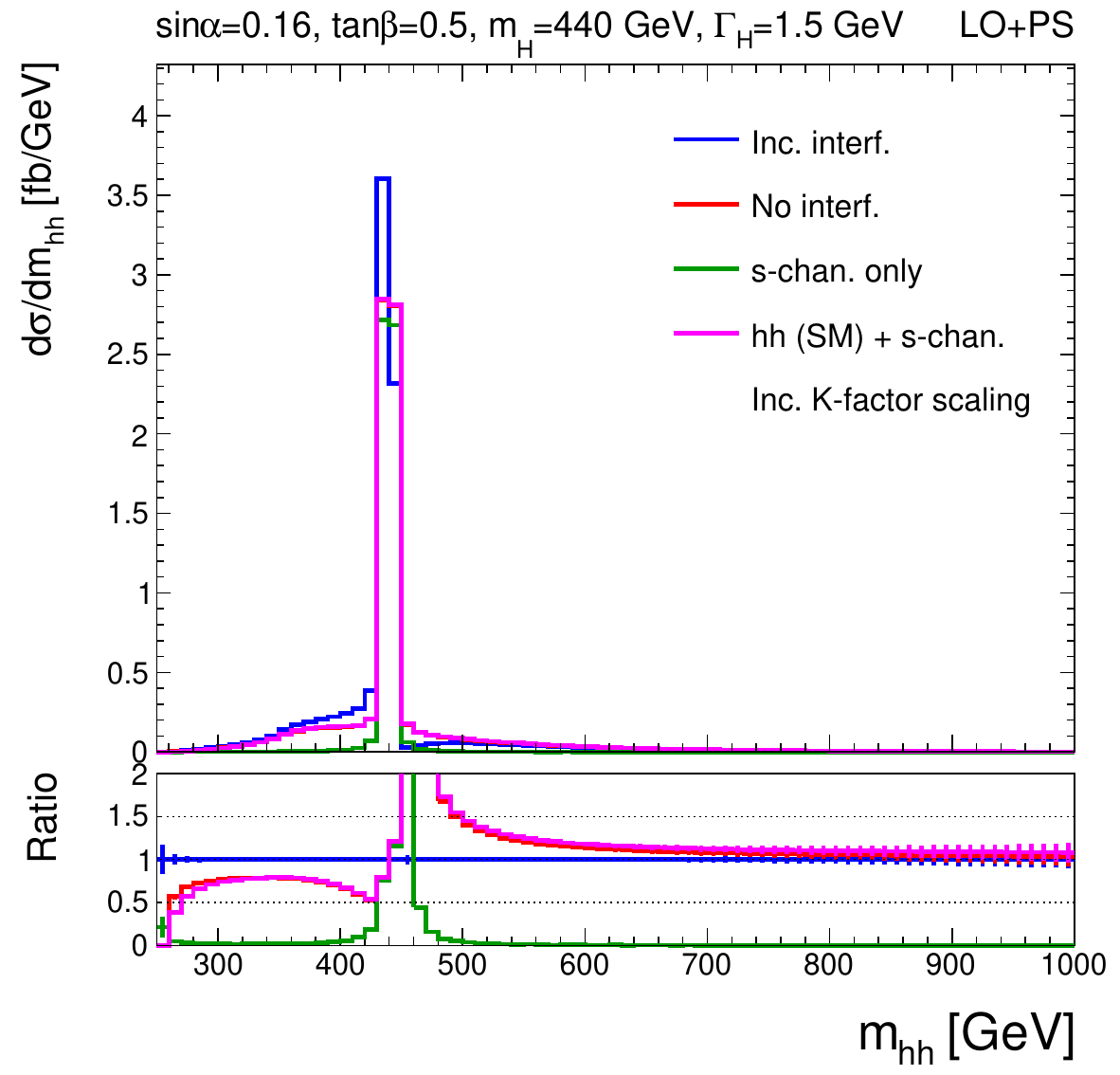}
  \includegraphics[width=0.48\textwidth]{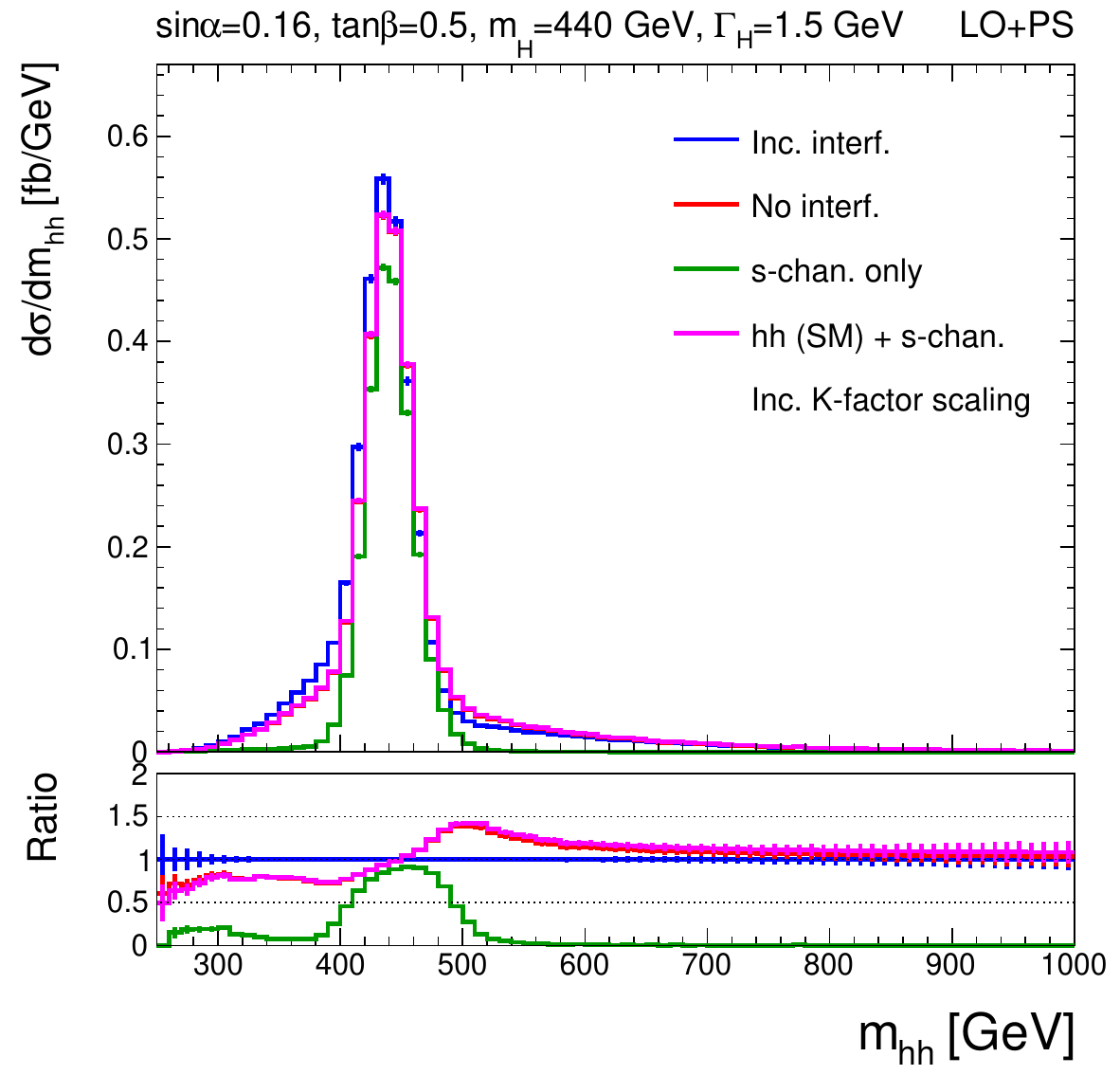}
\caption{The di-Higgs mass distributions for \BMb before (left) and after (right) experimental smearing.
The blue lines show the full di-Higgs spectrum including all interference terms. This is compared to the spectrum excluding the \SHBox and \SHSh interference terms shown in red, the \PS $s$-channel spectrum (\SH) shown in green, and the incoherent sum of \SH and the SM di-Higgs spectrum in magenta.}
\label{fig:mass_BMb}
\end{figure*}

\begin{figure*}[htbp]
  \includegraphics[width=0.48\textwidth]{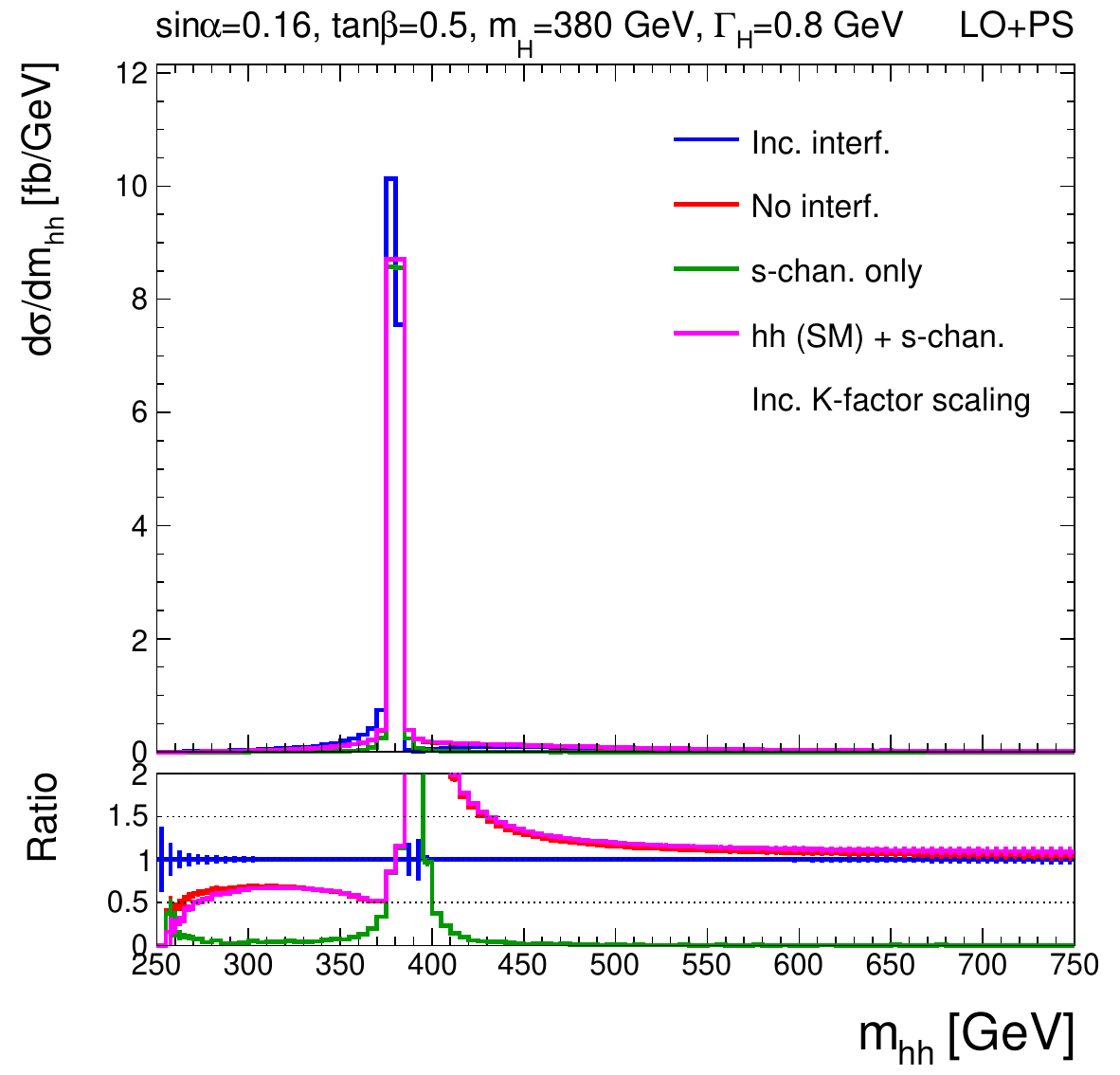}
  \includegraphics[width=0.48\textwidth]{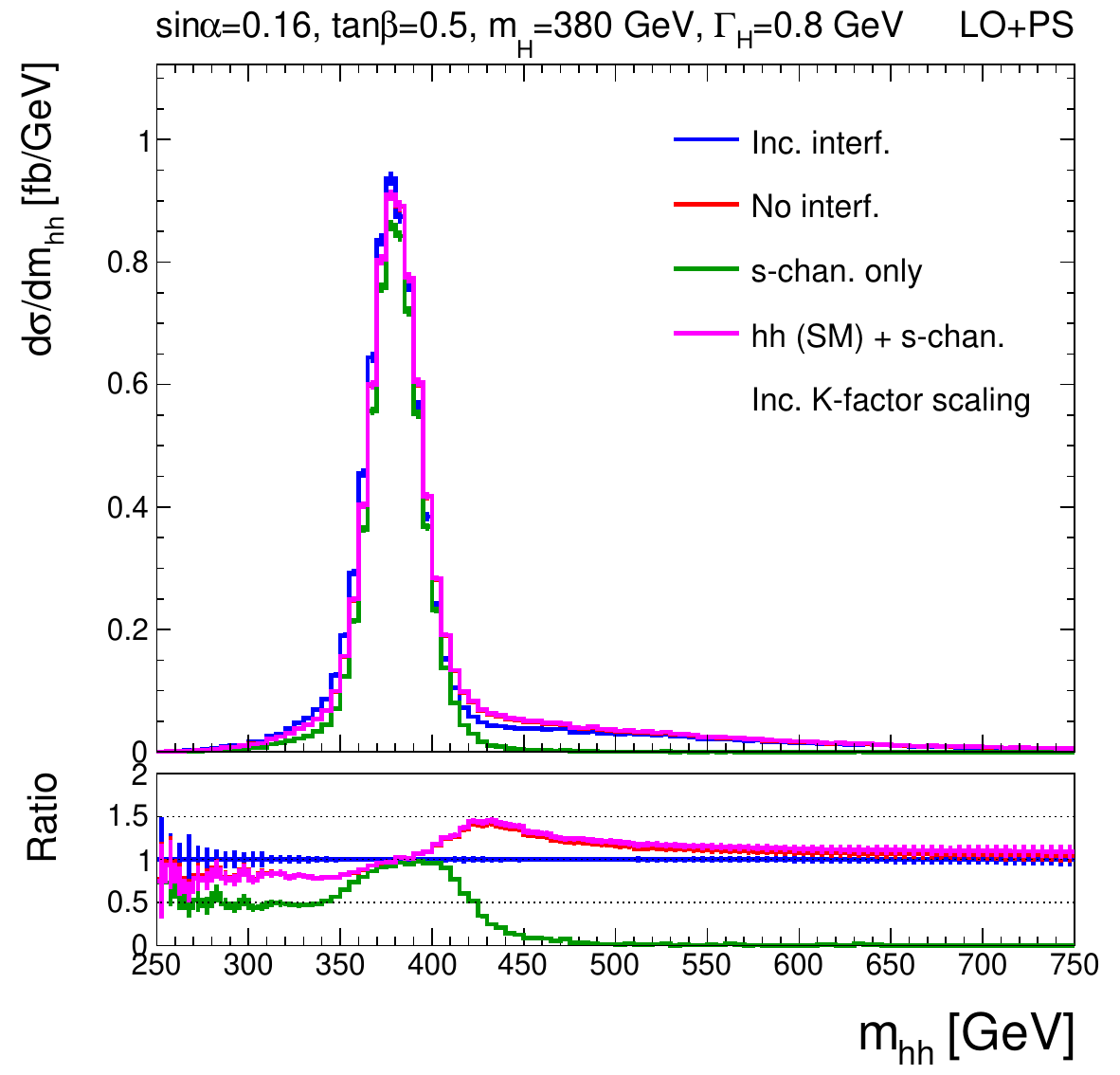}
  
\caption{The di-Higgs mass distributions for \BMc before (left) and after (right) experimental smearing.
The blue lines show the full di-Higgs spectrum including all interference terms. This is compared to the spectrum excluding the \SHBox and \SHSh interference terms shown in red, the \PS $s$-channel spectrum (\SH) shown in green, and the incoherent sum of \SH and the SM di-Higgs spectrum in magenta. }
\label{fig:mass_BMc}
\end{figure*}

The \mhh distributions for \BMb and \BMc are shown in Figures~\ref{fig:mass_BMb} and \ref{fig:mass_BMc}, respectively. In general, the effect of the interference in these BMs is visibly smaller than for \BMa, especially after the experimental smearing is applied. However, comparing the ratios in the figures we observe that the relative differences away from the resonant peak are actually quite similar to \BMa, but as the relative contribution of the non-resonant component compared to the \SH contribution is smaller the effect is less noticeable. 

For \BMc, we observe only a small impact from neglecting the interference, and even the \SH-only assumption is fairly close to the full distribution in this case. This is somewhat expected since the non-resonant contribution compared to the \SH component is small in this scenario  and so the $\intrel<1\%$ requirement imposed when selecting the point means that the interference manifests almost exclusively as a shift in the position of the resonant peak. The experimental smearing then largely washes out this shift reducing the impact of the interference on the distribution.

\begin{figure}[htbp]
  \centering\includegraphics[width=0.48\textwidth]{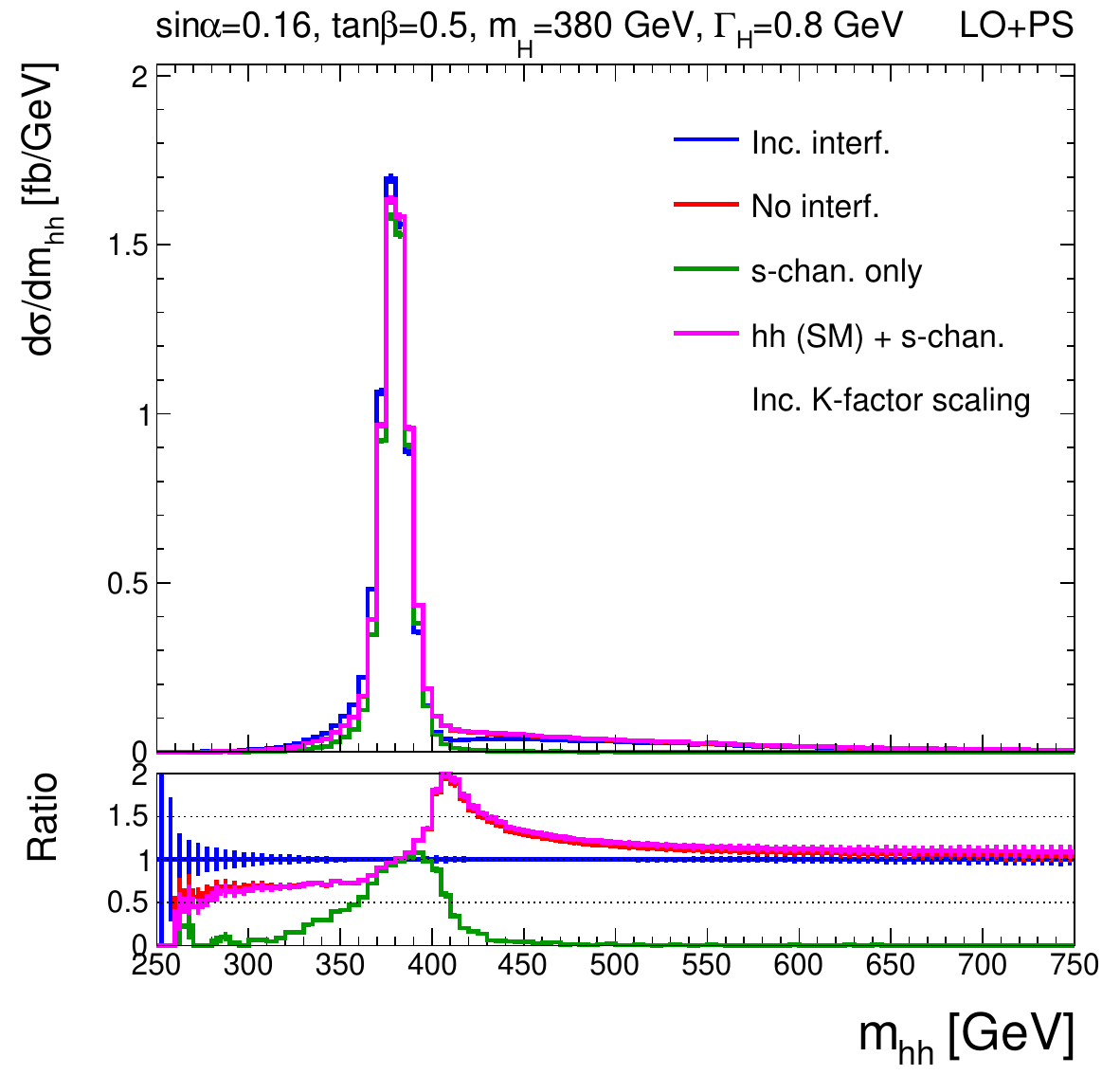}
\caption{The di-Higgs mass distributions for \BMc after experimental smearing while artificially improving the \mhh resolution by a factor of two compared to Figure~\ref{fig:mass_BMc}.
The blue lines show the full di-Higgs spectrum including all interference terms. This is compared to the spectrum excluding the \SHBox and \SHSh interference terms shown in red, the \PS $s$-channel spectrum (\SH) shown in green, and the incoherent sum of \SH and the SM di-Higgs spectrum in magenta.}
\label{fig:mass_BMc_optimistic}
\end{figure}

Since it is possible that the experimental collaborations could improve the \mhh resolution in the future, we investigated if such an improvement would significantly change the conclusion drawn in this case. To this end, we artificially improved the \mhh resolution by a factor of two and display the resulting \mhh distribution in Figure~\ref{fig:mass_BMc_optimistic}. In this case, we observe a slightly larger difference between the full spectrum and the no-interference scenario.

\subsection{\BMd and \BMe}

We define two BM points, \BMd and \BMe, to investigate scenarios where the non-resonant contribution to the di-Higgs spectrum is significant, independent of the interference. These BMs are selected by looking for the points where the relative fraction of non-resonant events close to the resonant peak is maximal. The fraction is defined as $f=\frac{\sigma_{\mathrm{woH}}}{\sigma_{\SH}+\sigma_{\mathrm{woH}}}$, 
where $\sigma_{\mathrm{woH}}$ is the cross-section for the $\Box+\Sh+\ShBox$ process requiring $0.9<\frac{\mhh^{\mathrm{truth}}}{\mH}<1.1$ to select events with $\pm 10\%$ of \mH.
This definition thus ignores the effect of the  interference of the \PS with the non-resonant terms. 
Two BMs are defined, since the scan returns different points depending on whether we use $L=400~\invfb$ or $L=3000~\invfb$. \BMd was found for the $L=400~\invfb$ scan which has $f=28.8\%$. The model parameters are $\mH=560\,\GeV$, $\sina=-0.16$, and $\tanb=0.5$.
\BMe was defined for the $L=3000~\invfb$ which estimates $f=54.5\%$. The model parameters in the case are $\mH=500\,\GeV$, $\sina=0.08$, and $\tanb=0.5$.

\begin{figure*}[htbp]
  \includegraphics[width=0.48\textwidth]{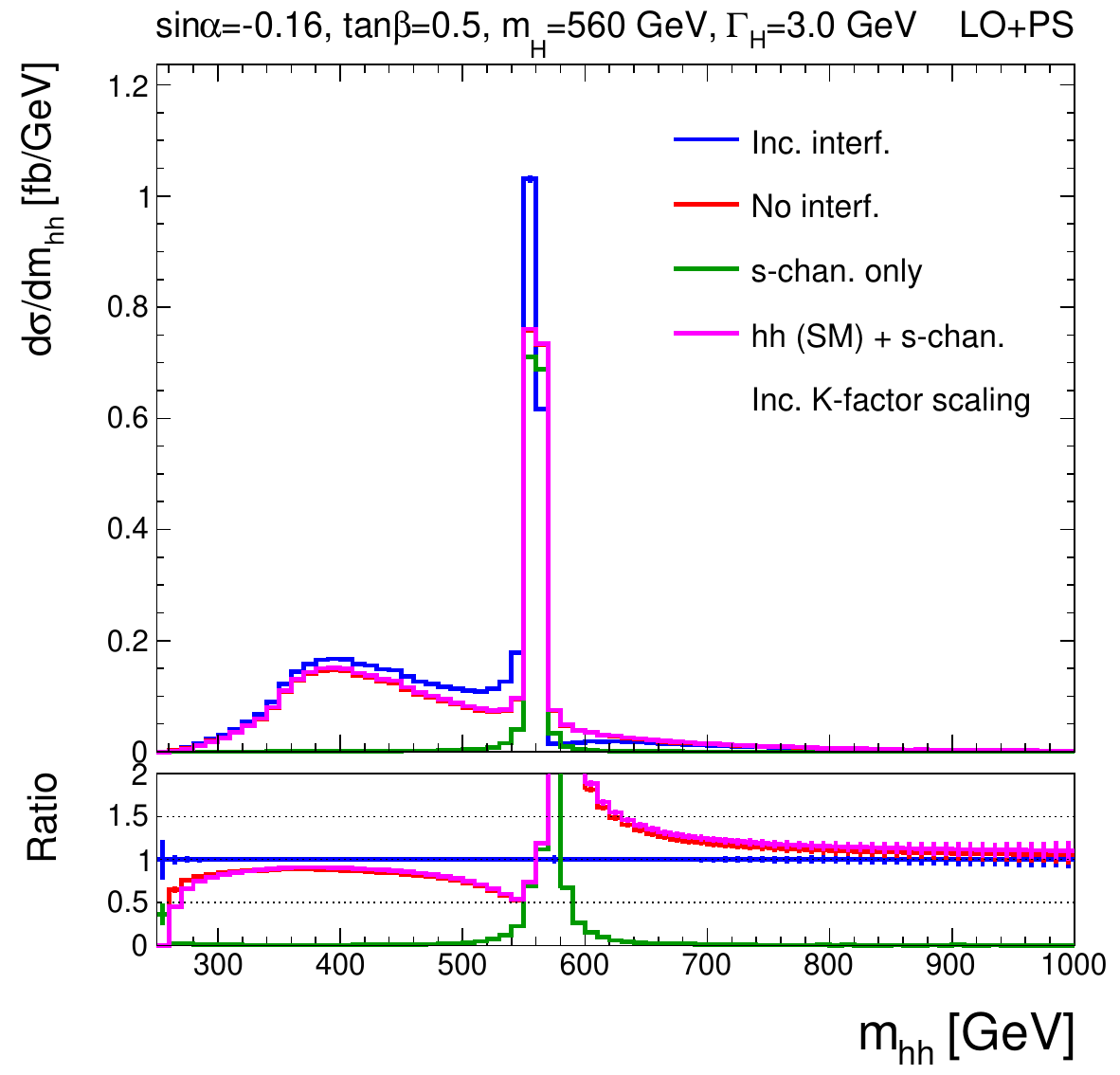}
  \includegraphics[width=0.48\textwidth]{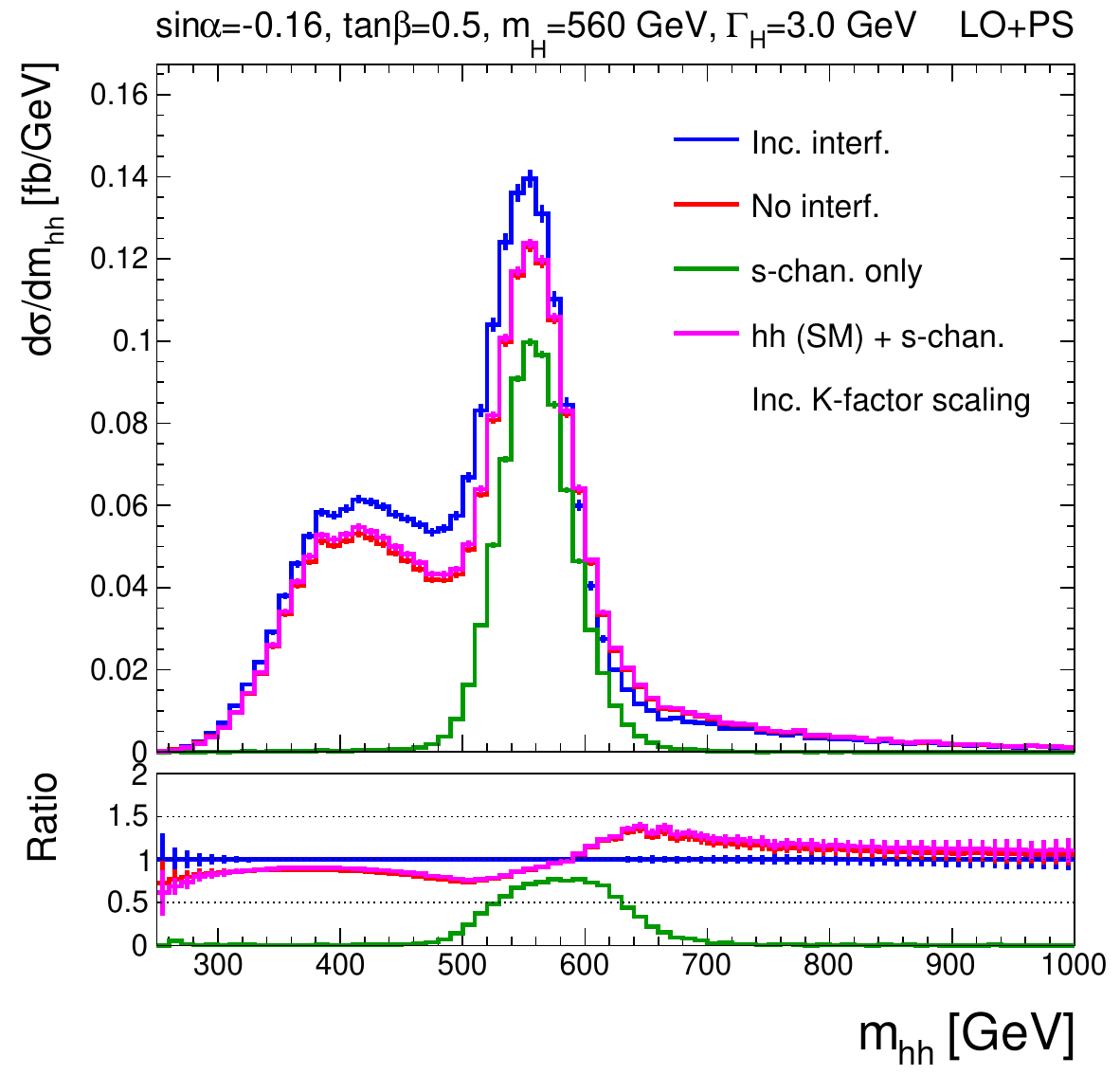}
\caption{The di-Higgs mass distributions for \BMd before (left) and after (right) experimental smearing.
The blue lines show the full di-Higgs spectrum including all interference terms. This is compared to the spectrum excluding the \SHBox and \SHSh interference terms shown in red, the \PS $s$-channel spectrum (\SH) shown in green, and the incoherent sum of \SH and the SM di-Higgs spectrum in magenta.}
\label{fig:mass_BMd}
\end{figure*}

\begin{figure*}[htbp]
  \includegraphics[width=0.48\textwidth]{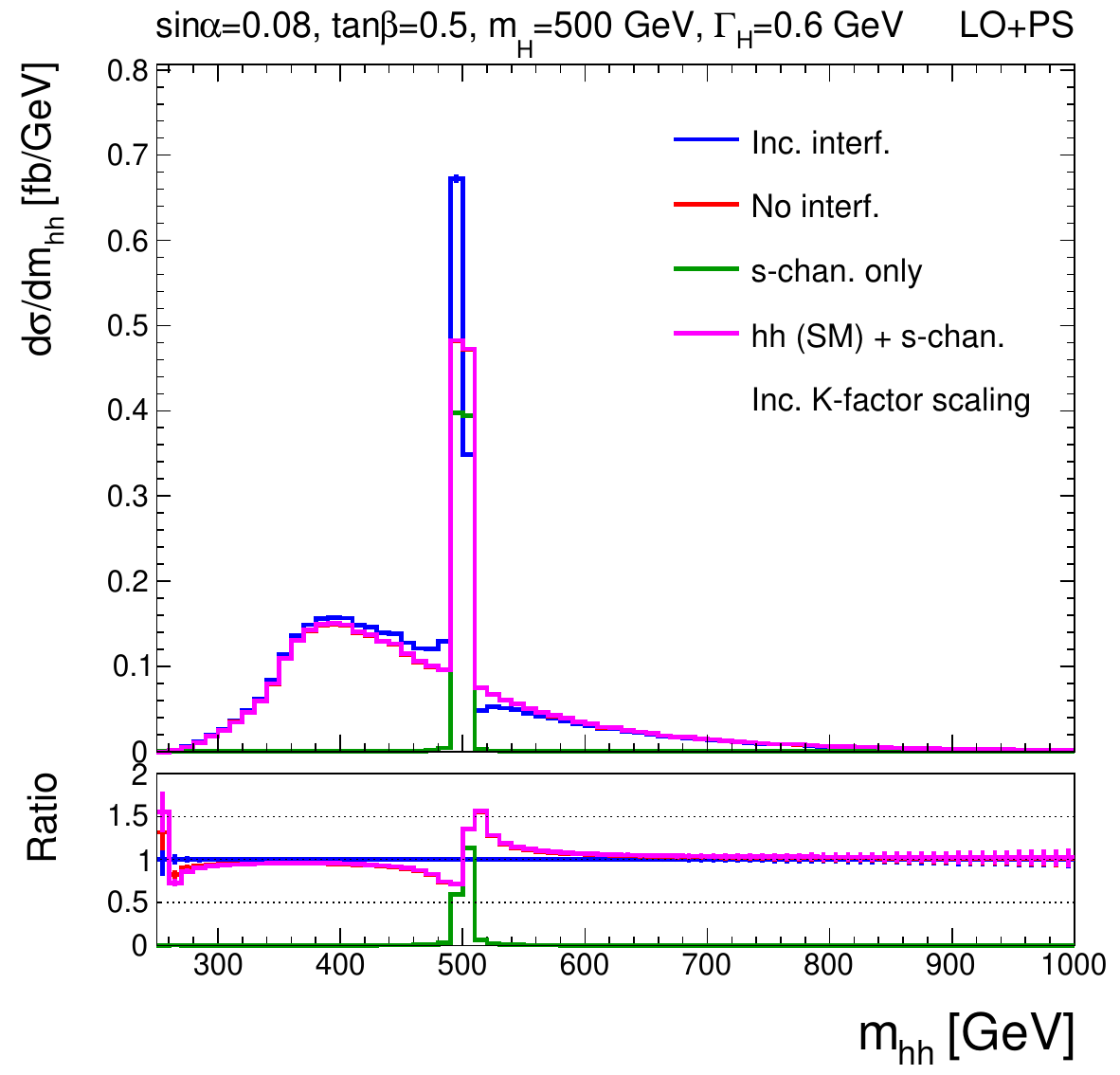}
  \includegraphics[width=0.48\textwidth]{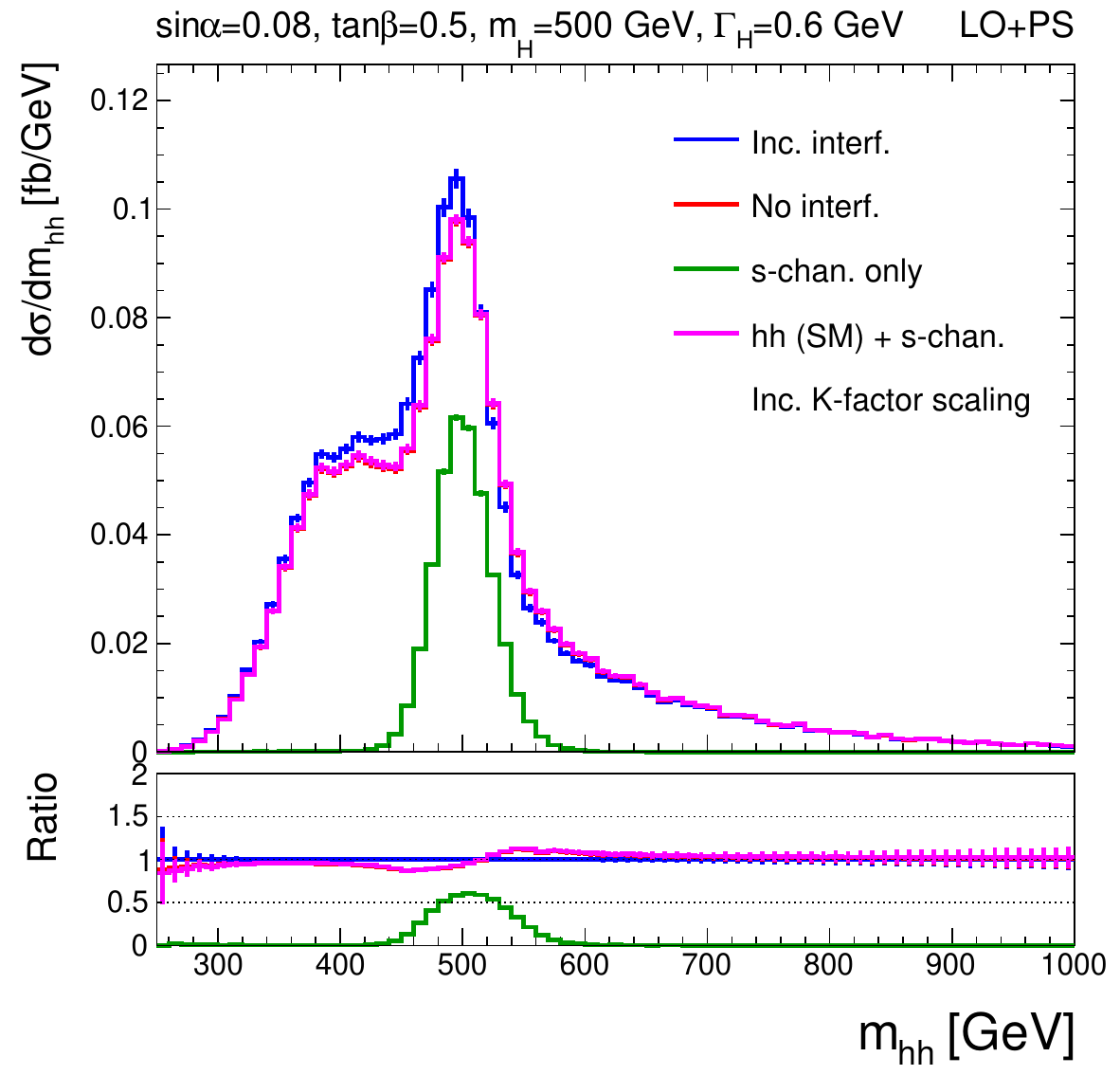}
\caption{The di-Higgs mass distributions for \BMe before (left) and after (right) experimental smearing.
The blue lines show the full di-Higgs spectrum including all interference terms. This is compared to the spectrum excluding the \SHBox and \SHSh interference terms shown in red, the \PS $s$-channel spectrum (\SH) shown in green, and the incoherent sum of \SH and the SM di-Higgs spectrum in magenta.}
\label{fig:mass_BMe}
\end{figure*}

Figures~\ref{fig:mass_BMd} and \ref{fig:mass_BMe} show the \mhh distributions in \BMd and \BMe, respectively. In both cases there is a significant difference between the \SH-only assumption and both the full model and no-interference scenarios. \BMd also shows a sizeable effect due to the interference. In contrast, the impact of the interference in \BMe is quite small. However, the non-resonant contribution is even more significant in this case and it clearly has to be taken into account by future experimental searches. 

\subsection{\BMf and \BMg}

We define two BM points to study the interference effects for large values of \mH. We define the BMs, \BMf and \BMg, as the points with maximal values of \mH within reach of the LHC for $L=400~\invfb$ and $L=3000~\invfb$, respectively. \BMf was found for $\mH=680\,\GeV$, $\sina=0.16$, and $\tanb=1.0$. The model parameters for \BMg are $\mH=870\,\GeV$, $\sina=0.15$, and $\tanb=1.1$. 

\begin{figure*}[htbp]
  \includegraphics[width=0.48\textwidth]{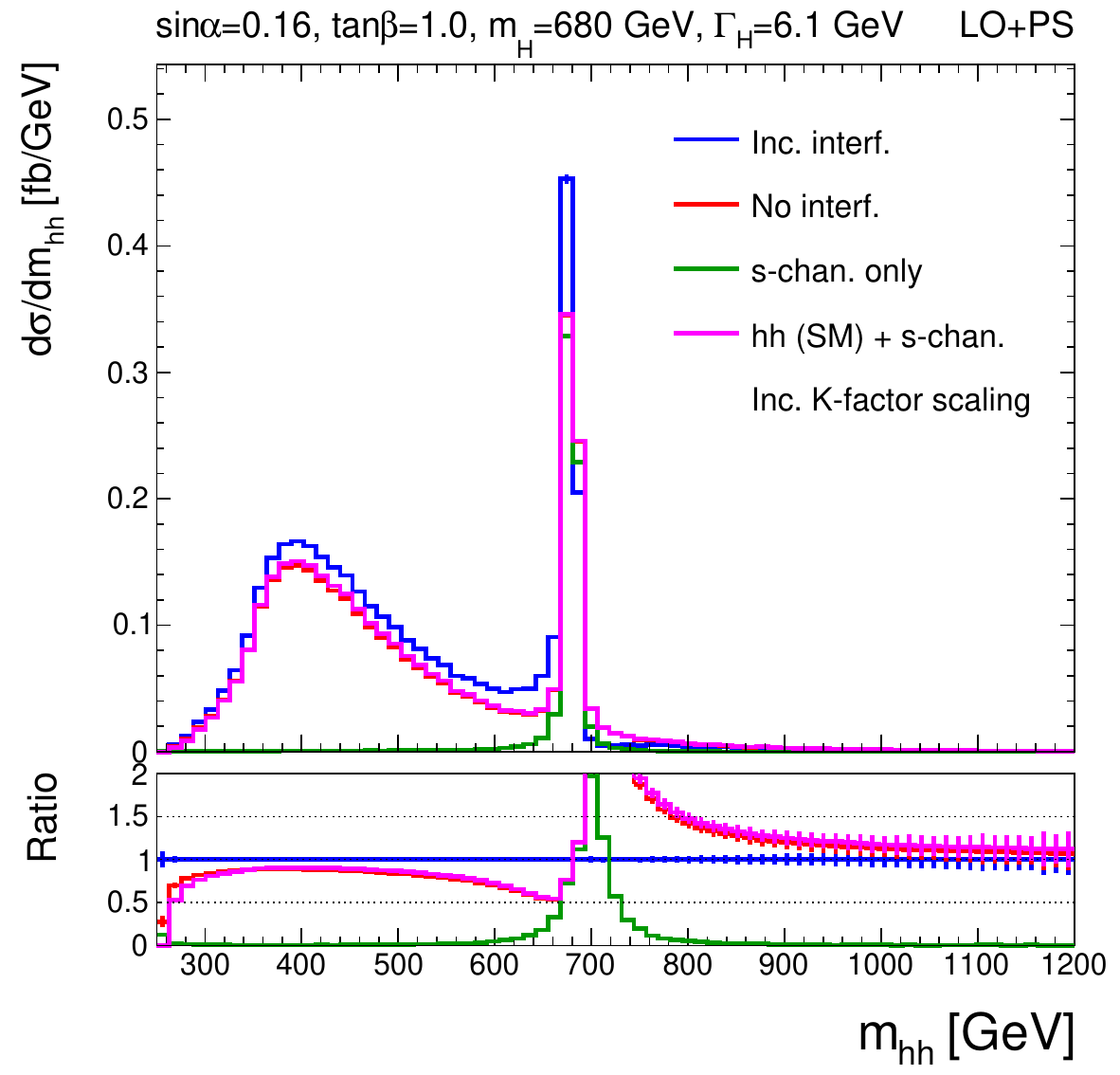}
  \includegraphics[width=0.48\textwidth]{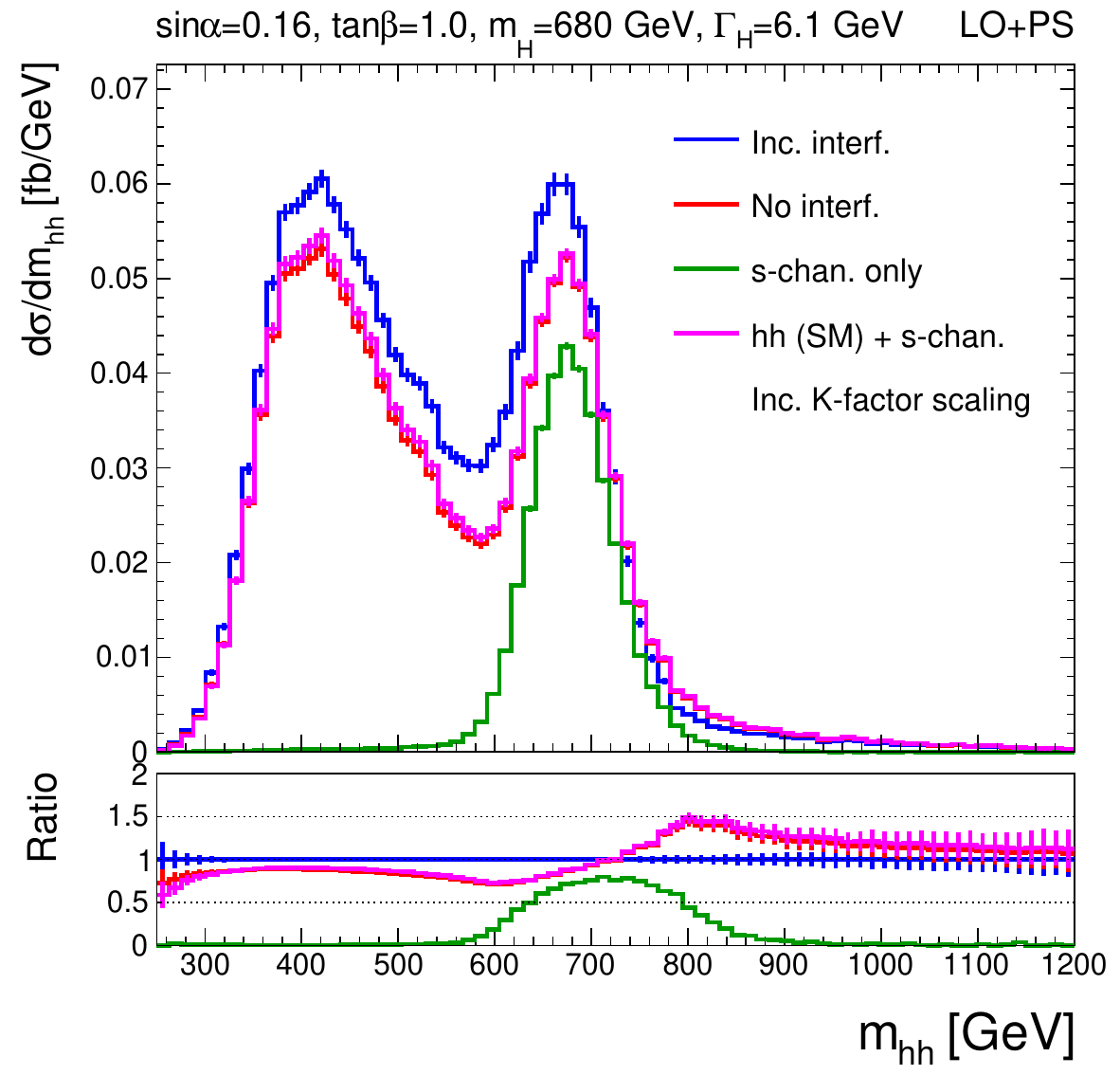}
\caption{The di-Higgs mass distributions for \BMf before (left) and after (right) experimental smearing.
The blue lines show the full di-Higgs spectrum including all interference terms. This is compared to the spectrum excluding the \SHBox and \SHSh interference terms shown in red, the \PS $s$-channel spectrum (\SH) shown in green, and the incoherent sum of \SH and the SM di-Higgs spectrum in magenta.}
\label{fig:mass_BMf}
\end{figure*}

\begin{figure*}[htbp]
  \includegraphics[width=0.48\textwidth]{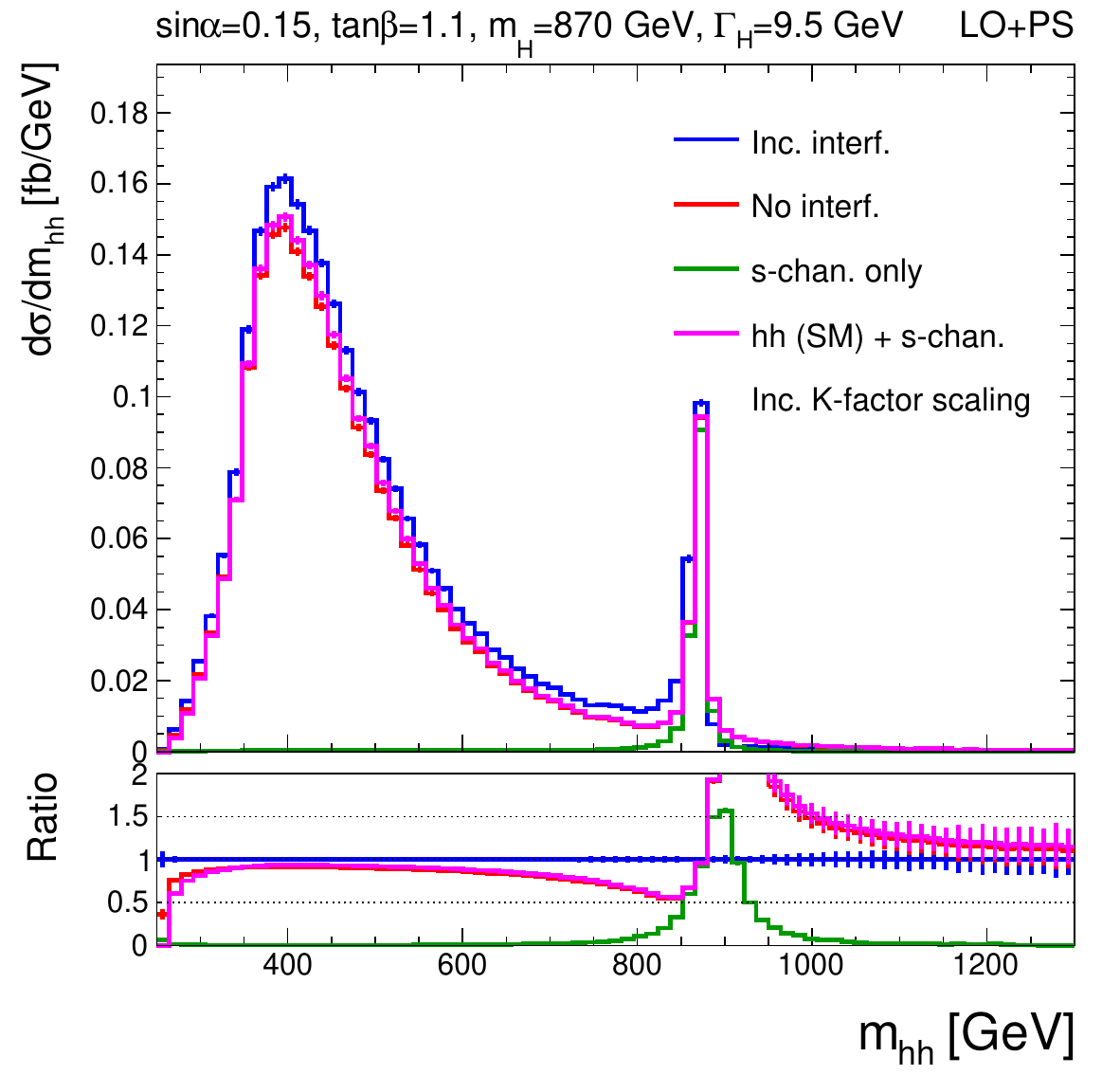}
  \includegraphics[width=0.48\textwidth]{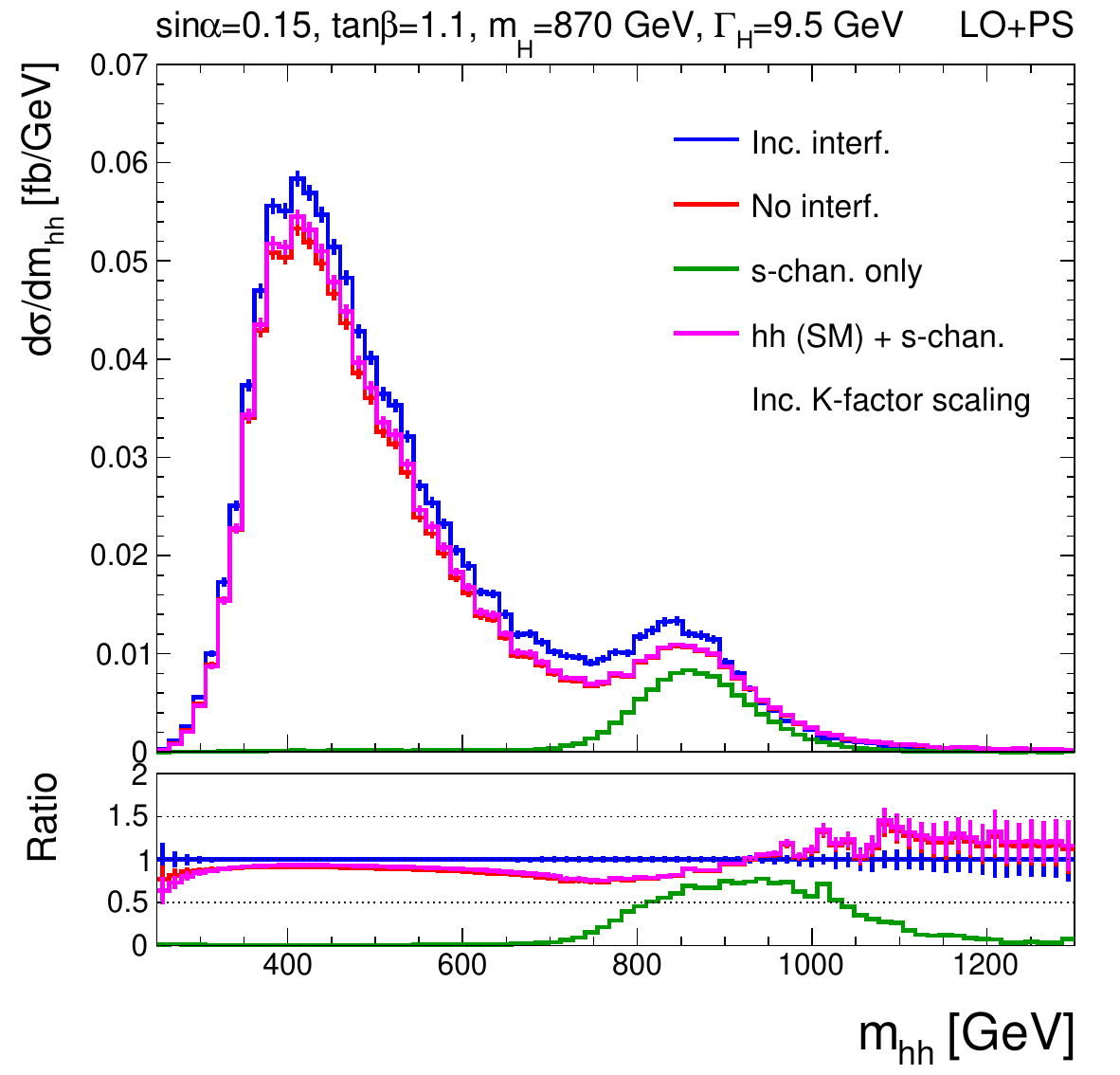}
\caption{The di-Higgs mass distributions for \BMg before (left) and after (right) experimental smearing.
The blue lines show the full di-Higgs spectrum including all interference terms. This is compared to the spectrum excluding the \SHBox and \SHSh interference terms shown in red, the \PS $s$-channel spectrum (\SH) shown in green, and the incoherent sum of \SH and the SM di-Higgs spectrum in magenta.}
\label{fig:mass_BMg}
\end{figure*}

The \mhh distributions are shown in Figures~\ref{fig:mass_BMf} and \ref{fig:mass_BMg}. Both BMs predict a more sizeable contribution to the total di-Higgs spectrum from the non-resonant contribution compared to \SH. However, we note that, as the resonant part peaks at higher mass, it may still drive the sensitivity for the experimental searches, as the backgrounds are probably also small for large \mhh. Once again, these distributions underline the need to take the non-resonant contributions into account in the experimental searches. 

\subsection{\BMh}

\BMh is defined as the point which has the largest deviation of \lamhhh from the SM value at LO. 
This main purpose of this BM is to investigate whether or not assuming SM-like couplings for the non-resonant contribution is a reasonable approximation. This may have practical implications for experimental searches, \ie fixing these couplings to SM values reduces the dimensionality of the parameter space that needs to be scanned.
The largest deviation for \lamhhh is found for $\mH=260\,\GeV$, $\sina=0.24$, and $\tanb=3.5$ which yields a value of $\kappalam=0.87$. 

\begin{figure*}[htbp]
  \includegraphics[width=0.48\textwidth]{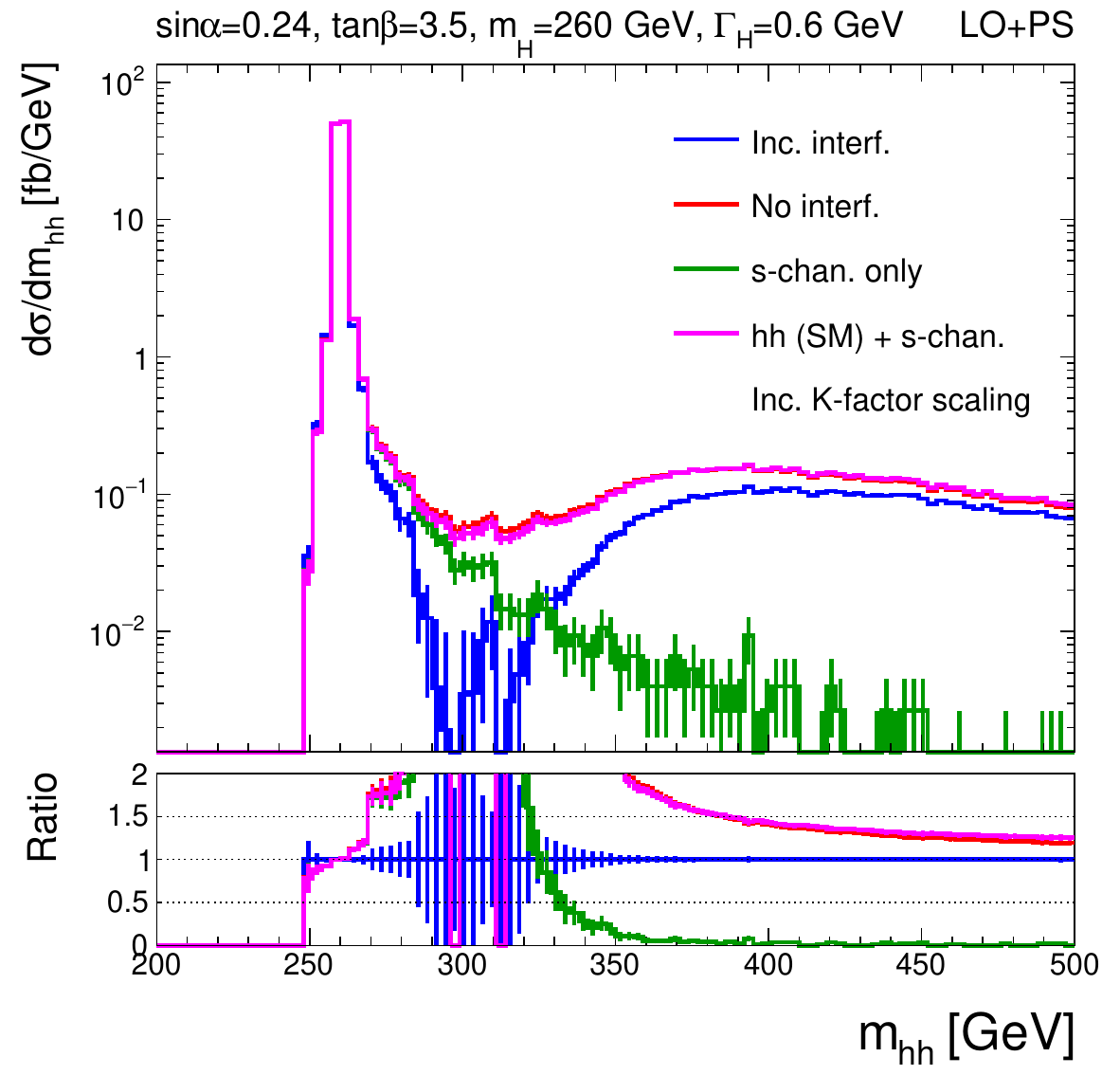}
  \includegraphics[width=0.48\textwidth]{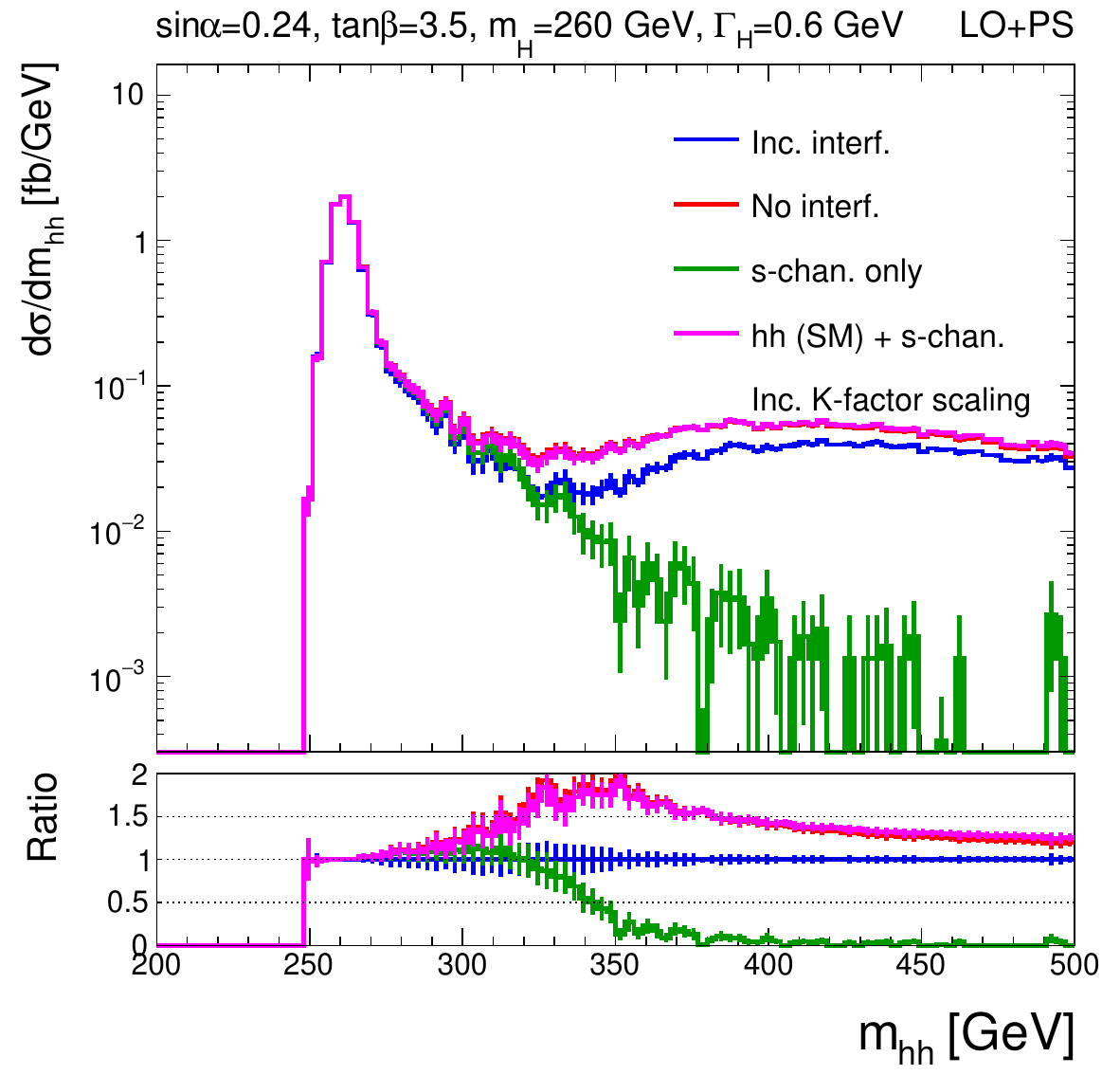}
  
\caption{The di-Higgs mass distributions for \BMh before (left) and after (right) experimental smearing.
The blue lines show the full di-Higgs spectrum including all interference terms. This is compared to the spectrum excluding the \SHBox and \SHSh interference terms shown in red, the \PS $s$-channel spectrum (\SH) shown in green, and the incoherent sum of \SH and the SM di-Higgs spectrum in magenta.}
\label{fig:mass_BMh}
\end{figure*}

\begin{figure}[htbp]
  \centering\includegraphics[width=0.48\textwidth]{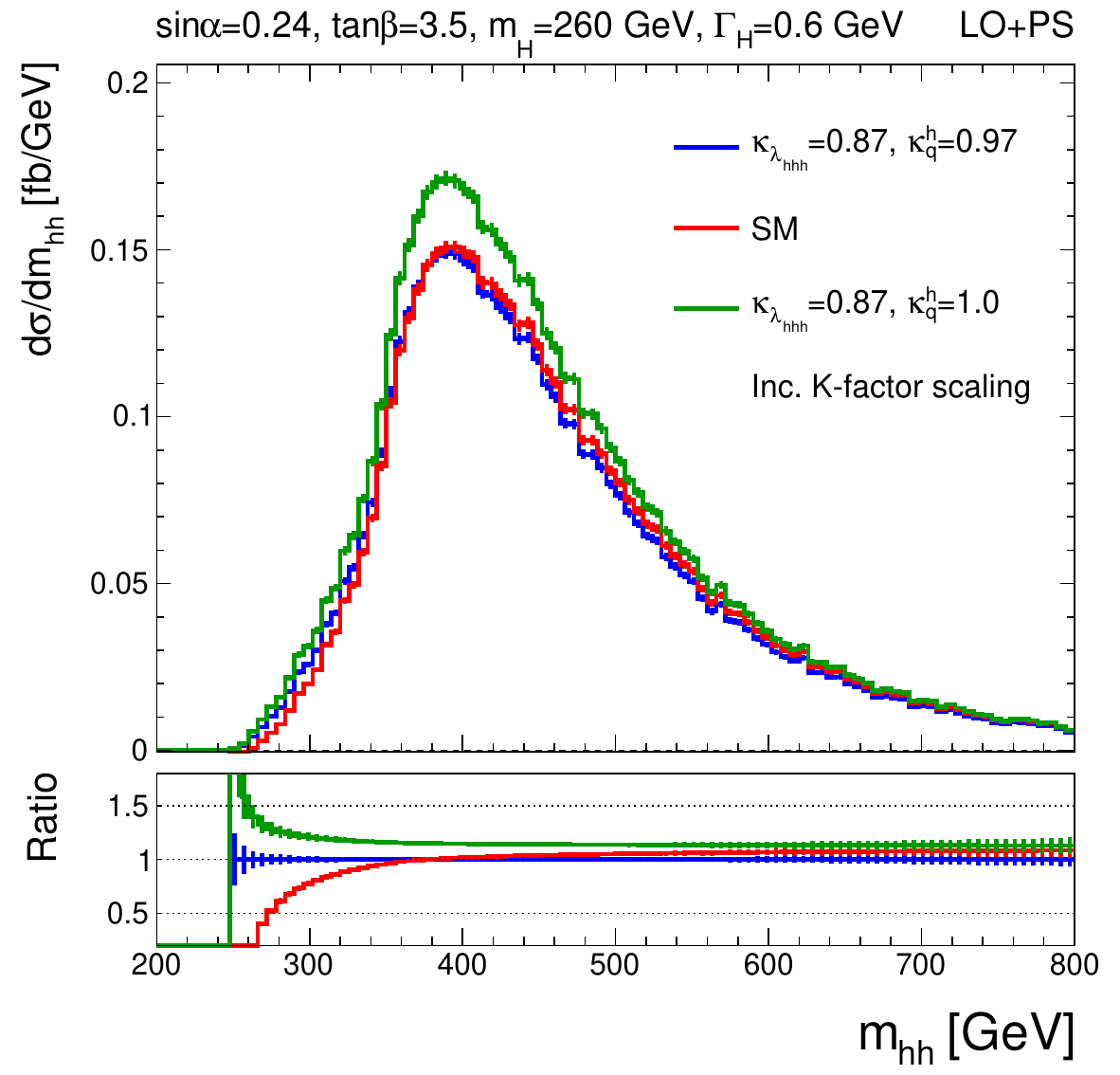}
\caption{The non-resonant contributions to the di-Higgs mass spectrum for \BMh before experimental smearing.
The blue line shows the case where both \kappalam and \kappat are set to the values predicted in \BMh ($\kappalam=0.87$, $\kappat=0.97$). This is compared to the SM spectrum shown in red and the spectrum when \kappat is set to the SM value and $\kappalam=0.87$ shown in green.}
\label{fig:nonres_comp_BMh}
\end{figure}

The \mhh distributions are shown in Figure~\ref{fig:mass_BMh}. We observe no major differences between the no-interference and SM+\SH distributions, implying that assuming SM couplings for the \Ph is a reasonable assumption in this case. We note that this conclusion does not imply that the interference effects are insignificant; it only indicates that fixing the \Ph couplings to the SM-values has a minimal impact on the distributions.
The non-resonant contribution to the spectrum is less noticeable in these plots as the resonant \SH peak is large. To study in more detail how the values of the \Ph couplings influence the non-resonant contribution, we remove the \SH, \SHBox, and \SHSh contributions to the spectrum, as shown in Figure~\ref{fig:nonres_comp_BMh}. The non-resonant spectrum for \BMh shown in blue is compared to the SM shown in red. We see only small differences between the two distributions, mainly impacting the low \mhh bins where the event yield is low. The differences in the distributions are somewhat smaller than one might expect given that \kappalam deviates from the SM value by a non-negligible amount. This is due to a small deviation in the value of \kappat, which is equal to 0.97 in this BM. This causes an accidental cancellation between the $\Box$ and \Sh components, which decrease in size for couplings $<1$, and  the \ShBox component which increases due to the destructive nature of the interference. The scenario where only \kappalam deviates from the SM and $\kappat=1$ is shown in green in Fig.~\ref{fig:nonres_comp_BMh} for comparison. Even in this scenario the deviation in the non-resonant yield is only $\approx 10\%$ so it is unlikely that the LHC experiments will be sensitive to the deviations in \kappalam in this model\footnote{We note that \BMh has since been excluded by the new combination presented by CMS in Ref.~\cite{CMS:2024phk} that is not yet included in \HiggsTools.  However, as the purpose of this benchmark is to study the most extreme cases where \kappalam deviates from the SM, we still consider this a useful scenario since the new bounds from CMS restrict \kappalam to be even closer to unity, and thus the conclusions drawn from this BM are still valid. We thank H. Bahl for useful discussions regarding this point. 
}.

\subsection{\BMi}

\begin{figure*}[htbp]
  \includegraphics[width=0.48\textwidth]{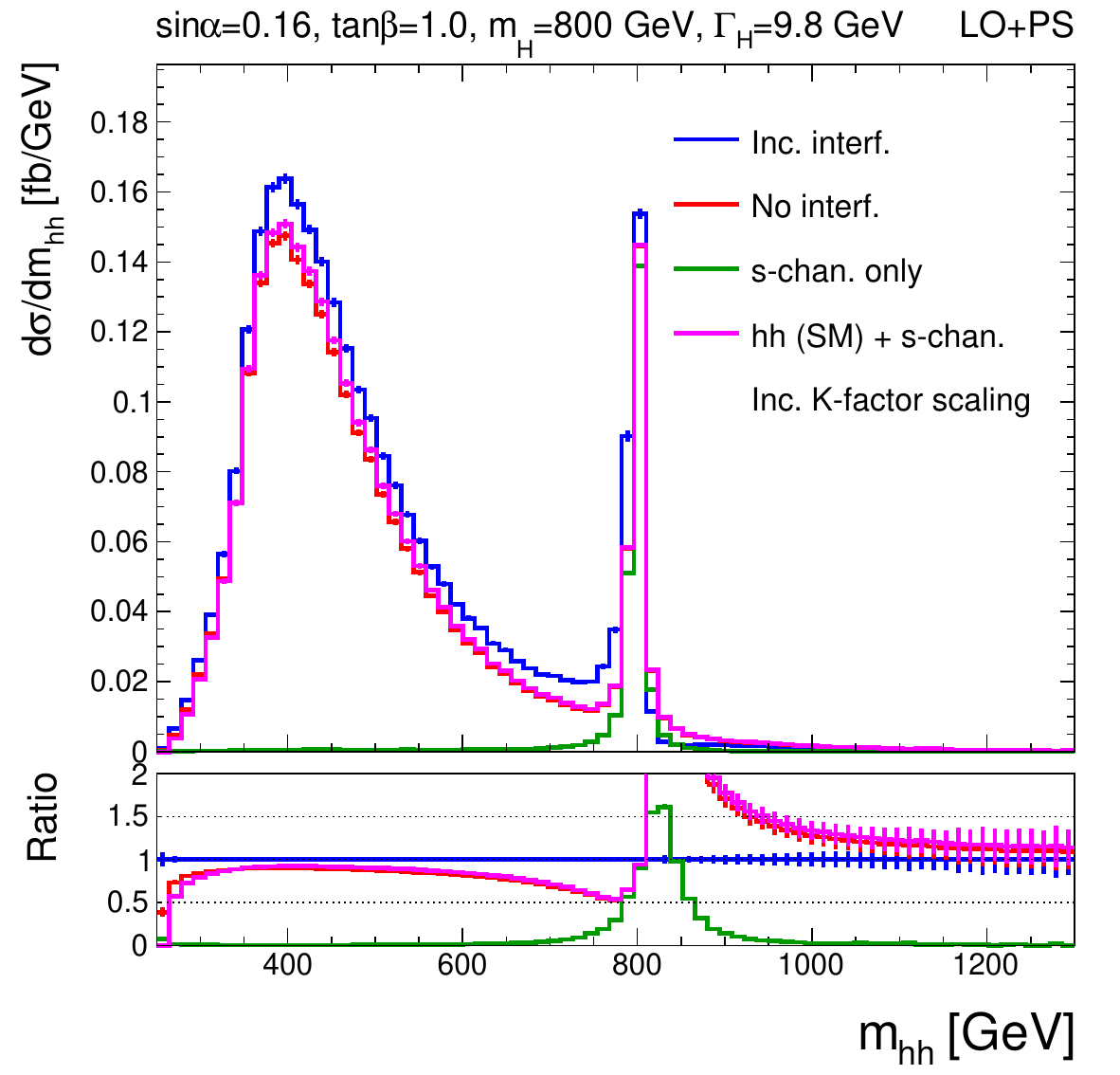}
  \includegraphics[width=0.48\textwidth]{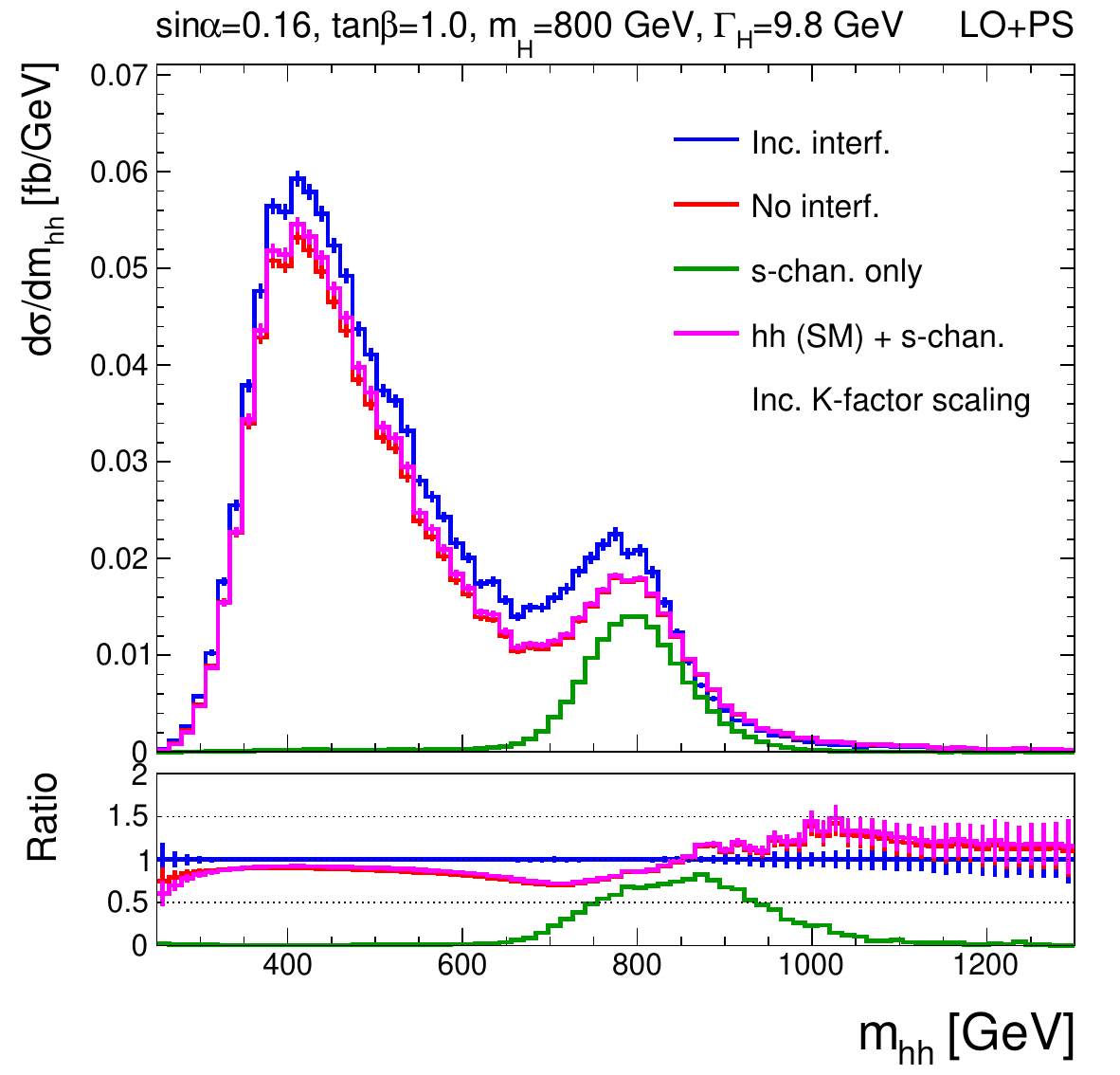}
  
\caption{The di-Higgs mass distributions for \BMi before (left) and after (right) experimental smearing.
The blue lines show the full di-Higgs spectrum including all interference terms. This is compared to the spectrum excluding the \SHBox and \SHSh interference terms shown in red, the \PS $s$-channel spectrum (\SH) shown in green, and the incoherent sum of \SH and the SM di-Higgs spectrum in magenta.}
\label{fig:mass_BMi}
\end{figure*}

The final BM, \BMi, is defined as the point which has the largest value of \relwH, which is found for $\mH=800\,\GeV$, $\sina=0.16$, and $\tanb=1.0$. The value of \relwH is 1.2\%. 
The \mhh distributions are shown in Figures~\ref{fig:mass_BMi}. 

\begin{figure*}[htbp]
  \includegraphics[width=0.48\textwidth]{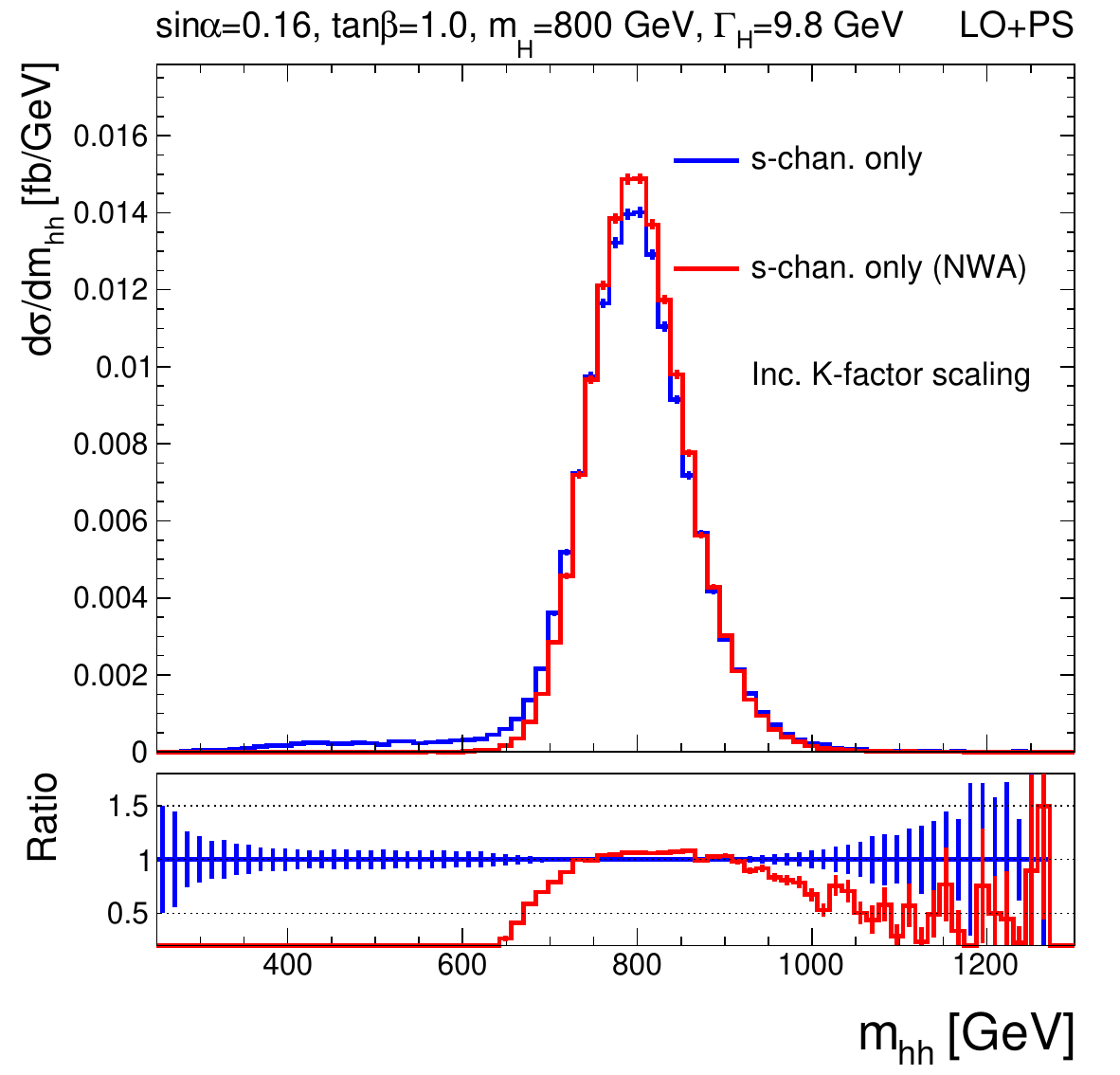}
  \includegraphics[width=0.48\textwidth]{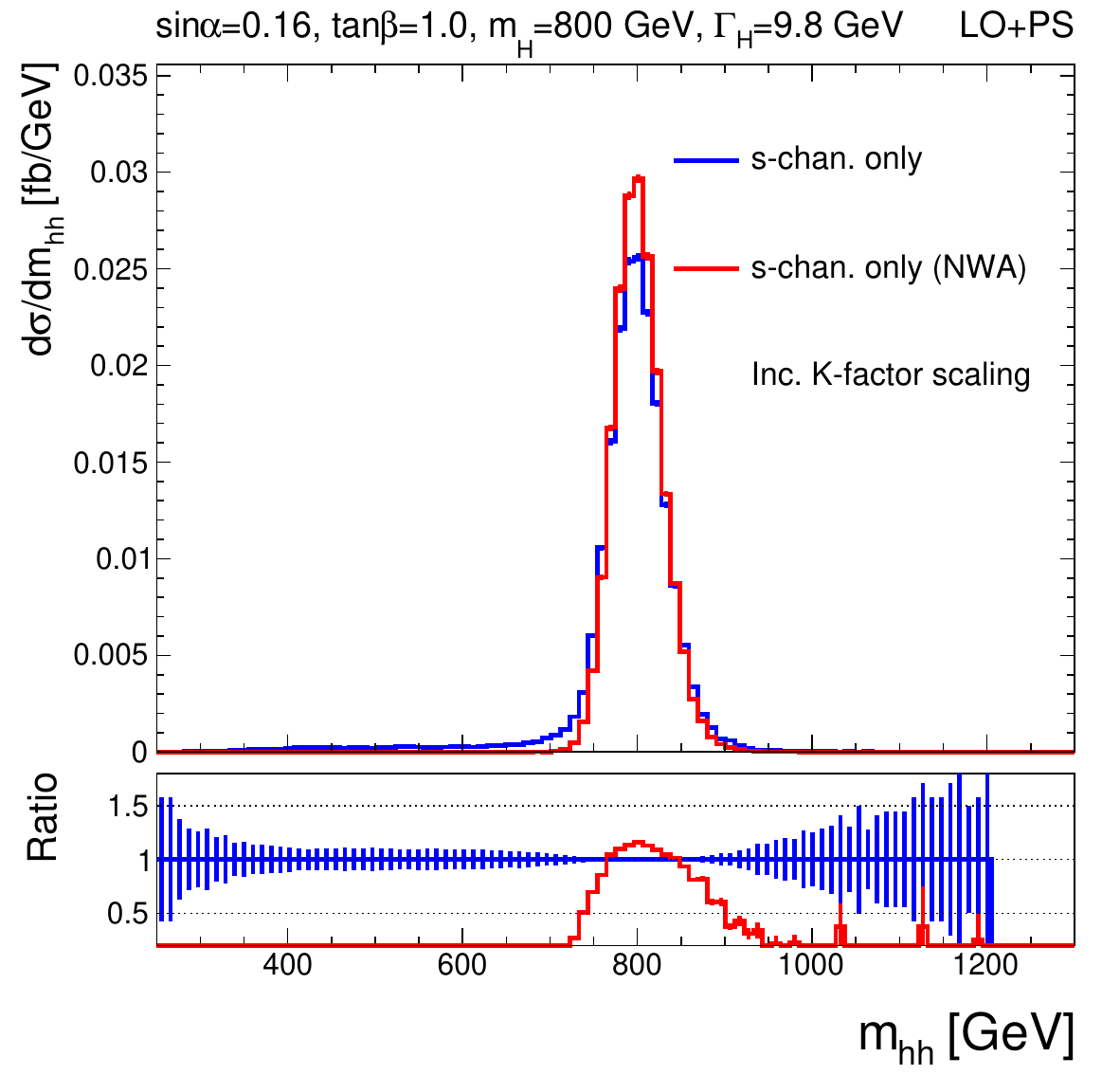}
\caption{The di-Higgs mass distributions for \BMi after experimental smearing are displayed for the $s$-channel only (\SH) contributions. 
The blue lines show  distributions obtained using the correct width ($\wH=9.8\,\GeV$), while the red lines display the distributions obtained using the NWA with $\wH=20\,\MeV$ and rescaling the branching fractions by $(20\,\MeV )/\wH$ as described in the text.
The left plot shows the distributions for the nominal assumed experimental resolution while the right plot shows the distributions after artificially improving the \mhh resolution by a factor of two. 
}

\label{fig:nwa_BMi}
\end{figure*}

The main purpose of this BM is to investigate whether finite width effects may invalidate the procedure followed by several experimental searches whereby a narrow width is assumed for the generated \PS signal. To investigate this question, we display the \mhh distributions for the \SH contribution in Figure~\ref{fig:nwa_BMi}. The figure compares the \mhh distributions predicted for samples generated with the correct width to samples generated assuming a narrow-width of $20\,\MeV$. For the latter, the sample yield is scaled by $(20\,\MeV )/\wH$ to account for the different branching fractions. The distributions are shown for our nominal assumption about the experimental resolution (left plot) and a more optimistic experimental resolution, which is improved by a factor of two (right plot). 

The \mhh distributions obtained for the nominal smearing are in quite good agreement. The difference between the two distributions is only around 7\% at the maximum of the peak, and thus the experiments would probably not be able to resolve the differences. 
The optimistic scenario is intended to investigate the impact of potential future improvements to the experimental resolution if the experimental collaborations are able to improve the \mhh resolution in the future and/or if a channel with a better resolution is used instead (\eg $bb\gamma\gamma$).  
As expected, when the resolution is bettered the experiments become more sensitive to the width effects, and the difference between the \mhh distributions can be as large as 14\% at the maximum of the peak. 
While this may lead to small modifications to the measured experimental limits, it would probably not significantly alter the overall conclusions of the searches and thus we conclude that the narrow-width assumption for generation of the \SH contributions is a reasonable approximation for the singlet model. 
However, we stress that this conclusion is specific to the singlet model since other beyond the SM scenarios can still accommodate more sizeable widths $\mathcal{O}(10\%)$, and these scenarios will require a proper treatment of finite width effects.  
It is also important to stress that, while the finite width effects for the \SH-only process can be accounted for via the branching fraction rescaling, this is not possible for the \SHBox and \SHSh interference terms. As demonstrated previously in Figure~\ref{fig:width_comp}, the width modifies these distributions in a non-trivial way (\eg differentially in \mhh), and thus it is not possible to account for this by the application of a constant scale factor as for the \SH contribution.

\section{Summary}
\label{sec:summary}

We have presented the latest experimental and theoretical limits on the parameter space of the real singlet extension of the SM Higgs sector. The tightest bounds on the model at high \mH come from the measurements of the $W$-boson mass, whereas the bounds at low \sina are driven by direct searches for $\PS\rightarrow VV/\Ph\Ph$ at the LHC. 

We have investigated in detail interference effects between resonant and non-resonant contributions, where the former originates from the production of a heavy additional Higgs boson which subsequently decays into two lighter SM-like Higgs bosons ($\PS\rightarrow\Ph\Ph$).  We have studied various kinematic observables, such as the commonly investigated di-Higgs invariant mass ($m_{hh}$), as well as a number of other variables that are relevant for the respective experimental searches. 
In particular, we find that the transverse momenta of the two SM-like Higgs bosons have a comparable sensitivity to interference effects. 

We have identified regions of the parameter space where interference effects between the non-resonant and resonant diagrams are sizeable. 
In order to account for NLO kinematics as well as interference effects, we have proposed a prescription that includes both, where we find good agreement with the correct NLO description when available. Using our methodology, we have defined benchmark scenarios exhibiting interesting features of the model. 
For several benchmarks scenarios, we find that the contribution of the non-resonant di-Higgs diagrams is significant and in some cases even dwarfs the resonant component. 
We have also observed that interference effects have a non-negligible effect on both the cross-sections and differential distributions. 
We strongly encourage the experimental collaborations to take these effects into account in future searches for additional Higgs bosons. 
Furthermore, the correct physical prediction of the total width needs to be used in realistic simulation as the width can crucially impact the rates as well as the relevance of the interference term. 
We have presented a publicly available tool, \texttt{HHReweighter}~\cite{reweight_repo}, for performing matrix-element reweighting that allows such interference effects to be taken into account in a computationally efficient way. 

Finally, we want to point out that the model studied here is already vastly explored by the experimental collaborations, and very often limits are set neglecting the effects discussed here. Therefore, even for such simple scenarios, results as currently presented should be taken with a grain of salt and in fact be reinterpreted using e.g. our reweighting tool. It would also be desirable to have experimental results presented e.g. in multi-dimensional grids, the simplest being additional results for varying widths for a given new scalar mass. Of course for other new physics scenarios the respective parameter space would have to be extended.

\begin{appendices}
\counterwithin{figure}{section}
\renewcommand{\thesection}{\Alph{section}}
\renewcommand{\thefigure}{\Alph{section}.\arabic{figure}}

\section{Interference evaluation in an additional observable}
\label{sec:AppendixA}

\begin{figure*}[htbp]
    \centering
    \subfloat[]{\includegraphics[width=0.46\textwidth]{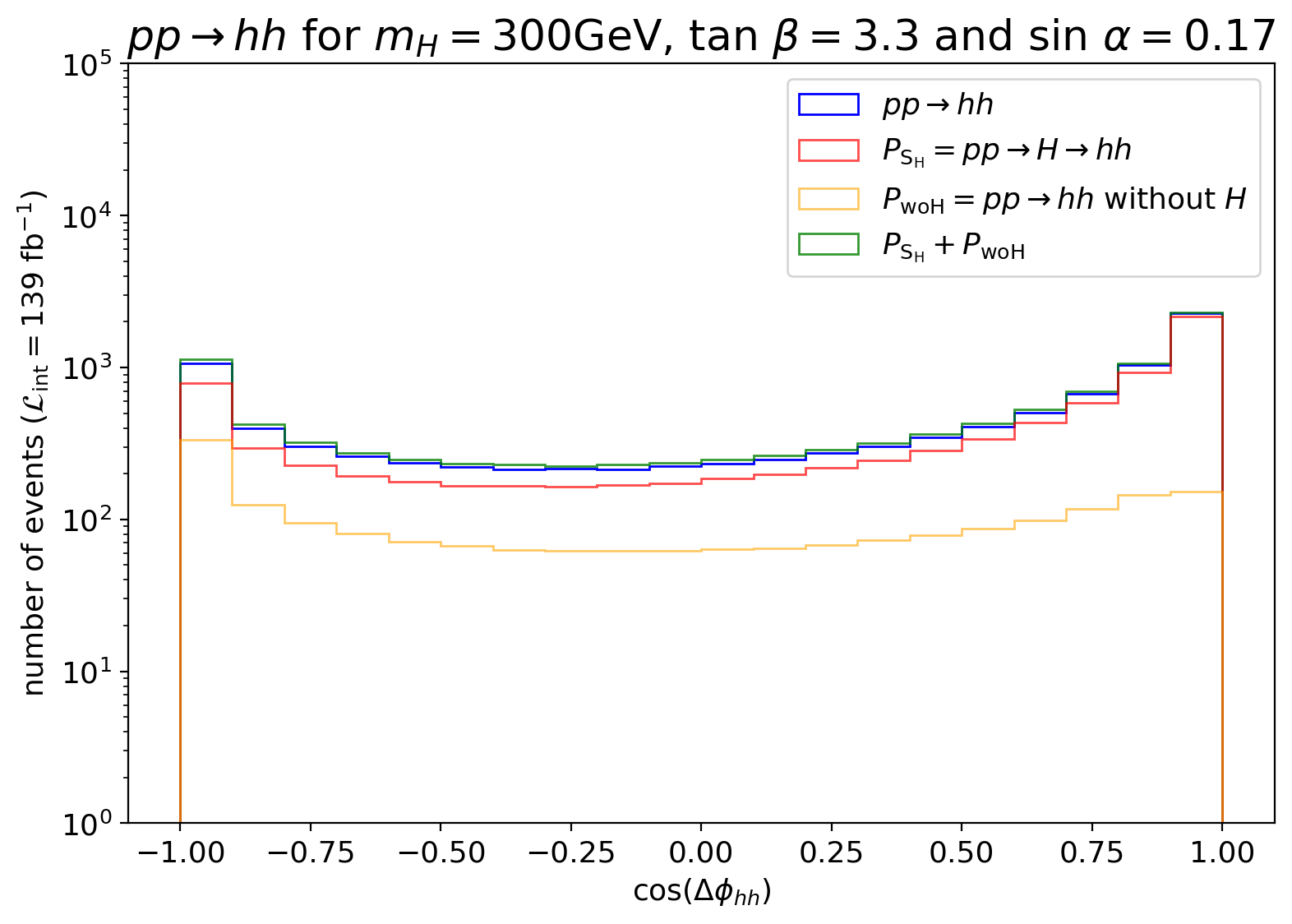}\label{figure_300_benchmark_varphi}}\quad
    \subfloat[]{\includegraphics[width=0.46\textwidth]{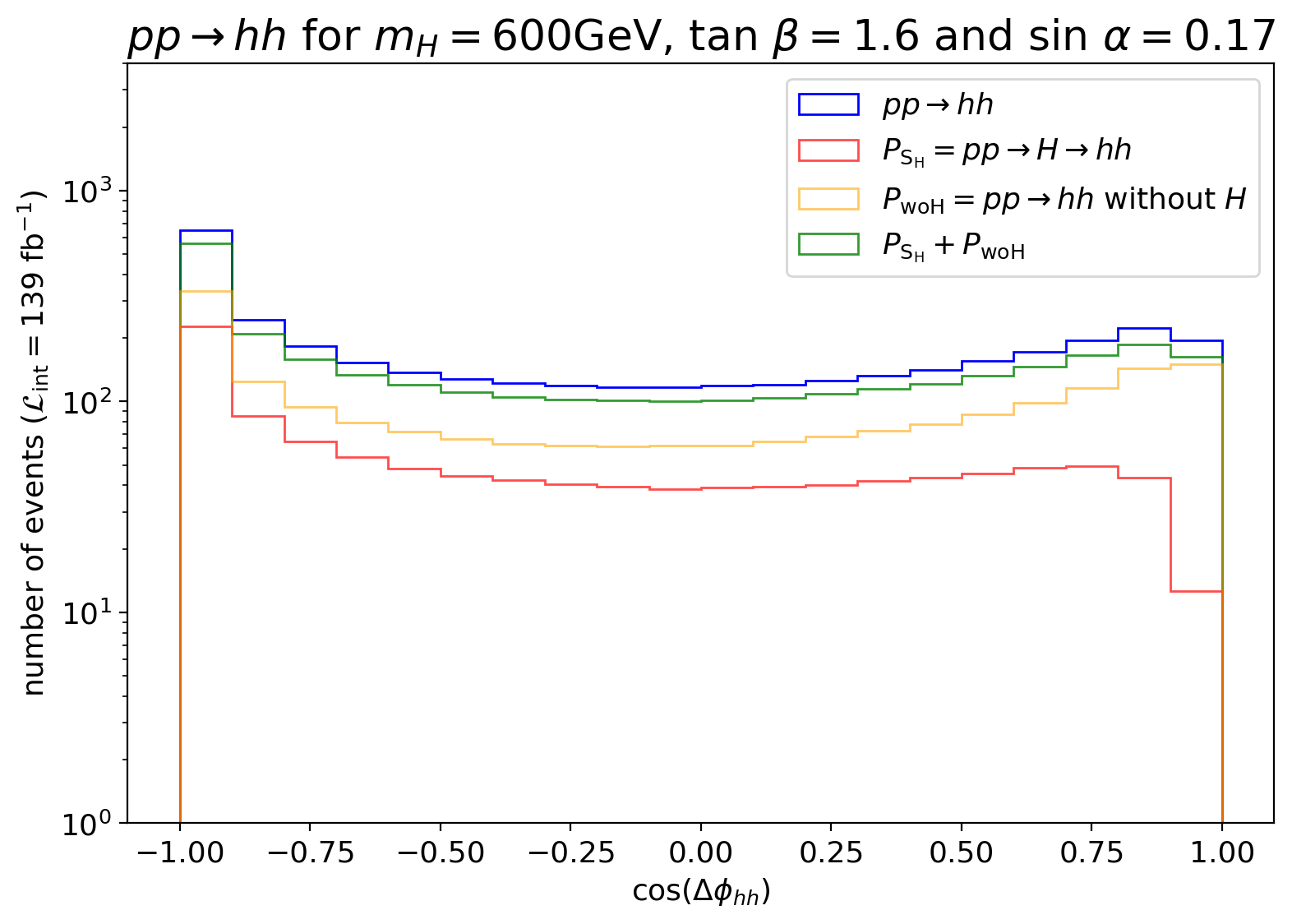}\label{figure_600_benchmark_varphi}}\quad
    \subfloat[]{\includegraphics[width=0.46\textwidth]{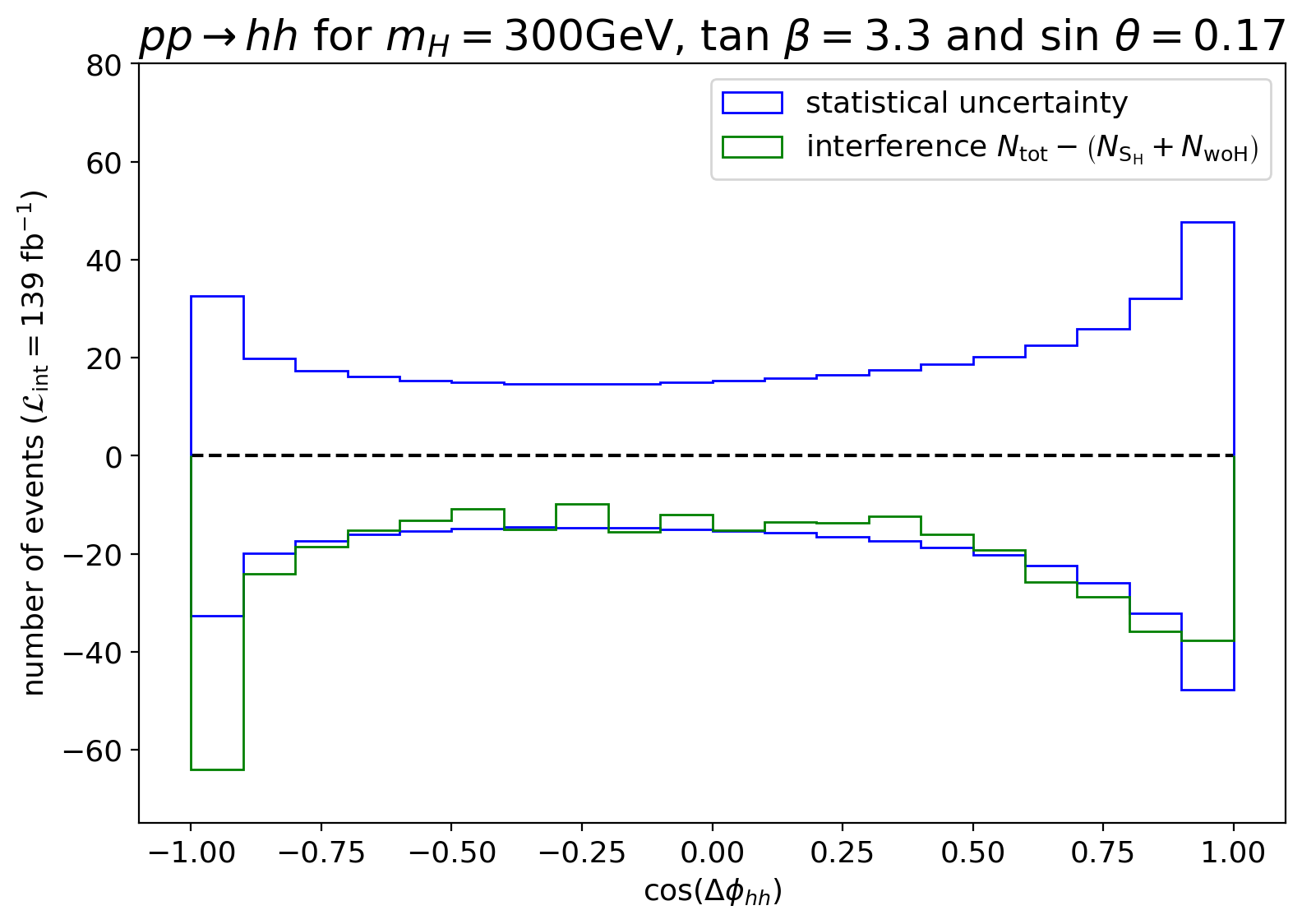}\label{figure_300_benchmark_varphi_diff}}\quad
    \subfloat[]{\includegraphics[width=0.46\textwidth]{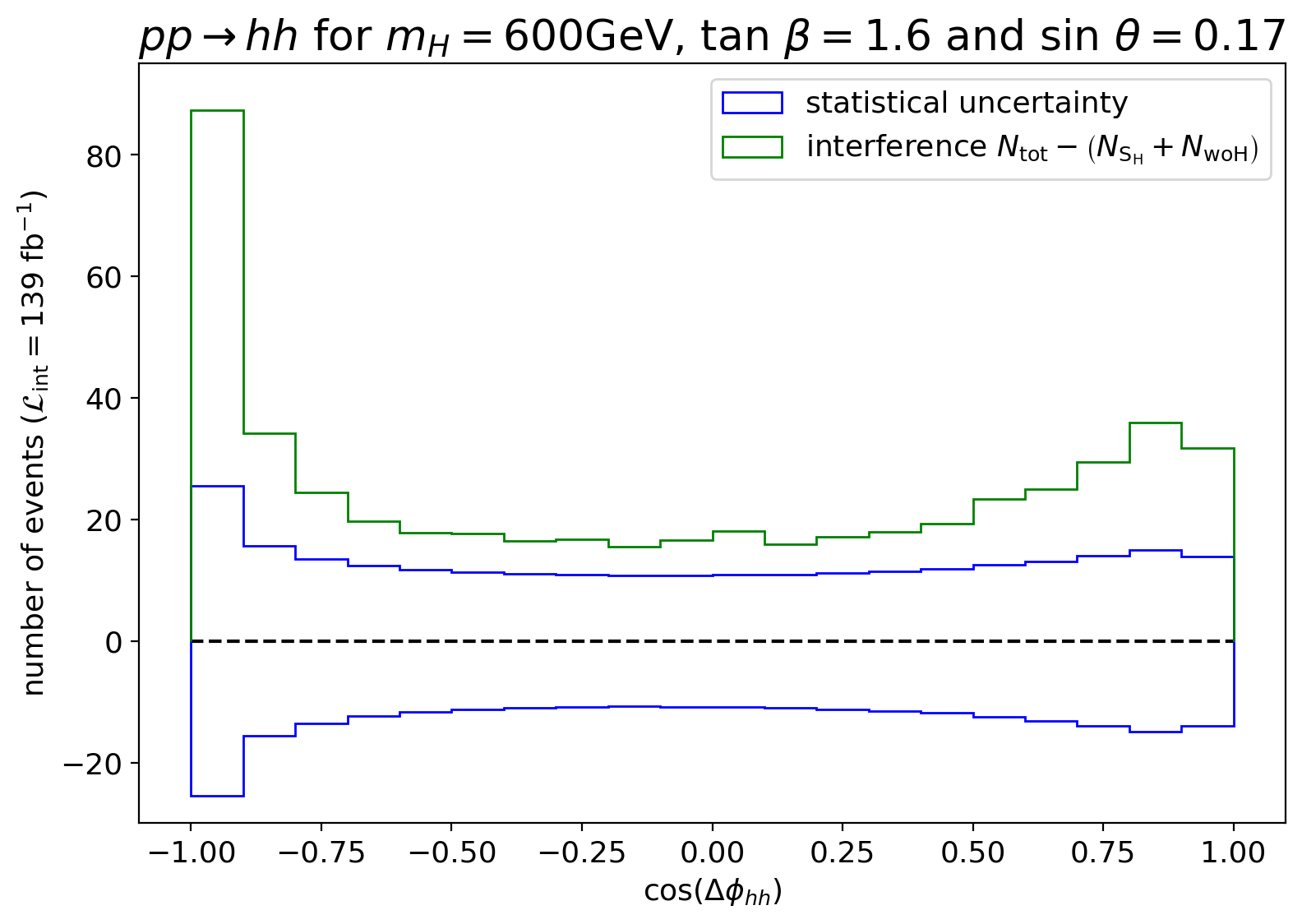}\label{figure_600_benchmark_varphi_diff}}
    \caption{Upper row: the angle between the Higgs bosons $\Delta\phi_{\Ph\Ph}$ 
    with $P_{\text{tot}}$ (blue), $P_{\SH}$ (red), $P_{\text{woH}}$ (orange) and the sum of $P_{\SH}$ and $P_{\text{woH}}$ (green). The difference between the blue and the green line corresponds to the interference term.
    Lower row: the interference for the two respective parameter points (left column vs. right column). }
    \label{figure_Observable_Plots_varphi}
\end{figure*}

In Figure~\ref{figure_Observable_Plots_varphi}, we display the differential distribution in $\Delta\phi_{\Ph\Ph}$ for both parameter points discussed in Section~\ref{subsec:parscan_interference}.
The similarity of the blue line (coherent sum of all contributions to $pp\to hh$) and the green line (incoherent sum of the contributions via $H$ and without $H$, i.e. without the interference term) indicates that the interference does not contribute significantly to this variable.

\section{Additional validation plots of the reweighting method}
\label{sec:AppendixB}

\begin{figure*}[htbp]
  \includegraphics[width=0.48\textwidth]{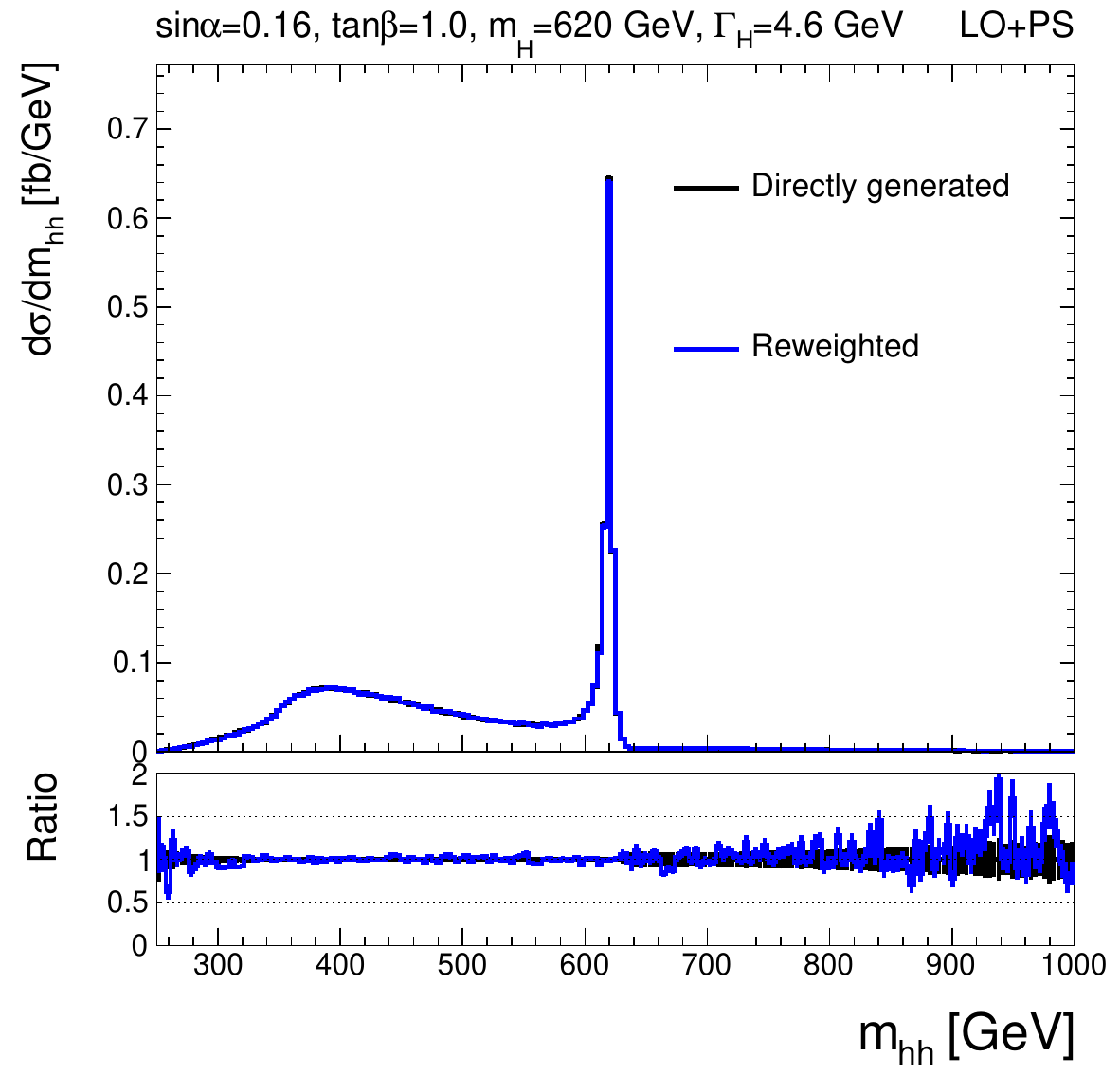}
  \includegraphics[width=0.48\textwidth]{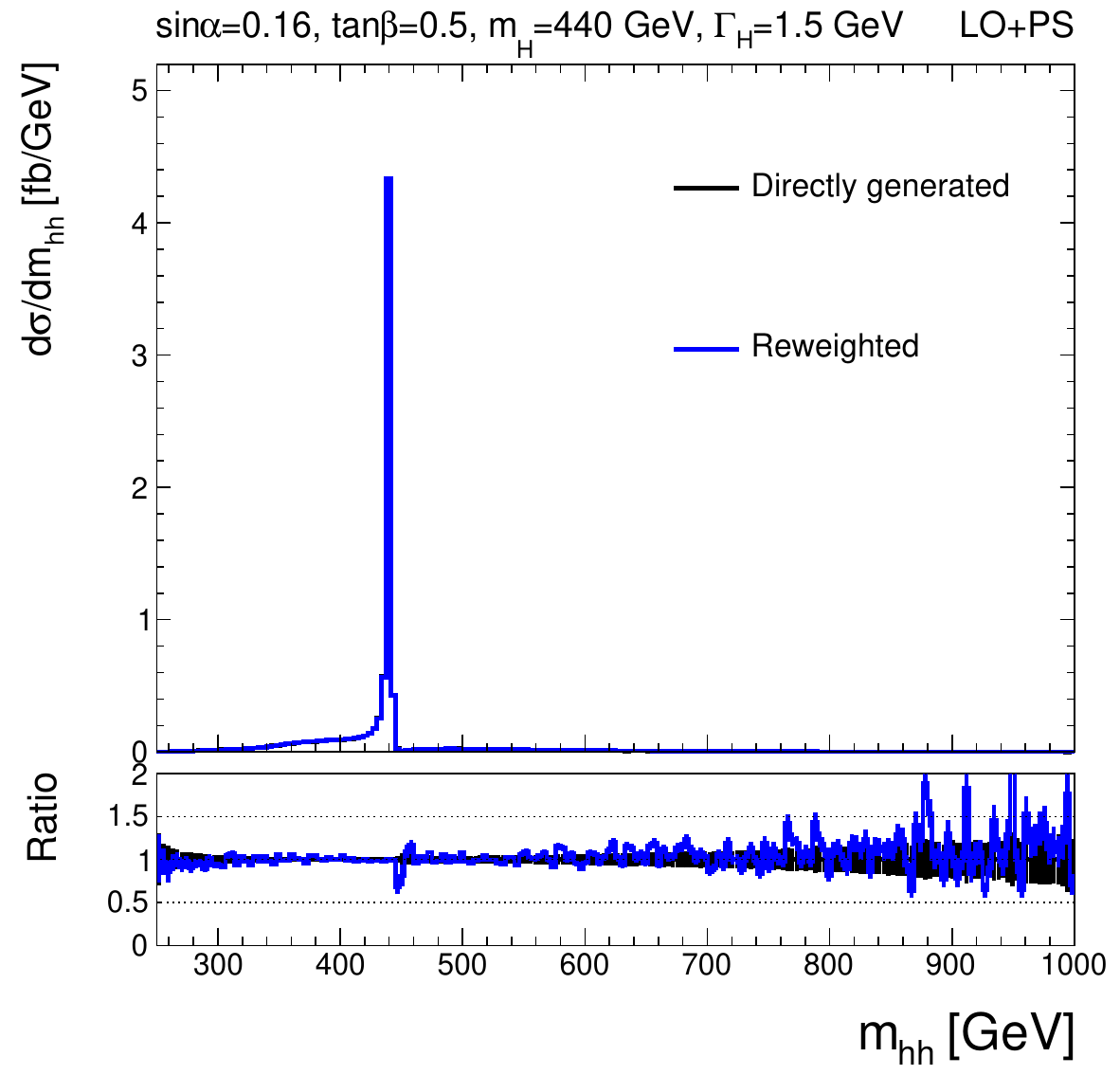}\\
  \includegraphics[width=0.48\textwidth]{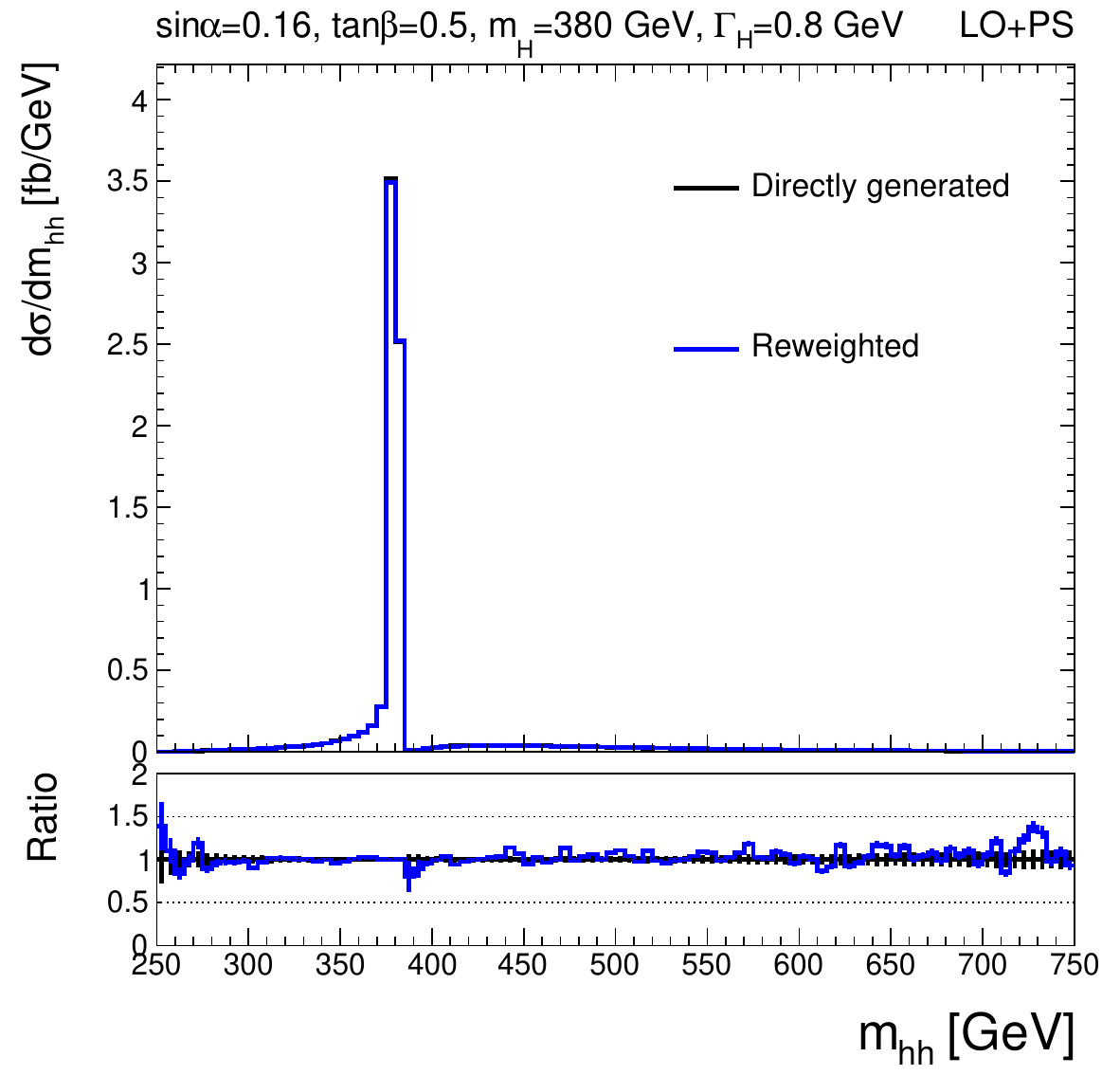}
  \includegraphics[width=0.48\textwidth]{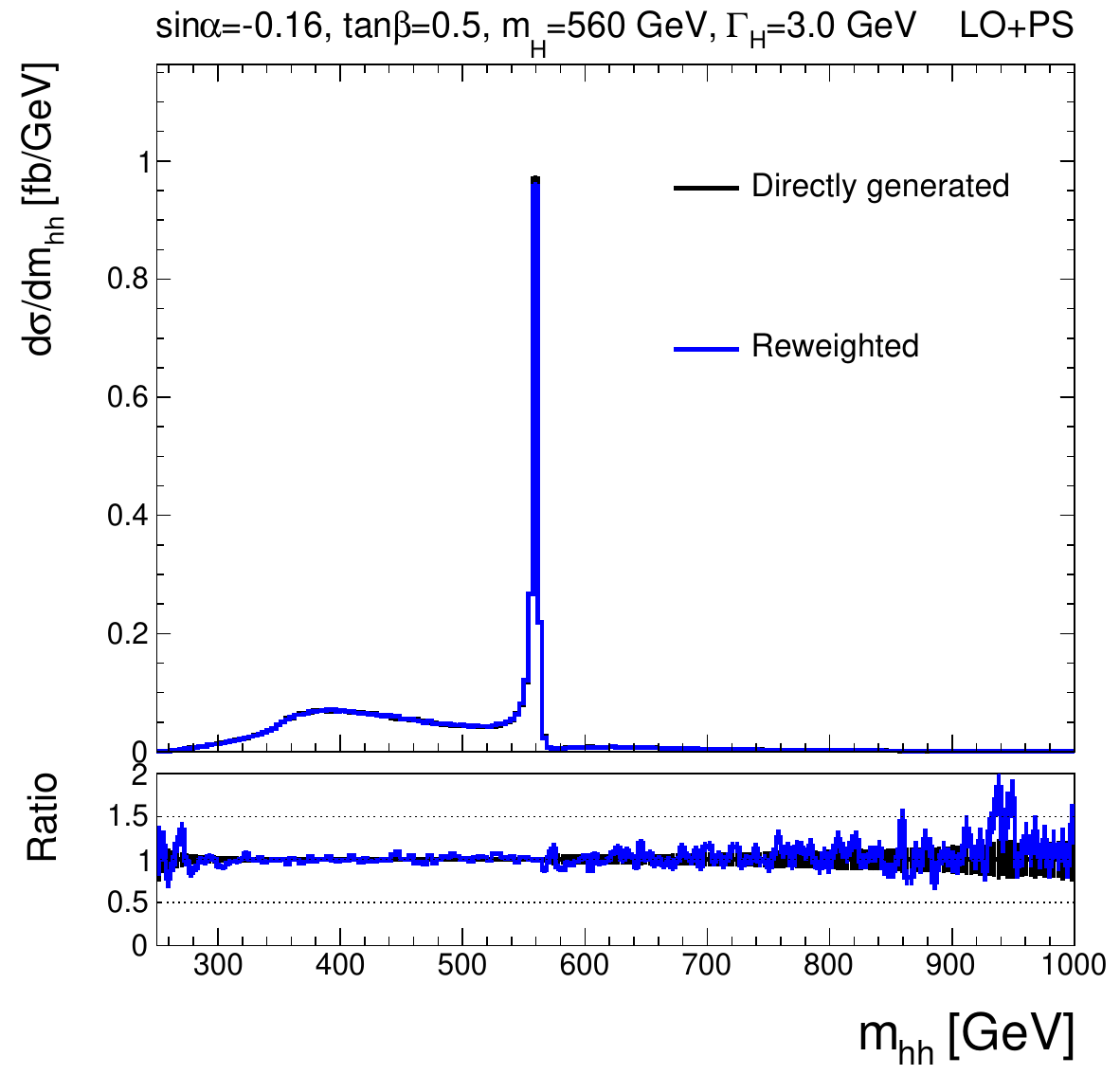}
\caption{The reweighting method is validating by comparing reweighted di-Higgs mass distribution (blue) to the distribution obtained by directly simulating the di-Higgs events for a set of parameter points (black). The distributions are shown for \BMa (upper left), \BMb (upper right), \BMc (lower left) and \BMd (lower right).
}
\label{fig:reweight_validations_allbms_1}
\end{figure*}

\begin{figure*}[htbp]
  \includegraphics[width=0.48\textwidth]{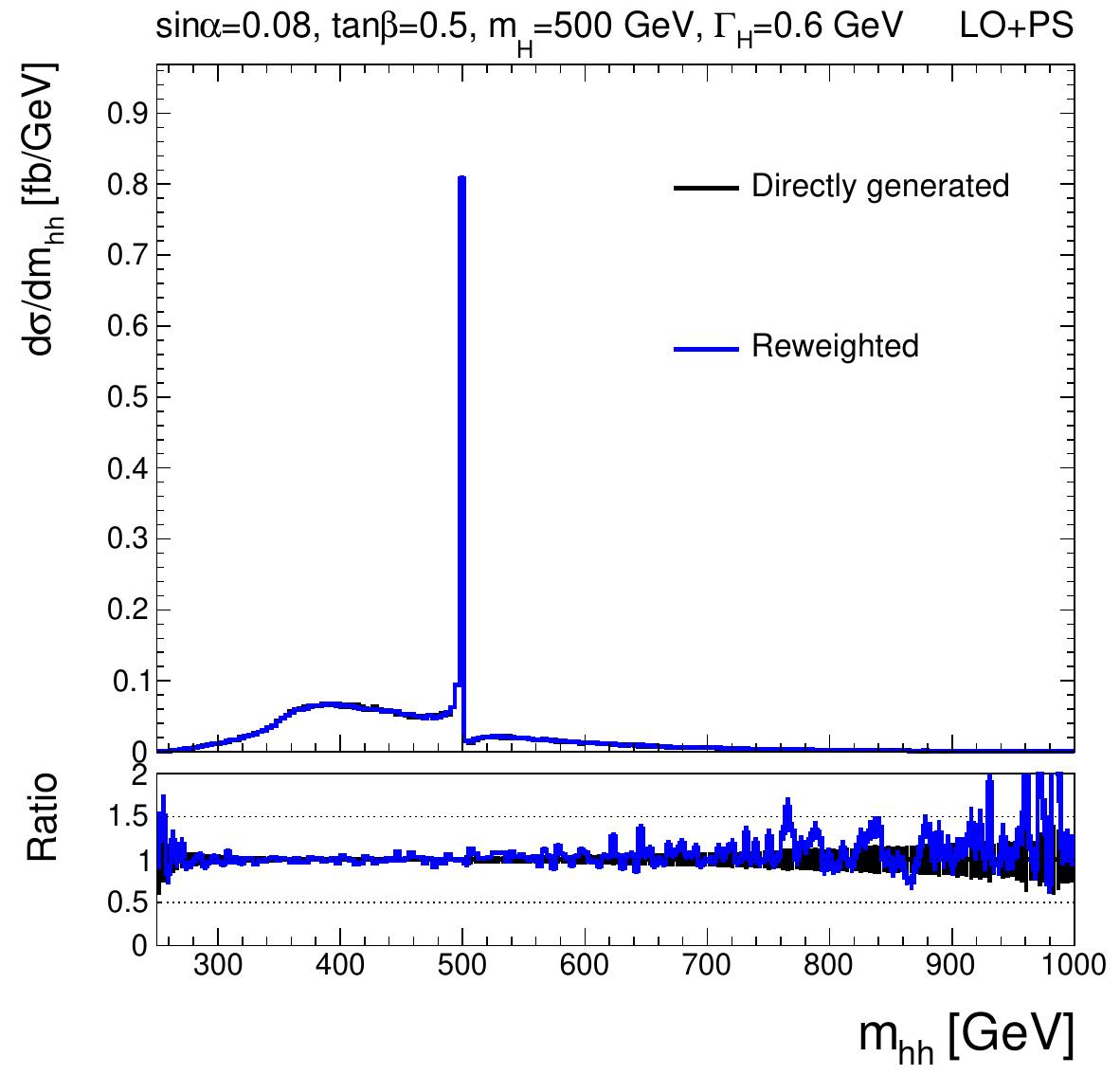}
  \includegraphics[width=0.48\textwidth]{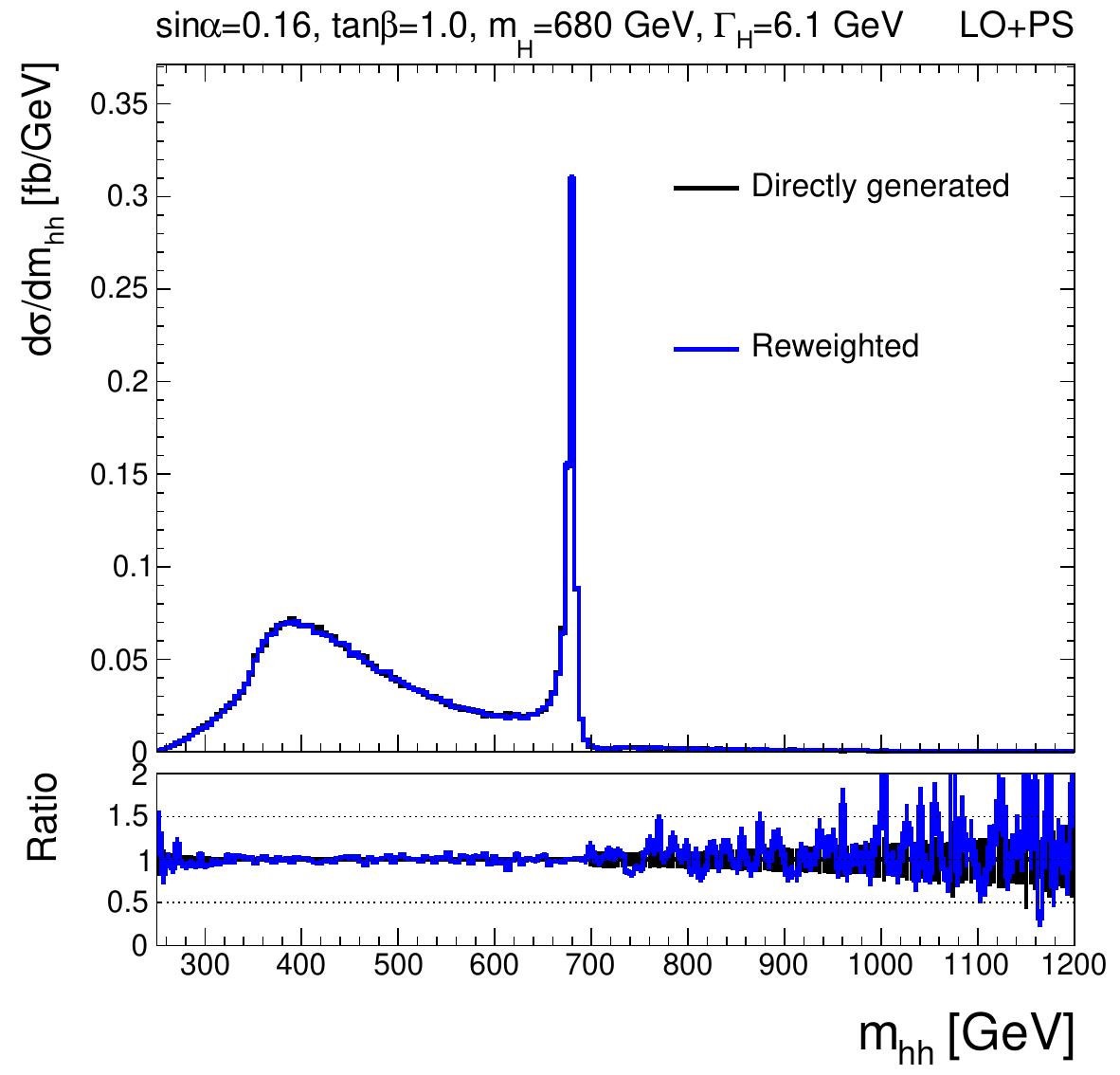}\\
  \includegraphics[width=0.48\textwidth]{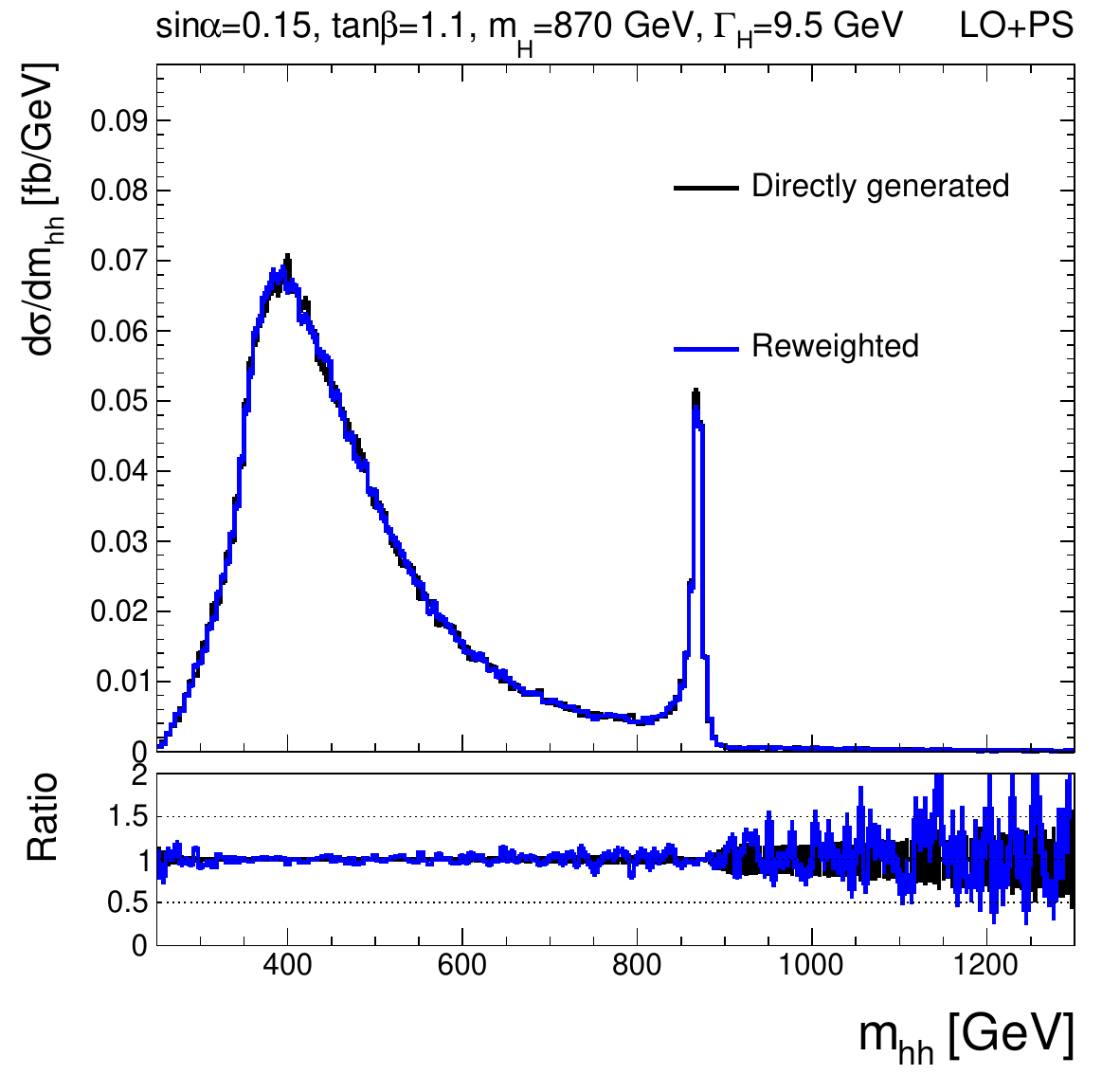}
  \includegraphics[width=0.48\textwidth]{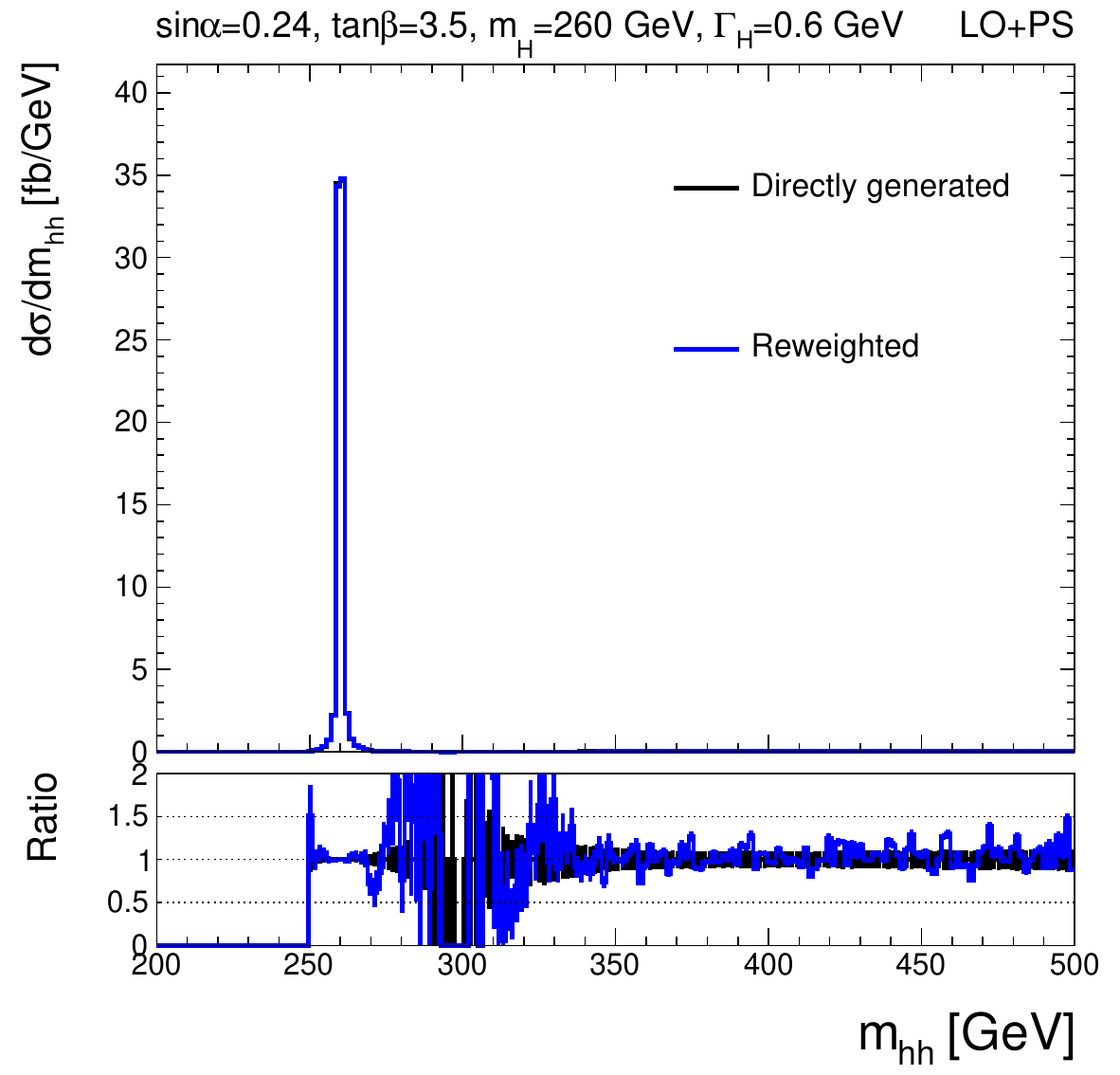}
\caption{The reweighting method is validating by comparing reweighted di-Higgs mass distribution (blue) to the distribution obtained by directly simulating the di-Higgs events for a set of parameter points (black). The distributions are shown for \BMe (upper left), \BMf (upper right), \BMg (lower left) and \BMh (lower right).}
\label{fig:reweight_validations_allbms_2}
\end{figure*}

\begin{figure*}[htbp]
  \centering\includegraphics[width=0.48\textwidth]{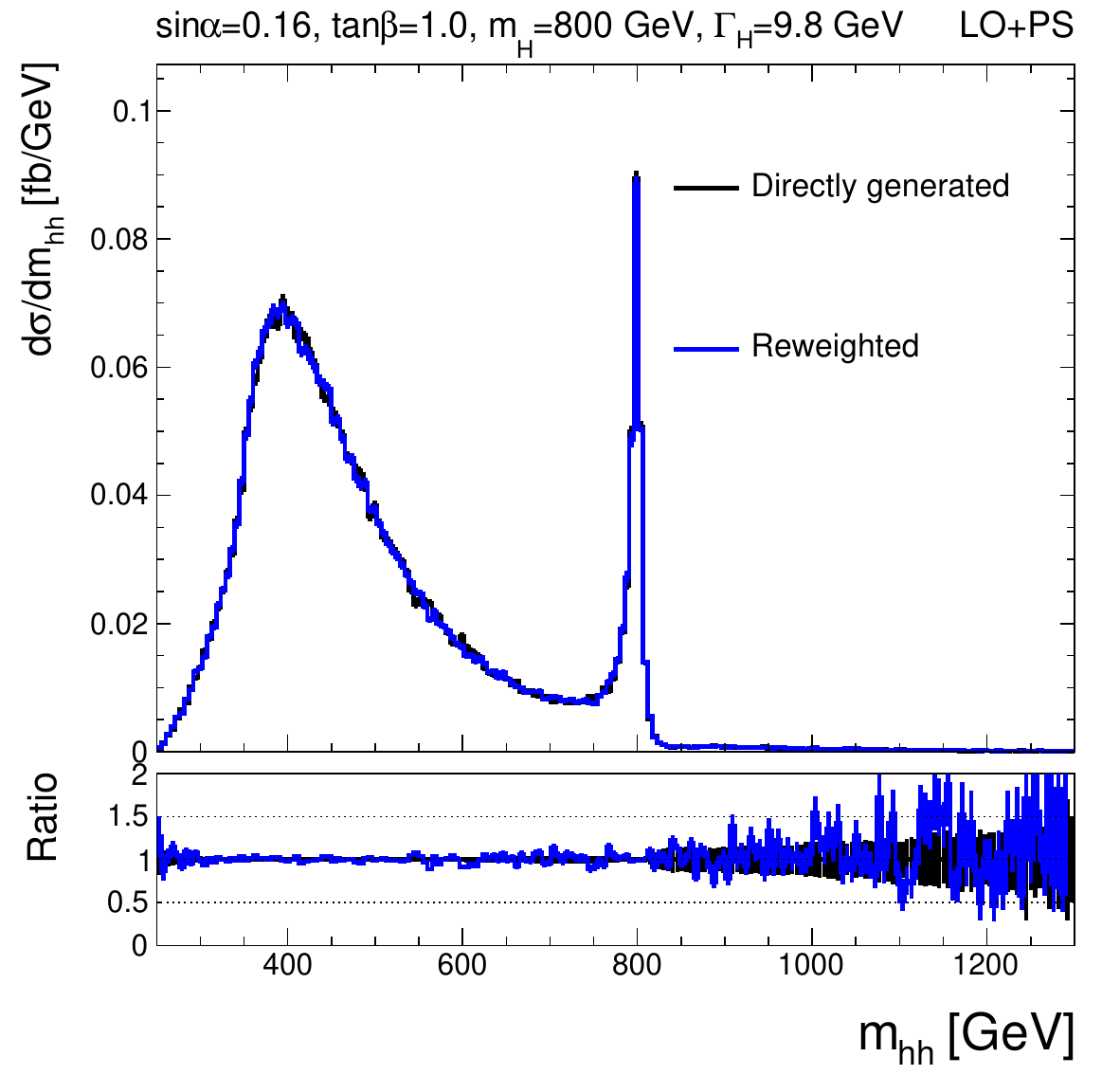}

\caption{The reweighting method is validating by comparing reweighted di-Higgs mass distribution (blue) to the distribution obtained by directly simulating the di-Higgs events for a set of parameter points (black). The distribution is shown for \BMi.}
\label{fig:reweight_validations_allbms_3}
\end{figure*}

The reweighting method described in Section~\ref{sec:method} is validated for all benchmark points given in Section~\ref{sec:benchmarks}. Figures~\ref{fig:reweight_validations_allbms_1}--\ref{fig:reweight_validations_allbms_3} show the reweighted \mhh distribution (blue) compared to the \mhh distribution obtained from a MC sample generated directly for the target model parameters (black).

\end{appendices}
\FloatBarrier
\section*{Acknowledgements}
TR is supported by the Croatian Science Foundation (HRZZ) under project HRZZ-IP-2022-10-2520.
EF and FF acknowledge funding by the Deutsche Forschungsgemeinschaft (DFG, German Research Foundation) under Germany’s Excellence Strategy – EXC-2123 QuantumFrontiers – 390837967.
DW was supported by the Science and Technology Facilities Council [grant number ST/W000636/1]. 
TR and EF also want to thank the CERN-TH group for their hospitality during parts of this work.
%%%%%%%%%%%%%%%%%%%%%%%%%%%%%%%
\bibliographystyle{JHEP}
\bibliography{lit}
\end{document}